\documentclass[preprint,authoryear,11pt]{elsarticle}
\usepackage{deluxetable}
\usepackage{amssymb}
\usepackage{mathrsfs}
\usepackage{amsmath}

\usepackage{epsfig}

\usepackage{amssymb}

\newcommand{\be}{\begin{equation}}
\newcommand{\ee}{\end{equation}}
\newcommand{\bea}{\begin{eqnarray}}
\newcommand{\eea}{\end{eqnarray}}

\newcommand{\mdm}{M_{dm}}

\newcommand{\sigmav}{\sigma_{\textrm{ann}} v}
\newcommand{\sigmavavg}{\langle \sigma_{\textrm{ann}} v \rangle}

\newcommand{\apj}{ApJ}
\newcommand{\apjl}{ApJ Lett}
\newcommand{\apjs}{ApJS}

\newcommand{\bain}{Bull. Astron. Inst. Netherlands} 
\newcommand{\mnras}{MNRAS}
\newcommand{\aj}{AJ}
\newcommand{\aap}{A\&A}
\newcommand{\jcap}{JCAP}

\newcommand{\nat}{Nature}

\journal{Physics Reports}

\topmargin = -\topskip
\advance\topmargin by -\headsep
\begin{document} 

\author{Louis~E.~Strigari}

\address{Kavli Institute for Particle Astrophysics and Cosmology, Stanford University, Stanford, CA 94305 USA}

\title{Galactic Searches for Dark Matter} 

\begin{abstract}

\par For nearly a century, more mass has been measured in galaxies than is contained in the luminous stars and gas. Through continual advances in observations and theory, it has become clear that the dark matter in galaxies is not comprised of known astronomical objects or baryonic matter, and that identification of it is certain to reveal a profound connection between astrophysics, cosmology, and fundamental physics. The best explanation for dark matter is that it is in the form of a yet undiscovered particle of nature, with experiments now gaining sensitivity to the most well-motivated particle dark matter candidates. In this article, I review measurements of dark matter in the Milky Way and its satellite galaxies and the status of Galactic searches for particle dark matter using a combination of terrestrial and space-based astroparticle detectors, and large scale astronomical surveys. I review the limits on the dark matter annihilation and scattering cross sections that can be extracted from both astroparticle experiments and astronomical observations, and explore the theoretical implications of these limits. I discuss methods to measure the properties of particle dark matter using future experiments, and conclude by highlighting the exciting potential for dark matter searches during the next decade, and beyond.

\end{abstract}

\maketitle

\newpage 

\tableofcontents


\newpage 

\section{Dark Matter Searches: Historical Context} 
\label{sec:problem} 

\subsection{Establishment of dark matter} 

\par When it first became possible to measure the motions of stars in the Milky Way, its structure and dynamics was revealed to a level not previously possible. On examination of the brightest stars in the disk of the Milky Way, Oort found that the stars were only able to account for about one-third of the total mass contained within the disk~\citep{Oort1932}. Oort deduced that the disk contained two-thirds ``dark matter" that was predominantly in the form of stars less luminous than the Sun, as well as gas and dust in the interstellar medium. Nearly thirty years later,~\cite{Kinman:1959} analyzed the velocities of Milky Way globular clusters, and deduced possible deviations from a pure bulge and disk mass model of the Galaxy, known at the time as ``Schmidt's Law." Though there were large error bars associated with his measurements, the data was consistent with a linearly rising mass distribution beyond the disk. Kinman suggested that the data be held in reserve due to uncertainties in the globular cluster distances and the local circular velocity. 

\par~\cite{Babcock:1939} was the first to measure the mass distribution of M31, the nearest galaxy with luminosity comparable to the Milky Way. From the measurement of radial velocities in the optical emission line region extending out to approximately $20$ kpc from the center of M31, Babcock determined that the total mass-to-light ratio increases in the outer regions. He suggested that absorption of light was the cause of this increasing mass-to-light ratio. Two decades later,~\cite{Kahn1959} were the first to connect the dynamics of local galaxies to the universe on larger scales. These authors noted that if the Milky Way and M31 were once receeding due to Hubble expansion, and are now approaching at the observed velocity, there must be ten times more unseen material within both of the galaxies and within the inter-galactic medium to reverse their initial separation. They made the suggestion that this material was in the form of intergalactic gas and dust. 

\par The results on the mass distribution of M31 were further corroborated by the 21-cm observations of~\cite{vandeHulst1957} and~\cite{Roberts1975}, which clearly indicated that the rotation curve of this galaxy is flat out to approximately $35$ kpc from the center, and did not fall-off in the Keplerian fashion as expected if the stars and gas dominated the gravitational dynamics. Flat rotation curves were identified in larger samples of nearby spiral galaxies through a variety of techniques at several observatories~\citep{Bosma1978,Rubin1980}. After several years of dealing with systematics, such as the effect of beam smearing, it became the widespread consensus amongst astronomers that the results were not due to observational limitations, and that rotation curves of many spiral galaxies remained flat in the outermost regions~\citep{Sofue:2000jx}. 

\par The seemingly linear rise in the mass distribution of the Milky Way beyond the disk, combined with the measured rotation curves of M31 and nearby spiral galaxies, provided a fresh perspective on well-known observations obtained from larger and more distant systems. About a year after Oort published his results on the kinematics of the stars in the Milky Way, Fritz Zwicky famously deduced that there was hundreds of times more mass in the Coma cluster of galaxies than was contained within the luminous galaxies~\citep{Zwicky}. Zwicky considered a few explanations for his data: the constituent galaxies contained much more mass than he ascribed to them, there was undetected intergalactic matter, or perhaps even the known physical laws were not appropriate for the mega-parsec scales he was considering. Upon noting that the cluster may be dominated by ``dark matter," Zwicky both used Oort's phrase for the first time in the context of extragalactic astronomy, and also in a manner that most closely resembles its meaning in the modern scientific vernacular. 

\par As incontrovertible evidence built indicating that galaxies and clusters of galaxies require much more mass that is contained in the luminous material, the question became eminently focused:  What is the nature of the dark matter that dominates the mass of astronomical systems? While there has been, and continues to be, a variety of theoretical of explanations to this question, two particularly compelling explanations have lent themselves to rigorous investigation. On the one hand there is the plausible conjecture, entertained by many of the aforementioned astronomers, that our inventory of astronomical bodies is incomplete, in particular we do not have a good census of bodies that are faint sources of radiation that escape detection. On the other hand, a theoretical picture has been developed that predicts that dark matter is a new elementary particle that is distinct from the known particles of nature. In order to set the stage for the results that are presented in this article, it is informative to review the historical establishment of these two classes of dark matter candidates, the predictions that they make, and the searches for signatures of them. 

\subsection{MACHOs: Review of searches and results} 
\par  We do not have a complete understanding of the population of low mass stars, stellar remnants, and planetary mass bodies in the Galaxy. Compact astronomical bodies that constitute a significant component of the mass of the Galaxy are historically referred to as Massive Astrophysical Compact Halo Objects (MACHOs).  Because of their intrinsic faintness, or because they in some scenarios emit no radiation at all, MACHOs typically cannot be identified from direct imaging. This leaves gravitational interactions, and in particular gravitational lensing, as the best available technique to search for MACHOs. 

\par If a massive body, in this case a MACHO, intervenes between the Earth and a background source star, the massive lensing body creates multiple images of the background star. Though the images are too small to be resolved by modern telescopes, the source star brightens on a timescale that is related to the mass of the intervening object and its velocity. The amplification of the source is the ratio of the area of the images that are created to the area of the unlensed source star. For a stellar mass object moving at a typical velocity through our Galaxy,  the characteristic brightening timescale of the star is in the range of a few weeks to a year. The above phenomenology of microlensing, initially predicted nearly a century ago~\citep{Einstein1936}, is now a standard phenomenological tool in modern astrophysics~\citep{Mao:2012za}. If the dark matter in our Galaxy is comprised of faint stars, stellar remnants, planetary mass objects, black holes, or some combination of compact objects thereof, ~\cite{Paczynski:1985jf} suggested that a long term monitoring program of stars in the nearby Large Magellanic Cloud (LMC) would reveal dark matter in this form, even though the objects could not be identified directly. 

\par During the past two decades, the MACHO project, the  Optical Gravitational Lensing Experiment (OGLE), and the Experience pour la Recherche DÕObjets Sombres survey (EROS) project have searched for MACHOs in the Milky Way via microlensing. Though microlensing events were indeed detected by these experiments, it is generally concluded that they do not have significant enough abundance to dominate the mass of the Galactic halo. As no conclusive signature for MACHOs was found, these experiments combined to place strong upper limits on the number of MACHOs in the Galaxy in the mass regime of approximately $10^{-7} - 30$ M$_\odot$~\citep{Alcock1998,Tisserand:2006zx,Wyrzykowski:2011}. 

\par For MACHOs greater than about a hundred solar masses, the timescale of the typical microlensing event for a lensing object in the halo is a few years, which is near the lifetime of the surveys themselves. Therefore the efficiency for detecting objects in this mass range drops significantly, and other methods must be utilized to identify them. Two such methods rely on studies of the population of wide binaries in the Galaxy~\citep{Bahcall1985}, and on the kinematics of the Galactic disk~\citep{Lacey1985}. The former method utilizes the fact that wide binaries are broken up by encounters with MACHOs, either by weak and diffuse encounters that occur during the entire lifetime of the binary system, or by catastrophic encounters that break up the binary during a single interaction with a MACHO. The disruption timescale for binaries with a separation $\gtrsim 0.1$ pc is less than approximately 10 Gyr for MACHOs with mass $\gtrsim 10$ M$_\odot$, so the number of wide binaries with this separation or larger provides a constraint on the number of MACHOs in this mass regime. The most conservative constraints find that MACHOs of mass $\sim 10 - 10^7$ M$_\odot$ comprise no more than $\sim 50\%$ of the Galactic halo~\citep{Yoo:2003fr,Quinn2009}. Above this mass scale, the indirect arguments from the velocity dispersion of the Galactic disk may be interpreted as placing strong constraints on MACHOs with mass $\gtrsim 10^7$ M$_\odot$~\citep{Lacey1985}.

\par Though microlensing studies find that MACHOs do not dominate the mass of the Galaxy, there have been hints that significant populations of stellar remnants and low mass bodies do indeed exist in the Galactic halo.~\cite{Oppenheimer:2001mx} find a substantial population of old, cool white dwarfs with high proper motions, and that these objects are abundant enough to comprise $\gtrsim 2\%$ of the Galactic halo. Subsequent analyses has however indicated that these objects are more likely members of the thick disk and the stellar halo, and not components of the more extended ``dark" halo of the Galaxy~\citep{Reyle2001,Flynn:2002au}. 

\par Though the analysis above is phrased in the context of constraints on low mass stars and stellar remnants, microlensing searches are sensitive to exotic objects such as primordial black holes if they constitute a component of the dark matter. Rather than forming as an end state of stellar evolution, or via growth from smaller seed black holes in the centers of galaxies, primordial black holes may have a wide range of masses. Primordial black holes may have formed in regions of the early universe that were sufficiently overdense that they halted expansion and recollapsed. They may also have formed during a phase transition in the early universe. Primordial black holes less massive than approximately $10^{15}$ g $\simeq 5 \times 10^{-20}$ M$_\odot$ would have evaporated via Hawking radiation. Though reasonable physical processes predict their formation, there is no strong theoretical motivation for primordial black holes to be as abundant as the observed dark matter. General astrophysical limits on the abundance of primordial black holes of a given mass result from searching for their gravitational interactions or for particles that may be produced during their evaporation~\citep{Josan:2009qn}. During the next few years, microlensing observations are expected to further improve the limits on primordial black holes with mass $\lesssim 10^{-7}$ M$_\odot$~\citep{Griest:2011av}. 

\par The above discussion indicates that Galactic searches for dark matter in the form of MACHOs have reached a significant level of maturity. These results have firmly established that dark matter does not dissipate energy to clump into objects as massive as stars or planets, indicating that the dominant matter component in Galactic halos is more diffusively distributed. This is probably the most direct empirical evidence that we currently have for the existence of particle dark matter. 

\subsection{WIMPs: Theoretical motivation and initial search results} 
\par The lack of MACHO detection was not surprising considering theoretical developments in cosmology and particle physics. Going back over thirty years observations of the abundances of light elements have indicated a baryon (neutron and proton) density in the universe considerably less than the critical density, implying that some form of non-baryonic matter must be cosmologically-significant~\citep{Gott:1974gy}. Even at the time of these earliest calculations, this was consistent with the determination that the extended mass distributions around galaxies comprised a significant fraction of the critical mass density of the universe~\citep{Ostriker1974}. Non-baryonic matter was also found to be very appealing from the perspective of large-scale structure formation, because density fluctuations grow much earlier as compared to a universe dominated by baryonic matter~\citep{Peebles1982}. This nicely accounts for fluctuations in the Cosmic Microwave Background (CMB) at the level of one part in $10^5$. These lines of evidence indicate that the dark component of spiral galaxies that was initially deduced through rotation curves may be connected to observations on larger cosmological scales, and a profound connection was developing between the production of dark matter in the early universe, large scale structure formation, and the mass that dominates galaxies and clusters of galaxies. 

\par As these cosmological calculations of the matter abundance in the universe improved and became more precise, an interesting tension came to the forefront between these calculations and the developing standard model of particle physics. In particular it was not clear that the standard model of particle physics contained a particle with the necessary properties to be significant on cosmological scales. The most natural particle to consider for the non-baryonic dark matter was the neutrino. A neutrino with mass of a few electron volts (eV) was appealing because if it was in equilibrium in the early universe, the present mass density of neutrinos would be near the critical density~\citep{Cowsik:1972gh}. However, under more detailed scrutiny neutrinos at this mass scale turn out to be disfavored as a candidate for the dark matter. On the one hand, phase space limits strongly constrain neutrino dark matter with mass less than a few hundred eV~\citep{Tremaine:1979we}. Further, numerical simulations of the large scale structure distribution in the universe revealed that neutrinos with mass below tens of electron volts were too ``hot" to explain the observed galaxy distribution; because they were relativistic when they decoupled they produced significantly fewer low mass galaxies in comparison to the observations~\citep{White:1984yj}. 


\par Though neutrinos with mass of a few hundred eV or less became disfavored as the dominant component of non-baryonic dark matter, this analysis within a cosmological context motivated a general theoretical framework for determining the abundance of a stable particle species that was in thermal equilibrium in the early universe. Neutrinos with mass greater than about $1$ GeV that fall out of equilibrium while non-relativistic were found to be cosmologically-significant~\citep{Lee:1977ua}, and that generically heavy leptons with similar interactions could be the dominant component of mass in galaxies and clusters of galaxies~\citep{Gunn:1978gr}. These results pointed to a preferred scale, the weak scale, to describe the interactions of a cosmologically-significant component of non-baryonic dark matter. For a standard thermal history in the early universe, the abundance of a particle is related to its thermally-averaged annihilation cross section times relative velocity as
\be 
\sigmavavg \approx \frac{3 \times 10^{-27} \, \textrm{cm}^3 \textrm{s}^{-1}}{\Omega_{\textrm{DM}} h^2} 
\label{eq:abundance}
\ee
~\citep{Zeldovich:1965,Chiu:1966kg,Steigman:1979kw,Scherrer:1985zt,Steigman:2012nb}.
Recent cosmological measurements find that the fraction of non-baryonic dark matter relative to the critical density is $\Omega_{\textrm{DM}} h^2 \simeq 0.11$~\citep{Komatsu:2010fb}. For this value of $\Omega_{\textrm{DM}} h^2$, for particles in the approximate GeV mass range, this scale for the annihilation cross section is characteristic of weak interactions. A particle with interactions at this scale is broadly referred to as a weakly-interacting massive particle (WIMP). 

\par Because of their weak scale interaction strength, WIMPs are detectable via non-gravitational methods. In addition to the prediction of the annihilation strength of WIMPs, Equation~\ref{eq:abundance} leads to a basic prediction for the WIMP scattering cross section on ordinary matter (i.e. quarks). Combining this scattering cross section with the estimation for the local number density of WIMPs, the interaction rate is large enough for them to have been detected over two decades ago in the first generation of low temperature germanium experiments that were primarily designed to search for signatures of neutrinoless double beta decay. While it was certainly possible that these experiments would detect WIMPs and render the non-gravitational detection of particle dark matter relatively straightforward, the first two experiments to systematically search for WIMPs reported null results~\citep{Ahlen:1987mn,Caldwell:1988su}. The reported constraints ruled out the simplest model for WIMP interactions with ordinary matter, and in the process first showed that if WIMPs comprise the dominant component of dark matter in galaxies, physics must be invoked to suppress the scattering cross section relative to that derived from the most basic cosmological arguments in Equation~\ref{eq:abundance}.

\subsection{Modern perspective on dark matter searches} 
\par The results from MACHO searches and the seminal WIMP searches have formulated our modern picture of particle dark matter. While MACHO searches have largely ended, WIMP searches are continuing to pick up incredible momentum, largely because there are still very compelling regions of theoretical parameter space left to explore. The size and scope of searches for particle dark matter have expanded considerably since the first set of experiments that were designed to search for them. For these reasons, it is important to have a better classification of the types of WIMP searches going on around the world at present, and what each of them will ultimately be able to tell us about the properties of dark matter. The focus of this article is on two broad classes of observations that are used to probe the particle nature of dark matter: searches with astroparticle experiments and colliders, and searches using astronomical data sets. 

\subsubsection{Astroparticle and collider searches}

\par There are three types of dedicated astroparticle and collider experiments actively searching for particle dark matter. Each type of experiment is sensitive to a different aspect of dark matter interaction with ordinary matter. 

\bigskip

$\bullet$ {\em Indirect Searches}: Indirect dark matter searches measure the annihilation and/or decay products of dark matter from astrophysical systems. Schematically, they measure the rate for DM DM $\rightarrow$ SM SM, where DM represents the dark matter particle and SM represents any standard model particle. In many instances, the particle represented by SM is unstable, and decays into particles (for example, photons or neutrinos) that are observable in detectors. Comparing to Equation~\ref{eq:abundance}, the cross section probed by indirect searches is most closely related to the process that sets the abundance of the dark matter in the early universe, assuming that the dark matter was once in thermal equilibrium. In order to best interpret the results from indirect searches, we must have a good idea as to both how the dark matter is distributed in halos, and what standard model particles the dark matter preferentially annihilates or decays into.  

\bigskip 

$\bullet$ {\em Direct Searches}: Direct dark matter searches, an example of which was described above, measure the scattering of dark matter off of nuclei in low background underground detectors, DM SM $\rightarrow$ DM SM. In these experiments, the collision of dark matter is deduced through energy input into particles in the detector. At the most fundamental level, the SM particle is represented by up and down quarks that make up the protons and neutrons within nuclei, so direct detection experiments ultimate measure the strength of the interactions between WIMPs and these types of quarks. Though there is no specific cosmological prediction for the WIMP-nucleus scattering cross section similar to the prediction that leads to Equation~\ref{eq:abundance}, as discussed throughout this article modern experiments are steadily improving limits and are probing well-motivated theoretical models.

\bigskip 

$\bullet$ {\em Collider Searches}: Collider searches measure the production of dark matter through the collision of high energy particles on Earth, SM SM $\rightarrow$ DM DM. Hadron colliders, such as the Tevatron at Fermilab and the Large Hadron Collider (LHC) at CERN, collide protons to produce new stable particles either directly or through processes that are initiated by quarks and gluons. New stable dark matter particles produced directly cannot be observed, so these events must be tagged by observable initial state radiation. However, the cross section for this process is typically small. On the other hand, the cross section is larger for the production of quarks and gluons, which eventually decay down into the lightest stable particle in the spectrum. In this process, much of the energy of the collision is not transferred to final state particles, so it is often not straightforward to reconstruct the kinematics of the collision and determine the mass of the stable particle that is produced. Lepton colliders, such as the Large Electron Positron (LEP) collider that operated from 1989-2000 at CERN, transfer all of the initial state energy into final state particles. LEP has produced a lower bound on the existence of stable charged particles of approximately $100$ GeV, and through the non-observations of perturbations to standard model cross sections motivates existence of new stable particles near the electroweak scale (see the arguments discussed in~\cite{Feng:2010gw}). 

\bigskip 

\par If dark matter is in the form of a new elementary particle, understanding its interactions requires measuring the cross section with standard model particles via each of the three methods described above. From the three sets of measurements, the theoretical challenge will be to determine whether the WIMP is a single stable particle, or it is embedded into a larger theory of high energy physics, perhaps one that contains a spectrum of particles with masses larger than that of the WIMP.

\par Though it is probably most prudent at this stage to treat the respective cross sections described above as independent, under some simplifying theoretical assumptions it is possible to connect the cross section as measured by one experiment with that measured in another. A straightforward example is if the WIMP annihilates pre-dominantly to standard model fermions, and there are no resonant interactions that serve to enhance the annihilation cross section. In this case a measurement of the WIMP-fermion coupling from the strength of the annihilation cross section translates directly to the elastic scattering rate off of nucleons in underground detectors. Typically in these models, the WIMP is the only particle the constitutes the dark matter, with all other particles in the theory at a significantly larger mass scale so they do not affect the interaction rate of WIMPs with standard model particles. Models along these lines are discussed in more detail in Sections~\ref{sec:indirect_detection_theory} and~\ref{sec:direct_detection_theory}. 

\par Given the compelling cosmological motivation for WIMPs, a vast number of theories have been developed that embed a particle (or particles) with these properties into a consistent framework. The most famous and well-studied example of WIMP dark matter is the neutralino in supersymmetry~\citep{Goldberg:1983nd}. Dark matter candidates also arise from theories of universal extra dimensions~\citep{Cheng:2002ej,Hooper:2007qk}, as well as many others~\citep{Duffy:2009ig,Feng:2010tg} (see the recent reviews of~\cite{Bertone:2004pz} and~\cite{Bertone:2010} for a more complete list of dark matter candidates). In many cases these theories can be appealing not only from the perspective of having a well-motivated dark matter candidate, but also because they are often linked to various other unsolved problems in high energy physics, such as the stabilization of the Higgs mass, the unification of the forces at the GUT scale, or the explanation of the matter-antimatter asymmetry. A stable dark matter particle is thus embedded into a larger theoretical edifice, that almost always contains a larger spectrum of particles at higher energy scales. It is these theories that we hope to study over the next several years with the three separate astroparticle and particle experiments that are outlined above. 

\subsubsection{Astrophysical searches} 

\par In modern astrophysics, two types of observations are particularly sensitive to the nature of particle dark matter:  

\bigskip

$\bullet$ {\it Abundance of low mass (dwarf) galaxies}: Very roughly, the minimum mass dark matter halo that can form in the universe is related to the velocity of the dark matter at the time it decoupled from the rest of the standard model particles in the early universe. In deriving Equation~\ref{eq:abundance}, the standard assumption was made that the dark matter mass is larger than the temperature at decoupling. If the opposite is true, and the dark matter mass is less than the decoupling temperature, then the dark matter decouples not as cold dark matter (CDM), but as so-called warm dark matter (WDM) (with hot dark matter in the form of standard model neutrinos already being ruled out). As the ratio of the dark matter mass to the decoupling temperature decreases, so too does the number of low mass dark matter halos. Though this argument provides a general guiding theoretical principle, numerical simulations are required to most accurately predict the abundance of dark matter halos as a function of halo mass given properties of the dark matter particle. 

\par In order to determine the minimum mass dark matter halo in the universe, a mapping between dark matter halo mass and galaxy luminosity is required for the faintest known galaxies. However, the uncertain theoretical mapping between galaxy luminosity and dark matter halo mass makes it difficult to interpret modern observations as a direct constraint on particle dark matter. Further, even though modern surveys now identify galaxies several magnitudes fainter than the Milky Way, these surveys are notoriously incomplete. In spite of these difficulties, steady progress is being made on both the observational and the theoretical fronts, and as discussed in significant detail in Sections~\ref{sec:satellites} and~\ref{sec:simulations}, interesting parameter space for dark matter particle models is well within reach. 

\bigskip

$\bullet$ {\it Central density of dark matter halos}: The properties of particle dark matter are also strongly linked to the mass distribution in the central regions of galaxies. The two most straightforward properties of the dark matter that these observations probe are 1) the phase space density of dark matter, and 2) the self-interaction, or scattering rate, DM DM $\rightarrow$ DM DM. Because of its low velocity at decoupling, the phase space density of the WIMP is large, many orders of magnitude larger than the phase space density of a standard model neutrino that decouples while it is relativistic. Very broadly, due to the low phase space density neutrinos favor constant density cores in the centers of dark matter halos, while perfectly cold WIMPs favor more steeply rising central densities. However, because of limited observational resolution, and because the effect of baryons on the central densities of dark matter in galaxies is not well-understood, mapping the measurements of the central densities of dark matter halos into a constraint on particle dark matter is not straightforward. 

\par The self-interaction property is relevant because in order to form extended halos that are stable over cosmological timescales, there should be minimal interactions between dark matter particles in halos through non-gravitational methods. For a power law density profile, significant amount of interaction leads to the mixing of the ``cold" (low velocity dispersion) central component of the galaxy with the ``hot"  (high velocity dispersion) component. It is straightforward to understand why, in the context of WIMP dark matter, interaction between dark matter is insignificant. From the weak scale interactions, the characteristic scale for DM-DM scattering is typically $\sim 10^{-40}$ cm$^2$. However, this cross section is so small that WIMP-WIMP scattering will effectively never happen in a dark matter halo. Assuming a density of $\sim$ GeV/cm$^3$, and a WIMP of mass $100$ GeV, for a characteristic WIMP velocity of $\sim 100$ km/s, the self-interaction rate of WIMPs is at least thirteen orders of magnitude longer than the Hubble time. 

\par The best modern limits on the self-interaction rate of dark matter do not arise from the observations of the central densities of galaxies, but rather result from the interpretation of the collisions of clusters of galaxies, the most famous being the so-called bullet cluster~\citep{Clowe:2006eq}. In this system, three separate components are observed: the galaxies within the clusters (which are assumed to be collisionless), the gas distribution as traced in X-rays, and the total mass as determined from a gravitational lensing map of the matter distribution. The collisionless galaxies are observed to follow the total mass distribution, while the gas components are observed to be more strongly interacting. From these observations, the upper limit on the self-interaction rate of the dark matter for a $100$ GeV WIMP is approximately $10^{-24}$ cm$^2$, which is many orders of magnitude larger than predicted by the theory of weak interactions. Thus any model that predicts significant interaction between dark matter should have a much larger cross section than that which characterizes WIMPs. 

\bigskip

\par Within the cosmological context of the standard Lambda Cold Dark Matter (LCDM) model, particle dark matter is approximately five times more abundant than baryons, about $22\%$ of the total energy density in the universe, and its energy density is a little less than a third of the dark energy density~\citep{Komatsu:2010fb}. Though the cold dark matter paradigm is fundamental to modern cosmology and crucial to the interpretation of many aspects of modern astrophysics, a clear validation of it has remained elusive. Detection of particle dark matter with the above properties will almost certainly be the best means to verify the cold dark matter paradigm, providing further validation of the LCDM cosmological model. Approaching a half of a century after the initial particle dark matter candidates were first postulated, and they were connected to the mass distribution around galaxies and to cosmology, within the next several years we are finally in a position to obtain a better understanding of the properties of WIMP dark matter, if it does indeed exist.

\newpage 
\section{Scope of this article} 
\label{sec:scope}

\par There are a number of excellent recent reviews on theoretical aspects of particle dark matter~\citep{Bertone:2004pz,Hooper:2008sn,Feng:2010gw,Bergstrom:2012fi} and experimental searches using astroparticle data~\citep{Gaitskell:2004gd,Porter:2011nv}. The former articles in particular highlight how a large focus of the theoretical research has revolved around developing extensions to the standard model that contain either a single dark matter particle candidate or a spectrum of them. 



\par Now that we have new experimental data sets in hand, and these experiments are beginning to encroach into the regime of cosmologically well-motivated dark matter candidates, it is important to understand what information can be extracted from experimental data. The goal of this article is to review both the theoretical and experimental developments in the field of direct and indirect searches for Galactic dark matter, and to understand the implications of these results. As motivated in the introduction, Galactic searches for dark matter not only involve interpretation of data from astroparticle experiments, but also data from more classical astronomical surveys that are sensitive to the microscopic properties of the dark matter. For this reason, I devote the beginning sections of this article to understanding what we have learned in recent years from astronomical surveys, and examine both their direct implications on particle dark matter, and also the interplay that they have with astroparticle searches that are reviewed in more detail in subsequent sections. 
 
\par This article is specifically organized as follows. In Section~\ref{sec:MW} I review the observational evidence for dark matter in the Milky Way, and what is known about its distribution from a variety of different astronomical observations. In Section~\ref{sec:satellites}, I examine the population of Milky Way satellites, which are exceedingly important for several aspects of dark matter searches. In Section~\ref{sec:simulations}, I connect these observations to the theoretical predictions of the dark matter distribution in Milky Way-mass galaxies from modern numerical simulations. 

\par The astroparticle analysis begins in Section~\ref{sec:indirect_detection_theory}, where I review theoretical developments in indirect dark matter detection. In Section~\ref{sec:indirect_detection_experiment}, I review experimental results in indirect detection, and review the implications for theoretical models. In Section~\ref{sec:direct_detection_theory}, I review theoretical developments in direct dark matter detection. In Section~\ref{sec:direct_detection_experiment}, I review experimental results in direct detection, and review the implications for theoretical models.

\par In Section~\ref{sec:future}, I outline future directions for direct, indirect, and astrophysical searches for Galactic dark matter. In this section I discuss some very recent results that connect aspects of direct and indirect detection experiments to searches for dark matter at the LHC. I close by examining future directions for astrophysical dark matter searches, and focus on what about the properties of dark matter can be extracted from these observations. These will become particularly important if WIMPs are not discovered during the next decade. Example observations I focus on are kinematic measurements of the stars in the Galaxy from wide-field surveys, searches for Galactic satellites, and gravitational lensing measurements in our Galaxy and beyond. I hope to bring to the forefront how astronomical observations can test, constrain, and possibly uncover the properties of dark matter.

\newpage 
\section{Dark matter in the Milky Way} 
\label{sec:MW}

\par The Milky Way (MW) is the best-studied of all galaxies. The three primary components of the MW in visible light are the bar/bulge, the disk which contains the spiral arms, and the stellar halo. The dominant component of the disk by mass is the thin disk, which has had a sustained star formation for nearly ten billions years, and thus has a diverse population of stars with a wide range of ages and metalliticies. The scaleheight of the thin disk perpendicular to the Galactic plane is approximately $300$ pc, and the total mass of the thin disk is approximately $5 \times 10^{10}$ M$_\odot$. The second component is the thick disk, which has a scaleheight of approximately $1$ kpc and is composed of older stars that are more metal-poor than the thin disk. By mass, about 90\% of the visible material is contained in stars, while the remaining 10\% is in the form of gas and dust. The total absolute visual magnitude of MW is estimated to be $M_V = -20.9$~\citep{vandenBergh:2000qf}, corresponding to a total luminosity of approximately a few times $10^{10}$ L$_\odot$. 

\par In the era of large scale surveys, extraordinary progress that has been made in understanding MW properties~\citep{Juric2008}. Yet because our position viewing it from the interior, many aspects of the MW remain a mystery. For example, relatively little is known about the properties of the spiral arms in comparison to external galaxies of a similar type. Further, we do not yet have a firm understanding of the nature of the bulge. There is a growing line of evidence that the MW does not contain a so-called classical bulge, but rather a ``pseudo-bulge" that may evolve in concert with the disk. On more theoretical grounds, we do not yet have a good understanding of how rare the MW is in terms of its number of satellites, globular clusters, merger history, and metallicity distribution in comparison to other galaxies of similar mass, luminosity, and morphological type. 

\par This section is primarily focused on using the distribution of stars and gas in the MW to extract information on the mass distribution, both visible and invisible. This information is extracted from a variety of different stellar populations, which act as kinematic tracers of the underlying potential. Though as most broadly classified, the MW stars can be grouped into bulge, disk, and halo components, it is possible to further divide these populations in sub-populations. These sub-populations can be classified according to such properties as their distance, velocity dispersion, age, or metallicity. The salient point is that the stars in each population had a similar formation history, so all, or a subset of these properties, are common to all stars within it. 

\par Because we are interested in uncovering the nature of the dark matter, there are, broadly, three particularly important components to separately consider: the Galactic center, the Solar neighborhood, and the distant outer part of the dark matter halo. The status of our knowledge of the dark matter distribution in each of these three regimes is discussed, paying specific attention to what quantities are best measured by the data, and the systematics that are incurred in the measurements. The dark matter distribution in each of these regimes is determined by utilizing different populations of stars and gas; it is important to review each of these observations in detail in order to gain an appreciation for the systematics that are propagated to particle dark matter searches discussed in the sections below. 

\subsection{Dark matter in the Galactic center} 
\par The Galactic center is a notoriously difficult region to study from our position within the MW disk. It is obscured by approximately 30 magnitudes in visual wavelengths, so, aside from a few windows of below average obscuration, the best method to observe it are through infrared and radio wavelengths. More recently it has becomes possible to resolve the properties of individual stars to determine the shape, ages, metallicities, and kinematics of bulge stars. 

\par Stars that belong to the bulge dominate the luminosity within approximately $1$ kpc of the Galactic center. Though often referred to as a bulge, both integrated photometry maps and studies of individual stars indicate that the bulge is actually shaped more like a bar. There is evidence from the kinematics of M giant stars that the bulge is cylindrically-rotating, with a peak rotation velocity of $\sim 100$ km/s at $\ell \simeq 10^\circ$~\citep{Howard2009}. At low Galactic latitudes, the velocity dispersion of the bulge varies from $\sim 130$ km/s at $\ell = 0^\circ$ to $\sim 90$ km/s at $\ell = \pm 6^\circ$~\citep{Kunder2012}. It is difficult to define a total mass for the bulge, because this requires a model-dependent separation of this component from the disk. The separation is even more difficult in light of recent observations that indicate the bulge is not a classical bulge that formed and evolves separately from the disk of the Galaxy~\citep{Shen2010}.

\par Closer in towards the Galactic center, the MW contains a nuclear star cluster (NSC) originally discovered with near-infrared imaging. The MW NSC has a spatial extent of approximately one parsec, with a total stellar mass of $\sim 10^6$ M$_\odot$. Like NSCs at the centers of many other galaxies, it contains a mix of old and young stars. Because of its proximity, it is possible to resolve the central region of the star cluster, where it is observed to have a cusp with a power law index of $r^{-1.2}$ within the central 0.22 pc, and a break to a steeper index of $r^{-1.7}$ for larger radii~\citep{Schodel2007}. Kinematic studies of the MW NSC find that the total mass contained within it can be fully-accounted for by the constituent stars of the cluster~\citep{Schodel2009}. 

\par From the astrometric and kinematic measurements of short period stars in the Galactic center, there is now conclusive evidence for a $\sim 4 \times 10^6$ M$_\odot$ supermassive black hole in the center of the Galaxy~\citep{Ghez2008,Gillessen2009}. From measured values of the velocity dispersion of stars near the Galactic center, the sphere of influence of the central black hole is $\lesssim 0.1$ pc. Beyond this radius and out to a few pc, the population of stars within the central NSC dominate the potential near the Galactic center. 

\par Within the central few parsecs, the contribution from the dark matter and from the mass of the disk are sub-dominant. Assuming for the former a smooth dark matter density profile of the form predicted in Section~\ref{sec:simulations}, the expected integrated dark matter mass is $\sim 1000$ M$_\odot$ within the central few parsecs. Explicitly including the disk component, the statistical constraints on the dark matter distribution in the Galactic center are extraordinarily weak~\citep{Iocco:2011jz}. From the current observations, it is not possible to tell if the dark matter profile is cuspy or is flat within the bulge region. Since the kinematics of this region are well-described by the observed populations of stars, and the kinematic data from which these results are derived is itself still relatively sparse, these results confirm that it is difficult to disentangle a subdominant dark matter component. At the stage, the conservative upper bound on the contribution of the dark matter is set by the upper limit deduced from the bulge contribution to the potential. This lack of constraint is somewhat unfortunate, because it is reasonable to expect that the brightest source of gamma-rays from dark matter annihilation comes from the Galactic center region (Section~\ref{sec:indirect_detection_theory}). 

\par Theoretical calculations find that the central black hole affects the dark matter distribution right around the black hole~\citep{Gondolo:1999ef}. In particular, the adiabatic growth of the central black hole may steepen the central cusp of dark matter due to phase space conservation. The calculation follows in a similar manner to the enhancement of the stellar density near a point mass. Assuming an initial dark matter profile in the Galactic center of the form $\rho \propto r^{-\gamma}$, it can be shown from both conservation of mass and conservation of angular momentum that the after the appearance of the black hole, the final mass profile scales as $\rho \propto r^{-A}$, where $A = (9-2\gamma)/(4-\gamma)$~\citep{Ullio:2001fb}. Using this formalism~\cite{Gondolo:1999ef} derive a lower bound of $\sim 0.24$ GeV cm$^{-3}$ of dark matter near the Galactic center. Though the theoretical framework is compelling, it relies on assumptions that are not well understood, for example the adiabatic nature for the growth of the black hole itself~\citep{Ullio:2001fb}. Dark matter may also interact with the constituents of the NSC and create a shallower profile~\citep{Gnedin:2003rj}. 

\subsection{Local dark matter in the Solar neighborhood}

\par As discussed in a historical context in Section~\ref{sec:problem}, methods for determining the local mass distribution from the kinematics of disk stars were pioneered by~\cite{Oort1932}. The methods of Oort were revisited and improved upon several decades later~\citep{Bahcall:1984,Kuijken:1991}, and there was significant controversy in the interpretation of these results. In particular, it was not conclusive whether the determinations of the local potential pointed to Galactic disk that contains a significant amount of dark matter, or one that was entirely described by the observed stellar populations~\citep{Kuijken:1991,Bahcall:1992}. All of these studies shared the primary motivation of searching for a dark component of the Galactic disk-- a more spherically-distributed mass component was included in the analysis, but the motivation for this component was either matter in the form of an extended stellar halo or a dark matter halo as we typically refer to it today. The fact that this extended halo component could contribute significantly to the local inventory of matter was not the driver for the scientific investigation. 

\par More recently, there has been a revived interest in the topic of local dark matter, motivated by several factors. The most important motivation is that we now have improved experimental limits on the dark matter scattering cross section with nuclei (Section~\ref{sec:direct_detection_experiment}), and also limits on the annihilation cross section from a variety of indirect detection experiments (Section~\ref{sec:indirect_detection_experiment}). Understanding the local distribution of dark matter is important for interpretation of these results. Further, there are now theoretical motivations for the presence of a dark matter disk (Section~\ref{sec:simulations}), that is kinematically-distinct from the dark matter halo component. Establishing the presence of this component first requires an understanding of the measurements of the local inventory of dark matter, and the systematics, both theoretical and experimental, in its determination. 

\par Begin by understanding the generic theoretical framework which all of these investigations rely on in one form or another. Fundamentally, all determinations of the potential from the kinematics of local stars are based upon measurements of the stellar {\it distribution function} (DF). This quantity, which will be denoted as $f$, is defined so that there are $f(\vec x, \vec v) d^3 \vec x d^3 \vec v$ stars in a volume of size $d^3 \vec x$ centered on position $\vec x$ and in a volume of size $d^3 \vec v$ centered on velocity $\vec v$. Though a measurement of $f$ is the fundamental goal of stellar dynamical studies, in practice only a fraction of the six phase space coordinates are typically accessible from observations. 

\par For a local population of stars, the accessible phase space coordinates are typically the position on the sky (two dimensions), the distance to the star, and the line-of-sight velocity. For the primary analysis in this section, assume that the distance to the star, $d$, along with the Galactic latitude and longitude $(\ell, b)$ are determined. Define a standard coordinate system in which $x$ points toward the Galactic center, $y$ points in the direction of Galactic rotation, and $z$ points towards the north Galactic pole. Define $v_{los}$ as the line-of-sight velocity, and $\mu_b$ as the proper motion of the star in the $b$-direction. The $z$-component of the velocity of the star is thus given as $v_z(b,d,v_{los})  = v_{los} \sin b + d \, \mu_b$. 

\par For the large sample of disk stars that have associated kinematical information, now numbering in the thousands, there are several systematics that must be dealt with when attempting to reconstruct the local potential. First is the uncertainty in the kinematic properties that can be extracted from observations, because the measurement uncertainties on the line-of-sight velocities are typically in the range of $5-10$ km/s~\citep{Yanny:2009kg}. For cold stellar populations with a velocity dispersion of $\sigma_z \lesssim 20$ km/s, these uncertainties must be folded into the analysis. Further, accurate proper motion measurements are required; for local stars the typical uncertainty from the proper motion is $3-4$ mas yr$^{-1}$~\citep{Munn2004}. It is straightforward to see that significant noise may be introduced from both the binning of the data and the observational uncertainties in the measurements of the positions and velocities.   

\par Because the lack of full phase space information and observational uncertainties, it is important to consider independent analysis methods, and to understand the strengths and weaknesses of these analysis methods. Theoretically the methods can be broadly broken up into two main categories: those that directly model $f$ based upon a set of assumptions for it, and those that rely on the moments of the DF and thus the more averaged properties of it. 

\bigskip 

$\bullet$ {\underline {Distribution Function-Based Methods}}

\bigskip 

\par Begin by considering the direct modeling of the distribution function. Assuming that the potential and the density distribution of the Galactic disk are axially-symmetric, the collisionless Boltzmann equation is
\be
v_R\frac{\partial f}{\partial R} + v_z\frac{\partial f}{\partial z} + \left( F_R + \frac{v_\phi^2}{R}\right)\frac{\partial f}{\partial v_R} 
- \frac{v_R v_\phi}{R}\frac{\partial f}{\partial v_\phi} + F_z \frac{\partial f}{\partial v_z} = 0,
\label{eq:boltzmann_axial}
\ee
where $F_R = -\partial \Phi(R,z)/ \partial R$ and $F_z = -\partial \Phi(R,z)/ \partial z$, and $\Phi(R,z)$ is the total potential for all of the mass components that contribute to it. 

\par As typically implemented, the key assumption of distribution function based methods is that $f$ is separable in the coordinates that describe the motion out of the plane, so that $f = f_{R\phi}(R,v_R,v_\phi) f_z(z,v_z)$. For stars near the Galactic plane, this is a reasonable approximation, because disk stars have low orbital eccentricity, and the $z$-component of the force on a star is nearly constant with $R$ for a stellar orbit. 

\par An implication of this separability assumption is that energy is conserved in the $z$-direction, so that $E_z =  v_z^2/2 + \Phi(z)$ is constant. From the definition of the distribution function, the number density of stars perpendicular to the plane is then
\be
\rho_s(z) = \int_{-\infty}^\infty f_z(z,v_z) dv_z = 2 \int_{\Phi(z)}^\infty \frac{f_z(E_z)}{\sqrt{2(E_z-\Phi(z))}} dE_z. 
\label{eq:nuz}
\ee
An Abel inversion of this equation provides a solution for the distribution function, 
\be 
f_z(E_z) = \frac{1}{\pi} \int_{E_z}^\infty \frac{-d \rho_s /d \Phi}{\sqrt{2(\Phi(z) - E_z)}} d\Phi. 
\label{eq:fz}
\ee
 Equation~\ref{eq:fz} is the main equation that is utilized for the distribution function based method. For a given stellar population, $\rho_s(z)$ can be determined directly from star counts above the plane (for a realistic data set there will of course be uncertainties in the density profile incurred in binning of the data). With the stellar distribution in place, all that is required is a model for the potential perpendicular to the disk, $\Phi(z)$. As a reminder this is the total disk potential, which includes contributions from the stellar population and the dark matter component. 
 
 \par~\cite{Kuijken:1989a} introduced a convenient parameterization for the potential, 
 \be 
 \frac{\Phi_z(z)}{2\pi G} = K \left( \sqrt{z^2 + D^2} -D \right) + F z^2,
 \label{eq:phiz} 
 \ee
 where $K$, $D$, and $F$ are constants. The first term in Equation~\ref{eq:phiz} is designed to represent the contribution from a disk with a scaleheight $D$, while the second term represents the contribution from the more extended dark matter halo. These three parameters are determined from the local kinematic data. 
  
\par With the model for $\Phi(z)$ in place, the total mass density, $\rho_T$, can be determined from Poisson's equation:
\be 
\rho_T(R,z) = -\frac{1}{4 \pi G} \left [ \frac{1}{R}\frac{\partial}{\partial R} (R F_R) + \frac{\partial F_z}{\partial z}  \right]. 
\label{eq:poisson}
\ee
The first term in Equation~\ref{eq:poisson} is related to the circular velocity through $V_c = F_R /R$. In the plane of the Galaxy it is safe to neglect this term for plausible potential models of the disk and halo~\citep{Kuijken:1989a}, however extending up to approximately $4$ kpc above the plane neglecting this term leads to an estimated $15\%$ error in the determination of $\rho_T(R,z)$~\citep{Bovy:2012tw}. So for measurements in the disk midplane, the total volume mass density is directly related to the gradient of the potential perpendicular to the plane of the disk. 

\par The set of parameters $K$, $D$, and $F$ can be determined through maximum likelihood or Bayesian methods. Equation~\ref{eq:fz} can be thought of as the conditional probability distribution for $E_z$ given the set of parameters that determine the potential, $p(E_z | F,D,K)$, or explicitly in terms of the observable position and velocity, $p(z,v_z | F,D,K)$. For the measured density distribution of stars, this can be re-expressed as $p(z,v_z | F,D,K) \propto p(v_z | z, F,D,K) \rho_s(z)$. The total probability for the set of model parameters $\{F,D,K\}$ from the measured set of $n$ stars is then
\be
P(F,D,K | z,v_z) = \frac{1}{N} P(F)P(D)P(K) \prod_{\imath=1}^n p(v_{z,\imath} | z_\imath, F,D,K) \rho_s(z_\imath),
\label{eq:prob_stars} 
\ee
where the normalization factor is, assuming that $\{F,D,K\}$ are independent, 
\be 
N = \int dF dD dK P(F)P(D)P(K)  \prod_{\imath=1}^n p(v_{z,\imath} | z_\imath, F,D,K) \rho_s(z_\imath). 
\label{eq:normalization} 
\ee
Equation~\ref{eq:prob_stars} can then be maximized for the parameters $\{F,D,K\}$ to determine the best-fitting model of the potential. Then, by comparing $\rho_T$ derived from Equation~\ref{eq:poisson} and $\rho_s$ as derived directly from the data, it is possible to measure the local density of dark matter contributed by the spherically-distributed halo component. 

\bigskip

$\bullet$ {\underline {Moment-based methods}}

\bigskip 

\par Though distribution function-based methods are the most direct route to constructing the local potential, their accuracy depends on the validity of the assumption for the shape of $f$. In the above case, the critical assumption is that of separability. Complementing the distribution function methods, an additional widely used method relies on taking moments of Equation~\ref{eq:boltzmann_axial}. This method has the benefit of not depending on the assumptions for the shape of the distribution function. 

\par Multiplying Equation~\ref{eq:boltzmann_axial} by $v_z$ and integrating over all velocities gives the jeans equation
\be
\frac{\partial (\rho_s \langle v_R v_z \rangle)}{\partial R} + \frac{\partial (\rho_s \langle v_z^2 \rangle )}{\partial z}
+ \frac{\rho_s \langle v_R v_z \rangle}{R} + \rho_s \frac{\partial \Phi}{\partial z} = 0. 
\label{eq:jeans_cyl}
\ee
Two additional independent equations can be obtained by similarly multiplying by $v_R$ and $v_\phi$, though for the purposes of the discussion in this section we will be primarily interested in the information that is extracted from Equation~\ref{eq:jeans_cyl}. 

\par As in the distribution function-based method, in order to implement Equation~\ref{eq:jeans_cyl}, we must have a model for $\Phi$ and a measurement of $\rho_s(z)$. An additional requirement is a measurement of the velocity dispersion perpendicular to the disk, $\langle v_z^2 \rangle$. This is a key difference between the distribution function and moment based methods-- in the distribution function based method all that is required is a measurement of the stellar velocities. In addition, from examination of Equation~\ref{eq:jeans_cyl}, it is clear that we need an estimate of the scale of the ``tilt" term, $\langle v_R v_z \rangle$, or the correlation between the radial velocity in the plane and the velocity perpendicular to the plane. However, from the discussion of, for example,~\cite{BT2008} this term is expected to be small. 

\par At a height $z$ above the disk, the solution to Equation~\ref{eq:jeans_cyl} for a single stellar population is 
\be 
\rho_s(z) = \rho_s (z=0) \frac{ \langle v_z^2(z=0) \rangle } {  \langle v_z^2 (z) \rangle }
\exp \left[ - \int_0^{z} \frac{1}{\langle v_{z^\prime}^2 \rangle} \frac{d\Phi}{dz^\prime} dz^\prime \right]. 
\label{eq:rhos_isothermal} 
\ee
Through an iterative procedure, Equation~\ref{eq:rhos_isothermal}, combined with Poisson's equation (Equation~\ref{eq:poisson}), allow for a solution of the density in the stellar population $\rho_s$, the total density, $\rho_T$, and thus the density in any component that is not associated to the observed population. 

\par The calculations above show how the volume mass density is determined over the regime probed by the tracer stellar population. A related quantity, that is also derived from the poisson equation (Equation~\ref{eq:poisson}) and is often used as a probe of the Galactic potential is the surface mass density. The integral surface mass density out to a distance $z_{max}$ from the Galactic plane is defined as 
\be
\Sigma(R,z_{max}) = 2 \int_0^{z_{max}} \rho_T(R,z) dz.
\label{eq:surface_mass_density}  
\ee
Depending on the scaleheight of the stellar population that is used, the surface mass density probes the potential on different scales relative to those probed by the volume mass density, so it can be viewed as a measurement complementary to the volume density. Note that as defined Equation~\ref{eq:surface_mass_density} is the surface mass density from all components, and not just the component that is associated with the stellar population. 

\bigskip 

$\bullet$ {\underline {Application to data sets}}

\bigskip 

\par With the theoretical frameworks established we are now in position to understand the results obtained from their application to data. In order to reduce the systematics in measurements it is best to first define a population of disk stars, and from this population obtain accurate distance and kinematic measurements. There are several important properties that an observed population must satisfy. From the observational perspective, the population of stars clearly must have both accurate distance measurements and must be large enough to obtain a good measurement of the density fall-off $\rho_s(z)$. From the theoretical perspective, it is beneficial for the stars to be old enough to be well-mixed, which in the context of disk stars means that they have undergone several vertical oscillations. 

\par Different measurements of the local dark matter density have relied on different populations of stars. The original measurements of~\cite{Oort1932} utilized bright K giants as tracers, as did subsequent measurements of~\cite{Bahcall:1984} and ~\cite{Kuijken:1989c}.~\cite{Kuijken:1989b} utilized a sample of approximately $2000$ K giant stars; the distances for this sample were re-calibrated by~\cite{Garbari:2012ff} in a more recent analysis.~\cite{Holmberg:2004fj} utilized a sample of K giants selected from the Hipparcos survey. Because of their typical age these stars have the benefit of being old and dynamically well-mixed. In an earlier study~\cite{Holmberg:1998xu} utilized a sample of A-F stars, which are typically not as old. The recent analysis of~\cite{Bidin:2012vt} utilized a sample of approximately $400$ thick disk red giants towards the South Galactic Pole. 

\par The main results of these various analyses are summarized in Table~\ref{tab:local_measurements}. Also indicated in the final column is whether the method used by the respective authors most closely resembles the distribution function based or moment based method. Measurements of the total surface mass density of the disk out to $1.1$ kpc are indicated, along with the contribution to the surface mass density from the visible mass that is reported by the respective authors. Note that although the visible surface mass density is less than the total surface mass density, this does not mean that the difference is contributed by the component associated with the extended dark matter halo. In the parameterization of Equation~\ref{eq:phiz}, the $F$ parameter most closely parameterizes the contribution from the extended dark matter halo. Some of the difference may be contributed by matter associated with the disk that is not included in the visible components. As discussed in more detail in Section~\ref{sec:simulations}, a dark matter disk is predicted to exist from cosmological simulations~\citep{Read:2008}, however using methods similar to those described above there is not yet kinematical evidence for such a structure in the Milky Way~\citep{Bidin:2010rj}. 

\begin{deluxetable}{l c l c}
\tablecolumns{4}
\tablecaption{Compilation of measurements of the local surface mass density and the local dark matter density. Above the horizontal line, the measurements of the surface mass density and dark matter density are derived directly from local measurements of different sets of stellar populations. The final column indicates the method used to obtain the results, either the distribution function method (DF) or the moment-based method (M). Below the horizontal line, the measurements are obtained from a combined analysis including additional measurements of Galactic model parameters. Note that $1$ GeV cm$^{-3}$ = 0.027 M$_\odot$ pc$^{-3}$. 
\label{tab:local_measurements}}
\tablehead{
\colhead{Authors} &  \colhead{Total (visible) surface mass density} & \colhead{Dark matter density} &  \colhead{Method}  \\
\colhead{} & \colhead{(M$_\odot$ pc$^{-2}$)} & \colhead{(M$_\odot$ pc$^{-3}$)} & \colhead{}
}
\tablewidth{0pc}
\startdata
~\cite{Kuijken:1991}  & $71 \pm 6$ \,($48 \pm 8$)&--& DF\\
~\cite{Holmberg:2004fj} &$74 \pm 6 \, (56)$&--& M\\ 
~\cite{Bienayme:2006} & $64 \pm 5 \,(53)$&--& DF\\  
~\cite{Holmberg:1998xu}~\tablenotemark{a} &--&$0.007$& DF\\
~\cite{Garbari:2012ff} &--& $0.022_{-0.013}^{+0.015}$ (90\%)& M\\
~\cite{Bovy:2012tw} &--& $0.008_{-0.003}^{+0.003}$ & M\\
\hline
~\cite{Catena:2009mf} &--&$0.011_{-0.0010}^{+0.0010}$&--\\
~\cite{Weber:2009pt} &--&$[0.005-0.010]$&--\\
~\cite{Salucci:2010qr} &--&$0.012_{-0.003}^{+0.003}$&--\\
~\cite{McMillan:2011wd} &--&$0.011_{-0.0011}^{+0.0011}$&--\\
\enddata
\tablenotetext{a}{Estimated by subtracting the total mass density from the mass density in visible matter.
}
\end{deluxetable}

\subsection{The mass of the Milky Way}
\par Dating back to the original work of~\cite{Kinman:1959}, the velocities of distant globular clusters and satellite galaxies have been used as kinematic {\it tracers} for the mass in the outer halo. Many authors have improved upon this analysis using both better theoretical models and observational samples~\citep{Little:1987,Zaritsky:1989,Kulessa:1992,Kochanek:1995xv,Wilkinson:1999hf,Sakamoto:2003,Watkins:2010}. More recently, thanks to large scale surveys such as the Sloan Digital Sky Survey (SDSS), it has become possible to measure the dark matter halo mass profile using the kinematics of a large sample of Blue Horizontal Branch (BHB) stars~\citep{Xue:2008se,Deason:2012wm}. These BHB stars are particularly useful because they are both intrinsically bright and have accurate distance measurements. 

\par Similar to the analysis of the dark matter distribution in the Solar neighborhood above, in measurements of the dark matter halo mass of the Galaxy there are two general theoretical methods to consider: those based on assumptions for the distribution function, and those based on the moments of the distribution function. In each case, a stellar population is observed, and the kinematic information on this population is used to study the underlying properties of the dark matter halo. Though similar in this aspect, the analysis in this subsection differs from the Solar neighborhood analysis in a couple of different ways. First, in this case not only are stellar populations utilized, but also populations of systems of stars, in particular globular clusters and dwarf spheroidals (dSphs). Second, the populations in the halo analysis make a negligible contribution to the total potential. Although this is not a necessity of the analysis, it is a physical statement that the influence of the dark matter halo continually increases at larger radii from the Galactic center. 

\bigskip 

$\bullet$ {\underline {Distribution function based models}}

\bigskip 

\par Following the sequence above begin by considering models based on distribution functions. Observations considered for this analysis typically provide at most four of six phase space coordinates; it is convenient to take these four coordinates as the latitude ($b$), longitude ($l$), distance ($d$), and line-of-sight velocity of the object ($v_{los}$). The distribution function is defined as above, though in this case we will not make an assumption on the separability of the distribution function. Rather break up the distribution function-based models into two broad classes of sub-models: those that assume that the velocity distribution for the tracers is isotropic, and those that assume that the tracer velocity distribution is anisotropic. The degree of anisotropy is determined through the standard parameter, 
\be 
\beta = 1 - \sigma_t^2 / \sigma_r^2, 
\label{eq:beta}
\ee
where a spherical polar coordinate system ($r,\theta,\phi)$ about the Galactic center is assumed, $\sigma_r$ is the radial component of the velocity dispersion, and $\sigma_t^2 = \sigma_\theta^2 + \sigma_\phi^2$ is the sum of the components of the velocity dispersion orthogonal to the radial component. The components of the velocity dispersion orthogonal to the radial component are taken to be equal, $\sigma_\theta^2= \sigma_\phi^2$. For $\beta < 0$, the orbits are more circular, while for $\beta > 0$, the orbits are more radial. For $\beta = 0$ the orbits are isotropic. 

\par For the isotropic model, the distribution function depends on a single integral of motion, which is the total energy. The mapping between a given density/potential model and the distribution function, $f(\epsilon)$, is given by the Eddington formula, 
\be
f(\epsilon) = \frac{1}{\sqrt{8}\pi^2}\frac{d}{d\epsilon} \int \frac{d\rho_s}{d\Psi}\frac{d\Psi}{\sqrt{\epsilon-\Psi}},
\label{eq:eddington}
\ee
where as above $\rho_s$ represents the density of the tracer population. Here $\Psi$ is the total potential from all of the components, and $\epsilon = \Psi -v^2/2$, where $v$ is the velocity. 

\par Equation~\ref{eq:eddington} is used to determine the velocity distribution of a population described by a set of theoretical parameters, and it is then to compared to the observed population. For the analysis in this sub-section, consider a simplified model for the Galactic potential, which includes contributions from the bulge, the disk, and the dark matter halo, so that $\Psi = \Psi_b + \Psi_d + \Psi_{dm}$. It will be most straightforward to assume that the potential components are spherically-symmetric. The bulge potential can be modeled as
\be
\Psi_b = \frac{G M_b}{(r+c_0)},
\label{eq:phi_bulge}
\ee
where $M_b \simeq 10^{10}$ M$_\odot$ is the total mass of the bulge and $c_0 = 0.6$ kpc is its scalelength. Though the potential from the bulge near the Galactic center deviates from a spherically-symmetric model, Equation~\ref{eq:phi_bulge} will be appropriate at much larger distances. The disk potential can be modeled as
\be
\Psi_d = \frac{G M_d[1-\exp(-r/b_d)]}{r},
\label{eq:phi_disk}
\ee
where $M_d \simeq 5 \times 10^{11}$ M$_\odot$ is the mass of the disk, and $b_d \simeq 3$ kpc is the disk scale length. For the dark matter halo, there are a variety of well-motivated models. In this case, we can start from the spherically-symmetric density profile. A popular modern model is the Navarro-Frenk-White (NFW) profile, 
\be 
\rho_{NFW}(r) = \frac{\rho_0}{(r/r_0)(1+r/r_0)^2},
\label{eq:NFW} 
\ee
which is motivated by the results from modern numerical simulations (See Section~\ref{sec:simulations} for more of a discussion on this profile). Then from poisson's equation, the potential corresponding to the NFW profile is 
\be 
\Psi_{NFW}(r) = 4 \pi G \rho_0 r_0^3 \frac{\ln(1+r/r_0)}{r}. 
\label{eq:NFW_potential} 
\ee

\par At large radius, the sum of the potentials above can be approximated by a pure power law that scales as $\propto r^{-\gamma}$. This is primarily because beyond the scale radius the potential is dominated by the NFW mass profile. For the tracer density $\rho_s$ it is standard to consider two limiting models. One model is described by a pure power law in density, $\rho_s \propto r^{-\alpha}$, which is appropriate for a stellar halo population~\citep{Bell2008} or probably for the dSph population~\citep{Evans:2000wd}. In these cases, $\alpha \simeq 2-4$. The second model assumes that the observed stellar population traces the underlying dark matter halo, $\rho_s \propto \rho_{dm}$. 

\par The final step involved in comparing the theory to the observational data requires transforming the distribution function in Equation~\ref{eq:eddington} to a function of observed line-of-sight velocities. This is done as
\be
f(v_r; \vec r) = \int d^3 v^\prime f(\epsilon) \delta (v_r^\prime - v_r), 
\label{eq:fvr}
\ee
where $v^{\prime 2} = v_r^{\prime 2} + v_\theta^{\prime 2} + v_\phi^{\prime 2}$, and $\vec r$ denotes the position in the halo. A similar relation can be derived for the tangential motion, though these are typically subject to much larger uncertainties. 

\par In a manner similar to Equation~\ref{eq:prob_stars} we can perform a maximum likelihood analysis to determine the best fitting parameters that describe the MW dark matter halo. Assuming that the anisotropy is fixed, $\beta = 0$, the likelihood can be maximized for the parameters $(\alpha, \gamma)$, so that
\be 
{\cal L}(\alpha,\gamma) \propto \prod_{i=1}^m f(v_{r,\imath}, \vec {r_\imath} | \alpha, \gamma),
\label{eq:halo_likelihood} 
\ee
where the product is over the positions and the velocities of the components of the population. 

\par The analysis becomes more model-dependent when considering anisotropic distribution function models with $\beta \ne 0$, for the tracer population. In this case, even when a single population contributes to the potential, the distribution function no longer depends on a single integral of motion, and there is also no longer a one-to-one mapping between the density and the potential. To make the problem tractable, it is standard to assume a distribution function that is separable in the energy and the angular momentum per unit mass, $F(\epsilon,L) = g(L)f_1(\epsilon)$. If it is assumed that $g(L) = L^{-2\beta}$, then the anisotropy $\beta$ for the tracer population is constant as a function of radius, and the distribution function splits up into even and odd components as~\citep{Watkins:2010}
\be
F_{even} \propto L^{-2 \beta} f_1(\epsilon),  
\label{eq:fbeta_eeven}
\ee
with 
\be
f_1(\epsilon) \propto \epsilon^{ [\beta (\gamma-2)] / \gamma + (\alpha / \gamma) - 3/2}, 
\label{eq:f1_even}
\ee
and 
\be
F_{odd} \propto (1 - \eta) \tanh ( L_z /\Delta) F_{even},
\label{eq:fbeta_odd} 
\ee
where $L_z$ is the $z$-component of the angular momentum, and $\Delta$ is a smoothing parameter. The value of $\eta$ determines the amount of rotation: for $\eta = 0$ there is maximum prograde rotation, for $\eta$ = 2 there is maximum retrograde rotation. Similar to the likelihood analysis in the case of isotropic models, parameters of the distribution may be determined from the data sets. These models form a representative set that are typically used to measure the mass profile of the Galaxy from discrete populations. 

\bigskip

$\bullet$ { \underline{Moment-based methods} } 

\bigskip

\par Now consider measurements of the Milky Way mass distribution that are based not on parameterizations of the distribution function, but rather on its moments. As above,  both isotropic and anisotropic models can be considered, and the potential of the Galaxy is assumed to be spherically-symmetric, with $(r,\theta,\phi)$ the coordinates describing the system. The spherical jeans equation is 
\be
r\frac{d(\rho_s \sigma_r^2)}{dr} + 2 \beta(r) \rho_s \sigma_r^2 =  \rho_s(r) \frac{GM(r)}{r}, 
\label{eq:jeans_spherical} 
\ee
which is derived by taking moments of the Boltzmann Equation, similar to the manner in which Equation~\ref{eq:jeans_cyl} is derived. 

\par The velocity dispersions that appear in Equation~\ref{eq:jeans_spherical} are with respect to a coordinate system centered on the Galaxy. To perform the analysis in this subsection we need to convert the measured velocity dispersion, $\sigma_{GSR}$, into the velocity dispersion in the Galactic coordinate system, $\sigma_r$ (as in e.g.~\cite{Dehnen:2006cm}). First assume that the Sun is on the $z$-axis, and $\theta$ is the polar angle. The distance to a star is 
\be
d^2 = r^2 + R_\odot^2 - 2rR_\odot \cos \theta.
\label{eq:distance} 
\ee
The time derivative of this quantity is the velocity in the direction of the star, 
\be 
\dot d = \left[ \frac{r - R_\odot \cos \theta}{d} \right] v_r + \frac{R_\odot \sin \theta}{d} v_\theta.
\label{eq:velocity}
\ee
 The velocity dispersion is then the RMS velocity over all solid angles, $\sigma_{GSR} = \langle \dot d \rangle$, and can be written as
\be  
\sigma_{GSR} = \sigma_r \sqrt{1-\beta(r) A(r)},
\label{eq:sigmaGSR} 
\ee
where
\be 
A(r) = \frac{r^2 + R_\odot^2}{4r^2} - \frac{(r^2-R_\odot^2)^2}{8r^3 R_\odot} \ln \left | \frac{r+ R_\odot}{r-R_\odot} \right |. 
\label{eq:Ar}
\ee
These equations show that for stars local to the Sun, there is an important correction factor to the measured line-of-sight velocity that must be accounted for. For large distances, which in this case means $r \gg R_\odot$, the correction factor is insignificant, so that $\sigma_{GSR} \simeq \sigma_r$. 

\par
\begin{figure*}
\begin{center}
\begin{tabular}{cc}
\includegraphics[width=0.65\textwidth]{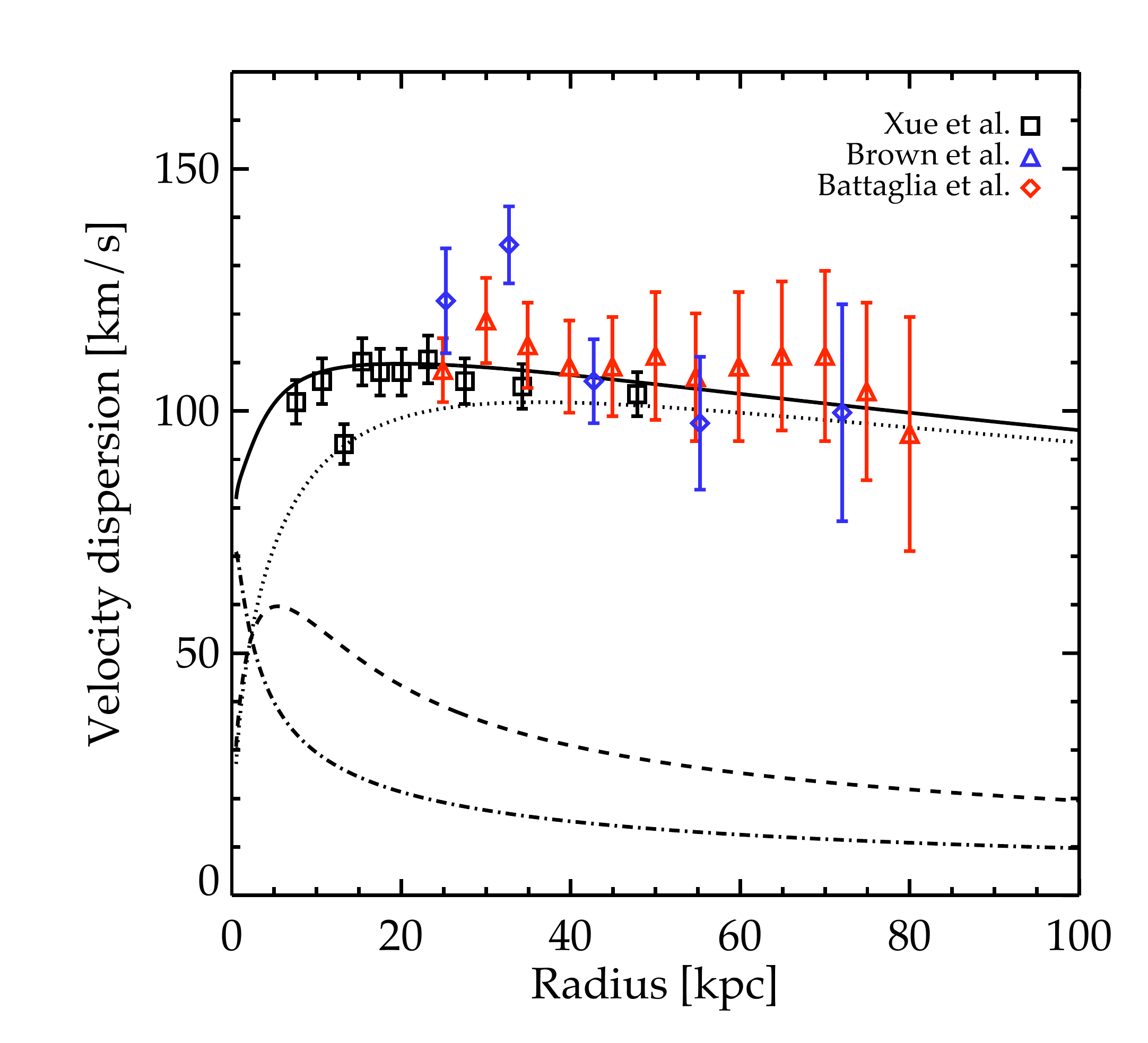} \\
\end{tabular}
\end{center}
\caption{Velocity dispersion versus Galactocentric radius from the data of~\cite{Xue:2008se},~\cite{Brown:2009nh}, and~\cite{Battaglia:2005rj}. The dot-dashed curve is the velocity dispersion from the bulge potential, the dashed curve is from the disk potential, and the dotted curve is from the dark matter halo. The solid curve is the sum of all three components.  The dark matter halo potential is calculated from an NFW profile with $\rho_0 = 6.5 \times 10^6$ M$_\odot$ kpc$^{-3}$ and r$_0 = 22$ kpc in Equation~\ref{eq:NFW}. The velocity dispersion is assumed to be isotropic. 
}
\label{fig:mw_dispersion}
\end{figure*}

 From the above formalism, Figure~\ref{fig:mw_dispersion} shows the velocity dispersion versus radius in comparison to three recent observational samples. Shown is the velocity dispersion due to the three primary dynamical components of the Galaxy, the bulge, disk, and dark matter halo. The dark matter halo clearly becomes the most relevant in the outer regions of the Galaxy. The potential is determined assuming that $\rho_s \propto r^{-3.5}$ and that the orbits of the tracers are isotropic. 

\bigskip 

\par Table~\ref{tab:mass_measurements} shows a compilation of recent measurements of the Milky Way mass using both distribution function and moment-based methods described above. In nearly all of the analyses reported, there is a ``direct" measurement of the mass on a scale that is directly probed by the tracer population. Nearly all of the measurements of the total halo mass require an extrapolation beyond the radius of the visible tracer population. 

\begin{deluxetable}{l c c c c c c}
\tablecolumns{6}
\tablecaption{Compilation of the measurements of the mass of the Milky Way dark matter halo. The third column gives the direct measurements within the radius $R$ in the second column using kinematic data. The fourth column gives measurements of the total Milky Way mass; in this column, all except the~\cite{Watkins:2010} results rely on an extrapolation using data from smaller radii. The fifth column gives the types of tracers that are used: halo stars (S), globular clusters (GC), and/or dwarf spheroidals (dSph). The sixth column gives the statistics of the orbits of the tracers that are implied from the best-fitting masses, either isotropic (I), circular (C), or radial (R). The final column indicates the most closely-resembled method used to obtain the results, either the distribution function method (DF) or the moment-based method (M).
\label{tab:mass_measurements}}
\tablehead{
\colhead{Authors} &  \colhead{$R$} & \colhead{Mass ($<R$)} & Total Mass & Tracers &Orbits & Method\\
\colhead{} & \colhead{(kpc)} & \colhead{($10^{11}$ M$_\odot$)} & \colhead{($10^{12}$ M$_\odot$)} & \colhead{} & \colhead{} & \colhead{} 
}
\tablewidth{0pc}
\startdata
~\cite{Wilkinson:1999hf} & 50 & $5.4_{-3.6}^{+0.2}$ & $1.9_{-1.7}^{+3.6}$ &GC,dSph&R&DF\\
~\cite{Sakamoto:2003} & 50 & $5.3_{-0.4}^{+0.1}$  &  -- &S& C&DF\\
~\cite{Xue:2008se} & 60 & $4.0_{-0.7}^{+0.7}$ & $1.0_{-0.2}^{+0.3}$ &S&I & M \\
~\cite{Gnedin:2010fv} &80&$6.9_{-1.2}^{+3.0}$ &  -- &S&R& M \\
~\cite{Watkins:2010} & 100 & $3.3_{-1.1}^{+1.1}$ & $0.9_{-0.3}^{+0.3}$ &S,GC,dSph&I& M\\
~\cite{Deason:2012wm} & 50 & $4.2_{-0.4}^{+0.4}$ & -- &S&R& M\\
~\cite{Battaglia:2005rj}~\tablenotemark{a} & -- & -- & $0.50_{-0.17}^{+0.25}$ &S&R& M \\
~\cite{Dehnen:2006cm}~\tablenotemark{a} & -- & -- & $1.5$&S&R& M \\
~\cite{Li:2007eg} &--&--&$2.43_{-0.65}^{+0.66}$  &-- & --\\
~\cite{Busha:2010sg} &--&--&$1.2_{-0.7}^{+1.0}$&--&-- \\
\enddata
\tablenotetext{a}{The~\cite{Dehnen:2006cm} results are from a reanalysis of the~\cite{Battaglia:2005rj} data. The quoted result from ~\cite{Battaglia:2005rj} is the truncated flat model that gives the best fit.} 
\end{deluxetable}

\par For the measurements that use the dSphs as the tracers, there are two systematics that are particularly important: first is the unknown density law $\rho_s$ due to the small number of objects that are measured, and second is the issue of whether or not Leo I is bound to the Galaxy. Because Leo I is moving at a high Galactocentric velocity and is at a Galactocentric radius of $260$ kpc, it is not yet clear whether this is an outlier that is bound to the halo or if it is unbound and is on its first pass through the Galaxy. Most recent proper motions indicate that it may be bound to the Galaxy~\citep{Sohn:2012xt,BoylanKolchin:2012xy}. 

\bigskip

$\bullet$ {\underline{Escape Velocity}}

\bigskip 

\par The escape velocity is directly related to the mass of the Milky Way dark matter halo. It is derived independently from a local population of high velocity stars, providing an important cross check on the mass calculations above. For this measurement, it is convenient to parameterize the high velocity tail of the stellar distribution function as
\be 
f(v) \propto (v_{esc} - v)^k.
\label{eq:ftail_vesc}
\ee
This formula can be theoretically-motivated by noting that at high energies the distribution function can be parameterized as a power law, $f(\epsilon) \propto \epsilon^k$ for isotropic orbits. Once the escape velocity is determined, the total potential of the Galaxy at the Solar radius can be derived as $\Psi_{total} = -v_{esc}^2/2$. The potential here is the sum of the respective components from the bulge, disk, and halo as defined above in Equations~\ref{eq:phi_bulge} and~\ref{eq:phi_disk}. 

\par~\cite{Smith:2006ym} have recently determined the escape velocity by combining this theoretical formalism with a maximum likelihood analysis on the sample of stars from the RAVE survey. From a sample of a couple of dozen stars they find that the escape velocity lies within the range 498 km/s $< v_{esc} < 608$ km/s at the 90\% c.l. There is a strong correlation between $v_{esc}$ and $k$, so that only a lower bound can be determined from the data on these parameters, which is at about $v_{esc} = 450$ km/s. Negative values of $k$ are ruled out in the analysis. 

\bigskip 

$\bullet$ {\underline{Full Milky Way Mass Model} }

\bigskip

\par In order to develop a full mass model of the Milky Way, the above measurements of the local dark matter distribution and the distribution in the outer halo are supplemented by several other sets of observations. 

\par The first of these parameters is the distance from the Sun to the center of the Galaxy. This has been measured recently to high precision using the orbits of short period stars around the central black hole, giving $R_0 = 8.33 \pm 0.35$ kpc~\citep{Gillessen2009} or $R_0 = 8.4 \pm 0.4$ kpc~\citep{Ghez2008}. A second measurement is the so-called terminal velocity, which is defined as the peak velocity of the interstellar medium along a line-of-sight in the Galactic plane. Relative to the Local Standard of Rest (LSR), the terminal velocity is defined as
\be
v_{term} = v_c(R_0 \sin \ell) - v_c(R_0) \sin \ell,
\label{eq:vterm}
\ee
where $v_c$ is the circular velocity. Peak velocities for several lines-of-sight from surveys have been compiled in~\cite{Malhotra1994,Malhotra1995}. A third observation often utilized are Galactic maser sources. High precision trigonometric parallaxes and proper motions of masers in high-mass star-forming regions across the Milky Way have been determined, providing three-dimensional location and velocity vectors relative to the Sun.~\cite{Reid2009} provide the measurements of 18 proper motions and line-of-sight velocities for masers. A fourth observation is the proper motion of Sgr A$^\star$, the source at the center of the Galaxy. Since Sgr A$^\star$ is fixed at the Galactic center to within approximately 1 km/s, its measured proper motion stems primarily from the motion of the Sun around the Galactic center, which is due both to the motion of the LSR and the peculiar motion of the Sun relative to the LSR. 

\par By combining all of these observations, several authors have recently reported measurements of the local dark matter density from the ensemble of data. Since the model parameterizations used in each of the analyses are somewhat different, it is important to clarify the differences in the assumptions.~\cite{Catena:2009mf} use two different parameterizations of the dark matter halo, the first is the NFW profile and the second is the Burkert model, defined by 
\be 
\rho(r) = \frac{\rho_b}{(1+r/r_b)(1+r/r_b)^2}. 
\label{eq:burkert}
\ee
Their motivation was to determine whether the shape of the density profile affects measurement of the local dark matter density. From their Bayesian modeling,~\cite{Catena:2009mf}  conclude that for an NFW profile the local dark matter density is $\rho_\odot = 0.389 \pm 0.025$ GeV cm$^{-3}$. The mean of this result is confirmed by~\cite{McMillan:2011wd}, who finds $\rho_\odot = 0.40 \pm 0.040$ GeV cm$^{-3}$ also using an NFW dark matter profile. 

\par Though several authors have converged on similar values for the mean dark matter density with the NFW halo based analysis, it is important to dissect this further to see if there are any systematics that may affect these results. The first obvious one is the assumption of spherical symmetry; breaking from this assumption makes the problem theoretically much less tractable. As discussed at the end of this section there is some evidence that the dark matter halo is aspherical. An additional systematic that was brought to light in~\cite{Weber:2009pt} is that there is a degeneracy between the scale radius of the Milky Way disk and the dark matter halo. As an example, using a generalized parameterization of the NFW dark matter density profile 
\be 
\rho(r) = \frac{\rho_0}{x^a(1+x^b)^{(c-a)/b}},
\label{eq:zhao}
\ee
where $r/r_0$, there is a strong correlation between the local dark matter density and $r_0$. 

\par In sum, all of the results presented above and summarized in Table~\ref{tab:local_measurements} and Table~\ref{tab:mass_measurements} show that astronomical measurements are providing interesting constraints on the local dark matter density and the Milky Way halo mass, though significant systematics still remain in these measurements. 

\subsection{Shape of the Milky Way halo} 
\par The discussion to this point has focused on spherical models for the Milky Way potential. The motivation for this focus was twofold. On the one hand, spherical models are the most straightforward to understand and interpret theoretically. Extending to non-spherical potentials certainly enlarges the model parameter space, and it makes the comparison between different theoretical models often less straightforward. A second motivation for restricting to spherical models is that the data sets considered to this point are well-described by them, meaning that they typically do not require more complexity. This section on the observational properties of the Milky Way dark matter halo closes by examining sets of observations that are, in principle at least, able to determine if the Milky Way potential deviates from a purely spherical model, and examines the state of these observations. 

\par There are three separate sets of observations that can determine the shape of the Milky Way dark matter halo. The first relies on the measurements of the velocity ellipsoid from different sets of stellar populations in the Galaxy. The second relies on the properties of the gaseous HI disk. The third relies on the spatial and velocity distribution of disrupted satellite galaxies.

\par In order to understand the analysis discussed below, it will be first necessary to define a set of ellipsoidal coordinates, and relate them to the spherical coordinates utilized above. For a given Galactic model, the elongation of the potential will generally be different than the elongation of the corresponding density distribution. The ellipsoidal coordinates, either for the potential or for the density, can be defined by transforming the spherical coordinate $r \rightarrow m$, where $m^2 = x^2/a^2 + y^2/b^2 + z^2/c^2$. The ellipsoid is thus characterized by the axes $a,b,c$, and from these define the axis ratios $q = c/a$ and $s = b/a$. From these ratios we can define the three classes of ellipsoids, which are prolate, $a > b \simeq c$, oblate, $a \simeq b > c$, and triaxial, $a > b > c$. A spherical halo has $a = b = c$. 

\par For the purposes of the discussion below it will also be necessary to define the standard spherical polar coordinate system, in which the radial velocity coordinate $v_r$ points away form the Galactic center, $v_\theta$ is the velocity in the direction of the zenith angle measured from the North Galactic pole, and $v_\phi$ is the velocity in the azimuthal direction opposite of Galactic rotation. 

\bigskip 

$\bullet$ {\underline{Shape of the Velocity Ellipsoid} }

\bigskip 

\par Now consider in more detail measurements of the shape of the halo from the velocity ellipsoid. For this measurement, begin by considering a sample of stellar velocities from an observed population of stars. From the sample of stellar velocities, the velocity dispersion tensor along any set of orthogonal coordinates is defined as
\be 
\sigma_{\imath \jmath}^2 = \langle ( v_\imath - \langle v_\imath \rangle ) ( v_\jmath - \langle v_\jmath \rangle ) \rangle.
\label{eq:velocity_dispersion_tensor} 
\ee
As above the average is over the phase space distribution function, i.e. $\langle v_\imath \rangle = \int v_\imath f(\vec x, \vec v) d^3 \vec v$. Since the velocity dispersion tensor is a second-rank tensor that is symmetric, it can always be diagonalized, and the resulting set of principal axes form the velocity ellipsoid. 

\par For any pair of coordinates, the tilt of the velocity ellipsoid is defined as
\be 
\tan ( 2 \alpha_{\imath \jmath} ) = \frac{2 \sigma_{\imath \jmath}^2}{\sigma_{\imath \imath}^2 - \sigma_{\jmath \jmath}^2}.
\label{eq:tilt} 
\ee
Defined in this manner, the tilt represents the angle between the $\imath^{th}$ the major axis of the ellipse formed by projecting the velocity ellipsoid onto the $\imath \jmath$-plane. 

\par The tilt angles in Equation~\ref{eq:tilt} have important theoretical properties. It has been long known that if all of the tilt angles are zero, then the potential is spherically-symmetric~\citep{Eddington1915,Lynden-Bell1962}. More generally, if the velocity ellipsoid is radially aligned everywhere, then the underlying potential of the system is spherically-symmetric~~\citep{Smith:2009yn}. Even if the velocity ellipsoid is aligned along the spherical polar coordinates and the potential of the dark matter halo is spherical, the density distribution of the population that is used to determine the velocity ellipsoid itself need not be spherical. 

\par Thus the measurements of the velocity ellipsoid provides a powerful probe of the potential of the Galaxy, and thereby the shape of the dark matter halo. Clearly the observational challenge is to obtain a large population of stars that have full six-dimensional phase space information associated with them, so that the shape of the potential may be determined by finding the tilt angles via Equation~\ref{eq:tilt}. Further, since we are particularly interested in understanding the properties of the outer region of the Galactic halo, it is most beneficial to obtain a sample of stars that are outside of the disk, where the potential is dominated by the dark matter halo.  

\par Several recent studies have measured the tilt of the velocity ellipsoid.~\cite{Siebert2008} used the phase space information from a sample of 763 red clump giant stars from the RAVE survey that are distributed between distances of $500$ to $1500$ pc below the Galactic plane. From their sample they measured tilt angle of $\alpha_{UW} = 7.3 \pm 1.8^\circ$ at $\sim 1$ kpc below the plane. Because of the sensitivity to the disk potential at this radius, the fact that the tilt is different than zero does not imply that the potential of the dark matter halo is non-spherical. Using an axisymmetric NFW dark matter halo model,~\cite{Siebert2008} show that the observed tilt may be explained by either a prolate dark matter halo with a disk scale length of $\sim 3$ kpc, or an oblate dark matter halo with a disk scale length of $\sim 2$ kpc. These results highlight the degeneracies between the halo and disk potential even for a fixed NFW model. 

\par~\cite{Smith:2009yn} utilize a sample of approximately $1800$ sub dwarfs in SDSS stripe 82 in 250 deg$^2$ to measure the tilt of the velocity ellipsoid in all three of the spherical coordinates $(r,\theta,\phi)$ defined above. Since these stars extend out to a distance of $5$ kpc from the disk, they are a more ideal sample to probe the dark matter potential than the above RAVE sample. Interestingly, they show that each of the angles $\alpha_{r \phi}$ and $\alpha_{\theta \phi}$ are consistent with zero, and that the third angle $\alpha_{r \theta}$ is consistent with zero at the three-sigma level, and its deviation from zero can be explained as a perturbation from the Galactic disk potential. 

\par Ideally an even more extended sample of stars would provide the most robust measurement of the shape of the halo; the sample of halo stars used above to determine the Milky Way circular velocity out to $60$ kpc particularly interesting. A measurement of the kinematics of this population will be possible with the GAIA satellite, which will measure the proper motions of bright  halo stars with an uncertainty $\lesssim 1$ mas. It is only possible at present to determine the shape of the distribution of the halo stars, and because they most likely formed via a different physical process than did the underlying dark matter halo, a direct measure of their density distribution cannot reliability constrain the shape of the dark halo. Studies of halo stars out to approximately $35$ kpc using the Canada-France-Hawaii Telescope indicate the oblateness of the stellar halo is $c/a = 0.70 \pm 0.01$~\citep{Sesar:2010fv}. The best-fitting Einasto fit to the stellar density is $n = 2.2 \pm 0.2$ and $r_e = 22.2 \pm 0.4$ kpc, which is less steep than the dark matter as is predicted in the LCDM theory of structure formation (Section~\ref{sec:simulations}). 

\bigskip

$\bullet$  {\underline {Thickness of HI  gas disk} }

\bigskip

\par The thickness of the HI gaseous disk of the Milky Way is observed to increase with radius, with an extent of $0.1$ kpc at approximately $5$ kpc to $1$ kpc at approximately 25 kpc~\citep{Wouterloot1990}. Since the HI gas disk is more extended than the stellar disk, it is more sensitive to the potential from the dark matter in the outer halo than the stellar disk, and it is interesting to determine whether dark matter halo models can reproduce the above observed ``flaring" of the HI disk. Several studies have shown that it is difficult to reproduce the flaring within the context of spherical halo models~\citep{Narayan2005,Kalberla2007}.~\cite{Banerjee2011} find that the flaring can be reproduced by a prolate dark matter halo that becomes the most prolate at $\sim 25$ kpc, $q \simeq 2$. 

\bigskip 

$\bullet$ {\underline{ Dynamics of Stellar Streams}} 

\bigskip 

\par The dynamics of stellar streams also provide a probe of the shape of the Galactic dark matter halo. In particular, the positions and the velocities of many of the stellar components of the Sagittarius stream are now well-determined from a combination of photometry from SDSS, 2MASS and high resolution spectroscopy. Assuming the parameterizations of the bulge, disk, and halo discussed above, however, it has proven to be difficult to obtain a fit to the detailed properties of the Sagittarius stream. Two particular difficult aspects to fit are first what appears to be a bifurcation of the leading stream, and second the position of the stream on the sky. Assuming that the bifurcation is not related to the physical dynamics of the stream,~\cite{Law:2010pe} find that the best-fitting dark matter halo is triaxial, which for the potential give $q = c/a = 0.72$ and $s = b/a = 0.99$. For the halo density the respective axis ratios are $q = c/a = 0.44$ and $s = b/a = 0.97$. This corresponds to a nearly oblate ellipsoid with the minor axis contained in the Galactic plane. The Sun does not lie along a principal axis, but rather is rotated about approximately $10$ degrees from the minor axis. 

\par Because of the difficulty in matching the Sagittarius stream to a halo model, and because in future data sets fainter streams are likely to be more prevalent, it is also possible to utilize fainter and thinner streams to study the shape of the halo. As an example,~\cite{Lux2012} show that the globular cluster tidal stream NGC 5466 favors an oblate or triaxial halo due to a feature on the westward side of the stream. Orbital integrations and theoretical modeling designed to match observations of faint streams and globular clusters will become more important in the coming years. 

\bigskip 

\par In sum, through several complementary observational techniques it is now possible to determine whether the Milky Way dark matter halo is non-spherical, and even if the halo is prolate, oblate, or triaxial. As discussed each of these methods have their respective strengths and weaknesses, and there are different lines of evidence that point towards spherical, prolate, oblate, and traixial models. In addition to the potential for improving these results with future large scale surveys, a variety of other interesting methods may become available to complement these methods, for example via the use of precision astrometry of hypervelocity stars~\citep{Gnedin:2005pt}. 

\newpage 
\section{Dark matter in Milky Way satellite galaxies}
\label{sec:satellites} 

\par The first galaxy ultimately identified as a dwarf satellite of the Milky Way was discovered by Harlow Shapley in the constellation Sculptor~\citep{Shapley:1938a}. Initially Shapley classified the system in Sculptor as a cluster of galaxies, however follow-up observations revealed that the constituents of the system were not galaxies, but were in fact individual stars, and the distance to the system was approximately 80 kpc. At the time, the extent of this system, and the fact that the stars were individually-resolvable, made the system more akin to a globular cluster than a known type of galaxy. Further observations by Shapley revealed a similar organization of stars in the constellation Fornax~\citep{Shapley:1938b}.  

\par Fast forwarding up to the end of the 20th century,  seven more systems with similar properties to Sculptor and Fornax were discovered. Those discovered prior to 1990 (Draco, Ursa Minor, Leo I, Leo II, Carina) were located by visual inspection of photographic plate data. Sextans, which was discovered in 1990, was the first satellite discovered via automated plate machine scans. In 1994, the Sagittarius galaxy was serendipitously discovered during the course of a survey of the outer regions of the Galactic bulge~\citep{Ibata:1994}. Though this system is more luminous than Fornax, it was not located earlier because it lies in a region of high foreground density behind the Galactic center (For a more detailed review of the discovery of the Milky Way satellites, see~\cite{Willman:2009dv}). 

\par These nine systems have come to be classified as dwarf Spheroidal (dSph) galaxies, and for reference from this point forward these nine dSphs will be called ``classical" satellites . The classical dSphs range in Galactocentric distance from approximately $15-250$ kpc, and their overall distance distribution in the Galactic halo is much more extended than the more centrally-concentrated Globular cluster population. The two brightest of the Milky Way satellite galaxies, the Large and the Small Magellanic Clouds, are of course visible to the naked eye and were known long before the discoveries of Shapley. These galaxies are also distinct from dSphs in that they contain HI gas, whereas the dSphs are devoid of gas up to the present observational limits. 

\par At the turn of the century, it was certainly reasonable to assume that our census of Milky Way satellites was incomplete. However, the nature of the incompleteness was unknown. From a systematic search of the Southern sky,~\cite{Kleyna:1997} reported that the census of Milky Way satellites is likely complete down to approximately $1/8$ of the luminosity of Sculptor, or approximately a few times $10^5$ L$_\odot$.~\cite{Willman:2004} estimated that there may be a few times more dSphs at the luminosity of Fornax that may have gone undetected in previous surveys due to systematic effects such as disk obscuration. Whatever the nature of the incompleteness, it was clear that large scale galaxy surveys coming online would be able to find more faint satellites, if they indeed were to exist. 

\par The Sloan Digital Sky Survey (SDSS) is a photometric and spectroscopic survey of $\sim 1/5$ of the northern Galactic sky. The SDSS is spectroscopically-complete from objects brighter than a magnitude $r = 17.7$; since this is brighter than the most luminous stars in any of the classical dSphs, it was expected that SDSS would be sensitive to the detection of new satellites as resolved overdensities of stars through imaging data. As of the publication of this article, searches of the SDSS data for overdensities of stars in color space have revealed a dozen new Milky Way satellites~\citep{Willman:2005cd,Belokurov:2006ph}. The distances to these objects are determined by comparing the population of stars to old stellar populations in globular clusters, and determining the position of the turn-off of the stellar main sequence. 

\par
\begin{figure*}
\begin{center}
\begin{tabular}{c}
\includegraphics[width=0.50\textwidth]{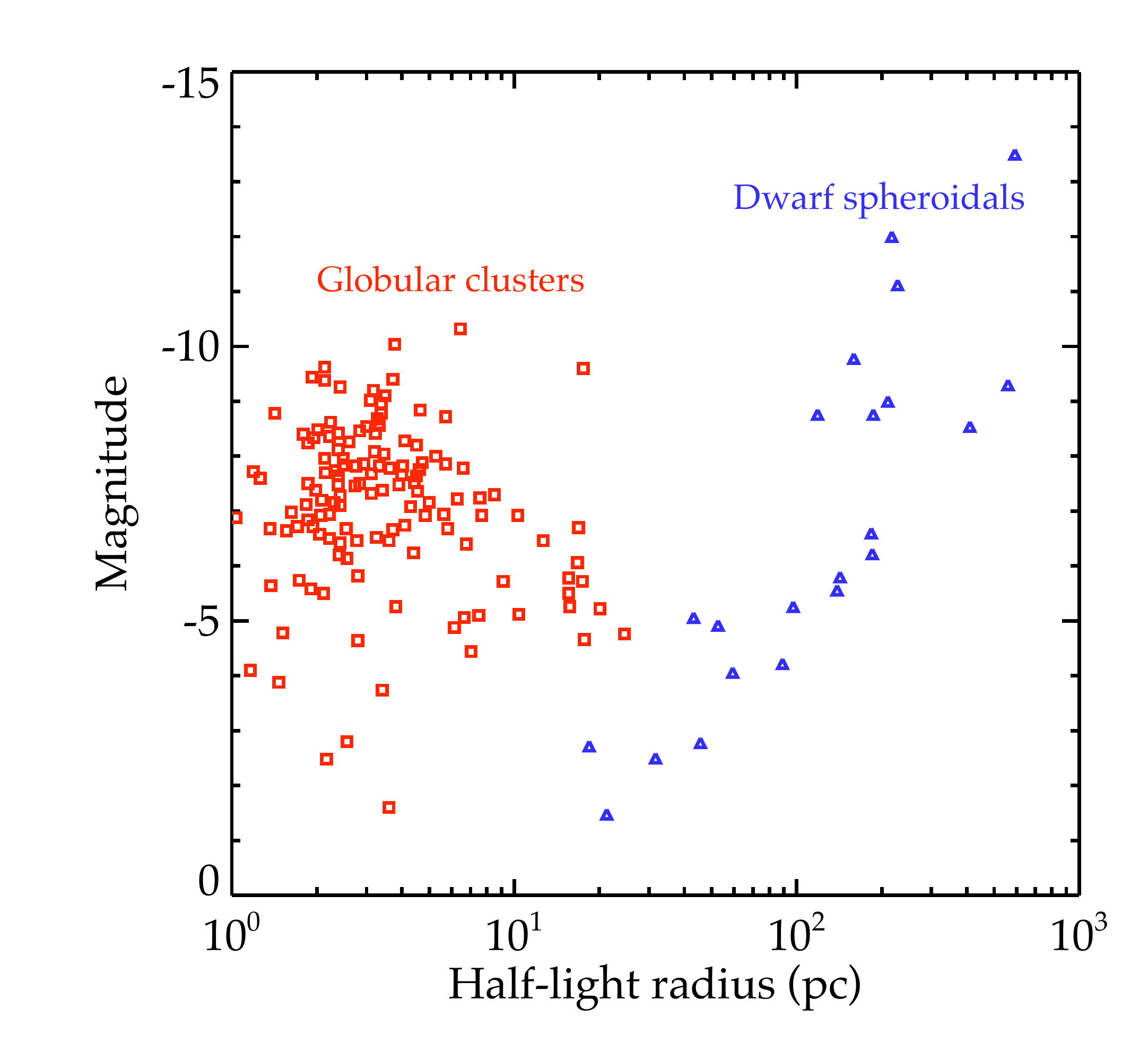} \\
\end{tabular}
\end{center}
\caption{Relationship between the magnitude and the half-light radius for Milky Way globular clusters (red squares, left) and dwarf spheroidals (blue triangles, right). The ultra-faint satellites appear here as the blue triangles in the lower middle portion of this figure. Globular cluster data from~\cite{Harris:1996}. 
}
\label{fig:mv_rh}
\end{figure*}

\begin{figure*}
\begin{center}
\begin{tabular}{c}
\includegraphics[width=0.75\textwidth]{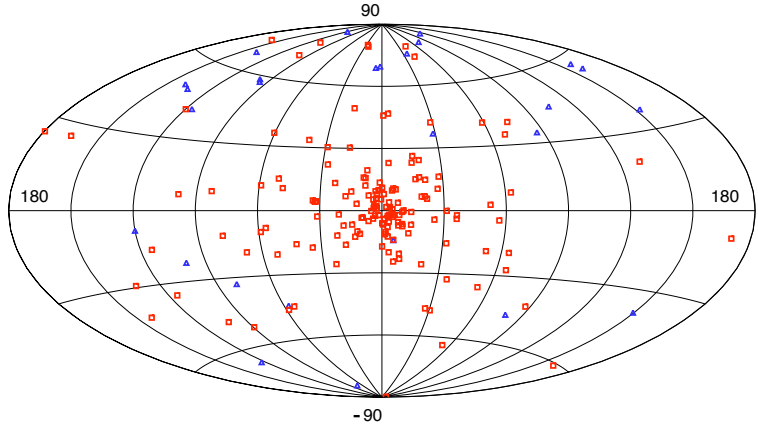} \\
\end{tabular}
\end{center}
\caption{Aitoff projection in Galactic coordinates of the globular clusters (squares, red) and dwarf spheroidals (triangles, blue). 
}
\label{fig:dsph_gc}
\end{figure*}

 From deeper photometric study of these objects, it was revealed that the satellites discovered in the SDSS are unique in many ways from the classical satellites. For starters, they all have total luminosities less than the faintest known classical satellite, so it indeed turns out that, at least in the part of the sky surveyed in detail by SDSS and for objects with surface brightnesses characteristic of the classical satellites, the census of classical satellites is complete. Examination of their metallicities indicates internal spreads that are more characteristic of galaxies that have had multiple episodes of star formation, as opposed to globular clusters, which largely formed its stellar population in a single burst. Further, several of these have half-light radii in the range $\sim 30-100$ pc, i.e. in between the respective half-light radii of typical globular clusters. This exceedingly faint population with intermediate half-light radii can only be detected out to distances of $\sim 50$ kpc, a small volume of the Galactic halo. The population of satellites discovered in the SDSS are typically referred to as ``ultra-faint" satellites. Figure~\ref{fig:mv_rh} shows the magnitude of the known population of globular clusters and dSphs as a function of their half-light radii. In this figure the ultra-faint satellites appear towards the bottom middle and right of the figure, while the classical dSphs appear largely in the upper right. Figure~\ref{fig:dsph_gc} shows a sky distribution in Galactic coordinates of the known Milky Way satellites; the satellites discovered in the SDSS are clearly clustered in the northern Galactic cap. 

\par The most relevant question about both the classical and ultra-faint satellites for this article is, do they contain dark matter, and if so, how much? About forty years passed from the discovery of the first satellites by Shapley before spectrographs were sensitive enough to measure velocities for constituent stars and study their kinematic properties. The first measurements of the velocities of stars in Draco and Ursa Minor were taken by~\cite{Aaronson:1983}. Even from his measurements of the velocities of only a few stars the~\cite{Aaronson:1983} observations hinted that these objects have total mass-to-light ratios of a hundred or greater, which if true would make them the most dark matter dominated objects known. Improved statistics on the sample of stellar velocities in the dSphs was clearly needed. In the following decade, the sample of velocities was improved upon by ~\cite{Mateo:1993};  these observations also indicated that the dark matter mass-to-light ratios of the dSphs was of order a hundred or greater, and there were additional intriguing hints that these objects had very similar masses. 

\par Concurrent with the early determinations of the dSph masses, it was recognized that they strongly constrain the nature of particle dark matter~\citep{Gerhard:1992}. Because of the extreme total mass-to-light ratios, it is unlikely that the luminous material has altered the distribution of dark matter in these systems, making them important testbeds for dark matter models~\citep{Hogan:2000bv}. In the following years, significant theoretical attention has been paid to the dSphs as perhaps the best astrophysical probe into the nature of particular dark matter. 

\par Modern kinematic data sets, which utilize multi-slit spectroscopy, now measure the velocities of hundreds to thousands of stars in both the classical and ultra-faint satellites; theoretical interpretation of these modern results is a main focus of this section. The theoretical framework in this section provides the basis for the analysis of satellites with indirect dark matter detection experiments, as discussed in Section~\ref{sec:indirect_detection_experiment}. 

\subsection{Theoretical modeling} 

\par The dark matter mass distributions of the dSphs are determined via methods similar to those discussed in Section~\ref{sec:MW}. In this case, the observations of bright stars in these systems provide three out of six phase space coordinates; these three coordinates are the two-dimensional projected position of the stars and the line-of-sight velocity from the stellar spectra. Similar to the analysis in Section~\ref{sec:MW}, reconstruction of the full stellar distribution function is difficult because of the lack of complete phase space information. It is therefore standard to use moments of the distribution function to analyze these observations. 

\bigskip

$\bullet${ \underline{Moment-based methods} }

\bigskip

\par Assuming that the stellar distribution and the dark matter distribution are spherically-symmetric, the analysis begins with the jeans equation, Equation~\ref{eq:jeans_spherical}. In this case, $\rho_s$ represents the three-dimensional density profile of the stellar distribution. The mass profile, $M(r)$, is the sum of the contribution from the dark matter and the stars. The three-dimensional density profile of the stellar distribution is related to the observed projected distribution, $I_s(R)$, via 
\be 
\rho_s(r) = -\frac{1}{\pi} \int_r^\infty \frac{d I_s(R)}{dR}  \frac{dR}{\sqrt{R^2 - r^2}}. 
\label{eq:abel_projection} 
\ee

\par The line-of-sight velocity of a star in a dSph is a mixture of the radial and tangential velocity components, and therefore depends on the intrinsic velocity anisotropy of the stars (Equation~\ref{eq:beta}). At a projected position $R$, the line-of-sight velocity dispersion is 
\be 
\sigma_{los}^2(R) = \frac{2}{I_s(R)} \int_R^\infty \left[1-\beta(r)\frac{R^2}{r^2}\right] \frac{\rho_s \sigma_r^2 r}{\sqrt{r^2-R^2}} dr. 
\label{eq:jeans_los} 
\ee
The three-dimensional radial velocity dispersion $\sigma_r$ is determined from Equation~\ref{eq:jeans_spherical}. From the examination of Equation~\ref{eq:jeans_los} it is manifest that there is a degeneracy between the mass profile and the velocity anisotropy, $\beta(r)$, if the mass profile is determined only from line-of-sight velocities. 

\par Because of the assumption of spherical symmetry, there is no equivalent separability condition that led to Equation~\ref{eq:fz}, so the stellar distribution function in dSphs does not simplify in the same manner as the distribution function of local disk stars. Perhaps the simplest theoretical approximation that can be made for the intrinsic line-of-sight velocity distribution is that it is gaussian with a variance given by $\sigma_{los}$. Define the true line-of-sight velocity of a star as $v_{true}$, the measured velocity as $v_{los}$, the mean line-of-sight velocity as $\bar v$, and the measurement uncertainty on a star to be $\sigma_m$. Take the measurement uncertainties to be gaussian-distributed. In this model, the line-of-sight velocity distribution for a star at a projected position $R$ is 
\begin{eqnarray}
f(v_{los}, R) &=& \int \frac{1}{\sqrt{ 2\pi \sigma_m^2 } } \exp\left[-\frac{(v_{los}-v_{true})^2}{2\sigma_m^2}\right]
\frac{1}{\sqrt{ 2\pi \sigma_{los}^2 } } \exp \left[-\frac{(v_{true}-\bar v)^2}{2\sigma_{los}^2}\right] d v_{true}  \nonumber  \\
&=& \frac{1}{\sqrt{ 2\pi (\sigma_m^2 + \sigma_{los}^2)} } \exp \left[-\frac{(v_{los}-\bar v)^2}{2(\sigma_m^2 + \sigma_{los}^2)}\right].
\label{eq:f_convolve}
\end{eqnarray}
In Equation~\ref{eq:f_convolve}, $\sigma_{los}^2$ reflects the intrinsic variance, as derived from the distribution function. Here no rotational component is assumed; the mean velocity $\bar v$ simply reflects the systemic motion of the dSph. The assumption of a small organized rotational component is valid because there are strong upper limits on the rotational velocity of dSphs, $\lesssim 1$ km/s, which implies that they are dominated by disorganized motion at all radii. 

\par Similar to the calculation presented in Section~\ref{sec:MW}, the best-fitting parameters for the mass distribution are determined by maximizing the likelihood function, 
\be
{\cal L}(\vec \theta) = \prod_{\imath=1 }^n f(v_{los,\imath}, R_\imath)
\label{eq:dsph_likelihood} 
\ee
where $n$ is the number of stars with line-of-sight velocities, and $\vec \theta$ represents the set of model parameters that describe the mass distribution. 

\par The approach outlined above is actually a hybrid method that utilizes both the moment-based modeling through the spherical jeans equation combined with an assumption for the gaussian nature line-of-sight velocity distribution. The observed velocity distribution function is a convolution of the intrinsic distribution with the distribution of measurement uncertainties. If the latter are gaussian and are of similar order to the intrinsic velocity dispersion then even if the true form of the distribution function differs from the gaussian model, the method outlined above is still likely to provide an accurate model for the distribution function, because it is dominated by the measurement uncertainties. This is the appropriate regime of parameter space for the ultra-faint satellites, as discussed below. However, if the measurement errors are small relative to the intrinsic velocity dispersion, as is the case with the classical dSphs, the line-of-sight velocity distribution may deviate from the gaussian model in Equation~\ref{eq:f_convolve}, so the approximation made in constructing Equation~\ref{eq:f_convolve} must be more rigorously tested. 

\par A related point is that for an assumed parameterization of the velocity anisotropy profile, the potential, and the stellar density distribution, there is no guarantee that a unique physical stellar distribution function exists. If a physical distribution function does not exist, then there may be a bias that is incurred when determining the mass, and other physical properties of the dark matter distribution. This provides an additional important motivation for testing the method outlined above with self-consisent distribution function models.  

\par It is possible to relax the assumption of spherical symmetry in the context of moment-based models. Indeed, in many ways non-spherical models are more appropriate, because despite their name, the dSphs are not spherical; typical ratios of the minor-to-major axes from the photometry are approximately $0.3$~\citep{Irwin:1995tb}. However non-spherical models do introduce a larger deal of complexity, and it is easy to lose sight of physical motivation for the parameters that describe them. For this reason, at this stage it is best to provide a basic outline for the ingredients of a non-spherical model for comparison to the spherical case. 

\par Take $(R,\phi,z)$ to be the standard cylindrical coordinates, and assume that the distribution function depends on the energy and the $z$-component of the angular momentum. With these assumptions $\langle v_R^2 \rangle = \langle v_z^2 \rangle$. From the axisymmetric jeans equations, the solutions for the velocity dispersions are 
\bea
\langle v_R^2 \rangle &=& \frac{1}{\rho_s(R,z)} \int_{z}^\infty dz^\prime \rho_s(R,z^\prime) \frac{\partial \Phi}{\partial z^\prime} \\
\langle v_\phi^2 \rangle &=& \langle v_R^2 \rangle 
+ \frac{R}{\rho_s(R,z)} \frac{\partial \rho_s  \langle v_R^2 \rangle}{\partial R} + R \frac{\partial \Phi}{\partial R} 
\eea
In order to compare to the measured dispersion profile, the $(R,\phi,z)$ coordinates must be converted to coordinates along the line-of-sight. The differences between the dark matter halo masses in spherical and certain non-spherical models using this modeling have been discussed in~\cite{Hayashi:2012si}. 

\bigskip 

$\bullet$ {\underline{Distribution function-based models} }

\bigskip 

\par The first modern set of self-consistent models for the dSphs were developed by~\cite{Wilkinson:2001ut}. These authors started with the premise that the stellar density distribution was described by a Plummer model, which is generally in good agreement with the measured stellar density profiles. The surface brightness for a Plummer model is
\be 
I_{pl} = \frac{4}{3} \frac{\rho_{pl,0} r_{pl}}{\left[ 1 + (R/r_{pl})^2 \right]^2}, 
\label{eq:plummer_surface_brightness} 
\ee
and the corresponding density is 
\be 
\rho_{pl}(r) = \frac{\rho_{pl,0}}{\left[ 1 + (R/r_{pl})^2 \right]^{5/2}}.
\label{eq:plummer_density} 
\ee
The Plummer model can then be fit to a dSph to obtain the Plummer radius, $r_{pl}$. With the parameters of the Plummer profile fixed for a given dSph,~\cite{Wilkinson:2001ut} developed a model for the distribution function specified by two parameters: one that controls the shape of the potential, $\alpha$, which is defined via 
\be
\psi(r) \propto \begin{cases} 
\left(1+r^2 \right)^{\alpha/2} 
& \mbox{if} \hspace{0.2cm} \alpha \neq 0 \\
\log \left(1+ r^2 \right) &
\mbox{if} \hspace{0.2cm} \alpha =0,
\end{cases}
\label{eq:wilkinson_potential}
\ee
and a second parameter that controls the anisotropy of the stellar velocity distribution. The parameters that describe the potential were chosen to vary between a pure mass-follows-light model (i.e. no dark matter) and a dark matter-dominated model with a harmonic potential that corresponds to a constant density core. From the solution to Eddington's formula (Equation~\ref{eq:eddington}), a self-consistent model for the velocity distribution function is obtained. 

\par Though these models are interesting because they allow for an analytic calculation of the distribution function, there is room for theoretical improvement. First, if anisotropic models are considered, the radial dependence of the velocity anisotropy is restricted in its functional dependence, so that the velocity anisotropy is an increasing function of radius. From numerical simulations however it is more probable that the velocity distribution of both the stars and dark matter are more circular because of the stripping of radial orbits due to interactions with the Galactic tidal field. Second, the Plummer profile is probably too restrictive of a model for the light profile, because it only allows for a core in the stellar distribution in three dimensions, and precludes the existence of cusped stellar density models. 

\par Though there is clear motivation for expanding the parameter space of distribution function-based models, because many are non-analytic it is difficult to gain a physical intuition for more general models of the potential and the stellar density. For example there is clear motivation to analyze both the photometry and the kinematics of the dSphs independently, in particular in order to allow for cusped central profiles of the stellar density. Though the distribution functions are typically non-analytic, it is straightforward to obtain it from a numerical solution to Eddington's formula (Equation~\ref{eq:eddington}). 

\par For both the moment-based and distribution function based modeling above, a functional parameterization for the velocity anisotropy is assumed, which ultimately leads to a degree of model dependence. Recently methods have been developed that bypass this issue and do not assume a form for the velocity anisotropy, but rather determine it directly in the analysis. This numerical method, originally developed by~\cite{Schwarzschild1979}, which has long been used to determine the masses of central black holes in more massive galaxies, is now being applied to discrete velocity samples of dSphs. 

\par In the context of dSphs, very broadly Schwarzschild modeling begins by assuming a form for the dark matter potential that is axially-symmetric to match the observed stellar distribution. With the assumed potential, orbits are launched and integrated for a large number of crossing times. Each orbit is then assigned a weight, and a set of weights are chosen so that the photometry and kinematics are matched. Because the orbits are directly determined in this method, the velocity anisotropy for the best-fitting model is automatically determined. On the down side, the computational time required to implement these models for the large modern data samples is significant. 

\bigskip 

\par In sum, there are now several theoretical methods that have been developed to analyze the kinematic data from dSphs. These models are very broadly classified into the following general categories: 1) moment-based methods in combination with a likelihood analysis described by Equation~\ref{eq:f_convolve}, and 2) distribution function models with either empirically or theoretically-motivated forms for the stellar density, potential, and velocity anisotropy. Below, these methods are compared to the most basic estimates for the mass obtained from the virial theorem. It is of course true that the dark matter masses as determined from each of these methods must be mutually consistent. 

\subsection{Classical satellites}
\par Classical Milky Way satellites are more extended and luminous than their ultra-faint counterparts discussed below. The half-light radii for these systems are typically of the order few hundreds of parsecs, and their luminosities spread over a range of nearly two orders of magnitude, approximately $10^5 - 10^7$ L$_\odot$. The photometric profiles as derived from star counts are typically consistent with a cored model in projection, followed by a turnover into an exponential fall-off in the outer region of the galaxy where the stellar density blends into the background star counts. For the classical dSphs there are hundreds, and in some cases thousands, of bright giant stars that have measured velocities to a precision of a few km/s or less~\citep{Walker:2008ax}. The distribution of stars with line-of-sight velocity measurements does not necessarily follow the same density distribution as the integrated star counts. 

\par An observed sample of velocities must be cleaned of interloping stars that are in the direction of the dSph, but are not physically associated with it. Several authors have discussed methods for cleaning interloping stars.~\cite{Klimentowski:2006qe} utilize a method, calibrated to simulations, that rejects stars with a velocity greater than a maximum velocity available to a star at a given projected radius. The maximum velocity is determined either by assuming that the star is on a circular orbit or is infalling freely into the galaxy potential.~\cite{Martinez:2010xn} develop a Bayesian method that may be applied to classical satellites that assigned a membership probability based on metallicity, distance, and velocity. 

\par Applying the above selection criteria to the data samples, a maximum likelihood analysis that utilizes Equation~\ref{eq:f_convolve} provides the integrated dark matter mass within a fixed physical radius $r$, for an assumed parameterization for the dark matter density profile~\citep{Strigari:2007at,Martinez:2009jh,Walker:2009zp}. Several analysis methods have shown that the line-of-sight velocity data sets constrain the integrated dark matter mass within the approximate half-light radius, independent of the parameterization of the velocity anisotropy~\citep{Strigari:2007vn,Walker:2009zp,Wolf:2009tu}. The formula for the mass within the half-light radius, $r_h$ is~\citep{Walker:2009zp,Wolf:2009tu}
\be 
M(R_{1/2}) \simeq 4\frac{ r_h \langle \sigma_{los}^2 \rangle}{G},
\label{eq:mhalf}
\ee
where here $\langle \sigma_{los}^2 \rangle$ is the luminosity weighted velocity dispersion. Because each of the dSphs have different half-light radii, the mass distribution is best constrained at different fixed physical radii. Table~\ref{tab:satellites} shows some properties of many important satellites whose masses can be readily derived. 

\begin{deluxetable}{l c c c c c}
\tablecolumns{5}
\tablecaption{Compilation of properties of Milky Way satellites. Classical satellites are above the horizontal line, while ultra-faint satellites are below the line. This table only shows dSphs with well-measured kinematic data, in which cases the mass can be determined using Equation~\ref{eq:mhalf}. In the fifth column the half-light radii are reproduced from~\cite{McConnachie:2012vd}. In the sixth column the luminosity weighted velocity dispersions are reproduced from~\cite{Wolf:2009tu}. 
\label{tab:satellites}}
\tablehead{
Name &  Distance [kpc] & Magnitude & Half-light radius ($r_h$) (pc)& $\langle \sigma_{los}^2 \rangle$ (km/s)}
\tablewidth{0pc}
\startdata
Sculptor  & 80 & -9.8&283&9.0\\
Fornax &138 & -13.1&710&10.7\\
Leo I  & 250  & -11.9&251&9.0\\
Leo II  &205 & -10.1 &176&6.6\\
Draco & 80 & -9.4 &221&10.1\\
Ursa Minor & 66 & -8.9&181&11.5\\
Carina  & 101 & -9.4 &250&6.4\\
Sextans & 86 & -9.5&695&7.1\\
\hline  
Willman 1& 38 & -2.7&25&4.0\\
Ursa Major I   &106 & -5.5 &319&7.6\\
Ursa Major II & 32 & -4.2&149&6.7\\
Hercules  & 138 & -6.6&330&5.1\\
Leo IV & 158 & -5.0&206&3.3\\
Canes Venatici I &  224 & -8.6&564&7.6\\
Canes Venatici II & 151& -4.9&74&4.6\\
Coma Berenices &  44 & -4.1&77&4.6\\
Segue 1 & 2006 & -1.5&29&4.3\\
Bootes I & 2004 & -6.3&242&9.0\\
\enddata
\end{deluxetable}

\par
\begin{figure*}
\begin{center}
\begin{tabular}{cc}
\includegraphics[width=0.35\textwidth]{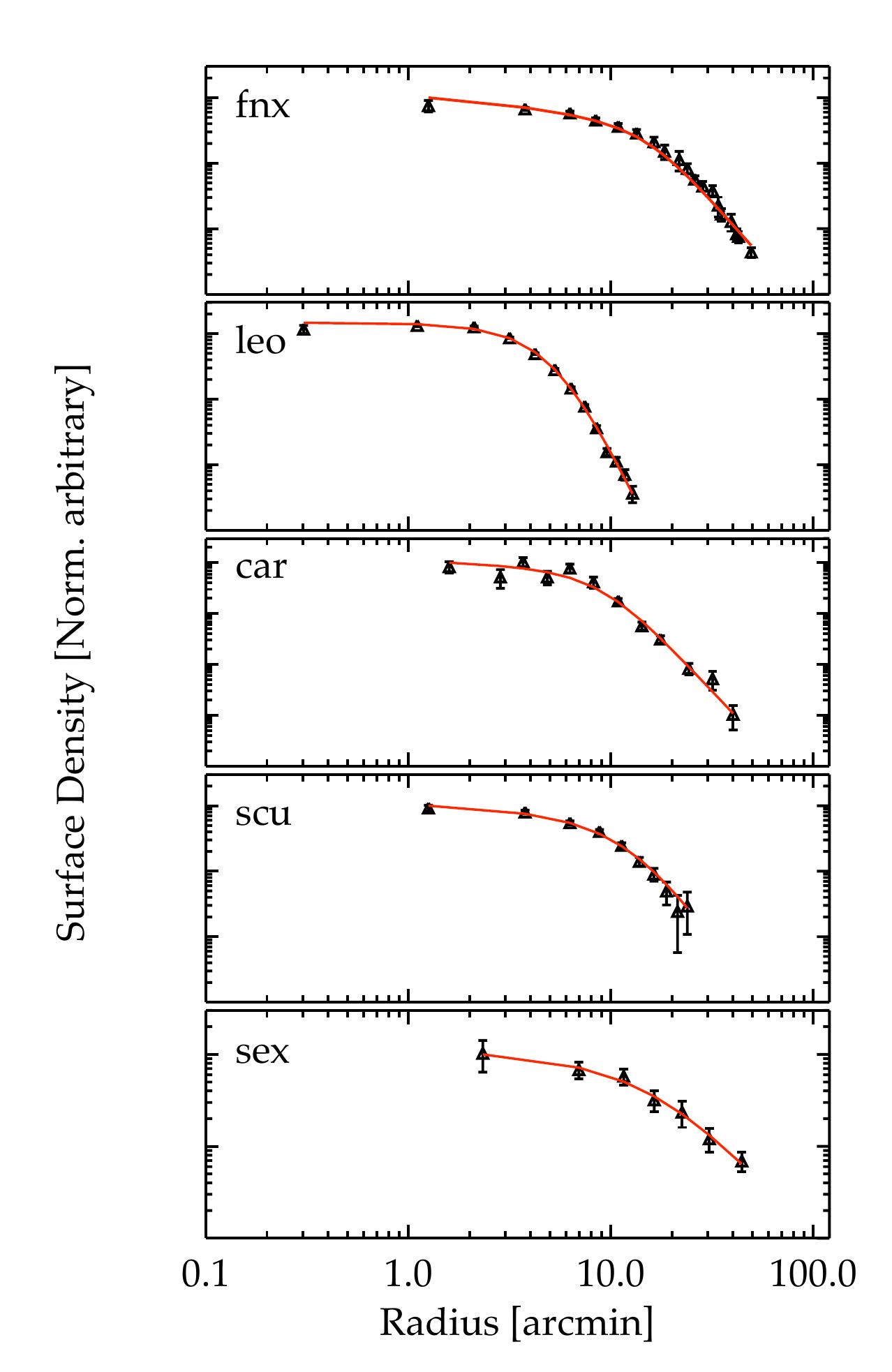} & 
\includegraphics[width=0.35\textwidth]{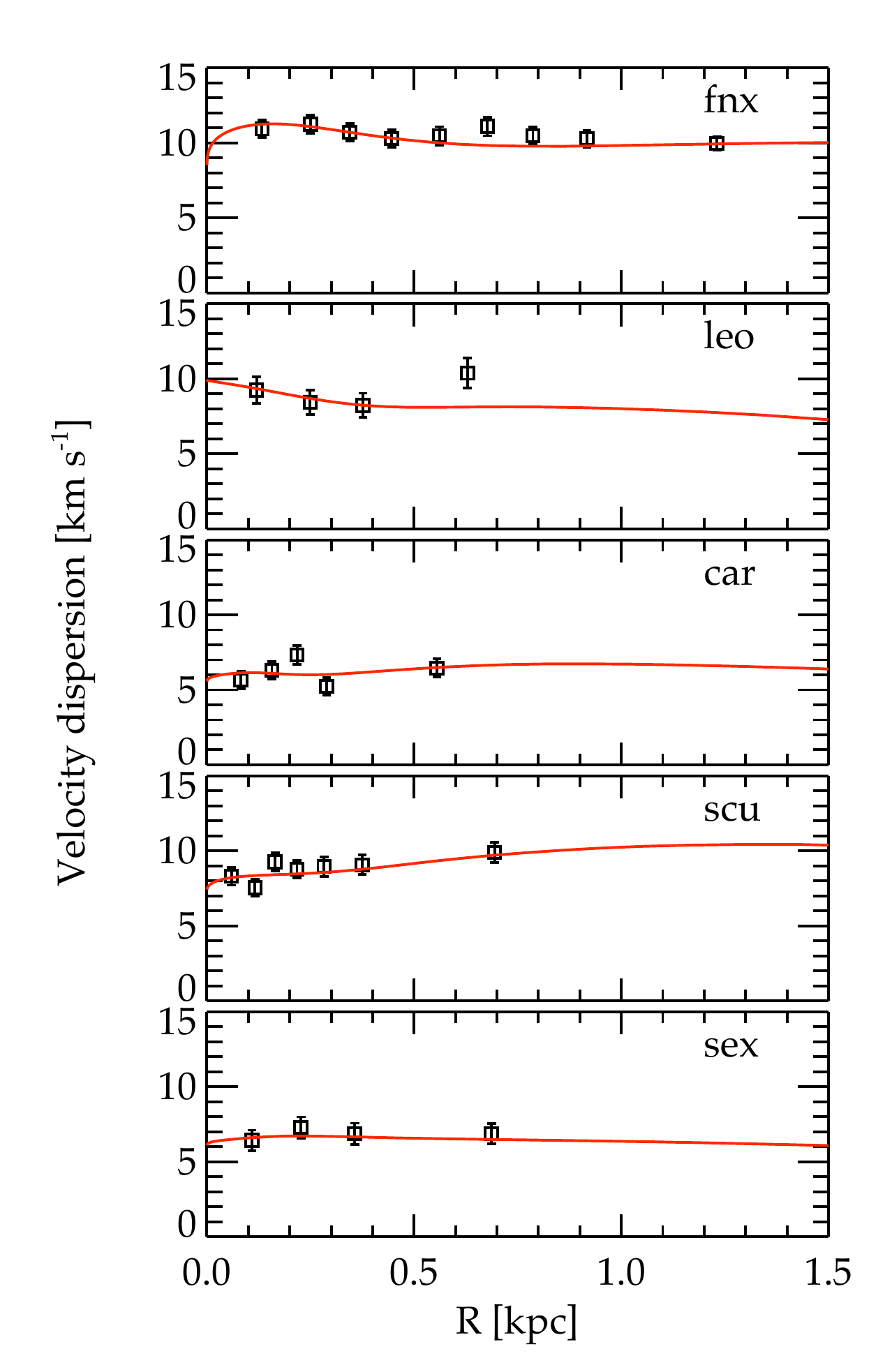} \\
\end{tabular}
\end{center}
\caption{Photometric profiles ({\it left}) and velocity dispersion profiles ({\it right}) for five classical dSphs, using LCDM-based models for the dark matter potentials. From~\cite{Strigari:2010un}. 
}
\label{fig:sigma_classical}
\end{figure*}

 Though the mass at the half-light radius does clearly indicate that the dSphs are dark matter-dominated and provides a measurement of the central dark mater density, it does not give a handle on the total mass of their respective dark matter halos. Unfortunately it is difficult to determine these total halo masses, primarily because it is not certain how far the halos extend beyond the observed stellar distributions. Therefore, any estimate of the halo mass necessarily introduces a significant degree of model dependence.~\cite{Strigari:2010un} examined isotropic velocity models, utilizing the light profiles, kinematic data, and dark matter potentials that are predicted in LCDM models (Section~\ref{sec:simulations}) to show that the total mass distributions of the brightest satellites are consistent with being in the range $10^8 - 10^9$ M$_\odot$. Figure~\ref{fig:sigma_classical} shows the velocity dispersion profiles and the best-fitting photometric and kinematic models that correspond to these dark matter halo masses. These results are broadly consistent with other determinations of the mass distribution that utilize moment based methods~\citep{Lokas:2009cp}.
 
\par Schwarszchild mass estimates have now been published for three dSphs, Fornax, Sculptor, and Draco. Using a cored model for the stellar light profile,~\cite{Jardel:2011yh} find a mass within the half-light radius that is consistent with those deduced from moment and distribution function-based methods.~\cite{Breddels:2012cq} determine that the mass of Sculptor within 1 kpc is $\sim 10^8$ M$_\odot$, which is again in agreement with the above methods.~\cite{Jardel:2012} determine a lower bound to the mass of Draco of a few times $10^8$ M$_\odot$ within a physical radius of about $500$ pc where kinematics of stars are measured. 

\subsection{Ultra-faint satellites} 
\par Measuring the velocity dispersion, and thus the mass, of ultra-faint satellites poses different sets of challenges in comparison to measuring the velocity dispersions of classical satellites. First, by their very nature, the constituent stars are fainter, with typical target stars having a magnitude of $r=20-21$. For a realistic exposure level, the Keck/DEIMOS spectrograph provides a signal-to-noise on a star of this magnitude of approximately $15$~\citep{Simon:2007dq}. Second, the measured uncertainties derived from the stellar spectra are approximately $2-3$ km/s, which, because it is within about a factor of two of the intrinsic velocity dispersions of the systems, complicates the extraction of the intrinsic velocity dispersion that arises from the distribution function (Low velocity dispersions at this level have been measured in globular clusters~\citep{Cote:2002hw}). Third, the measured line-of-sight velocities may contain a component that is due to the motion of the star around a binary companion. Early studies of classical dSphs indicated that this binary contamination was $\sim 1-2$ km/s, so it is a small systematic to classical satellite mass measurements~\citep{Kleyna:2001us}. However, as measurements of low velocity dispersion globular clusters clearly indicate that binaries do significantly contaminate the velocity dispersion of bound stellar systems~\citep{Bradford:2011aq}, detailed understanding of this effect is required in ultra-faint satellites. 

\par Another interesting difference between the ultra-faint and the classical satellites relates to how their total luminosities are derived. Determination of the total luminosity is important because, in addition to the kinematics, it provides a measurement of the total dark matter-to-luminous matter ratio in the system. The stellar density in ultra-faint satellites is low enough that it is not possible to obtain photometry from integrated unresolved starlight~\citep{Martin:2008wj,Munoz:2009hj}. Further, because of the low numbers of stars, removal of interloping stars is particularly important, because a single bright star not belonging to the satellite may significantly inflate the total luminosity (recall that the faintest of the ultra-faint satellites are about as luminous as a single red giant star). With the interloper population removed, photometric density profiles are derived assuming that the measured stars are an unbiased sampling of an underlying density profile. It is then assumed that the stars in the object are comprised of a single population, in this case an old, $\gtrsim 8$ Gyr, and metal-poor, $[Fe/H] < -2$, population. If the sample of stars in the system is complete brighter than a given magnitude, from the stellar population model the total number of stars at fainter magnitude can be determined, and from this the total integrated luminosity of the system. 

\par Similar to the classical dSphs, the ultra-faint satellites are not spherical systems. From the isodensity contours, the ellipticities of the systems, again measured in terms of the ratio of the minor to the major axis, are in the range $\sim 0.3-0.7$~\citep{Martin:2008wj}, and as a mean population are more elongated than the classical dSphs. Given the small samples of stars that belong to these systems, and the uncertainty in the precise properties of the dark matter halos that they inhibit, it is difficult to determine if they are elongated by tidal effects, in particular because their direction of elongation is  random with respect to the Galactic center~\citep{Munoz:2009hj}.

\par Before applying the theoretical models discussed above to the ultra-faint satellites, and answering the important question of their dark-to-luminous mass ratio, it will be useful to obtain an estimate of the velocity dispersion for a system with the total luminosity of an ultra-faint satellite that is bound and in equilibrium. For this example consider the system Segue 1, which is the least-luminous of the ultra-faint satellites. For a simple estimate, we can appeal to the virial theorem, which states that the total mass, $M_{vir}$, of a system with a half-light radius of $r_h$ is related to the velocity dispersion as $\sigma^2 \simeq 2 M r_h/G$ (note that here we have approximated the total extent of the system as the half-light radius). If we assume a mass-to-light ratio for the stars as $M_\star/L = 3$, as is appropriate for a stellar population as observed in the ultra-faint satellites~\citep{Maraston:2004em}, from the total luminosity the velocity dispersion of this system would be approximately $0.5$ km/s. For the brightest of the ultra-faint satellites, the corresponding velocity dispersion is $\lesssim 1$ km/s. A measured velocity dispersion that is significantly larger than this reveals that a dark matter component dominates the dynamics of the system. 

\par The first measurements of stellar line-of-sight velocities in ultra-faint satellites were obtained a few years ago~\citep{Martin:2007ic,Simon:2007dq,Geha:2008zr}. All of these measurements determined that the velocity dispersions are in the range of 4-5 km/s; the velocity dispersion for each satellite, averaged over the entire radius of the satellite, is shown in Table~\ref{tab:satellites}. From the estimates above, these velocity dispersions indicated that the masses of these systems are significantly greater than what is contributed by the luminous population of stars. Interpreting these results at face value, the ultra-faint satellites contain dark matter mass-to-light ratios in the range of hundreds to thousands, making them the most dark matter dominated objects known. Combined with the measurements of the internal spreads in metallicities, these systems are clearly more akin to galaxies than they are to the less extended globular cluster systems that do not contain appreciable dark matter. 

\par
\begin{figure*}
\begin{center}
\begin{tabular}{cc}
\includegraphics[width=0.45\textwidth]{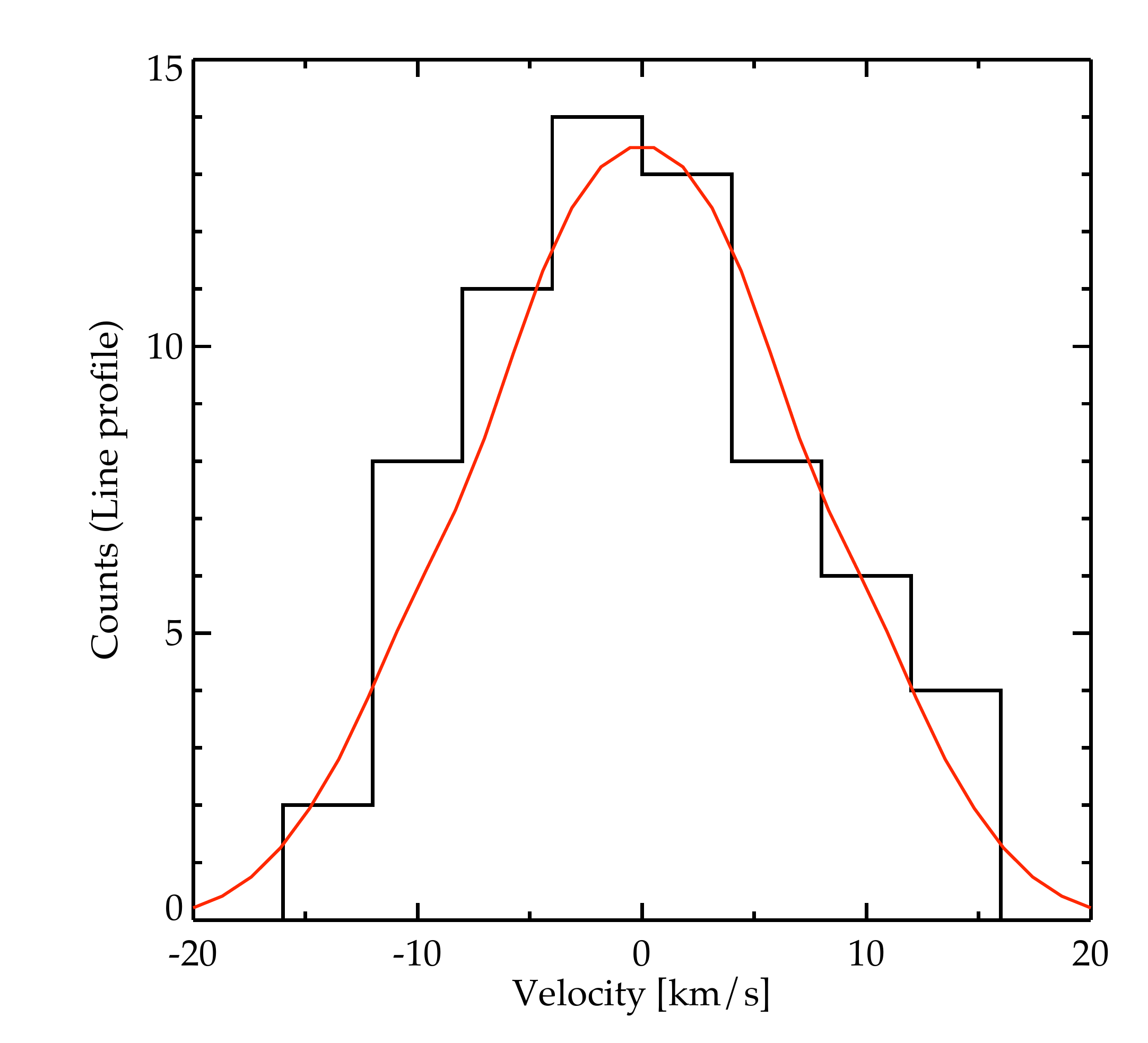} & 
\includegraphics[width=0.45\textwidth]{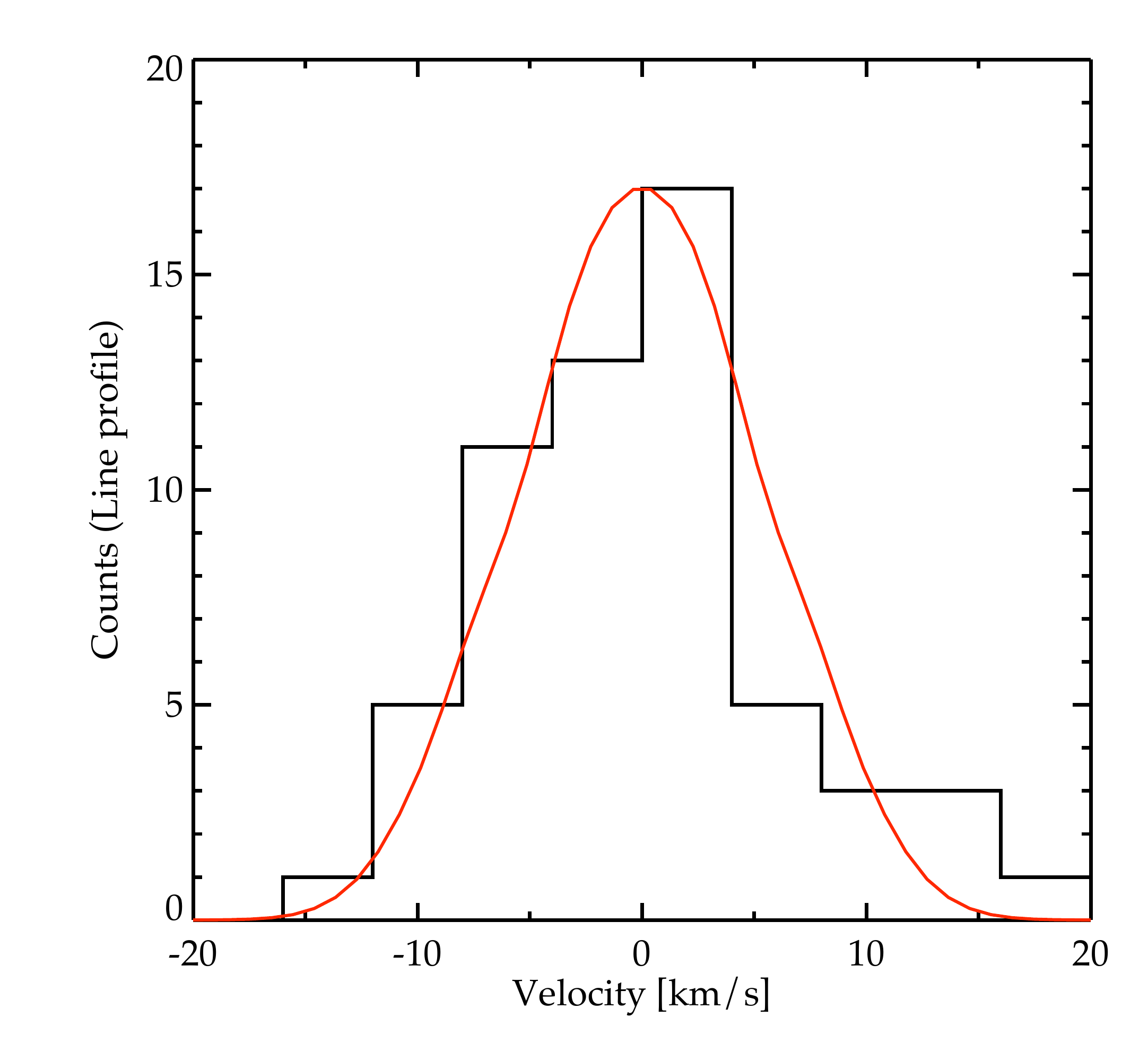} \\
\end{tabular}
\end{center}
\caption{Histogram of line-of-sight velocities of Segue 1 ({\it left}) and Coma ({\it right}), compared to a model with an NFW density profile that provides a fit to the data. 
}
\label{fig:hist_segue1}
\end{figure*}

 Figure~\ref{fig:hist_segue1} shows an example velocity distribution for the ultra-faint satellites Segue 1 and Coma Berenices. Also shown are fits to the velocity distributions assuming cusped NFW models for the dark matter distribution and using Eddington's formula in Equation~\ref{eq:eddington}. Given the relatively small data samples, it is not surprising that good fits can be obtained to the data for an input potential. Segue 1 has a luminosity of a few hundred solar luminosities and is at a distance of 23 kpc, and Coma Berenices has a luminosity of a few thousand solar luminosities and is at a distance of approximately 44 kpc.~\cite{Geha:2008zr} originally derived in detail the mass distribution of Segue 1 from a sample of 24 line-of-sight velocities, concluding that it was the most dark matter dominated system known. Even though the sample of stars was small, amongst ultra-faint satellites Segue 1 is a unique system given that it has a systemic velocity of $\sim 205$ km/s, which nicely separates it from the velocity distribution of Milky Way foreground stars. Folllow up studies of Segue 1 with three times the sample of stars has shown that this result holds~\citep{Simon:2010ek}, and that it is robust to systematic effects such as binary stars~\citep{Martinez:2010xn}. 

\par Though as a population there is now strong evidence that the ultra-faint satellites are dark matter-dominated, there are outlier systems with unique observational properties, so it is still necessary to examine each system on an individual basis. An example of such an intriguing system is Willman 1, which presents a unique set of challenges from a few perspectives. Though Willman 1 was the first ultra-faint satellite discovered, and was hinted to be a dark matter dominated object from early studies, the newest results present a less clear picture~\citep{Willman:2010gy}. With a systemic velocity of -11 km/s, there is significant contamination from Milky Way foreground stars in the direction of Willman 1, so determining a clean membership sample is difficult. While this object does appear to have properties more in common with that of a galaxy than a globular cluster, the velocity distribution shows a double peaked structure that is difficult to interpret.

\par Given the relatively small kinematic samples from the ultra-faint satellites, it is not possible to determine whether they have cored or cusped central dark matter distributions. However, similar to the classical satellites, the uncertainty on the mass profiles is minimized at about the half-light radius. The best constrained integrated half-light radius mass for an ultra-faint satellite is determined to approximately $20\%$ (Canes Venatici I)~\citep{Wolf:2009tu}. This can be compared to the best constrained classical dSph, Fornax, which has a half-light radius mass determined to approximately $5\%$. 

\bigskip

\par
\begin{figure*}
\begin{center}
\begin{tabular}{c}
\includegraphics[width=0.60\textwidth]{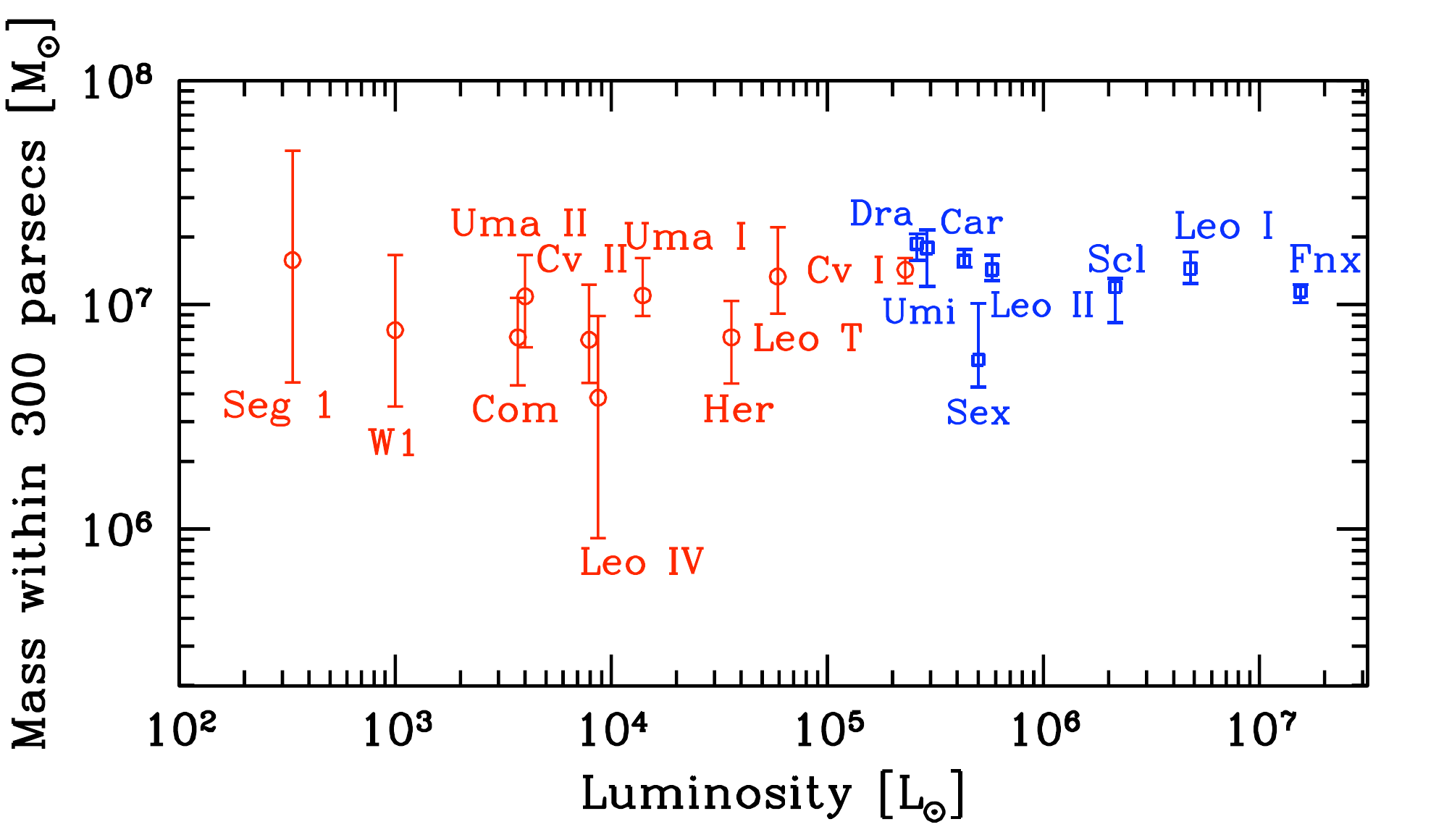} \\
\end{tabular}
\end{center}
\caption{Mass within $300$ parsecs versus luminosity for Milky Way satellite galaxies. From~\cite{Strigari:2008ib}. 
}
\label{fig:m300}
\end{figure*}

 The above discussion details how the dark matter distributions in the classical and ultra-faint satellites are determined; it is appropriate to put these determinations into a broader context. As outlined above, it is not straightforward to compare the classical and ultra-faint satellite populations because of their wide range of observed properties, in particular their approximate order of magnitude range in half-light radii. Since the total dark matter halo masses are unknown from the kinematic data samples, a better observational proxy is the mass within a fixed physical radius~\citep{Strigari:2007ma,Strigari:2008ib}; a convenient radius may be chosen as the approximate average half-light radius for the satellite population. Phrased in this context,~\cite{Strigari:2008ib} highlighted how there exists a common mass scale for all of the Milky Way satellites of $10^6$ M$_\odot$ within 100 pc, and $10^7$ M$_\odot$ within 300 pc (Figure~\ref{fig:m300}). The intriguing aspect of this result is that it highlights how the Milky Way satellites share a mean integrated central dark matter density, which is independent of the luminosity of the satellite. This is particular striking considering that the luminosities of the Milky Way satellites span a range of nearly five orders of magnitude. 

\par Though there have been subsequent revisions to the velocity dispersions of the ultra-faint satellites~\citep{Aden:2009vx}, the salient point still remains from the~\cite{Strigari:2008ib} analysis. It is certainly plausible that this scale could be set by dark matter physics that differs from the CDM framework~\citep{Lee:2008jp}, though it can also be viewed as a prediction of CDM-based models of galaxy formation~\citep{Maccio':2008qt,Li:2008fx,Stringer:2009ta,Busha:2009qn}. In addition, in the context of CDM, it has important implications for indirect dark matter detection, as will be discussed in more detail in Sections~\ref{sec:indirect_detection_theory} and~\ref{sec:indirect_detection_experiment}. In particular, if the dark matter annihilates to provide an appreciable gamma-ray flux, the luminosity of a galaxy in gamma-rays is independent of its optical luminosity. 

\subsection{Core versus cusp issue} 

\par The discussion above has focused on determining integrated masses of dSphs. Given the significant improvement in the kinematic data sets, it is also prudent to now ask if the data are able to distinguish between cored or cusped central dark matter density profiles. This question is of direct relevance for Galactic dark matter searches, because different dark matter particle models make different predictions for the central density of dark matter halos. As an illustration, it is possible to define a phase space density for the dark matter as $Q = \rho / \sigma^3$, where here $\sigma$ is the three-dimensional velocity dispersion and $\rho$ is the density~\citep{Hogan:2000bv}. This quantity can only decrease from its primordial value as a result of collisional shocks or coarse graining. For a neutrino-like thermal relic particle, the phase density is given in terms of the particle mass, $m$, as
\be 
Q = 5 \times 10^{-4} \left( \frac{m}{1 \textrm{keV} } \right)^2 \frac{ M_\odot / \textrm{pc}^{-3} }{ (\textrm{km} \, \textrm{s}^{-1})^3 }. 
\label{eq:Q} 
\ee
The key assumption in the derivation of this Equation is that the particle freezes out while it is still relativistic. It is also possible to obtain phase space densities of this order with non-thermal particles produced from decays~\citep{Kaplinghat:2005sy,Strigari:2006jf}. Cold dark matter particles have $Q$ values that are orders of magnitude larger, because their velocity dispersion is negligible at freeze-out. 

\par In order to probe the phase-space properties of the dark matter, the ideal systems to study are those that are the most dark matter dominated, so that central dark matter density is set by phase space properties, and not by interactions with baryons. The most dark matter-dominated systems known, whose core sizes are measurable given the scale of modern observations, are the dSphs and the low surface brightness galaxies~\citep{Simon:2004sr,Oh:2010ea}.  It is certainly true that a measured deviation from a cuspy profile in these galaxies would be the most significant clue from astrophysics that dark matter is different than is assumed in the LCDM model. 

\par This motivates a more detailed examination of the central density profiles of dSphs. Before examining them from the perspective of the line-of-sight velocity data, it will first be worthwhile to review related observations that test for the existence of a dark matter core or cusp. Indeed, several authors have suggested that observations not directly related to the analysis of the line-of-sight velocity data indicate that the dSphs may favor cored dark matter density profiles. 

\par As a first example, in Fornax dSph there are five surviving globular clusters, making it, along with Sagittarius, the only dSphs with surviving globular cluster systems.~\cite{Goerdt:2006} determined that, based on dynamical friction arguments, the globular clusters would not have survived to the present day if the dark matter density profile rises as $1/r$ towards the center of the galaxy. These authors surmise that the only explanation for their existence is a dark matter core in Fornax, of the scale $\lesssim 1$ kpc. Though this result is intriguing, it is based on the projected positions of the globular clusters; three-dimensional measurements of the globular cluster distances from the center of Fornax are required for more precise calculations. 

\par There are also arguments in favor of a core in the Ursa Minor dSph based upon observation of kinematically-cold stellar substructure~\citep{Kleyna:2003zt,Gilmore:2007}. A stellar substructure that is this old, $\gtrsim 10$ Gyr, is not expected to have survived in a galaxy of this size if the dark matter distribution is significantly cuspy, though may survive in a system with a dark matter core $\lesssim 0.85$ kpc.~\cite{Pace:2012ru} corroborate the presence of a kinematic substructure in Ursa Minor, though these latter authors find a different magnitude for the velocity dispersion of the substructure, and speculate that this substructure may have its own dark matter halo associated with it. 

\par In order to test these observations, it is best to turn to the line-of-sight velocity dispersion data. The likelihood analysis discussed above provides a means for determining whether a cored or a cusped dark matter distribution provides a better description of the data. However, significant degeneracies preclude robust conclusions. The most significant degeneracies are between the central slope of the dark matter density, the central slope of the stellar density, and the velocity anisotropy. Before addressing the degeneracies between these quantities in a more rigorous numerical manner, it is first worthwhile to set up an analytic description, for which we again appeal to the spherical jeans equation.  

\par For isotropic models, the spherical jeans equation reduces to a form of hydrostatic equilibrium, 
\be 
\frac{d(\rho_s \sigma_r^2)}{dr} = - \rho_s \frac{d\Phi}{dr}, 
\label{eq:hydrostat}
\ee
where the quantity $\rho_s \sigma_r^2$ plays the role of the pressure. As a further simplification take the velocity dispersion to be constant, which is a good approximation to the dSph data, and $\sigma_r = \sigma_{los}$. Taking the potential to be dominated by the dark matter, combining Equation~\ref{eq:hydrostat} with poisson's equation then gives 
\be
\gamma_s + r \frac{d\gamma_s}{dr} = \frac{4\pi G}{\sigma_{los}^2} \rho r^2,
\label{eq:gamma_s}
\ee
where $\rho$ represents the density of the dark matter, and $\gamma_s = - d(\ln \rho_s)/d \ln r$ is the log-slope of the stellar density profile. In the limit that $r \rightarrow 0$, assuming that $\gamma_s = 0$ Equation~\ref{eq:gamma_s} implies that the central dark matter density must also be cored. If $\gamma_s > 0$, then the central density of the dark matter must be cusped as $\rho \propto r^{-a}$, with $a < 2$. In other words, assuming isotropy, if the stellar density is cusped then the dark matter density is cusped, while if the stellar density is cored then the dark matter density is cored~\citep{Evans:2008ik}. 

\par With this background we can now go into more detailed modeling with the spherical jeans equation. In order to implement the jeans equation we need an assumption for the dark matter density profile; take this to be of the form Equation~\ref{eq:zhao}. The log-slope of this profile as a function of radius is 
\be
\frac{d \log \rho}{d \log r} = a - (a-c)\left(\frac{r}{r_0}\right)^b \left[ 1 + \left(\frac{r}{r_0}\right)^b \right]. 
\label{eq:zhao_logslope} 
\ee
Note that this is a five parameter model for the density profile. Also including a model for the velocity anisotropy, $\beta$, implies at least six parameters must be input when fitting to the line-of-sight velocity data. Despite the quality of the data sets described above, theoretical analysis shows that the degeneracies amongst parameters are so significant that it is not possible to distinguish between central cores and cusps. This can be stated on more statistical grounds for hypothetical data sets that are even larger than the current sets~\citep{Strigari:2007vn}. The addition of stellar proper motions hold promise for breaking these degeneracies~\citep{Wilkinson:2001ut,Strigari:2007vn}, though the required accuracy of approximately a few km/s per star will likely be challenging for modern observatories. 

\par The core versus cusp issue can also be studied via distribution function and Schwarzschild based methods outlined above. As an example,~\cite{Strigari:2010un} study five classical dSphs (Fornax, Sculptor, Carina, Sextans, and Leo I) with high-quality kinematic and photometry data. They assume an isotropic stellar velocity dispersion, allow for cusps in the stellar density $\rho_s$, and utilize cusped central dark matter density models that are predicted in LCDM simulations. The line-of-sight velocity distribution is determined from the isotropic Eddington formula. Though the models presented are in general non-analytic, numerical solutions find that the cuspy LCDM-based profiles may fit the full line-of-sight velocity distributions, even under the assumptions of isotropic stellar orbits. This result generalizes the simple analytic formula presented in Equation~\ref{eq:gamma_s} for line-of-sight velocity profiles that are not strictly constant. 

\par More recently, Schwarzschild modeling has been applied to Fornax~\citep{Jardel:2011yh}, Draco~\citep{Jardel:2012}, and Sculptor. For Sculptor,~\cite{Breddels:2012cq} use the parameterization of the dark matter profile presented in Equation~\ref{eq:zhao_logslope}, and confirm the results from moment-based methods that the kinematic data set from Sculptor is unable to constrain the central slope $a$. Interestingly these authors do find that the velocity anisotropy parameter $\beta$ transitions from positive, implying radially-biased orbits, to negative, implying circularly-biased orbits, outside of approximately the half-light radius. For Draco,~\cite{Jardel:2011yh} find that the data are consistent with a cusped central density profile with $a = 1.0 \pm 0.2$ for $r \ge 20$ pc. 

\par The results above seemingly indicate that even with high quality modern data sets, the line-of-sight velocity data is unable to distinguish between cored and cusped dark matter central density profiles, possibly a fundamental limitation of the information that is extractable from the data. However, more recently a new approach has come to light that looks at the data from a different perspective. It does not revolve around larger data sets, but rather around a reinterpretation of the data sets that already exist. The idea is straightforward, and goes back to the discussion of the previous section on multiple stellar populations. Because stellar populations can be separated based on properties such as their metallicity, velocity, and spatial distribution, they may be viewed as distinct kinematic data sets. This is of practical interest for the dSph data because a given data set can be reinterpreted as multiple data sets that probe the same underlying potential~\citep{Walker:2011zu}. Depending on the nature of the data sets, degeneracies may be broken between the stellar density, velocity anisotropy, and the dark matter density, possibly pinning down more precisely the central dark matter distribution. 

\par The Sculptor dSph has received the most attention, because it is observed to have two distinct populations of stars: a metal rich component with $[Fe/H] > -1.5$, and a metal poor component with $[Fe/H] < -1.7$. The half-light radius for metal rich component is $\sim 230$ kpc, and the half-light radius for the metal-poor population is 350 kpc~\citep{Battaglia:2008jz}. From jeans-based modeling with the assumption of cored Plummer profiles in each case,~\cite{Walker:2011zu} indicate that a cored model is favored for Sculptor, ruling out a cusped NFW model at $\gtrsim 96\%$ c.l. Because of the implications of this result, it is imperative to scrutinize the assumptions in more detail, in particular the assumptions of the slope of the central stellar density profile, which was shown above to be important in determining the central density slope~\citep{Strigari:2010un}. 

\par A simplified analytical model has recently been presented to expand upon the results obtained in~\cite{Walker:2011zu}, that derives from the projected virial theorem~\citep{Agnello:2012uc}. In general, it is difficult to implement the projected virial theorem to dSph data sets, because it involves an integral over the line-of-sight velocity dispersion, weighted by the spatial distribution of the stars. Performing this integration requires a fit to the line-of-sight velocity dispersion, so it is a weighted average of the second moment of the distribution. It is clear to see how this method loses information relative to using the information that is directly contained in the line-of-sight velocity distribution. Because of these issues there are still significant degeneracies that preclude the measurement of the mass distribution from a single population, however, for the case of multiple populations there is a twist on the argument. 

\par For a spherically-symmetric system, the projected virial theorem provides a simple relation between the line-of-sight kinetic energy tensor, $K_{los}$, and the potential projected along the line-of-sight, $W_{los}$, $2 K_{los} + W_{los} = 0$. In terms of the observable properties of a galaxy, these quantities are given by,
\be 
K_{los} = \pi \int_0^\infty I_s(R) \sigma_{los}^2 R dR 
\label{eq:klos}
\ee
\be 
W_{los} = - \frac{4\pi G}{3} \int_0^\infty r \rho_s(r) M(r) dr. 
\label{eq:wlos}
\ee
The key idea brought to light in~\cite{Agnello:2012uc} is that if there are two populations of stars in a dSph, each with a distinct surface brightness profile and velocity dispersion then the ratio of the virial quantities may be taken. For two populations, labeled 1 and 2, the virial theorem then implies that $K_{los,1}/K_{los,2} = W_{los,1}/W_{los,2}$. So from the surface brightness profile and the velocity dispersion for each population, a mass distribution that generates the underlying potential can be assumed, and it is then possible to determine if this mass distribution satisfies the ratio of the virial quantities. Though it is a more simplified theoretical argument than the jeans-based analysis presented above, perhaps the most intriguing aspect of this method is that it depends only on the projected light profile, and is not subject to the degeneracies incurred from deprojection of the light profile. 

\par Though theoretically-interesting, the projected virial theorem is probably still not the final word on the issue of multiple stellar populations in dSphs. Systematic issues that pertain to the use of the projected virial theorem include testing the assumption of spherical symmetry, testing the robustness to the variation in the projected light profile, and understanding if significant information is lost when performing the integrals in Equations~\ref{eq:klos} and~\ref{eq:wlos}. 

\par Though there are some intriguing hints of dark matter cores in dSphs, a more definitive answer will likely have to wait until proper motions for individual stars are obtained, as well as higher quality photometric data sets in the central regions. Both of these observations will help in breaking the degeneracy between the velocity anisotropy and the central dark matter slope. 

\subsection{Streams and disrupted satellites} 
\par To this point this section has focused on understanding bound, self-gravitating populations of stars and dark matter around the Milky Way and the extraction of their dark matter distributions. Though we now have measurements of over two dozen dark matter-dominated satellites, and approximately six times more globular clusters, it is very likely that many of these objects are transients on cosmological timescales. Interactions with the Galactic potential may induce many of them to dissipate their stars and dark matter into tidal streams and debris that will eventually phase mix into the Galaxy. Understanding the dynamics of these debris streams is not only important as a point of comparison to the predicted distribution of disrupted satellites in LCDM simulations (Section~\ref{sec:simulations}), it is of direct relevance for direct dark matter searches, for which event rates are significantly affected by dark matter streams passing through the Solar neighborhood (Section~\ref{sec:direct_detection_theory}). 

\par There is now significant evidence that tidal disruption is a common phenomenon that is an integral part of the formation of the Galaxy. Several streams have now been identified as distinct features in phase space. The most spectacular example is the Sagittarius dSph; this object was previously utilized in Section~\ref{sec:MW} as a probe of the shape of the Milky Way dark matter halo. The major axis of stars associated with Sagittarius extends well over $10^\circ$, and a nearly constant velocity dispersion of $\sim 10$ km/s is now measured in the core region of $\sim 2$ kpc~\citep{Frinchaboy:2012ur}. The core of Sagittarius has a luminosity that is similar to Fornax, and also like Fornax, Sagittarius has several globular clusters associated with it. Modeling indicates that the dark matter mass of Sagittarius before it fell into the Milky Way halo was in the range $10^8 - 10^9$ M$_\odot$~\citep{Law:2004ep}. 

\par Another intriguing stream in the Milky Way is the Magellanic stream, which is a neutral hydrogen structure that appears to trail behind the Magellanic Clouds~\citep{Mathewson1974}. Previous modeling has associated this stream with the Magellanic clouds, relying on the hypothesis that the Magellanic clouds have undergone multiple close passages with the Galaxy, and the stream is a result of either ram pressure stripping or tidal stripping. However, recent proper motions have questioned this multiple passage model, showing instead that the clouds are most likely on their first passage through the Galaxy~\citep{Kallivayalil:2005mm}. More recent modeling has shown that it is possible to connect the Magellanic stream to the clouds under the first passage scenario~\citep{Besla:2010ws,Diaz:2011dq}. 

\par In addition to the Sagittarius and the Magellanic streams, several other streams have recently been identified in the SDSS~\citep{Belokurov:2006kc,Duffau:2005ta,Grillmair:2008fv}. These streams may have originated either from a dwarf galaxy or a globular cluster. Table~\ref{tab:streams} lists prominent streams in the Milky Way halo that may be plausibly associated with a tidally-disrupted dwarf galaxy (several other known streams exist in the Milky Way, but likely originated from the disruption of a globular cluster). The detection of a multiple streams, either from disrupted dwarf galaxies or globular clusters, lends more support to the theory that the stellar halo of the Galaxy was formed via the chaotic accretion model~\citep{Searle1978}, as opposed to the theory of formation via monolithic collapse~\citep{Eggen1962}. 

\begin{deluxetable}{l c l}
\tablecolumns{3}
\tablecaption{Prominent stellar streams in the Milky Way halo that may be associated with tidally-disrupted dwarf galaxies. 
\label{tab:streams}}
\tablehead{
Name & Distance [kpc] & Reference}
\tablewidth{0pc}
\startdata
Magellanic Stream & -- &~\cite{Mathewson1974} \\
Sagittarius & --& ~\cite{Ibata:1994}\\
Orphan Stream & -- & ~\cite{Belokurov:2006kc}\\
Styx & 45 &~\cite{Grillmair:2008fv}\\
Bootes III & 45 &~\cite{Grillmair:2008fv,Carlin:2009cu}\\
\enddata
\end{deluxetable}

\par Probably the simplest means to determine whether a stream originated from a dark matter-dominated dwarf galaxy or from a globular cluster or star cluster is through the width of the stream. For example,~\cite{Grillmair:2008fv} identified four streams in the SDSS data ranging in distances from 3 to 15 kpc (Acheron, Cocytos, Lethe, and Styx), and indicates that three of them have widths $\lesssim 100$ pc, while the fourth, Styx, has a width of approximately 2.6 kpc. This conclusion is strengthened by measurements of the velocity dispersion of $\sim 14$ km/s of the Bootes III object, which is a clump of stars that appears to be in the initial phases of tidal disruption and may be the progenitor of the Styx stream~\citep{Carlin:2009cu}. An additional diagnostic of the origin of a stream comes from understanding the metallicity of the constituent stars.~\cite{Boer:2012sm} show that dSphs are typically deficient in Ni and Na relative to globular clusters, and use this diagnostic to conclude that the progenitor of the recently-discovered Aquarius stream is a globular cluster. 

\par In addition to their role in understanding the population of accreted dwarf galaxies, streams are also excellent probes of the population of dark matter substructures in the Galaxy.~\cite{Carlberg:2011xj} showed that gaps in streams are created by dark matter substructure, and that the high density of gaps in narrow streams may be indicative of a population of $\gtrsim 10^5$ substructures greater than approximately $10^6$ M$_\odot$. If the orbit of a stream can be reconstructed, it is also possible to constrain the shape of the Milky Way potential that gives rise to that stream. This was evident in the discussion of the Sagittarius stream in Section~\ref{sec:MW}. However, due to the typically sparse data associated with the streams, it is difficult to reconstruct the orbital trajectory and obtain unique fits to the data for a given potential model~\citep{Eyre:2009}. 

\par Finally, a discussion of the population of Galactic satellites and stellar streams cannot be complete without mentioning the unique spatial distribution of the satellites. It was noted many years ago that the observed Milky Way satellites are distributed to a good approximation along an inclined plane relative to the Milky Way disk~~\citep{Lynden-Bell1976}. As discussed in the next section, this may be a reflection of the fact that dark matter satellites accrete in an anisotropic manner along filaments into the Milky Way~\citep{Libeskind:2005hs,Zentner:2005wh}. It has also been hypothesized that the observed population of satellites are so-called tidal dwarf galaxies, which form when a gas-rich galaxy merges into a larger host halo, and fragments to produce distinct populations of stars~\citep{Metz:2007cg}. 

\par The spatial distribution of satellites, as well as the other dark matter properties of Milky Way satellites and streams that were discussed in this section, comprise a series of compelling issues that must be understood in a larger cosmological context. This now leads into the subject of the next section, which lays out the predictions of the structure, and substructure, in the Milky Way within a cosmological context.

\newpage  
\section{Simulations of Galactic halos: Predictions and observational connections}
\label{sec:simulations} 

\par The observations discussed in the previous sections have provided a wealth of information on the structure of the Galaxy and its satellites. They provide measurements of the local dark matter density, the mass of the Galactic dark matter halo, and the masses of the satellite galaxies. These quantities will be important for interpreting results from direct and indirect dark matter searches (Sections~\ref{sec:indirect_detection_theory} and~\ref{sec:direct_detection_theory}). It is important to recognize, however, that there is still some limitation to the scope of these observations, and that dark matter searches require information not directly measured from modern astronomical data. Two particularly important examples are the dark matter velocity distribution, and the properties  of dark matter substructure.  In order to model properties such as these that are not directly determined from modern observations, we must appeal to theory, which now mainly revolves around the use of numerical simulations. The primary goal of this section is to advance the theoretical description of the dark matter properties in the Galaxy, using both analytic models and N-body simulations, focusing on those that are of most practical use for dark matter searches. 

\par The first analytic models for the formation of dark matter halos were based on the theory of self-similar accretion of material around a seed perturbation~\citep{Gunn:1972,Fillmore:1984wk,Bertschinger:1985pd}. Though a great deal of theoretical insight has been gained from this theory, it is becoming clear that this picture of self-similar accretion must be considered together with a more violent and chaotic picture for the formation of dark matter halos that involves hierarchical merging of smaller mass halos. In the hierarchical merging scenario, dark matter halos of a given mass can have significantly different formation histories; for example some halos may experience a violent late merger, while others may experience a quiescent merger history. 

\par In recent years, it has become widely appreciated that the ``smooth" component of a dark matter halo is only part of its phase space structure. One of the more profound theoretical insights in dark matter over the past couple of decades is that halos formed in a LCDM universe are not smooth, but rather contain an abundance of substructure. This substructure contains a wealth of information on the nature of particle dark matter. Indeed, if dark matter has an annihilation or an elastic scattering cross section that is too low to be detectable by indirect or direct methods, probably the best method for studying the nature of particle dark matter will be through the properties of satellites and substructure. As a result there is clear motivation for a detailed theoretical understanding of the mapping between a particle dark matter model and its substructure properties, and how different models can be studied observationally. 

\par This section begins with a description of numerical simulations that are used to study the structure of dark matter halos. It then goes into more detail on the output of the simulations, in particular the important quantities such as the mass and the velocity distributions. The theoretical predictions for dark matter substructure in the Galaxy are then discussed, and the connection is made to the observations of satellite galaxies discussed in Section~\ref{sec:satellites}. 

\subsection{Simulations of Milky Way-like galaxies} 

\par Since this section will frequently quote the results from numerical simulations, it is beneficial to have a general understanding of the most important properties of the simulations. Standard N-body (i.e. dark matter-only) simulations assume that dark matter is the dominant component of mass in the universe, and that the particles responsible for setting the formation of structure interact only gravitationally, so their direct encounters can be ignored on cosmological scales. In order to run the computations in a reasonable amount of time on modern computers, a single dark matter particle is made to represent a much larger number of smaller mass particles. 

\par In a simulation the unperturbed universe is represented by a uniform grid of particles. From seed linear density perturbations on this smooth model, structure in the universe forms as overdense regions grow and eventually break away from the expansion of the universe to form self-bound gravitating dark matter halos. The number and the structure of the halos depends sensitivity on both the initial conditions for the density perturbations and the nature of the dark matter. In the cosmological volume within which modern numerical simulations are run, the mass function of dark matter halos is well-determined in the range of approximately $10^8$ $M_\odot$ to scales as massive as clusters of galaxies, approximately $10^{14}$ $M_\odot$. 

\par Halos at the mass scale of the Milky Way, $\sim 10^{12}$ $M_\odot$, are identified in a larger volume N-body cosmological simulation at redshift $z = 0$ according on their mass and a density criteria on the matter distribution in the local vicinity of the halo. Once the halo is identified, a Lagrangian region within which the halo formed is determined. Within this region, a larger number of lower mass particles are generated according to a new set of initial conditions. Regions of larger and larger distance outside this Lagrangian volume are populated with more and more massive particles. Since the sampling is much higher in the Lagrangian region than in the initial larger scale simulation, not all of the particles within this region wind up in the final halo, and a negligible fraction of more massive particles in the outer regions end up in the final halo~\citep{Springel:2008cc}.

\par The above procedure for simulating the formation and the subsequent identification of Milky Way mass dark matter halos in an expanding universe works well for particles that travel a small distance compared to the grid size at the redshift when the simulation is started. This is a good approximation for the cold and warm dark matter models considered here. However, in warm dark matter models, the sharp cut-off in the power spectrum gives rise to artificial halos that form along filaments on scales that are characteristic of the inter-particle separation~\citep{Lovell:2011rd}. This turns out to be an important effect in the regime of interest for both observations and theory. 

\subsection{Mass distribution} 
\par The mass profiles of dark matter halos in N-body simulations are of great interest for many aspects of this article. In combination with the observational results discussed in Section~\ref{sec:MW}, the results from N-body simulations are used for predicting signals in indirect dark matter experiments (Section~\ref{sec:indirect_detection_experiment}). Because the formation of halos and the resulting mass profiles are directly related to the properties of dark matter, before delving into the results from the modern simulations, it is worthwhile to take a step back and understand different aspects of theoretical predictions for mass profiles of halos in an expanding universe. 

\par The first predictions for the mass profiles of galaxies that form in an expanding universe were based on the self-similar accretion model. The self-similar model begins by defining a spherical region, and assumes a perturbation in this region of the form $\delta M/M \sim M^{-\epsilon}$, where the case $\epsilon = 1$ corresponds to accretion about a point mass~\citep{Gunn:1972}. Assuming spherical symmetry and entirely radial orbits, self-similar models predict single power-law models for the density profile, of the form $\rho \propto r^{-2}$ for $\epsilon \le 2/3$, and $\rho \propto r^{-9\epsilon/(1+3\epsilon)}$ for $\epsilon \ge 2/3$~\citep{Fillmore:1984wk}. These results predict that density profiles depend on the halo mass, and the density profiles are always steeper than $r^{-2}$. Because the density at a given position is set by a sum over mass shells that contribute to the density, the interior of the density profile is smooth as a result of the summation of a large number of mass shells, whereas in the outer region of the halo, only a small number of shells contribute to the density profile. The power law shape of the density profile breaks at the transition to the infall regime, where the density profile is characterized by sharp caustic features~\citep{Bertschinger:1985pd}. 

\par Self-similar halo models have now been extended to include non-radial motions and deviations from spherical symmetry, and solutions for the density profile have been obtained. The addition of angular momentum increases the range of slopes allowed, predicting a pure power law that scales as $r^{-1.3}$ for Milky Way-mass halos~\citep{Nusser:2000xn}. When torques are introduced into the spherical infall model, the central slope is mass dependent, with larger mass halos being described by steeper density profiles~\citep{Zukin:2010gx}. It is also interesting to note that even in the simplest case of pure radial orbits, numerical simulations with spherical initial conditions show that the radial orbit instability turns the resulting halo into a non-spherical structure~\citep{Vogelsberger:2010eh}. 

\par Understanding self-similar halo models is not just important because it provides physical intuition for halo formation, it also brings up the important question of the dependence of density profile on halo mass. A related question is the scatter in the density profile at a fixed halo mass. If density profiles are found to not depend on mass, or if there is significant scatter in the density profile at a fixed mass, it would provide direct evidence that the physics of halo formation is not entirely captured in the self-similar model. 

\par In numerical simulations, the formation of halos does not occur through a smooth and nearly-spherical accretion, but rather through a combination of smooth particle accretion and the merging of lower mass halos. It is still a matter of debate what fraction of the mass of a halo comes from mergers and what fraction comes from smooth accretion; some studies find that the merger component constitutes nearly the entire mass of the halo~\citep{Angulo2010}, while others find that approximately 50\% is due to mergers~\citep{Stewart:2007qv,Genel:2010pb}. 

\par Whatever the origin of the mass is that comprises dark matter halos, it is clear that halos that form via hierarchical clustering are not spherical. Analysis of the shapes of halos over a mass range of approximately $10^{10} - 10^{14}$ M$_\odot$ indicate that halos become less spherical as the mass increases~\citep{Allgood:2005eu,Schneider2012}. Halos are found to be more preferentially prolate, and the major axis is aligned with the large scale structure distribution, indicating an intriguing interplay between the properties of the halo and the local environment. 

\par In order to best understand whether the density profiles of halos are universal, it is most convenient to work in the context of spherically-averaged profiles. Initial studies of the mass distribution of dark matter halos in cosmological simulations indicated that the central slope is shallower than is predicted by the secondary infall model, with the central slope reaching a constant asymptotic inner value $\rho \sim r^{-1}$~\citep{Dubinski:1991}. Further, it was found that the outer regions of the density profile fall-off more steeply, as $r^{-3}$, and that this double power law model, the Navarro-Frenk-White (NFW) profile, is nearly universal in halos over a wide range of mass scales~\citep{Navarro:1995iw}. 

\par Though these results put forth the profound idea that density profiles are the same for any mass halo, they still suffered from insufficient resolution of the central regions. The recent sample of Aquarius simulations, which resolve down to less than one percent of the halo virial radius for approximately $10^{12}$ M$_\odot$ halos, now find that this asymptotic slope is not yet achieved, and that the profile becomes progressively more shallow towards the center of the halo~\citep{Navarro:2008kc}. These recent results favor the Einasto profile, 
\be
\rho_e(r) = \rho_{-2} \exp [-( (r/r_{-2})^\alpha -1) )], 
\label{eq:einasto_profile}
\ee
where $r_{-2}$ is the radius at which a log slope of $-2$ is achieved and $\rho_{-2}$ is the density at that radius. For Milky Way-mass halos, $\alpha = 0.17$~\citep{Navarro:2008kc}. This value of $\alpha$ also works well for subhalos~\citep{Springel:2008cc}, though recent results suggest that a modification may work better (see discussion of substructure below). In addition to the fact that the density profile does not achieve a pure power law, there is a small mass trend that is seen in the density profiles as a function of halo mass. However, neither of these trends are as significant as is predicted in the self-similar halo formation model as described above. These simulations also show that different halos at a fixed mass cannot be scaled to look identical. 

\par In addition to the density profiles, simulations predict additional important trends for the structure of dark matter halos. It is found that lower mass halos are more centrally-concentrated than halos of larger mass, where the concentration can be defined as the ratio of the radius at which the density reaches 200 times the critical density, defined as $r_{200}$, to the characteristic radius of the halo. For the Einasto profile, the concentration is $c = r_{200}/r_{-2}$, while for the NFW profile the concentration is $c = r_{200}/r_0$, where $r_0$ is defined in Equation~\ref{eq:NFW}. For the updated values of cosmological parameters and for halo masses above approximately $10^{11}$ M$_\odot$, the concentration-mass relation for the Einasto profile is~\citep{Duffy:2008pz}
\be
c_{vir} (M_{vir,z}) = \frac{8.82}{(1+z)^{0.87}} \left(\frac{M_{vir}}{2 \times 10^{12} h^{-1} \textrm{M}_\odot} \right)^{-0.106}. 
\label{eq:concentration_mass} 
\ee

\par It is important to ask whether the halo profile in Equation~\ref{eq:einasto_profile} and its corresponding properties are specific to the LCDM cosmological model.~\cite{Maccio':2012uh} have recently examined the density profiles in universes in which a component of the dark matter is warm. Due  to the cut off in the matter power spectrum in warm dark matter models, halos typically form later, so their concentrations are lower than in pure cold dark matter universes. Though the concentrations are lower, the NFW model still is a good description of the density profile, which points to the intriguing idea that the NFW-like profile results from a general dynamical processes in halo formation. 

\par All of the results discussed to this point concern pure dark matter-only simulations, and do not include the effect of baryonic physics. Inclusion of baryons leads to a great deal of challenges from the perspective of numerical simulations, so understanding halo structure in more realistic universes will likely remain elusive for some time. Higher resolution simulations of single halos, in combination with analytic models, have recently indicated that the central densities of dark matter halos become less steep than those found in pure N-body simulations because the baryons induce repeated epochs of feedback due to star formation activity. Cusps are turned into cores because energy is repeatedly input into the collisionless particles due to fluctuations in the gravitational potential~\citep{Pontzen2012}. This model provides an appealing explanation for low surface brightness galaxies, for which there is now good indication that the central densities are flatter than the Einasto model~\citep{Oh:2010ea}. 

\subsection{Velocity distribution}
\par The density profiles of dark matter halos are only a part of the phase space information that is of interest for dark matter searches. A second important quantity is the dark matter velocity distribution function (VDF) at the Solar radius. While there has been a significant amount of literature devoted to understanding the universality of halo mass profiles, by comparison relatively little work has been dedicated to understanding the global properties of the VDF in N-body simulations. The VDF is of particular importance for direct dark matter searches (Sections~\ref{sec:direct_detection_theory} and~\ref{sec:direct_detection_experiment}), which are most sensitive to the particles in the high velocity tail of the distribution due to the kinematics of the collision between the dark matter and nucleus. Indirect searches for high energy neutrinos from dark matter particles that get captured in the Sun (Sections~\ref{sec:indirect_detection_theory} and~\ref{sec:indirect_detection_experiment}) are sensitive to the lowest velocity component of the VDF. 

\par Following the discussion above on halo density profiles, it is first informative to gain some theoretical intuition to the nature of the VDF. For this we can compare to the observations of collisionless stars in elliptical galaxies. For stars in elliptical galaxies, it is straightforward to show that the two-body relaxation scale is much longer than the Hubble time~\citep{BT2008}. However, since their density profiles are near isothermal, a physical mechanism beyond two-body relaxation must set the dynamics of galaxies, unless for some reason galaxies form in a state near equilibrium. The above paradox led to the development of the theory of `violent relaxation' by~\cite{Lynden-Bell:1967}. The main premise of this theory is that particles in a galaxy do not respond to interactions with individual objects, but instead to changes in the form of groups of objects, which causes the potential of the system to vary rapidly during its formation. The stars continually respond to different groups in a statistical manner; for completely random interactions the central limit theorem dictates that the velocity distribution approaches a maxwell-boltzmann distribution and thus an isothermal density distribution. It is straightforward to see why this logic may translate over to dark matter in halos, which form via a combination of mergers and accretion. 

\par Though the broad arguments above suggest that the VDF should be close to a maxwell-boltzman distribution, it is important to recognize that different VDFs are predicted in other theoretical models. For example the self-similar model described above predicts that the density profile at each location is comprised of distinct flows of particles. These flows manifest themselves in the VDF as a series of delta-functions that are normalized by the density of particles in each flow relative to the total density at that location. Also the disruption of a dark matter satellite generates a VDF that is represented by a single flow of particles. A significant amount of modern theory is devoted to understanding the relative contribution between these non-thermalized flows and a more thermalized component 

\par In modern N-body simulations, the VDF is studied in two separate statistical regimes.  On the one hand, the resimulation of dark matter halos in a full cosmological volume via the methods described above provides a large number of particles in a relatively small number of halos. In this case, although single halos are well-sampled, it is difficult to obtain a measure of the scatter in the VDF between halos of a given mass. On the other hand, the VDF is analyzed in a large sample of halos taken from a large cosmological volume. While this later method does give a better estimate of the scatter in the VDF between halos of a fixed mass, accurately measuring the VDF in the central region of the halo is difficult due to lack of numerical resolution. In both cases some statistical information is lost, either in the total number of halos analyzed or in the number of particles per halo.   

\par To understand this matter more concretely, consider the relevant scales of modern simulations. The two largest volume cosmological simulations to date that resolve a statistical sample of Milky Way-mass halos identify approximately $10^4$ halos with mass near $10^{12}$ M$_\odot$~\citep{BoylanKolchin:2009nc,Klypin:2010qw}. Due to the particle resolution of a few times $10^8$ M$_\odot$, halos at this mass contain $10^3 - 10^4$ particles per halo. For comparison, the Aquarius~\citep{Springel:2008cc} and Via Lactea-II~\citep{Diemand:2008in} simulations of Milky Way-mass halos contain nearly a billion particles within the virial radius of their respective halos. 

\par In order to interpret the VDF from these simulations, it is best to establish a point of comparison. With this motivation define the maxwell-boltzmann velocity distribution, 
\be 
f(v) = \frac{1}{\sqrt{2\pi \sigma^2}} \exp [-v^2/2\sigma^2],
\label{eq:maxwell_boltzmann}
\ee
where $\sigma$ is the velocity dispersion, which as discussed below in section~\ref{sec:direct_detection_theory} can be related to the circular velocity of the halo. The maxwell-boltzmann velocity distribution is isotropic in that it only depends on the magnitude of the velocity, and may be self-consistently derived from an isothermal density profile. Though it is convenient and motivated in part by theory, there is no a priori reason why we should expect halos to have this exact distribution. 

\par
\begin{figure*}
\begin{center}
\begin{tabular}{cc}
\includegraphics[width=0.49\textwidth]{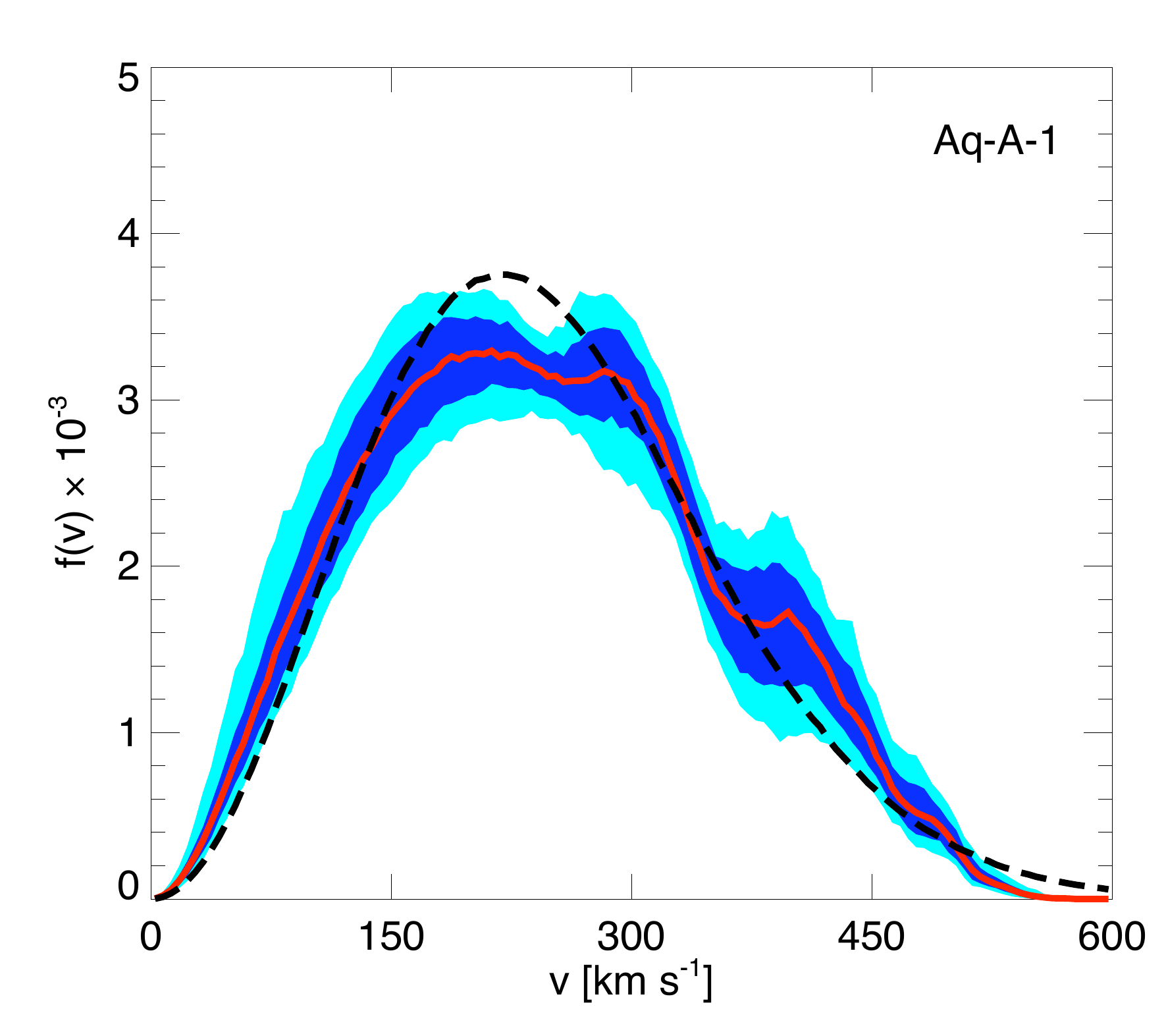} & 
\includegraphics[width=0.55\textwidth]{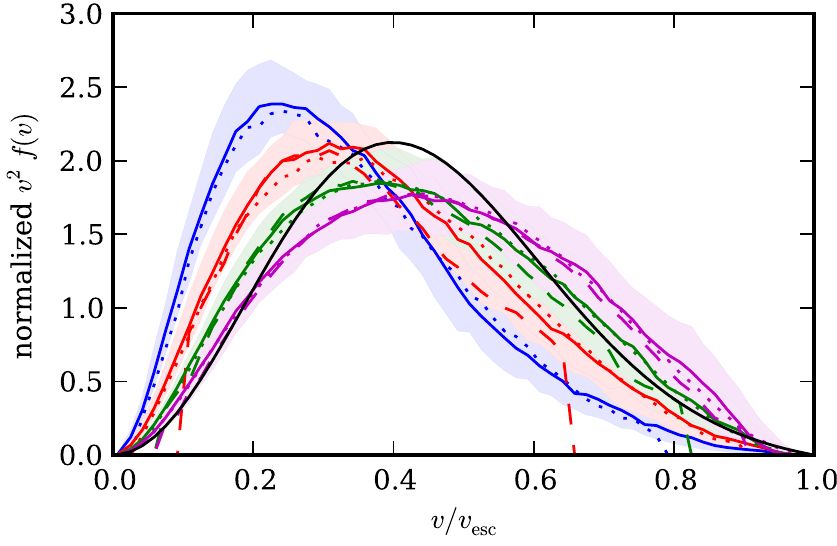} \\
\end{tabular}
\end{center}
\caption{Velocity distribution function in dark matter only simulations. The left panel shows the distribution from the dark matter-only Aquarius simulations (Figure reproduced from~\cite{Vogelsberger:2008qb}). The right panel shows the stacked velocity distribution function for halos from Milky Way mass to cluster mass scales (Figure reproduced from~\cite{Mao:2012hf}). In the right panel, from left to right, the VDF is shown at increasing radius relative to the scale radius of the halo. The solid dashed-line in the left, and the solid line in the right give the best-fitting maxwell-boltzmann distributions. 
}
\label{fig:simulations_vdf}
\end{figure*}

 As shown in Figure~\ref{fig:simulations_vdf}, the VDF at the Solar radius differs from a pure maxwell-boltzmann distribution in the highest resolution dark matter only simulations~\citep{Vogelsberger:2008qb,Kuhlen:2009vh}. In both cases the deviation from the corresponding best fitting maxwell-boltzmann model from Equation~\ref{eq:maxwell_boltzmann} is similar. First, the distributions have a larger fraction of particles at velocities less than its peak velocity. Second, the peak of the distribution is broader than the maxwell-boltzmann model. Third, there are fewer particles in the extreme high energy tail of the distribution. For the individual halos studied, these results incorporate the scatter in the VDF at the Solar radius by measuring it at different locations within an ellipsoidal radius that is determined from the halo density distribution. 

\par The Aquarius and Via Lactea-II simulations also reveal the presence of sharp features that are due to substructure. These features are only recognizable when a halo has a significant fraction of its mass, $\gtrsim 50\%$, in an individual substructure. To a good approximation, the VDF is dominated by the ``smooth" component, which reflects particles that have been both smoothly accreted and those that were once part of a subhalo that has been tidally-destroyed. The VDF also reveals broader and more extended features that are related to the energy distribution~\citep{Vogelsberger:2008qb}. These features are not seen in the individual velocity components, which are well-described by gaussians of different widths; the triaxiality of the VDF is clearly indicated in~\cite{Vogelsberger:2008qb} and~\cite{Abel:2011ui}. 
 
\par Turning now to the other regime of identifying a larger sample of halos with lower resolution,~\cite{Mao:2012hf} have recently studied the properties of the VDF in halos from Milky Way mass up to the cluster mass scale. From a stacked analysis of the VDF from thousands of halos, the general results from the individual high-resolution halos are corroborated.~\cite{Mao:2012hf} find that the VDF is well described by the form
\be
f(v) = \exp(-|v|/v_0)(v-v_{esc})^p, 
\label{eq:universal_VDF} 
\ee
where $v_0$ and $p$ are correlated fit parameters that depend on the scale radius of the halo. The key aspect of this VDF is that it is broader at the peak and steeper at the tails than the maxwell-boltzmann model, with the power-law index $p$ controlling how steep the distribution falls off beyond the peak. The right hand panel of Figure~\ref{fig:simulations_vdf} shows the VDF at different radii in a large sample of halos from cosmological simulations. 

\par The above methods rely on direct analysis of the output from cosmological simulations. Complementary methods for studying the VDF combine these results with various analytical methods. One such method involves sampling halos from cosmological simulations, merging them with a halo that is similar in mass, and evolving the merged system until it is relaxed. Here relaxed means that the system is in equilibrium, so that evolving it further in time does not change its phase space distribution. This methodology was utilized in~\cite{Hansen:2005yj}, who started from an initial condition of two halos with an NFW profile, and used them to generate a distribution function according to the isotropic Eddington model (Equation~\ref{eq:eddington}).~\cite{Hansen:2005yj} et al. suggested that the best-fitting VDF of the merged structure after many dynamical timescales is the Tsallis distribution, 
\be
f(v) = \left[ 1 - (1-q) \left( \frac{v}{\kappa \sigma_r}\right)^2 \right]^{\frac{q}{1-q}},
\label{eq:tsallis} 
\ee
where $q \simeq 1$ is the Tsallis index and $\kappa$ is a free parameter. The maxwell-boltzmann distribution is recovered in the limit that $q \rightarrow 1$. 

\par Numerical simulations also show a correlation between $\beta$, velocity anisotropy parameter (Equation~\ref{eq:beta}), and the log-slope of the density profile such that $\beta$ increases with radius from the center of the halo. This behavior of the anisotropy can be qualitatively understood in terms of the infall properties of particles in the halo, because infalling particles with radially-biased orbits tend to dominate the outer region of the halo. The Ospikov-Merritt model predicts behavior of the anisotropy of this form, 
\be
\beta(r) = \frac{r^2}{r^2 + r_\beta^2},
\label{eq:beta_ospikov}
\ee
where $r_\beta$ is the scale radius of the velocity anisotropy profile. Though it does qualitatively provide the correct behavior for $\beta$, the slope of the rise of $\beta$ is much steeper than what is seen on average in simulated dark matter halos.~\cite{Hansen:2004qs} more accurately quantify the relation between the anisotropy and the log-slope, $\gamma$, as
\be 
\beta = 1 - 1.15 \left( 1 + \frac{\gamma}{6} \right). 
\label{eq:beta_gamma}
\ee
Recent analysis of cluster mass galaxies corroborate this relation, although they find that there is a significant amount of scatter~\citep{Lemze:2011ud}. 

\subsection{Galactic substructure}
\par The discussion to this point has focused on the smooth component of dark matter halos. One of the great insights into dark matter that has been deduced from N-body simulations is that the mass distribution within Galactic dark matter halos is in fact not entirely smooth, but that a significant fraction of it is bound up in substructure, or subhalos. Modern simulations of Milky Way-mass galaxies are now complete in their measurement of subhalos down to a mass of approximately $10^6$ $M_\odot$, corresponding to approximately $10^{-6}\%$ of the total mass of the halo. For masses $\gtrsim 10^6$ $M_\odot$, the mass function of subhalos is measured to be a power law that scales as $dN/dM \propto M^{-\alpha}$, with $\alpha = 1.9$. Because $1 < \alpha < 2$, subhalos at the low mass end of the mass function dominate the distribution by number, while the subhalos at the high mass end of the mass function dominate the total mass in substructure. The fraction of the total mass of the Milky Way halo in substructure is in the range $10\%-50\%$~\citep{Diemand:2008in,Springel:2008cc}. 

\par The most massive subhalos in a typical Milky Way-mass dark matter halo are approximately $10^{10}$ $M_\odot$. Subhalos near this upper end of the mass range are preferentially found in the outer regions of the halo, because dynamical friction effects destroy massive subhlaos as they approach the central regions of the halo. Less massive subhalos are found to more faithfully track the smooth dark matter density profile~\citep{Springel:2008cc}. The predicted radial distribution of subhalos in primary halos is of particular importance for indirect dark matter searches (section~\ref{sec:indirect_detection_experiment}). 

\par The formation of substructure down to the mass scale resolved in N-body simulations is a unique prediction of cold dark matter theory. For standard cold dark matter particles with mass around the weak scale and freeze-out temperatures $\gtrsim$ GeV, the minimum mass subhalo is predicted to be somewhere in the range of an Earth mass~\citep{Green:2003un,Loeb:2005pm,Profumo:2006bv,Martinez:2009jh}. It is not yet clear if subhalos at this low mass scale would survive tidal disruption in the Milky Way halo though (see~\cite{Diemand:2005vz}). In warm dark matter models or its variants, free streaming induces a cut-off in the power spectrum, and there is expected to be a corresponding cut off in the mass function of substructure at a significantly larger mass scale. 

\subsubsection{Connection to Galactic satellites} 

\par The bound subhalos that survive within Milky Way-mass dark matter halos are expected to be related to the dSph galaxy population; understanding this mapping provides a unique probe of cosmology and the properties of dark matter. The correspondence between subhalos and dSph satellite galaxies is by no means straightforward; it has been known for two decades that there are orders of magnitude more dark matter subhalos predicted in simulations of Milky Way-mass halos than there are observed dSph satellite galaxies~\citep{Kauffmann:1993gv,Klypin:1999uc,Moore:1999}. This discrepancy, often known as the ``missing satellites problem," has received a great deal of attention in the literature (for recent reviews of different aspects of this issue see~\cite{Kravtsov:2009gi} and~\cite{Bullock:2010uy}), and it has inspired a wide range of theoretical variations of the LCDM model~(e.g.~\cite{Kamionkowski:1999vp,Zentner:2003yd,Hooper:2007tu}). 

\par At the time when it first received a great deal of attention in the astronomical community, it was unclear what to make of the missing satellites issue. For example, it was well-appreciated that the census of Milky Way satellites was incomplete, even for satellites that are as luminous as the brightest dSphs~\citep{Willman:2001qj}. In addition, kinematic data sets from dSphs were not as complete as they are today, so there was a great deal of uncertainty on their dark matter masses. Because of these uncertainties, there were well-motivated theoretical solutions to the satellite discrepancy. First, and perhaps most simply, it could be that the observed dSphs reside in the most massive dark matter subhalos. This would imply a sharp cut off in the formation of galaxies below a fixed present day mass scale, which could result from, for example, supernova feedback processes~\citep{Stoehr:2002ht}. It was also hypothesized that the dSphs do not reside in the most massive subhalos at present, but rather in the most massive subhalos that formed before a redshift near reionization. This solution directly connects the luminous fraction of satellites to the properties of the intergalactic medium at the time when the halos formed~\citep{Bullock:2000wn,Benson:2001au}. Combinations of the above solutions were developed, which hypothesized that the dSphs reside in subhalos that were the most massive at the time of accretion into the Milky Way~\citep{Kravtsov:2004cm}. 

\par
\begin{figure*}
\begin{center}
\begin{tabular}{c}
\includegraphics[width=0.65\textwidth]{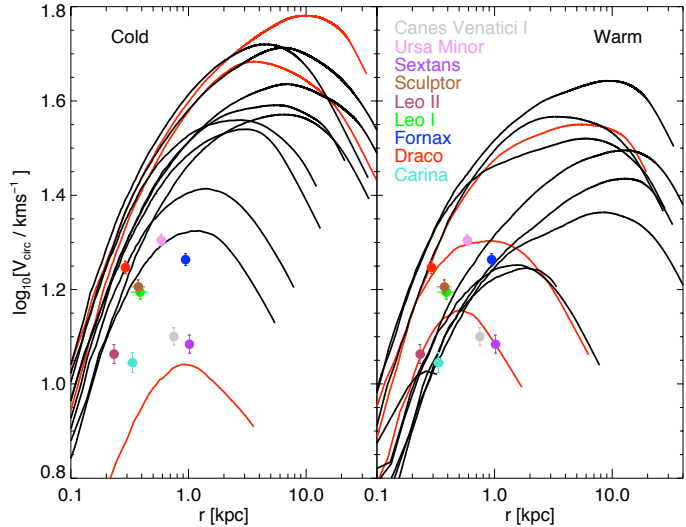} \\
\end{tabular}
\end{center}
\caption{The circular velocities for the subhalos with the most massive progenitors, in a CDM simulation ({\it left}) and a WDM simulation ({\it right}). The data points are the masses within the half-light radius for the brightest dwarf spheroidals. Figure from~\cite{Lovell:2011rd}; see larger sets of related figures in~\cite{BoylanKolchin:2011de}.
}
\label{fig:RrVr}
\end{figure*}

 Within the past decade, advances in the census of Milky Way satellites from the SDSS and in their kinematic data (see Section~\ref{sec:satellites}) have served to better focus the understanding of the missing satellites problem. Assuming that the subhalos in the simulations host the dSphs, data and theoretical modeling indicates that the peak circular velocities for the dSphs are in the range 10-30 km s$^{-1}$~\citep{Strigari:2010un}, confirming earlier estimates based on sparser data samples~\citep{Zentner:2003yd,Kazantzidis:2003hb}. This now clearly implies that there are dozens of dark subhalos that are devoid of stars entirely, or host stellar populations that are too faint to resolve in modern surveys~\citep{BoylanKolchin:2011de}. This is explicitly shown in the left hand panel of Figure~\ref{fig:RrVr}, which compares the measured circular velocities of the dSphs to the circular velocities of the most massive subhalos in the Aquarius simulation. Given that the satellite population is complete down to at least a luminosity of the classical satellites~\citep{Walsh2009}, in the regime surveyed by SDSS it is no longer possible to postulate that a large population of undetected satellites exist at the same luminosity as the classical satellites, so long as they have surface brightnesses similar to the classical satellites. 

\par Now that the abundances of the Milky Way satellites and subhalos are more secure, does the new experimental data point to a solution of the missing satellites problem? Combining both the theoretical and the observational advances, the possible solutions can be broadly classified into three main categories. First, the solution may be that our knowledge of galaxy formation at the scales of the faintest known galaxies in the universe is still incomplete. It is certainly true that modern simulations, which are unable to appropriately account for baryonic physics, do not include all of the relevant physical processes in galaxy formation. Second, the solution may be that our observational knowledge is still incomplete, either in the form of our census in the Milky Way of ultra-faint satellites, or in our understanding of the distribution of satellites around a typical galaxy like the Milky Way. In other words, the Milky Way may be an outlier in terms of its population of luminous satellites. Third, as mentioned above, the solution may be that the cosmological model differs from the standard LCDM model. This could result in either a reduction of the number of subhalos, a reduction in their central dark matter densities, or both. Any one of these three effects, or a combination of them, may be a part of the solution. Each of them are now considered in turn. 

\paragraph{Inclusion of baryonic physics} 

\par To better address the first hypothesis, there are now numerical simulations that have examined the effect of baryonic physics on the formation and structure of Milky Way-mass dark matter halos and satellite galaxies~\citep{Wadepuhl:2010ib,Parry:2011iz}. The physical processes included in these simulations are supernova feedback, radiative cooling, cosmic ray pressure, and energy input from the supermassive black hole at the center of the host galaxy. Because of the low density of baryons in the satellites and the scales that the simulations resolve, these cosmological simulations do not point to a significant change in the dark matter properties of the satellites. This picture has been challenged recently in simulations of isolated satellite galaxies, which indicate that baryonic feedback can significantly affect the central densities of dark matter halos~\citep{Brooks:2012vi}. The reduced central densities affect the population of satellites in two ways. First, it alters the mass distribution of the most massive satellites, putting them seemingly more in line with the kinematics of the observed satellites. Second, because of the reduced central densities of the satellites, it lends them more susceptible to tidal stripping, reducing the overall number of satellites. 

\paragraph{Satellites of external galaxies} 
\par Now addressing the second hypothesis, a complete assessment of the satellites discrepancy requires a better understanding and identification of faint satellites around the Milky Way and external galaxies. Observations of the Milky Way satellite population were covered in section~\ref{sec:satellites} and will be discussed again in section~\ref{sec:future}. Here two modern methods are discussed for examining the satellite population around external galaxies: the first is the direct counting of satellites from large scale surveys, and the second utilizes the method of gravitational lensing. 

\bigskip 

$\bullet$ {\underline{Direct counting methods}} 

\bigskip 

\par Recently a great deal of progress has been made on searches for faint satellite galaxies around systems that have physical properties similar to the Milky Way. These properties include absolute magnitude, color, and isolation. From modern galaxy surveys, in particular the SDSS, it is now possible to build up a statistical sample of Milky Way-like galaxies along with their respective satellite populations. To understand better how these studies work, begin by considering satellites brighter in absolute magnitude than the Magellanic Clouds. Modern kinematic measurements indicate that the Magellanic Clouds have maximum circular velocities $\gtrsim 60$ km/s~\citep{vanderMare2002,Stanimirovic2004}. For comparison, N-body simulations find that $\lesssim 5\%$ of Milky Way-mass halos in cosmological simulations contain two satellites with maximum circular velocities greater than the accepted maximum circular velocities of the Magellanic Clouds~\citep{BoylanKolchin:2009an,Liu:2010tn}. So the Milky Way is somewhat uncommon in that it has two satellites with masses greater than the mass of the Magellanic Clouds, but probably not uncomfortably so.  

\par
\begin{figure*}
\begin{center}
\begin{tabular}{c}
\includegraphics[width=0.50\textwidth]{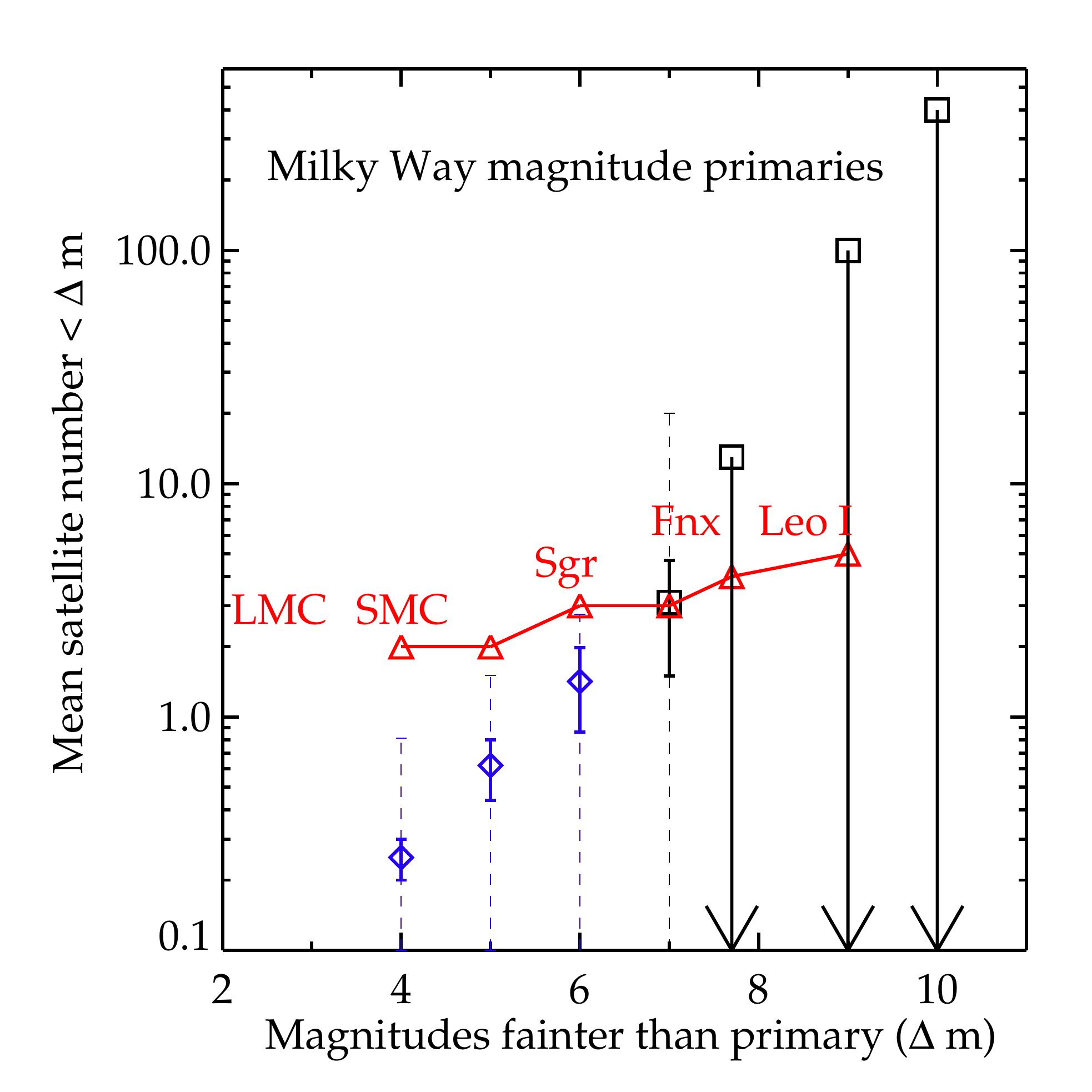} \\
\end{tabular}
\end{center}
\caption{Mean number of satellites brighter than an apparent magnitude around Milky Way-like galaxies, versus the satellite apparent magnitude. Daimonds (blue) and squares (black) are from SDSS data, and triangles (red) are Milky Way data. From~\cite{Strigari:2011ps}.
}
\label{fig:satellite_abundance}
\end{figure*}

 In order to better probe the satellites discrepancy, it is necessary to extend this measurement down to fainter satellite magnitudes. However, because it becomes increasingly more difficult to reliably assign redshifts to satellites that are several orders of magnitude fainter than the Milky Way, the sample of host galaxies that can be used in the measurement dramatically decreases. Using SDSS data release 7,~\cite{Guo:2011qc} measure the satellite luminosity function down to the magnitude scale of Fornax, which is about eight magnitudes fainter than the Milky Way. These authors show that Milky Way-like galaxies have on average about a factor of two fewer satellites than our Galaxy or M31. Using SDSS data release 8, and a combination of photometric and spectroscopic redshifts,~\cite{Strigari:2011ps} translate this into an upper limit of $\sim 13$ satellites brighter than the luminosity of Fornax around Milky Way-like galaxies (Figure~\ref{fig:satellite_abundance}). At this point, these results broadly indicate that, down to at least the magnitude of Sagittarius, and probably down to the magnitude of Fornax (see Table~\ref{tab:satellites}), the Milky Way and M31 are not significant outliers in terms of their respective numbers of satellites. As discussed again in section~\ref{sec:future}, in the next several years these measurements will significantly improve because of new larger scale galaxy surveys, thereby allowing us to study the missing satellites issue in a complete statistical framework. 

\bigskip 

$\bullet$ {\underline {Gravitational Lensing}}

\bigskip 

\par Strong gravitational lensing is an interesting technique that may be utilized to measure the distribution of substructure in galaxies and to study the nature of dark matter. In a strong lensing system, there are two basic observables, the positions of the images that are created, and the ratios of the fluxes between the images. If the lensing potential is smooth, then relations can be established between the fluxes of the separate images. A deviation from a smooth model for the potential may then reveal the presence of a dark matter subhalo, even though the galaxy associated with it is not identified in optical surveys. 

\par The largest sample of strong lensed systems comes from the Cosmic Lens All-Sky Survey (CLASS)~\citep{Myers2003}. CLASS identifies lenses where multiple images of background radio sources are formed; radio sources are beneficial because they are free from dust extinction and they do not suffer from stellar microlensing. Interestingly, almost all of these lensing systems have images that violate the expected flux ratios for a smooth potential model. Several authors have argued that substructure, as is expected in LCDM models, causes the flux ratios to differ than expected in the smooth model~\citep{Mao:1997ek,Metcalf:2001es}. 

\par Utilizing simulations of dark matter halos that have similar properties to the lensing galaxies observed, several authors studied whether it is possible to reproduce the flux ratios in the LCDM cosmological model~\citep{Xu:2009ch,Xu:2011ru}. The general conclusion of these studies is that there is {\em too little} substructure in LCDM halos to account for the observed flux anomalies. These authors also find that, since most of the lensing signal is due to subhalos greater than approximately $10^6$ $M_\odot$, extrapolating the mass function to lower subhalo masses is unlikely to affect the conclusion. The addition of baryons in the subhalos, and the corresponding contraction in the dark matter distribution that makes the subhalos more dense, also does not significantly affect the results. Current studies are now focused on the contamination of low mass galaxies along the line-of-sight towards the lensing system. 

\par Detailed modeling of individual lensing systems now uncover galaxies that are too faint to be seen in optical light.~\cite{Vegetti:2012mc} report the detection of a $\sim 10^8$ $M_\odot$ mass satellite in the Einstein ring lens system JVASB1938+666, which is an elliptical galaxy at redshift $z=0.88$ that is lensing a bright infrared background source galaxy at redshift $2.059$. This follows up on the eariler detection of a $\sim 3 \times 10^9$ $M_\odot$ substructure associated with the lens system using a Bayesian reconstruction method~\citep{Vegetti:2009cz}. In the former case, the authors utilize this detection to make the first measurement of the substructure mass function beyond the Local Group, finding that $dN/dM \propto M^{-\alpha}$, with $\alpha = 1.1_{-0.4}^{+0.6}$. They determine the fraction of the total mass in substructure to be $\sim 3.3\%$. Given the uncertain mapping between mass and luminosity at this mass scale, it is difficult at this stage to predict the luminosity of this galaxy, though it does sit in the same mass range as the classical dSphs of the Milky Way. 

\par Assimilating the results presented above, direct satellite counting experiments seem to be indicating that the Milky Way is relatively normal amongst galaxies of its kind in terms of its population of satellites. While promsing, the gravitational lensing observations are still probably too sparse at this stage to deduce more general conclusions about the population of dark matter substructure around galaxies. Taking the former results at face value, the evidence is that the satellite abundance issue points to an inadequacy of modern theory, either in the understanding of galaxy formation, the properties dark matter, the standard cosmological model, or some combination. 

\paragraph{Variation of the LCDM Model} 

\par Finally, it is interesting to ask in more detail what modifications to the physics of dark matter provide the best solution to this problem. As mentioned above, the most basic solution involves giving the dark matter larger velocities at freeze-out than in the CDM framework. In this case, the formation of dark matter halos with mass scale below the corresponding free-streaming scale of the particle is significantly suppressed. These particles may be produced thermally or non-thermally, and well-motivated models typically require the particle mass to be approximately the keV scale~\citep{Abazajian:2005gj,Abazajian:2012ys}. Recent N-body simulations of Milky Way-mass galaxies find that a particle of this mass is effective in reducing both the total population of Milky Way subhalos and their central densities. As a result, the most massive dark matter subhalos are now able to host the observed dSph satellite galaxies, as indicated in the right hand panel of Figure~\ref{fig:RrVr}~\citep{Lovell:2011rd}. A second class of models involves introducing a larger-than-standard self-interaction cross section, as discussed in Section~\ref{sec:problem}~\citep{Spergel:1999mh,Loeb:2010gj}. Several groups have shown via N-body simulations that subhalos in this model are also less dense, and it is also more likely for the dSphs to populate the most massive subhalos of the Milky Way~\citep{Vogelsberger:2012ku,Rocha:2012jg,Peter:2012jh}. 

\subsection{The mass of the Milky Way} 

\par Very recently, an additional aspect of the missing satellites issue has come to the forefront that is related to our uncertainty on the mass of the Milky Way. Specifically, if the Milky Way mass is approximately $10^{12}$ $M_\odot$, the number of subhalos with maximum circular velocity $\gtrsim 30$ km/s can be made consistent with observations~\citep{Wang:2012sv}. Though this ``low Milky Way mass" solution is appealing, there are still some aspects to it that require further investigation. One of which is that reducing the mass of the Milky Way to this level seems to make the existence of the Magellanic Clouds even less probable than the $\sim 5\%$ chance discussed above. Further direct measurements are clearly required. 

\par Cosmological simulations discussed in this section have been utilized to estimate the mass of the Milky Way. These estimates rely on identifying a Milky Way or Local Group-like system in a large volume cosmological simulation, and then applying a version of the ``timing argument"~\citep{Kahn1959}. The mass of the Local Group is determined by identifying Milky Way-M31 like systems in the simulation, while the mass of the Milky Way is determined by identifying Milky Way-Leo I like systems. Using the millenimum simulation,~\cite{Li:2007eg} use this method to determine a total Local Group mass of approximately $5 \times 10^{12}$ $M_\odot$, and a 95\% c.l. lower limit on the Milky Way mass of $8 \times 10^{11}$ $M_\odot$, which is consistent with the low Milky Way mass solution to the satellite abundance problem. This result is compared to the mass measurements from the kinematics of halo stars, globular clusters, and dSphs in section~\ref{tab:satellites}. 

\par The mass of the Milky Way may also be estimated from properties of satellites other than Leo I. ~\cite{Busha:2010sg} recently construct the probability distribution for the Milky Way mass by locating the number of dark matter halos that have two satellites that are as massive as the Magellanic Clouds. Their results indicate a 68\% c.l. lower limit on the mass of approximately $5 \times 10^{11}$ $M_\odot$, which is also both consistent with many other mass measurements and the low-mass solution. While it would be ideal to extend this analysis to less massive satellites, both the lack of resolution in large scale N-body simulations and the uncertainty in the statistical mapping between the luminosity of a satellite and its dark matter mass preclude an analysis at this time. 

\subsection{Caustics} 
\par This section closes out with a brief discussion of an additional property of dark matter halos that is intriguing from both the perspective of their phase space properties and for the detection of particle dark matter. For cold and collisionless dark matter, since the phase space density is conserved along the particle trajectory, the collapse of halos is expected to give rise to caustics. For completely cold particles with zero velocity dispersion, particles effectively occupy a three-dimensional sheet in the total six-dimensional phase space, and the projection of the sheet into configuration space results in infinite density singularities. In reality these singularities are smoothed out because particles are cold and do not have precisely zero velocity dispersion. 

\par Caustics have been studied a great deal in the context of the spherical infall model~\citep{Fillmore:1984wk,Bertschinger:1985pd}. Within the context of hierarchical structure formation, caustics are also expected to form as a tidal stream folds back upon itself. It is not yet possible to resolve caustics in modern N-body simulations due to numerical resolution~\citep{Diemand:2008gf}, however analytical schemes have been developed that allow for the prediction of annihilation radiation from caustics~\citep{Hogan:2001xa}, including the smoothing of them through the finite velocity dispersion~\citep{White:2008as}. The issue of detection of annihilation radiation from caustics will be addressed again in the next section, in particular comparing the expected signal to the more often considered signals from the smooth halo and from halo substructure.

\newpage 
\section{Indirect Detection: Theoretical Developments}
\label{sec:indirect_detection_theory} 

\par As discussed in Section~\ref{sec:problem}, particles with roughly weak scale interactions that were in thermal equilibrium in the early universe may decouple with an abundance to constitute much if not all of the observed dark matter in the universe today. Via the same processes that set their relic abundance, dark matter particles annihilate today in regions of high density, for example in the centers of galactic halos over a wide range of mass scales. The particles that are produced in the annihilation processes may be detectable through a variety of modern ground and space-based observatories (Section~\ref{sec:indirect_detection_experiment}). 

\par Discussion of detection of standard model particles produced in the annihilation of dark matter dates back more than three decades~\citep{Gunn:1978gr}. Since then, the idea that particle dark matter may be detected through astrophysical observations has inspired an immense amount of theoretical work. From the theoretical perspective, the most pertinent questions are: Where is the dark matter? What standard model particles does it produce, and what is the rate of standard model particle production? The answer to the former question may be addressed through the observations and analysis methods discuss in Sections~\ref{sec:MW}-~\ref{sec:simulations}. The latter question is addressed by appealing to a specific particle physics framework, several of which are introduced for the first time in this article in this section. 

\par On the theoretical side there has also been interest in models beyond the standard WIMP framework. For example, dark matter particles that are unstable, such as sterile neutrinos or lightest particles in supersymmetric theories that violate R-parity, provide unique phenomenological signatures both from the perspective of astrophysics and particle physics. It has also been understood that dark matter may be detected indirectly in searches over  a wider spectrum of energies than are considered in searches for standard WIMPs.

\par This section reviews recent theoretical developments in the area of indirect dark matter detection, bringing together the astrophysical and particle physics frameworks that are the most relevant for interpretation of modern experimental results. In introducing a representative set of particle models, the goal is to make the most straightforward possible connection between the cross section that is relevant for dark matter production in the early universe and the cross section that is most relevant for detecting dark matter via indirect searches. 

\subsection{Dark matter annihilation}
 
\par Along a direction $\ell$, the observed flux of photons from dark matter annihilation above an energy $E_{min}$ per solid angle is 
\be
\frac{d{\cal L}(\Psi)}{d\Omega}= N_\gamma \frac{1}{2} \frac{\sigmavavg}{\mdm^2}  \frac{1}{4\pi} \int_{\ell_{-}}^{\ell_{+}} \rho^2 [r(\ell)] d\ell,
\label{eq:indirect_detection_flux_omega}
\ee
where $\Psi$ represents the angular separation from the center of the galaxy, $D$ is the distance to the center of the galaxy, so that $r^2 = \ell^2 + D^2 - 2 \ell D\cos \Psi$. The upper and lower boundaries to the integral are $\ell_\pm = D \cos \Psi \pm \sqrt{r_t^2 - D^2 \sin^2 \Psi}$, where $r_t$ is the tidal radius of the dark matter halo. The number of photons produced is the integral over the energy spectrum of decay photons, 
\be
N_\gamma = \int_{E_{min}}^{\mdm} \frac{dN}{dE} dE. 
\label{eq:Ngamma} 
\ee
Equation~\ref{eq:indirect_detection_flux_total} neatly separates into a component that depends on the properties of the dark matter particle, $N_\gamma \sigmavavg/(2 \mdm^2)$, and a component that depends on the distribution of the dark matter in the halo. Integrating over a solid angle $\Delta \Omega$ gives the total flux within the solid angle
\be 
{\cal L}({\Delta \Omega}) = N_\gamma \frac{1}{2} \frac{\sigmavavg}{\mdm^2} \frac{1}{4\pi} 2\pi
\int_0^{\Delta \Omega} \sin \Psi d\Psi \int_{\ell_{-}}^{\ell_{+}} \rho^2 [r(\ell)] d\ell.
\label{eq:indirect_detection_flux_total}
\ee
Explicitly separating out the component that depends on the dark matter distribution we have
\be 
J(\Delta \Omega) = \frac{1}{2} \int_0^{\Delta \Omega}  \sin \Psi d\Psi \int_{\ell_{-}}^{\ell_{+}} \rho^2 [r(\ell)] d\ell.
\label{eq:Jvalue}
\ee
Equation~\ref{eq:Jvalue} will be referred to here as the ``$J$-value."

\par Annihilation radiation is expected to arise from three primary components of a halo: the smooth component, subhalos, and caustics. The smooth component is straightforward to model under the assumption of a dark matter density profile (e.g. Equations~\ref{eq:NFW} and~\ref{eq:zhao}). The contribution from halo substructure and caustics requires a more detailed theoretical treatment. 
 
\bigskip

$\bullet$ {\underline{Substructure} }

\bigskip

\par If there is significant substructure in the dark matter distribution, as is predicted by N-body simulations, there is a corresponding ``boost" to the predicted flux over this smooth component~\footnote{Note that this factor is to be distinguished from increases in the annihilation cross section, such as from Sommerfeld enhancements}. There are several papers in the literature that provide a theoretical treatment of the substructure boost factor~\citep{Strigari:2006rd,Pieri:2007ir,Martinez:2009jh,Charbonnier:2011ft}, which complement the results derived from N-body simulations above their resolution limit~\citep{Springel:2008zz,Kuhlen:2008aw}. The following discussion closely follows the methods outlined in~\cite{Strigari:2006rd} and~\cite{Martinez:2009jh}.

\par Start from the definition of the luminosity of a source in Equation~\ref{eq:indirect_detection_flux_total}. Define the total flux from a galaxy, including the substructure boost factor, as
\be 
{\cal L}(M_h) = [1+B(M_h,M_{min})] \tilde {\cal L}(M_h), 
\label{eq:boost_definition}
\ee
where $B$ is defined as the boost factor, which is zero for a totally smooth dark matter distribution, and $\tilde {\cal L}$ is the contribution from the smooth halo. Here the boost is defined to depend on the total halo mass, $M_h$, and the minimum mass in the substructure hierarchy, $M_{min}$. While the former quantity can be estimated from observations of a given source, the later is of course only weakly theoretically-constrained. 

\par Due to the mass-dependence of the radial distribution of substructure discussed in Section~\ref{sec:simulations}, there is predicted to be a radial dependence of the substructure boost factor, which depends strongly on how efficiently subhalos of a given mass are destroyed by tidal interactions within the host halo. Simulations show that the substructure boost factor increases towards the virial radius of the halo~\citep{Springel:2008zz}, reflecting the distribution of the most massive subhalos in a host galaxy. Smaller mass subhalos contribute progressively more to the annihilation luminosity in the inner regions, though because of the resolution of modern simulations at this time the boost factor within the approximate  scale radius of a galaxy can only be determined from extrapolation of the results in the outer region of the host galaxy. Because of this theoretical uncertainty in the radial dependence of the boost factor for subhalos below the mass limit of modern simulations, and also because the angular resolution of modern experiments implies that they are likely not sensitive to the radial dependence of the boost factor (except perhaps for a couple of targets such as M31 and nearby clusters), the primary calculations in this section focus on the total enhancement from subhalos integrated over the entire primary halo. 

\par To determine the value of the boost, annihilation luminosity factors ${\cal L}(m)$ for subhalos within mass $m$ of the host halo are integrated, $B{\cal L}(M_h) = \int (dN/dm) {\cal L}(m) dm$. With the above assumptions for no radial dependence of the boost factor, from the definition in Equation~\ref{eq:boost_definition} this boost factor can be explicitly written out as
\begin{eqnarray}
B(M_h) & = & \frac{1}{{\cal L}(M_h)} \int_{M_{min}}^{{q M_h} } \frac{dN}{d m}  {\cal L}(m) d m \nonumber \\ 
& = & \frac{1}{{\cal L}(M_h)} \int_{M_{min}}^{{ q M_h} } \frac{dN}{d m} [1 + B(m)] \tilde {\cal L} (m) d  m \nonumber \\ 
& = & \frac{AM_h^{\alpha^\prime}}{ {\cal L} (M_h)} \int_{M_{min}}^{ {q M_h} } [1 + B(m)] \tilde {\cal L} (m) \, m^{-1-\alpha^\prime} dm. 
\label{eq:boost_integrated}
\end{eqnarray}
where a subhalo mass function of $dN/d \ln m = A(M_h/m)^{\alpha^\prime}$ is assumed, and $q$ reflects the fact that the subhalo mass function continues up only to a fraction of the mass of the host galaxy, typically $q \simeq 0.1$. From the results in Section~\ref{sec:simulations}, $\alpha^\prime = 0.9$. For a fixed $\alpha^\prime$, the normalization factor $A$ is determined by demanding that a fraction of the total mass in substructure is within a given mass range. 

\par To proceed in the evaluation of Equation~\ref{eq:boost_integrated} we need a model to relate the luminosity of a subhalo to its mass. For the purposes of this analysis approximate this ${\cal L}-m$ relation as an analytic relation, though its probably more realistic that there is scatter in this relation that can ultimately only be probed through higher resolution numerical simulations~\citep{Springel:2008zz,Kuhlen:2008aw} or Monte Carlo analysis~\citep{Martinez:2009jh}. Motivated by the concentration mass relation (e.g.~\cite{Bullock:2001}), the subhalo luminosity-mass relation has been shown to be well-approximated by a power-law relation, ${\cal L} \propto M^{0.87}$~\citep{Strigari:2006rd,Kuhlen:2008aw}. 

\par It is clear to see that the substructure boost factor is a strong function of the halo mass $M_h$. For example, for a typical galaxy cluster, the boost factor may be in the range $B \simeq 10^2 - 10^3$ depending on the minimum mass cut. This significantly improves the detection prospects of clusters. For galaxies about as massive as the Milky Way, such as M31, for an extrapolation of $M_{min}$ down to Earth mass, $B \sim 10^2$. On the other hand, for dSphs, $B$ is likely only a factor of few at the most, which in the context of the above calculation is a reflection of the small host halo mass for dSphs. If the radial distribution of the boost is also considered, it is also likely that the boost is insignificant for dSphs, because it is plausible that subhalos in the outskirts of the dSphs are efficiently stripped off by the Milky Way halo. However at this stage these the arguments are heuristic because theoretical knowledge of dark matter substructure in Milky Way-mass galaxies, much less sub-substructure distributions that may be present in dSphs, is subject to significant uncertainty. Though the above arguments do not improve the detection prospects of dSphs, they do motivate more of a clean interpretation of the annihilation cross section limits that are derived from the data, which are discussed in Section~\ref{sec:indirect_detection_experiment}.

\bigskip

$\bullet$ { \underline{Caustics} }

\bigskip 

\par As discussed in Section~\ref{sec:simulations} in the self-similar halo model caustics will form in regions where the mapping between phase space and physical space is singular. Caustics are also expected to form when a satellite is tidally-destroyed upon accretion into a parent halo. Because of the high density achieved at caustics, the dark matter annihilation rate is expected to be enhanced relative to the smooth dark matter halo contribution. This enhancement can be quantified by noting that, on one side of the caustic, the density diverges as $\rho \sim A/\sqrt{x}$, where $x$ is the distance from the center of the caustic and $A$ is the fold coefficient. 

\par Though the above relation shows that the density diverges for purely cold (i.e. zero velocity dispersion) particles, in practice the dark matter is not purely cold but has a finite velocity dispersion. For a WIMP of mass 100 GeV, conservation of phase space density gives the velocity dispersion today for particles that are just infalling into the halo as $\sim 0.03$ cm/s. This finite velocity dispersion cuts off the divergence of the caustic, so that the dark matter annihilation rate per unit surface area is $\int \rho(\ell) d\ell \simeq \int_{\delta \ell}^R A^2/\ell \propto \ln (R/\delta \ell)$, where $R$ is the radius of the halo. This predicts that the increase in the annihilation rate over the smooth halo due to the caustic is a factor of several~\citep{Hogan:2001xa,Natarajan:2007tk}. 

\subsection{Annihilation cross section models} 
\par With the input from astrophysics in place for calculating the gamma-ray flux from dark matter annihilation, it is next necessary to evaluate the annihilation cross section in some representative particle models. On general theoretical grounds, particle dark matter candidates can be classified according to their intrinsic spin.  Scalar particles have spin 0 and fermions have spin $1/2$ (Higher spin dark matter particle models have also been studied~\citep{Barger:2008qd,Yu:2011by}). Fermion dark matter candidates can be further classified as either Dirac or Majorana particles, which differ from one another in the interactions that are allowed. With these definitions interaction operators can be constructed between the dark matter and standard model particles that transform as odd or even under charge, parity, and time reversal transformations. This approach is valid under some simplifying theoretical assumptions, most importantly that the dark matter and standard model interactions are mediated by new and unknown physics above a certain energy scale. Models along these lines are sometimes called effective theory models, and have recently been examined in the context of WIMP direct and indirect detection by several authors~\citep{Beltran:2008xg,Cao:2009uw,Goodman:2010yf,Bai:2010hh,Zheng:2010js,Yu:2011by}.

\par To define a representative set of models, assume that the WIMP couples only to standard model fermions. For a scalar WIMP, the interactions between the WIMP, represented by the field $\phi$, and a fermion, represented by $f$, are
\begin{eqnarray}
\textrm{Scalar} &&\frac{F_S}{\sqrt 2} \bar \phi \phi \bar ff,
\label{eq:scalar:scal} \\
\textrm{Vector} &&\frac{F_V}{\sqrt 2} \bar \phi \overleftrightarrow
{\partial_\mu}\phi \bar f \gamma^\mu f,
\label{eq:scalar:vect} \\
\textrm{Scalar-Pseudoscalar} &&\frac{F_{SP}}{\sqrt 2} \bar \phi \phi \bar f \gamma_5 f,
\label{eq:scalar:s_Ps} \\
\textrm{Vector-axialvector} &&\frac{F_{VA}}{\sqrt 2}(\bar \phi \overleftrightarrow
{\partial_\mu}\phi) \bar f \gamma^\mu \gamma_5 f. 
\label{eq:scalar:v_Av}
\end{eqnarray}
For fermionic WIMPs, represented by $\chi$, the corresponding interactions are
\begin{eqnarray}
\label{eq:fermionic}
\textrm{Scalar} && \frac{G_S}{\sqrt{2}}
\bar \chi \chi \bar f f 
\label{eq:fermion:scal} \\
\textrm{Pseudoscalar} &&\frac{G_P}{\sqrt{2}} 
\bar \chi \gamma^5 \chi \bar f \gamma_5 f \\
\textrm{Vector} &&\frac{G_V}{\sqrt{2}}
\bar \chi \gamma^{\mu} \chi \bar f \gamma_{\mu} f 
\label{eq:fermion:vector}\\
\textrm{Axialvector} &&\frac{G_A}{\sqrt{2}}
\bar \chi \gamma^{\mu}\gamma^5 \chi \bar f \gamma_{\mu} \gamma_5 f 
\label{eq:fermion:axial} \\
\textrm{Tensor} &&\frac{G_T}{\sqrt{2}}
\bar \chi \sigma^{\mu \nu} \chi \bar f \sigma_{\mu \nu} f.
\end{eqnarray}
Note that the various couplings that appear in these equations have different dimensions, and in general the couplings to fermions could be universal (independent of fermion mass), or dependent on the mass of the standard model fermion. It is also possible to define mixed interactions for fermionic dark matter, i.e. scalar-pseudoscalar interactions of the form $\bar \chi \chi \bar f \gamma_5 f$, in which the lagrangian is invariant under the combination of charge, parity, and time reversal transformations. 

\par From the interactions above it is possible to define the annihilation cross sections, and expand them as $\sigmav \simeq a + b v^2$. For scalar dark matter these are given by~\citep{Beltran:2008xg,Zheng:2010js,Yu:2011by}
\bea
(\sigmav)_S &\simeq& \frac{1}{4\pi}
\left(\frac{F_{\mathrm{S}}}{\sqrt 2}\right)^2 c_f
\left(1-\frac{m_f^2}{\mdm^2}\right)^{3/2}
\label{eq:sv_scalar_scal}\\
(\sigmav)_V &\simeq& \frac{1}{2\pi}
\left(\frac{F_{\mathrm{V}}}{\sqrt 2}\right)^2 c_f  \mdm^2 \frac{1}{3}
\sqrt{1-\frac{m_f^2}{\mdm^2}} \left(2+\frac{m_f^2}{\mdm^2} \right)v^2,
\label{eq:sv_scalar_vect}\\
(\sigmav)_{SP} &\simeq& \frac{1}{4\pi}
\left(\frac{F_{\mathrm{SP}}}{\sqrt 2}\right)^2 c_f
\sqrt{1-\frac{m_f^2}{\mdm^2}}
\label{eq:sv_scalar_s_Ps}\\
(\sigmav)_{VA} &\simeq& \frac{1}{\pi}
\left(\frac{F_{\mathrm{VA}}}{\sqrt 2}\right)^2 c_f
\frac{1}{3} \mdm^2 \left(1-\frac{m_f^2}{\mdm^2} \right)^{3/2} v^2  
\label{eq:sv_scalar_v_Av}
\eea
and for fermonic dark matter, 
\bea
(\sigmav)_S&\simeq&\frac{1}{8\pi}
\bigg(\frac{G_S}{\sqrt{2}}\bigg)^2c_f
\bigg(1-\frac{m_f^2}{\mdm^2}\bigg)^{3/2}
\mdm^2v^2\label{sigma_v-S}\\
 (\sigmav)_P &\simeq&\frac{1}{2\pi}
\bigg(\frac{G_P}{\sqrt{2}}\bigg)^2c_f
\sqrt{1-\frac{m_f^2}{\mdm^2}}\mdm^2 
\label{eq:sigma_v-P}\\
(\sigmav)_V&\simeq&\frac{1}{2\pi}
\bigg(\frac{G_V}{\sqrt{2}}\bigg)^2c_f\sqrt{1-\frac{m_f^2}{\mdm^2}}
(2M_\chi^2+m_f^2) 
\label{eq:sigma_v-V}\\
 (\sigmav)_A &\simeq&\frac{1}{2\pi}
\bigg(\frac{G_A}{\sqrt{2}}\bigg)^2c_f
m_f^2 \left( \frac{8 \mdm^2/m_f^2 - 28 + 23 m_f^2/\mdm^2}{24\sqrt{1 - m_f^2/\mdm}} v^2 \right)
\label{eq:sigma_v-A}\\
(\sigmav)_T &\simeq&\frac{2}{\pi}
\bigg(\frac{G_T}{\sqrt{2}}\bigg)^2c_f
\sqrt{1-\frac{m_f^2}{\mdm^2}}(\mdm^2+2m_f^2) 
\label{eq:sigma_v-T}\\
\eea
In the formulae above for the annihilation cross section, only leading order terms ($a$ or $b$) are shown. The color factor is $c_f = 3$ for quarks and $c_f = 1$ for leptons. The above expressions show that scalar dark matter with scalar interaction and psuedo-scalar interactions with fermions annihilate through s-wave interactions, while scalar dark matter with vector and vector-axialvector interactions with fermions annihilate through p-wave interactions. Fermionic dark matter with pseudoscalar, vector, and tensor interactions annihilate through s-wave interactions, and fermonic dark matter with scalar or axial vector interactions annihilate through p-wave interactions. For majorana fermions, the vector current interaction vanishes. 

\par The above models represent only a subset of effective models because they focus on interactions with fermions. Models similar in spirit with couplings to gauge bosons are considered in, e.g.~\cite{Cao:2009uw}. A simple model that is related to those presented above is the singlet scalar model, in which a single dark matter particle represents the only beyond the standard model physics~\citep{McDonald:1993ex,Burgess:2000yq}. Depending on its mass, the dark matter can annihilate either through a Higgs, gauge bosons, or fermions. For each of these different models, the annihilation cross section is
\be 
\sigmav = \frac{16\lambda^2 v_{ew}^2}{(4\mdm^2 - m_H^2)^2 + \Gamma_H^2 m_H^2}\frac{\sum_\imath \Gamma(\tilde h \rightarrow X_\imath)}{2\mdm},
\label{eq:sigmav_darkon} 
\ee
where $v_{ew}$ is the vacuum expectation value of the Higgs, $m_H$ is the mass of the Higgs, $\Gamma_h$ is the total Higgs decay width, and $\Gamma(\tilde h \rightarrow X_\imath)$ is the partial decay width into the final state $X$. Note that, in contrast to the scalar WIMP with scalar interactions with fermions in Equation~\ref{eq:sv_scalar_scal}, in the singlet scalar model the annihilation cross section to fermions is proportional to the fermion mass because of the coupling to the Higgs. 

\par In supersymmetric models, neutralinos can annihilate into fermions through exchanges of Higgs, $Z^0$ bosons, and through sfermions. In all these cases the annihilation cross section is proportional to the mass of the outgoing fermion. There is a significant amount of literature devoted to computing neutralino annihilation cross sections; formulae for the annihilation cross section for neutralinos into fermions, as well as for gauge bosons and Higgs, have been compiled in~\cite{Bertone:2004pz}. In addition to the continuum spectra, neutralinos may also annihilate to a two-photon final state or a final state with a photon and a $Z^0$. In the later case, the energy of the photon is mono-chromatic, with an energy given by
\be 
E_\gamma = \mdm \left( 1- \frac{m_{Z}^2}{4 \mdm^2} \right). 
\label{eq:energy_line} 
\ee
The annihilation cross sections for these processes are dominated by the s-wave contributions, and for the neutralino parameter space maximize at approximately $10^{-29}$ cm$^{3}$ s$^{-1}$~\citep{Bergstrom:1997fh,Ullio:1997ke,Bern:1997ng}, nearly three orders of magnitude below the thermal relic scale. 

\par To calculate the relic density of dark matter and the flux from particles in halos today, the annihilation cross sections must be averaged over the velocity distribution, $\sigmavavg$. For the relic abundance calculation, this is performed via the methods outlined in, for example,~\cite{Gondolo:1990dk},~\cite{Griest:1990kh}, and~\cite{Edsjo:1997bg}. If the annihilation cross section does not depend on the relative velocity ($a \gg b$), as is the case for scalar dark matter with scalar and pseudoscalar interactions, and fermionic dark matter with pseudoscalar, vector, axialvector, and tensor interactions, then the averaged annihilation cross section in the early universe is the same as it is in halos today. In the opposite limit ($b \gg a)$, the annihilation cross section is lower in halos today than it was in the early universe, because the typical velocity in a halo is $v/c \simeq 10^{-3}$. In a halo the averaged annihilation cross section depends on the assumed halo density profile and velocity distribution, and also on the position within the halo~\citep{Campbell:2010xc,Frandsen:2012db}.

\subsection{Photon and charged particle production} 

\par The detectability of a model characterized by the annihilation cross sections presented above depends on the spectrum of photons produced through the decay or fragmentation of the fermions or gauge bosons in the final state. The main photon production channel is through production and subsequent decay of neutral pions. The photon spectra are typically determined numerically through programs such as Pythia 6.4.19~\citep{Sjostrand:2006za} and there are fitting formulae that describe the output~\citep{Fornengo:2004kj,Cembranos:2010dm}. For continuum spectra, the number of photons produced in the annihilation peak at an energy approximately an order of magnitude below the dark matter mass. 

\par In addition to continuum spectra, for dark matter that annihilates into final state fermions, photons from final state radiation (FSR) are produced~\citep{Beacom:2004pe,Bergstrom:2005ss}. FSR photons are produced whenever there are charged particles in the final state. The spectrum of photons from FSR is distinct from the line spectrum and the continuum spectrum in that it peaks near the mass of the dark matter~\citep{Birkedal:2005ep}. When there is a fermion in the final state, the photon energy spectrum from final state radiation is~\citep{Essig:2009jx}
\be 
\frac{dN_\gamma}{dy} = \frac{\alpha}{\pi} \left[ \frac{1+(1-y)^2}{y}\right] \left[ \ln \frac{s(1-y)}{m_f^2} - 1 \right].
\label{eq:FSR} 
\ee
Here $\alpha = 1/137$, $y = E_\gamma/\mdm$, $s = 4 \mdm^2$, and the maximum photon energy is $\mdm$. Note that for FSR, the number of photons produced is suppressed by a factor of $\alpha$ relative to the continuum emission.

\par If electrons and positrons are produced from the dark matter annihilation process, secondary photons that are produced by these $e^+ e^-$ pairs may also be detectable. The electron-positron population is non-thermal, and loses energy through a variety of physical processes, including: synchrotron from magnetic fields, inverse Compton radiation from interaction with both starlight and the cosmic microwave background, and classic bremsstrahlung via interactions with interstellar gas. Theoretical predictions for the corresponding photon fluxes, as well as the fluxes of electrons, positrons, and protons, are reviewed in~\cite{Cirelli:2010xx} and~\cite{Profumo:2010ya}. 

\subsection{Dark matter decay} 
\par In Section~\ref{sec:indirect_detection_experiment}, experiments that are sensitive to photons produced from particle decays will also be discussed. Though there is no scale corresponding to the preferred thermal relic scale in the case of annihilations to preferentially set the abundance of dark matter, a large class of well-motivated models predict observable signals from dark matter decays~\citep{Arvanitaki:2009yb,Abazajian:2012ys}. 

\par For decaying dark matter models, the flux corresponding to Equation~\ref{eq:indirect_detection_flux_total} is 
\be
\frac{d{\cal F}(\Psi)}{d\Omega} = N_\gamma \frac{1}{\mdm \tau} \frac{1}{4\pi} \int_{\ell_{-}}^{\ell_{+}} \rho [r(\ell)] d\ell,
\label{eq:flux_decay}
\ee
where $\tau$ is the lifetime of the particle. As in the case of annihilation, Equation~\ref{eq:flux_decay} nicely separates into a piece that depends on the properties of the dark matter particle, $1/\mdm \tau$, and the remainder which depends on the properties of the host dark matter halo. Because in this case the flux is proportional to the density and not the square of the density, boost factors from halo substructure are not relevant. The total flux within a solid angle $\Delta \Omega$ can be then calculated in a manner similar to Equation~\ref{eq:Jvalue}. 

\par The fact that the signal scales as the density and not as the density squared implies that the signals are more spatially-extended. The decaying dark matter signal from our Galaxy and from nearby galaxies is rivaled in some instances by the signal from diffuse background radiation from extragalactic sources, which in many instances constitute a background to the signal that is to be observed~\citep{Cembranos:2007fj,Cirelli:2012ut}. 

\subsection{Neutrinos} 
\par In addition to the above discussed signatures from high energy photons, indirect signals of dark matter annihilating or decaying may be detected through other channels. One such interesting channel is neutrinos. By their nature, neutrinos are much more difficult to detect than photons, however, they represent a complementary probe to photons for two reasons. First, they are produced similar to photons in the fragmentation or hadronization products of annihilating dark matter. Second, they propagate from the source without interacting, so they are able to probe dark matter produced in different environments than photons. There are two primary methods utilized to search for neutrinos from dark matter annihilation. First, through their production in the Galaxy and external galaxies as discussed above. Second, they are produced by dark matter that gets captured and annihilates in the Earth and the Sun. Since the astrophysical phenomenology of neutrinos produced in the Galactic halo or external galaxies is similar to that discussed above in the context of gamma-rays, the remainder of this section briefly reviews the production of neutrinos from dark matter that is captured in the Sun or the Earth, anticipating results to be discussed in Section~\ref{sec:direct_detection_experiment}. 

\par The capture rate, $C$, in a differential volume element of the Sun can be written as~\citep{Gould:1987ir}, 
\be
\frac{dC}{dV} = n \int_0^{u_{max}} du \frac{f(\vec u)}{u} w^2 
\int_{E_{min}}^{E_{max}} dE \frac{d\sigma}{dE}, 
\label{eq:capture} 
\ee
where $n$ is the number density of dark matter near the Sun, and $\sigma$ is the cross section for a dark matter particle interacting off of a nucleus. The velocity in a given shell, $w$, is related to the velocity at infinity $u$ and the escape velocity, $v_{esc}(r)$, in the shell at radius $r$, as $ w = \sqrt{u^2 + v_{esc}^2(r)}$. The upper limit to the integral in Equation~\ref{eq:capture} is defined as
\be 
u_{max} = \frac{2 \sqrt{\mdm m_A}}{\mdm - m_A} v_{esc},
\label{eq:umax} 
\ee
where $m_A$ represents the mass of a nuclear species. The upper and lower limits to the energy integral are, 
\bea 
E_{min} &=& \frac{1}{2} \, \mdm \, u^2 \nonumber \\
E_{max} &=&  \frac{4\mdm m_A}{(\mdm + m_A)^2}\frac{1}{2} \,\mdm \,w^2, 
\label{eq:e_neutrino}
\eea
where $E_{max}$ is the maximum energy for two-body elastic scattering, and $E_{min}$ is the minimum energy required for capture. The minimum energy corresponds to the capture of a WIMP into an orbit at any radius; requiring the WIMP to be captured within the radius of Jupiter changes the value of $E_{min}$~\citep{Kumar:2012uh}. 

\par The flux depends on a competition between the capture rate of WIMPs in the core of the Sun, determined from Equation~\ref{eq:capture}, and the annihilation rate times the relative velocity divided by the effective volume of the core of the Sun. For dark matter in equilibrium, the annihilation rate is one-half the capture rate. The time to reach equilibrium for dark matter in the Sun is
\be 
t_\odot \simeq 8.6 \times 10^{-2} \, \textrm{yr} \left[\frac{\Gamma_C}{\textrm{s}^{-1}}\right]^{1/2}
\left[\frac{\sigmav}{\textrm{pb}}\right]^{1/2}\left[\frac{\mdm}{10 \, \textrm{GeV}}\right]^{3/4},
\label{eq:equilibrium_sun}
\ee
where $\Gamma_C$ is the dark matter capture rate. Assuming that dark matter annihilation and capture are in equilibrium, neutrinos from the Sun are primarily sensitive to the scattering cross section. The spectrum of neutrinos produced per annihilation is generated from Pythia; fitting formula for dark matter annihilation into quarks, leptons, and gauge bosons are presented in~\cite{Cirelli:2005gh}. Low-energy neutrinos that are produced from annihilation and decay products that lose energy through interactions in the solar medium may also be detectable~\citep{Rott:2012qb,Bernal:2012qh}. 

\par From the perspective of Galactic astrophysics, the capture rate in the Sun is sensitive to the lowest velocity of WIMPs in the velocity distribution. Note that the situation is much different for capture in the Earth, which is more strongly affected by diffusion and interactions with the planets in the Solar System and acceleration from the Sun. This is particularly important at low velocity~\citep{Lundberg:2004dn}. This is in contrast to direct searches, which are primarily sensitive to the dark matter in the high velocity tail of the distribution. So, though the neutrino rates are a more complicated convolution between the cross section and the dark matter velocity distribution (Equation~\ref{eq:capture}), neutrino searches and direct dark matter searches also give complementary probes of the dark matter velocity distribution in the Solar neighborhood.

\newpage 
\section{Indirect Detection: Experimental Results}  
\label{sec:indirect_detection_experiment} 

\par The previous section has reviewed the theoretical aspects of indirect dark matter detection, highlighting the role of both astrophysics theory and particle dark matter theory.  The most exciting challenge confronting research in indirect dark matter detection involves understanding how to test theoretical models with modern experimental data. Coming at it from the other direction, it is also important to gain an understanding of what information can realistically be extracted from the modern experimental data sets that is most relevant for dark matter. With these general themes in mind, this section moves on to discussing the status indirect dark matter searches from a variety of modern astronomical data sets. 

\par Before going into details, it is necessary to take a step back and establish the scope of indirect dark matter searches. By their nature, these searches typically do not reflect ``controlled" experiments, in the sense that often it is difficult to simulate experimental backgrounds and get repeated measurements of them. In nearly all cases, backgrounds or foregrounds to the sought after signal result from physical processes that are poorly understood. 

\par Since indirect dark matter searches are typically undertaken by experiments motivated by other astrophysics topics outside of the realm of dark matter, it is useful to think generally about their strategies when it comes to searching for dark matter. A first type of search strategy is designed to identify dark matter from either a known type of target or from a distinct feature in the dark matter spectrum. In other words, we a priori know what type of signal we are looking for, and/or where to find it. There are two examples that illustrate this search strategy. First, gamma-ray studies of dSphs, which have no intrinsic gamma-ray backgrounds and relatively well-understood dark matter distributions. Also in this category are gamma-ray line searches in regions of expected high dark matter density. Since no known astrophysical process can produce sharp lines at energies of hundreds of MeV and above, this is a distinct spectral feature that is difficult to mimic without dark matter. A defining characteristic of these types of searches is that they are based upon existing theoretical predictions. For example in the case of the dSph searches they are based on determinations of the dark matter distributions from stellar kinematics, combined with predictions for the annihilation cross section from the freeze-out arguments. Or in the case of the line searches, they rely on the determination of the intrinsic strength of the line signal from a theoretical calculation. 

\par A second type of indirect search strategy relies on more of an empirical approach. It revolves around analysis of astronomical data sets, identification of features in these data sets that may be interpreted as a dark matter signal, and building a theoretical model around it. This line of thinking has given birth both to surprising experimental results and creative theoretical models, and is an important part of the process because it ultimately highlights our theoretical ignorance as to the nature of particle dark matter. It is also important because, as we have learned from many areas of astronomy in the past, the most ground breaking discoveries are often those not at all anticipated. An example that defines this category is the interpretation of the recent results from the PAMELA satellite, which finds a rise in the local positron fraction up to energies of hundreds of GeV. An additional more recent example is the possible existence of a feature in the Fermi-LAT data near the Galactic center at approximately 130 GeV. Though this has generated a great deal of theoretical interest, it was certainly not predicted a priori from a theoretical model. 

\par The primary motivation for the above discussion is that it provides a context to interpret the results of indirect dark matter searches, and allows us to make better sense of the myriad of results that now flood the literature in this topic. As for the results that are covered in this section, the discussion begins by focusing on the most updated results from gamma-ray observatories, in particular on the Fermi-LAT and modern Air Cerenkov Telescopes (ACTs). Since we now have a tremendous amount of data from these instruments over all regions of the sky, we are able to study essentially all classes of astrophysical sources from which a dark matter signal is expected. Following up on the gamma-ray analysis, the results and status of modern indirect searches that utilize neutrinos, cosmic-rays, and X-rays are discussed. This section closes with an interpretation of the most important experimental results, and the reach of future experiments in improving upon these results. 

\subsection{Overview of gamma-ray experiments: Fermi-LAT and ACTs}
\par The Fermi gamma-ray space telescope was launched in June 2008 into low Earth orbit~\citep{Atwood:2009ez}. It was designed as the successor of the successful EGRET mission, with an order of magnitude better energy and angular resolution than EGRET. Fermi consists of two experiments, the Large Area Telescope (LAT), which is the primary instrument aboard Fermi, and the gamma-ray burst monitor. The LAT is sensitive to photons in the range of approximately $20 \, \textrm{MeV} -300 \, \textrm{GeV}$, while the gamma-ray burst monitor is sensitive to photons in the range of hundreds of keV. This article will focus solely on the results obtained by the LAT. 

\par The LAT is an imaging, wide field-of-view pair conversion telescope that measures electron and position tracks that result from the pair conversion of an incident high-energy gamma-ray in converter foils~\citep{Atwood:2009ez}. In between the converter planes are silicon strip detectors that are sensitive to charged particles that measure the tracks that result from the pair conversion. The calorimeters measure the electromagnetic energy shower produced by the electron-positron pair and the development of the shower. The single photon on-axis angular resolution varies from approximately $0.1^\circ$ at energies greater than 10 GeV to approximately $3.5^\circ$ at 100 MeV. The energy resolution, $\Delta E/E$, over the energy range of interest in approximately 10\%. The effective area of the LAT is approximately $10^4$ cm$^2$. 

\par
\begin{figure*}
\begin{center}
\begin{tabular}{c}
\includegraphics[width=0.70\textwidth]{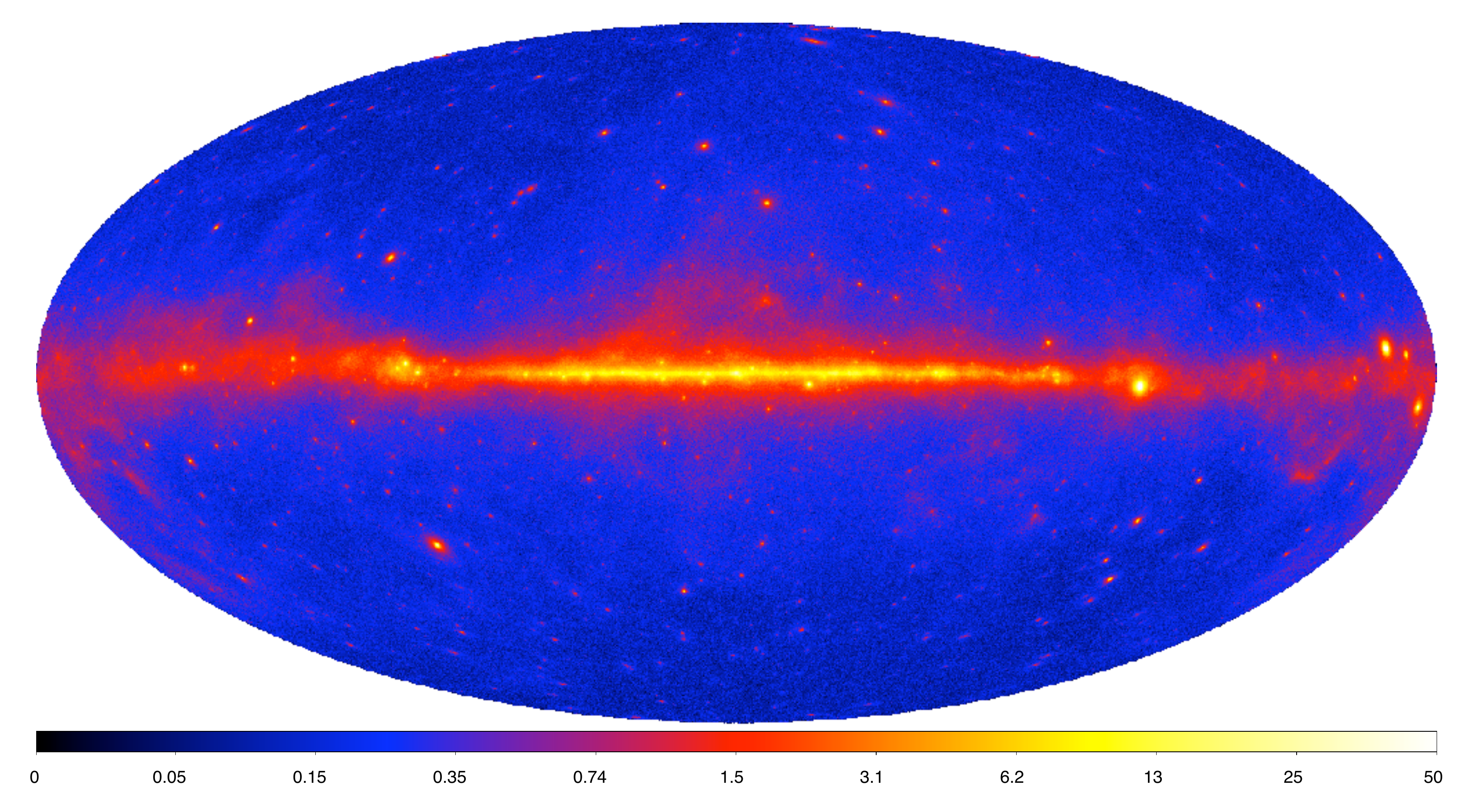} \\
\end{tabular}
\end{center}
\caption{Two year all sky Fermi-LAT map of the energy flux. The bottom legend is the flux in units of $10^{-7}$ cm$^{-2}$ s$^{-1}$. From~\cite{2012ApJS..199...31N}. 
}
\label{fig:fermi_allsky} 
\end{figure*}

 Figure~\ref{fig:fermi_allsky} shows an all-sky gamma-ray map in Galactic coordinates obtained from two years of Fermi-LAT data for photons in the energy range 100 MeV -300 GeV. Because of the steep nature of the power law spectrum of gamma-ray sources, most of the photons visible are in the energy regime of hundreds of MeV. As is clear from Figure~\ref{fig:fermi_allsky}, the brightest source of gamma-rays in the sky is the diffuse emission along the Galactic plane. Diffuse gamma-rays from the Galactic plane in the Fermi-LAT energy bands are created by cosmic ray interactions in the interstellar medium. Over the energy range of hundreds of MeV to tens of GeV, the dominant component of the gamma-rays results from $\pi_0$ decays. Above $\sim 1$ GeV, the next largest contribution to the diffuse gamma-ray flux arises from inverse Compton scattering of electrons from the interstellar radiation field. Below $\sim 1$ GeV, the flux from bremsstrahlung is near the flux from inverse Compton~\citep{Abdo:2009mr}. 

\par Bright sources in the Galactic plane, such as pulsars and supernova remnants, are clearly visible, as well as extragalactic sources at high latitude, in particular Active Galactic Nuclei (AGN). The two-year point source catalog of the Fermi-LAT contains over 1800 point sources, with over 100 of them firmly identified in other wavelengths, and over 1100 associated with sources at other wavelengths~\citep{2012ApJS..199...31N}. Firm identification with sources at other wavelengths is typically done through timing and spectral information. This leaves over 30\% of the point sources in the two-year catalog as unidentified at other wavelengths, with the fraction of unidentified sources increasing towards the Galactic plane, where confusion with diffuse Galactic radiation becomes an important issue. As discussed in more detail during the development of this section, understanding both the diffuse Galactic radiation and the properties of the sources in the two-year catalog will be important in our quest for dark matter signals. 

\par Complementing the Fermi-LAT searches for dark matter, there are now several ground-based imaging Air Cerenkov Telescopes (ACTs) that have rapidly improved in both size and sensitivity, giving them an important role in indirect dark matter searches. Generally, ACTs detect Cerenkov light from air showers that are produced in the atmosphere by cosmic rays. The shape and the orientation of the shower allows for a discrimination between those events generated by gamma-rays and those induced by other types of cosmic rays. The typical effective area for a modern ACT is much larger than that of the Fermi-LAT,  approximately $10^5$ m$^2$, and the angular resolution of approximately $0.1$ degrees over the appropriate energy regime is much better than that of the LAT. The lower energy thresholds are approximately hundreds of GeV, while the high energy threshold is much greater than that of the Fermi-LAT, up to tens of TeV. 

\par There are now hundreds of gamma-ray sources that have been detected and studied by modern ACTs, including blazars and pulsars. Most interestingly from the perspective of dark matter studies, ACTs are pointed instruments that are able to integrate for a long period of time on an individual dark matter source, as long as that source is within its field-of-view. However, relative to the LAT they are less sensitive to studies of the nature of the diffuse radiation over large regions of the sky. Further, as discussed in more detail below, given modern exposures (i.e. the integration time multiplied by the effective area) the sensitivity to a given dark matter cross section is typically a couple of orders of magnitude weaker than the sensitivity of the LAT. At present ACTs require tens to hundreds of hours of integration per source in order to obtain even this level of sensitivity. Table~\ref{tab:gamma_ray_experiments} provides some basic information on modern gamma-ray experiments that have presented dark matter results.  

\par With the above background, the remainder of this section is devoted to reviewing and interpreting results from gamma-ray experiments. The results are organized according to the location and the classification of the source: 1) Galactic center, 2) dwarf spheroidals, 3), Galactic halo and extragalactic, 4) Galactic substructure, and 5) clusters of galaxies. 

\begin{deluxetable}{l c c c c l}
\tablecolumns{4}
\tablecaption{Modern gamma-ray experiments that are active in indirect dark matter searches. Experiments below the horizontal line are ACTs. 
\label{tab:gamma_ray_experiments}}
\tablehead{
Experiment & Effective Area $[m^2]$& Energy Range [GeV] & Location & Reference}
\tablewidth{0pc}
\startdata
Fermi-LAT & $1$ & 0.1 - 300 & Space &~\cite{Atwood:2009ez}\\
\hline
VERITAS & $10^5$ & $50 - 50000$ & Arizona &\cite{Holder:2006gi}\\
MAGIC & $2 \times 10^4$ & $50-30000$& Canary Islands &~\cite{Aleksic:2011bx}\\\
HESS& $5 \times 10^4$ &$10-10000$& Namibia &~\cite{Aharonian:2006pe}\\
\enddata
\end{deluxetable}

\subsection{Galactic center}

\par Due to its relative proximity, it is reasonable to assume that the largest flux of gamma-rays from dark matter annihilation comes from the Galactic center. However, as discussed in Section~\ref{sec:MW}, the dark matter distribution in the Galactic center is poorly constrained from direct kinematic measurements, because in this region the stars in the bulge and the central nuclear star cluster appear to dominate the dynamics. Unfortunately this means that there is no reliable way to estimate the dark matter density in the central region of the Galaxy, and therefore there is no way to a priori predict the precise photon flux from dark matter, and test this with the observations (even if we allow for the fact that we don't know the dark matter particle mass). 

\par To get an estimate of a theoretically-plausible prediction for the gamma-ray flux from the Galactic center region, consider an Einasto dark matter profile (Equation~\ref{eq:einasto_profile}) with $\rho_{-2} = 3.7 \times 10^5$ $M_\odot$ kpc$^{-3}$ and $r_{-2} = 16$ kpc~\citep{Navarro:2008kc}. A dark matter halo with these parameters is consistent with both the measurements of the local mass density and the total Milky Way halo mass (Section~\ref{sec:MW}). With this set of parameters, within a region of radius of one degree around the Galactic center from Equation~\ref{eq:indirect_detection_flux_total}, $J(1^\circ) \simeq 1.5 \times 10^{19}$ GeV$^2$ cm$^{-5}$. For $\sigmavavg = 3 \times 10^{-26}$ cm$^3$ s$^{-1}$, a dark matter mass of 100 GeV, and a $b\bar b$ spectrum, above energies of 1 GeV, $N_\gamma \simeq 14$. From Equation~\ref{eq:indirect_detection_flux_total}, the photon flux above 1 GeV is $F(1^\circ) = 6.3 \times 10^{-10}$ cm$^{-2}$ s$^{-1}$. For comparison, the two year Fermi-LAT source catalog list five point sources within one square degree from the Galactic center. For energies greater than 1 GeV, the total flux from these point sources is $\simeq 1.6 \times 10^{-7}$ cm$^{-2}$ s$^{-1}$. So even though the predicted flux from a standard dark matter model is above threshold for the LAT, the point sources in this regime represent a significant background for the signal.  

\par At higher energies than is accessible by the LAT, there is now firm evidence from HESS~\citep{Aharonian:2004wa} and MAGIC~\citep{Albert:2005kh} for a point source, HESS J1745-290,  that is coincident with the position of Sgr A*. The energy spectrum of this source is $dN/dE \simeq E^{-2.1}$, and there is a cutoff in the spectrum at approximately 15 TeV. In the two-year Fermi-LAT source catalog, there is a source coincident with HESS J1745-290, with a best-fit power law of $dN/dE \simeq E^{-2.3}$ over the range of 100 MeV to 100 GeV. Above 1 GeV, the flux from this source is $7.7 \times 10^{-8}$ cm$^{-2}$ s$^{-1}$, making it the brightest source over these energies observed by the Fermi-LAT near the Galactic center. 

\par In addition to the above point sources, diffuse gamma-ray emission has been detected from the Galactic center region. HESS finds that this emission extends in both directions along the Galactic plane, and that it is spatially-correlated with giant molecular clouds in the central 200 pc of the Galaxy. The emission is attributed to the decays of neutral pions produced in the interaction of hadronic cosmic rays in the molecular clouds~\citep{Aharonian:2006au}. Characterization of the diffuse emission from the Galactic center measured by the Fermi-LAT is more challenging because of the aforementioned number of point sources. After carefully subtracting the emission associated with them, the diffuse emission is determined by assuming a model of the sources of cosmic rays, their propagation, and the properties of the interstellar medium. It is now standard to calculate these distributions is with the GALPROP code~\citep{Strong:1998fr,Moskalenko:2001ya,Strong:2004de,Porter:2008ve,Vladimirov:2010aq}. With this modeling in place, it is then left to understand if any residuals remain that may be due to a dark matter signal. 

\par The combined effects of the point sources, diffuse emission, and uncertainty in the dark matter distribution make extraction of a signal from dark matter difficult. In order to avoid these uncertainties, the HESS collaboration has recently searched for gamma-ray emission between angles of $0.3-1.0^\circ$ from the Galactic center, corresponding to a physical distance range of 45-150 pc. In addition to avoiding contamination from gamma-ray sources in the Galactic center and the plane, this observational strategy is also beneficial because there is expected to be less theoretical uncertainty in the dark matter distribution over this region. For example, normalizing to the local density and the total halo mass, there is less than a factor of two difference between the corresponding densities of the Einasto profile and the NFW profile. However, the difference is two orders of magnitude when assuming a cored dark matter profile that matches both the local density and the halo mass. Assuming a NFW or Einasto profile, HESS finds that the upper limit on the annihilation cross section is $\sigmavavg \simeq 10^{-25}$ cm$^3$ s$^{-1}$ for masses greater than 1 TeV and annihilation into quark anti-quark pairs.~\cite{Abazajian:2011ak} extend the limits on the HESS data into the Galactic center, finding an upper bound of $\sigmav \simeq 10^{-25}$ cm$^3$ s$^{-1}$ at a WIMP mass $\gtrsim 1$ TeV for cusped dark matter profiles. 

\par
\begin{figure*}
\begin{center}
\begin{tabular}{c}
\includegraphics[width=0.70\textwidth]{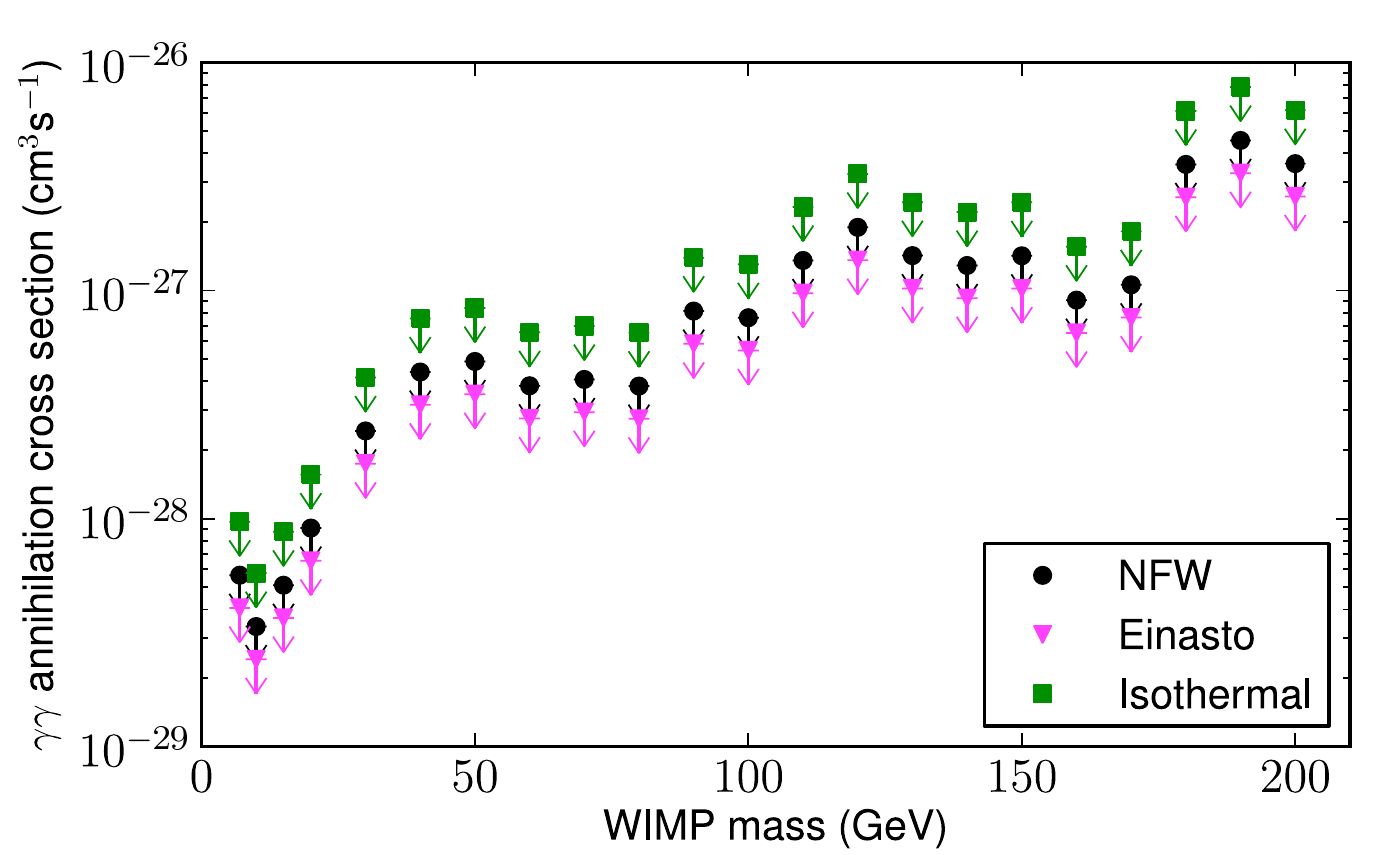} \\
\end{tabular}
\end{center}
\caption{Upper limits on the annihilation cross section into lines. The different sets of data points are for different assumptions on the dark matter density profile. From~\cite{Ackermann:2012qk}. 
}
\label{fig:fermi_lines} 
\end{figure*}

 The Fermi-LAT collaboration has published searches for gamma-ray lines within a $20^\circ \times 20^\circ$ region around the Galactic center~\citep{Abdo:2010nc,Ackermann:2012qk}. To search for lines in the gamma-ray data, a sliding window is used over energies in order to account for the energy resolution. Over the different energy bins the data is fit to a power law plus signal model, where the index of the power law is allowed to vary and the signal represents the line feature. In the Galactic center, the Fermi-LAT collaboration reports no anomalous gamma-rays lines, and places upper limits on the annihilation cross section to lines that vary from approximately a few times $10^{-29}$ cm$^3$ s$^{-1}$ at 10 GeV to a few times $10^{-26}$ cm$^3$ s$^{-1}$ at 200 GeV (Figure~\ref{fig:fermi_lines}). There are factors of few uncertainty reported reflecting the uncertainty in the dark matter density. Gamma-ray lines from the Galactic center will be discussed again at the end of this section. 

\par Though the precise nature of the diffuse gamma-ray emission in the Fermi-LAT energy range remains to be characterized, an interesting morphological structure has come to light in recent analysis of the data. After subtracting out templates for the neutral pion, inverse Compton, and bremsstrahlung emission,~\cite{Dobler:2009xz} first reported a detection of the ``Fermi Haze," which extends 50$^\circ$ above and below the Galactic plane. Two characteristic features of this gamma-ray haze are that it has a long-to-short axis ratio of two, and the emission in this haze region is spectrally-harder than the other gamma-ray components in the Galaxy. The gamma-ray haze is consistent with the microwave haze discovered in the WMAP data by~\cite{Finkbeiner:2004us}, which is now believed to result from synchrotron emission from the same population of relativistic electrons that are the source of the gamma-ray haze. A similar haze has also recently been announced by the Planck collaboration~\citep{2012arXiv1208.5483P}. Further analysis has revealed that the LAT data is more consistent with a ``bubble" like morphology~\citep{Su:2010qj} that has sharp edges. Though the emission is extended in a manner similar to would be expected from emission from particle annihilation in the dark matter halo~\citep{Finkbeiner:2004us}, both the morphology and the edges seem to argue against a dark matter interpretation (see however~\cite{Dobler:2011mk}). 

\par Though there has been great progress in recent years in our understanding of gamma-ray astrophysics from the Galactic center, it may leave readers of this article somewhat unsatisfied that a definitive limit (or detection) of dark matter has not been advocated for. While it is certainly true that cross sections much larger than the thermal relic scale are now clearly ruled out for any reasonable dark matter distribution in the Galactic center, reliably pushing down to the thermal relic scale with Galactic center observations will certainly require better understanding of the systematics that have been discussed above. 

\subsection{Dwarf spheroidals}

\par The dSph galaxies, whose properties were discussed at length in Section~\ref{sec:satellites}, provide excellent targets for indirect dark matter searches, and in particular gamma-ray searches, for several reasons. First, there are no gamma-ray point sources and there is no intrinsic diffuse emission associated with them, as is the case of the Galactic center. Indeed, the very strict upper limits on star formation processes and the gas content of these systems predicts a negligible gamma-ray flux from these sources~\citep{Grcevich:2009gt}. Second, they have well-understood dark matter distributions that are derived from the stellar kinematics, as was shown in detail in Section~\ref{sec:satellites}. Third, since the boost factor is predicted to be negligible in these systems, interpretation of a null-flux detection as a limit on $\sigmavavg$ is much more straightforward than it is for the cases of the Galactic center and larger-mass galaxies (Section~\ref{sec:indirect_detection_theory}). 

\par
\begin{figure*}
\begin{center}
\begin{tabular}{c}
\includegraphics[width=0.70\textwidth]{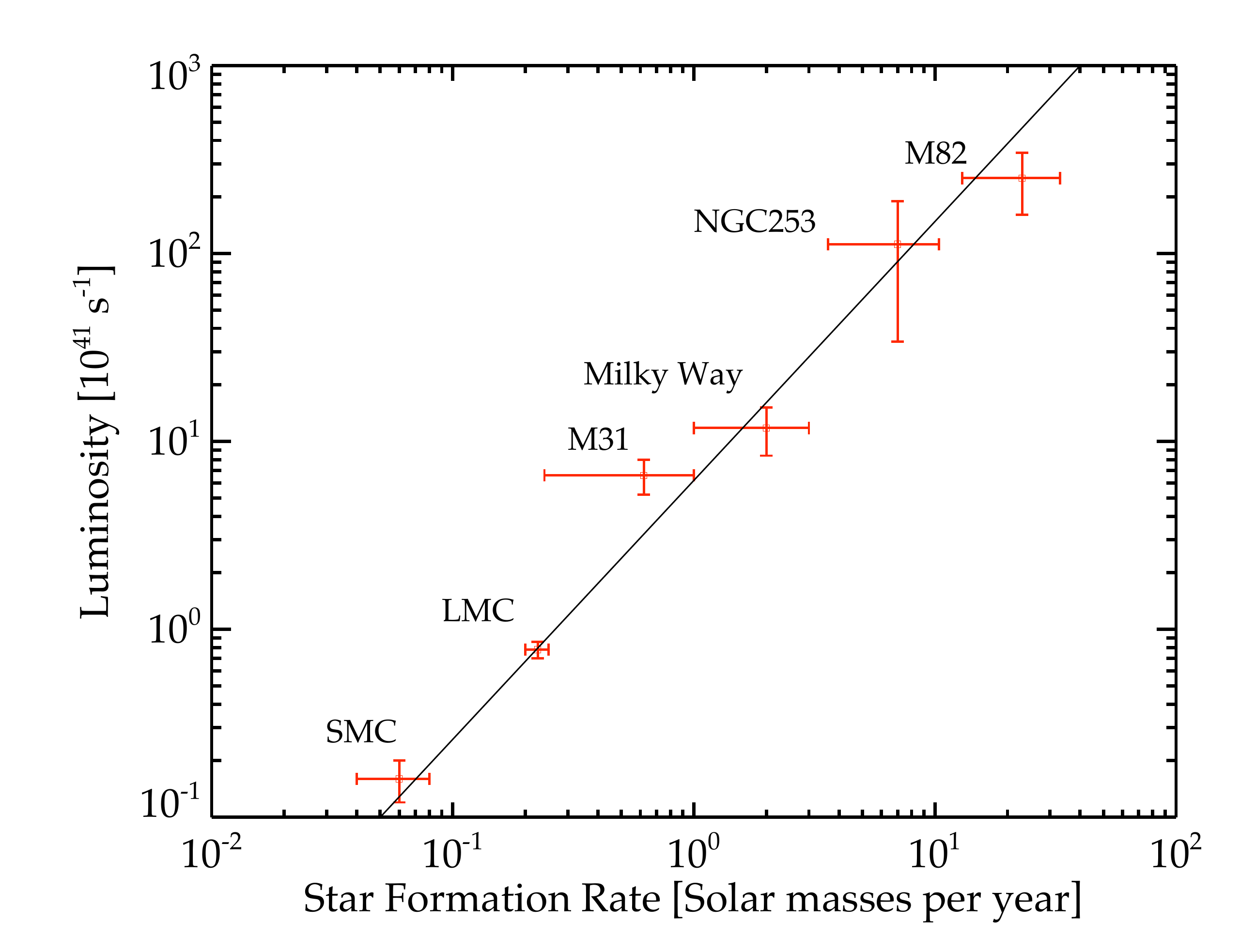} \\
\end{tabular}
\end{center}
\caption{Gamma-ray luminosity as a function of star formation rate for galaxies detected by Fermi-LAT. Adapted from~\cite{Ackermann2012a}.
}
\label{fig:SFR_gamma}
\end{figure*}

 The lack of gamma-ray emission from cosmic-ray processes or from stellar remnants is an important point, so it is informative to be as quantitative as possible with this discussion. The most salient point can be seen in Figure~\ref{fig:SFR_gamma}, which shows the relation between the star formation rate and the gamma-ray luminosity for star forming galaxies~\citep{Ackermann2012a}. Though it is not yet clear what this relation should be for galaxies with lower star formation rate than is shown, we can get a rough idea by extrapolating the linear relation to lower star formation rates. The extrapolation down to the scale of the only dSph with observed star formation activity, Leo IV, predicts a negligible gamma-ray flux from this object. So considering that other dSphs have negligible star formation, it is very likely that any gamma-rays observed from these objects arise from dark matter annihilation (Note that it is possible that several more star forming galaxies could appear as Fermi-LAT sources before the end of the mission, with the next most likely candidate being the M31 satellite galaxy M33~\citep{Abdo:2010ip}).

\par It is also insightful to compare the gamma-ray emission from local star forming galaxies to the emission from nearby globular clusters. The Fermi-LAT has now detected approximately ten globular clusters in gamma-rays, with the emission most likely arising from milli-second pulsars~\citep{Abdo2010AA}. For globular clusters there is an interesting observed correlation between the density of the system and the gamma-ray luminosity, which may be reflective of the fact that milli-second pulsars are born in stellar captures in the dense clusters. In contrast, since the dSphs are much more extended than globular clusters (Figure~\ref{fig:mv_rh}), the stellar capture rate in them is much lower, so the capture of stars and remnants into binary systems does not occur. Only one dSph, Sculptor, has a population of X-ray binaries associated with it, which likely formed from an entirely different physical mechanism~\citep{Maccarone:2005de}. 

\par In order to precisely predict the gamma-ray signal from dSphs, we must return to understanding the dark matter distributions within them. It is informative to begin with some heuristic arguments that will be important in the context of these measurements. First, note that the dark matter halos of the dSphs likely do not extend beyond approximately a few kpc. This limit can be motivated by determining the Roche limit for a dSph for a minimum estimate of dark matter it contains, the known distance $D$ to it, and the assumption that it is on a circular orbit. The Roche limit is specifically given by $R_t \simeq D\left[M_\textrm{dsph}/M_\textrm{mw}(D)\right]^{1/3}$, where $M_\textrm{mw}(D)$ is the mass of the Milky Way out to the distance of the dSph. So in order to obtain a conservative lower estimate on $R_t$ we must have a reliable estimate of the minimum mass of the dSph. This in turn depends on the largest radius out to which we find stars that are associated with the dSph, and understanding if these stars are tidally-bound to the system. For all of the classical satellites, it is certainly robust that within spherical radii of a few hundred parsecs the stars are faithfully tracing the local potential~\citep{Strigari:2008ib}. 

\par As discussed in Section~\ref{sec:satellites}, the modern kinematic data strongly constraints the integrated dark matter masses of the dSphs within their respective half-light radii of approximately a few hundred parsecs. This statement is weakly dependent on whether there is a central core or cusp in the dSph, so long as the log-slope is $-d (\log \rho) /d \log r \lesssim 1.5$. The importance of this statement cannot be overestimated, because the constraints on the mass profile directly translate into constraints on the integral over the density-squared within the same region~\citep{Strigari:2007at}. Indeed, it can be shown that in a manner similar to the calculation for the integrated mass in Section~\ref{sec:satellites}, the $J$-value is best constrained within an integrated physical radius that strongly correlates with the half-light radius~\citep{Walker:2011fs}. To better appreciate this, consider that the nearest classical satellites are at distances of approximately $70-80$ kpc. For a dSph at this distance, the half-light radius corresponds to less than approximately one degree, which is about the angular resolution of the Fermi-LAT over a large energy range of interest. This is the region within which the integrated density and the integrated density-squared are the best constrained from the kinematic data sets. Thus {\it the assumption of a core or a cusp for the density profile does not significantly affect the gamma-ray flux predictions for the Fermi-LAT}.  As discussed more below, however, for instruments with better angular resolution than the Fermi-LAT, the assumption of a core or the cusp is much more relevant. 

\par The theoretical developments outlined above have significantly improved the determinations of the $J$-values of the dSphs since the time when they were first determined over a decade ago~\citep{Baltz:1999ra,Tyler:2002ux,Evans:2003sc,Bergstrom:2005qk}.~\cite{Strigari:2007at} and ~\cite{Martinez:2009jh} have developed a maximum likelihood method to determine $J$-values from stellar kinematical and photometric data using the likelihood in Equation~\ref{eq:f_convolve}. More recently groups have extended this analysis though in all cases the calculations are generally in good agreement~\citep{Charbonnier:2011ft}. 

\begin{figure}
\begin{center}
\begin{tabular}{c}
\includegraphics[width=0.45\textwidth]{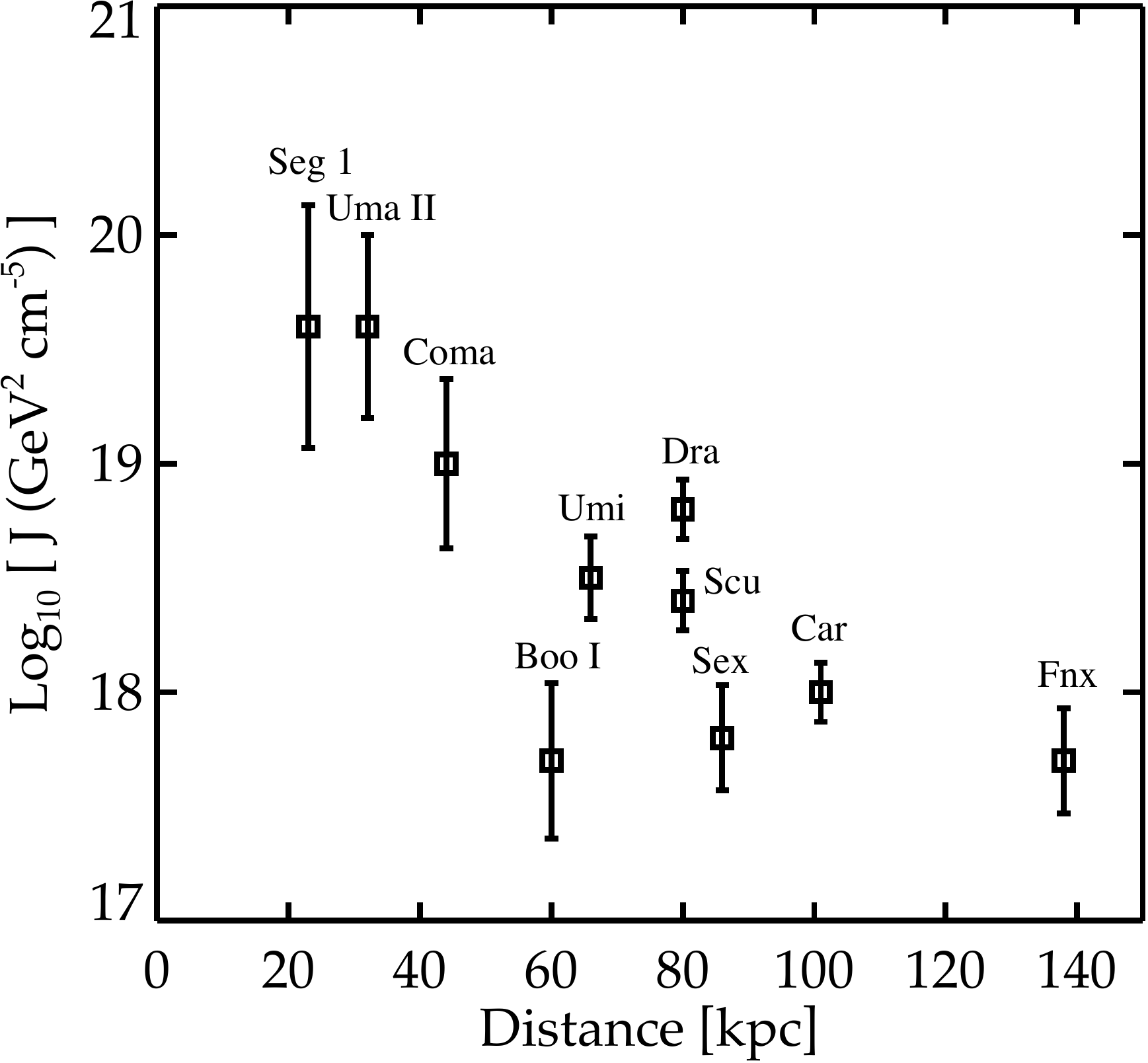} \\
\end{tabular}
\end{center}
\caption{J-values for dSphs within a radius of $0.5^\circ$ as a function of their Galactocentric distance. In this case NFW profiles are assumed for the dark matter density, though the results are very weakly dependent on the assumed central dark matter profile for dSphs with large data samples (compare to Figure~\ref{fig:J_pdfs}). 
}
\label{fig:Jvalues}
\end{figure}

\par
\begin{figure}
\begin{center}
\begin{tabular}{cc}
\includegraphics[width=0.45\textwidth]{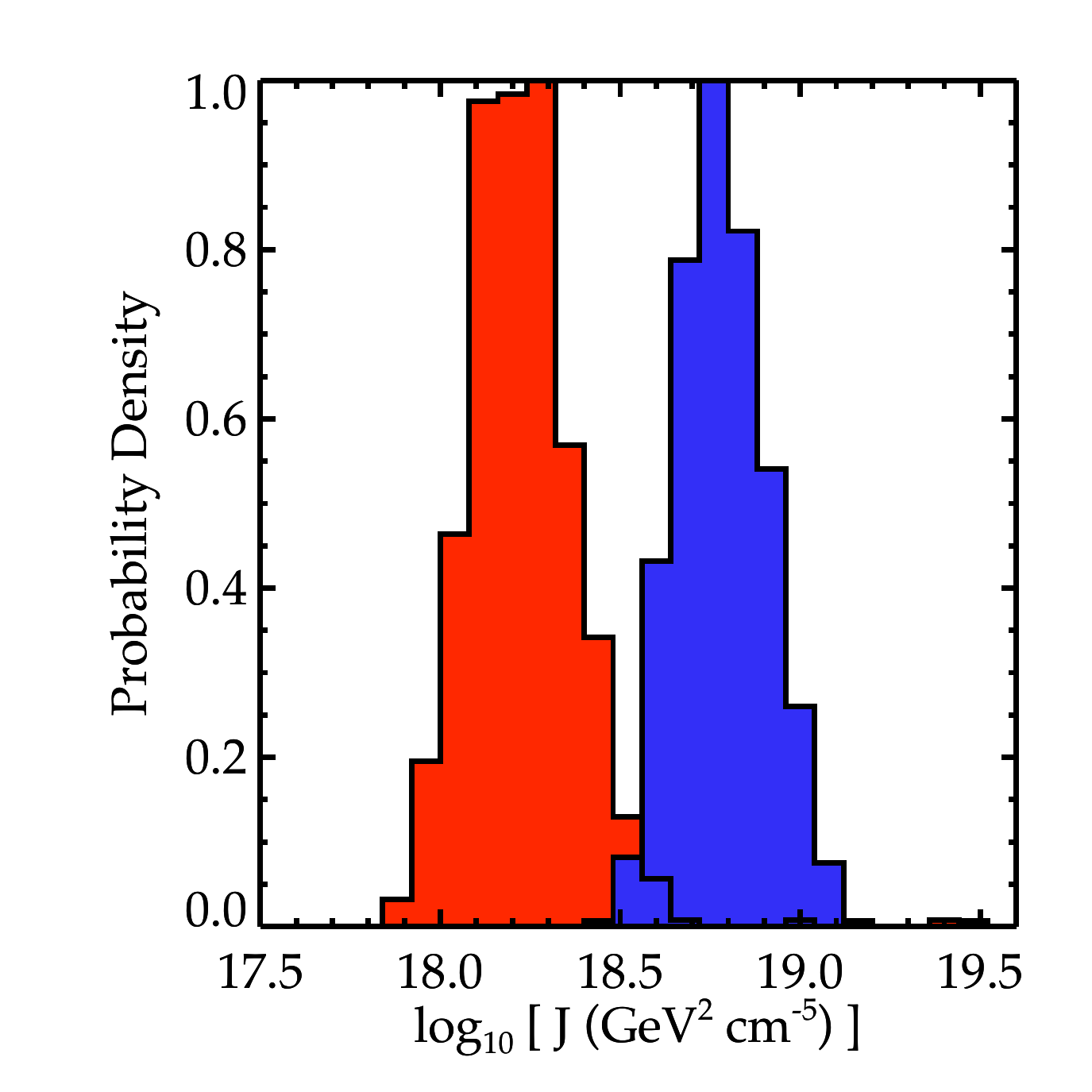} & 
\includegraphics[width=0.45\textwidth]{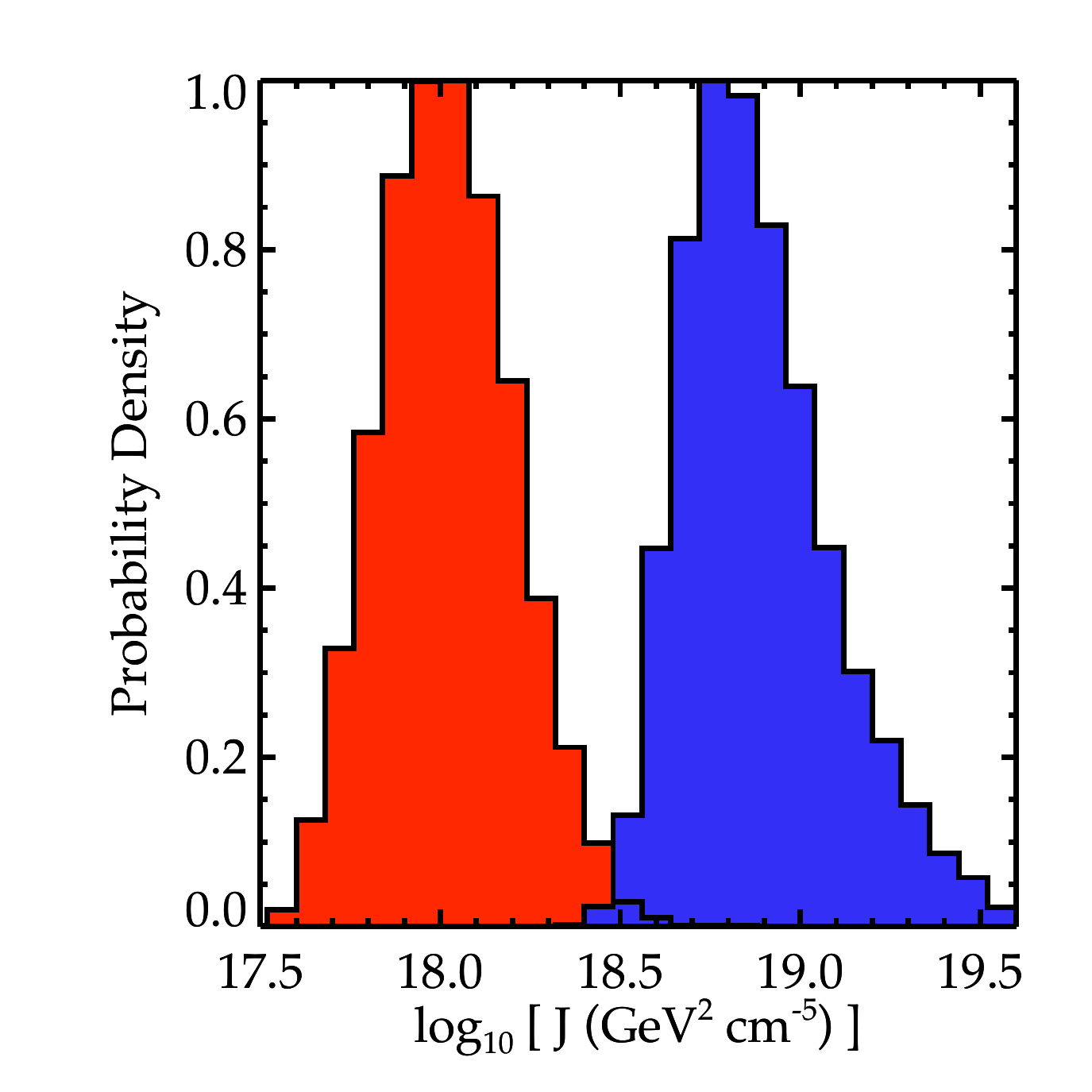} \\
\end{tabular}
\end{center}
\caption{J-values for Draco within $0.1^\circ$ (left histograms) and $0.5^\circ$ (right histograms). The left panel assumes an NFW profile, while the right panel marginalizes over inner slopes of the dark matter density profile from a range of $0$ (core) to $1$ (NFW cusp). Comparison of the peaks of the right-hand (blue) histograms shows that the $J$-value is weakly-dependent on the assumed central slope of the dark matter density profile within $0.5^\circ$. 
}
\label{fig:J_pdfs}
\end{figure}

 For nearby dSphs that are most relevant for gamma-ray observations, the most updated determinations of the $J$-values are shown in Figure~\ref{fig:Jvalues}. Here an NFW profile is assumed for the dark matter density profile, as in~\cite{Ackermann:2011wa}. However, as is shown in the probability density in Figure~\ref{fig:J_pdfs} the results are weakly dependent on whether a cored or cusped central density profile is assumed for the dark matter. Figure~\ref{fig:Jvalues} clearly indicates which dSphs are the most interesting targets for indirect dark matter detection experiments. The two dSphs with the largest $J$-values, Segue 1 and Ursa Major II, are ultra-faint satellites with sparse samples of stars associated to them. Specifically, the $J$-values for Segue 1 and Ursa Major II in Figure~\ref{fig:Jvalues} were derived from samples of 66 and 20 stars, respectively. Though these dSphs have the largest mean flux, they also have the greatest uncertainty due to the small stellar samples. After Segue 1 and Ursa Major II, the dSphs with the next largest $J$-values are Ursa Minor and Draco, at 66 and 80 kpc, respectively. These $J$-values are determined from samples of 150 and 320 stars, respectively, and as is seen uncertainties are much lower than the uncertainties on the $J$-values for Segue 1. In all cases, the one-sigma uncertainties are approximately log-normal; Figure~\ref{fig:J_pdfs} shows example probability densities from which the $J$-values in Figure~\ref{fig:Jvalues} are derived.

\par
\begin{figure*}
\begin{center}
\begin{tabular}{cc}
\includegraphics[width=0.51\textwidth]{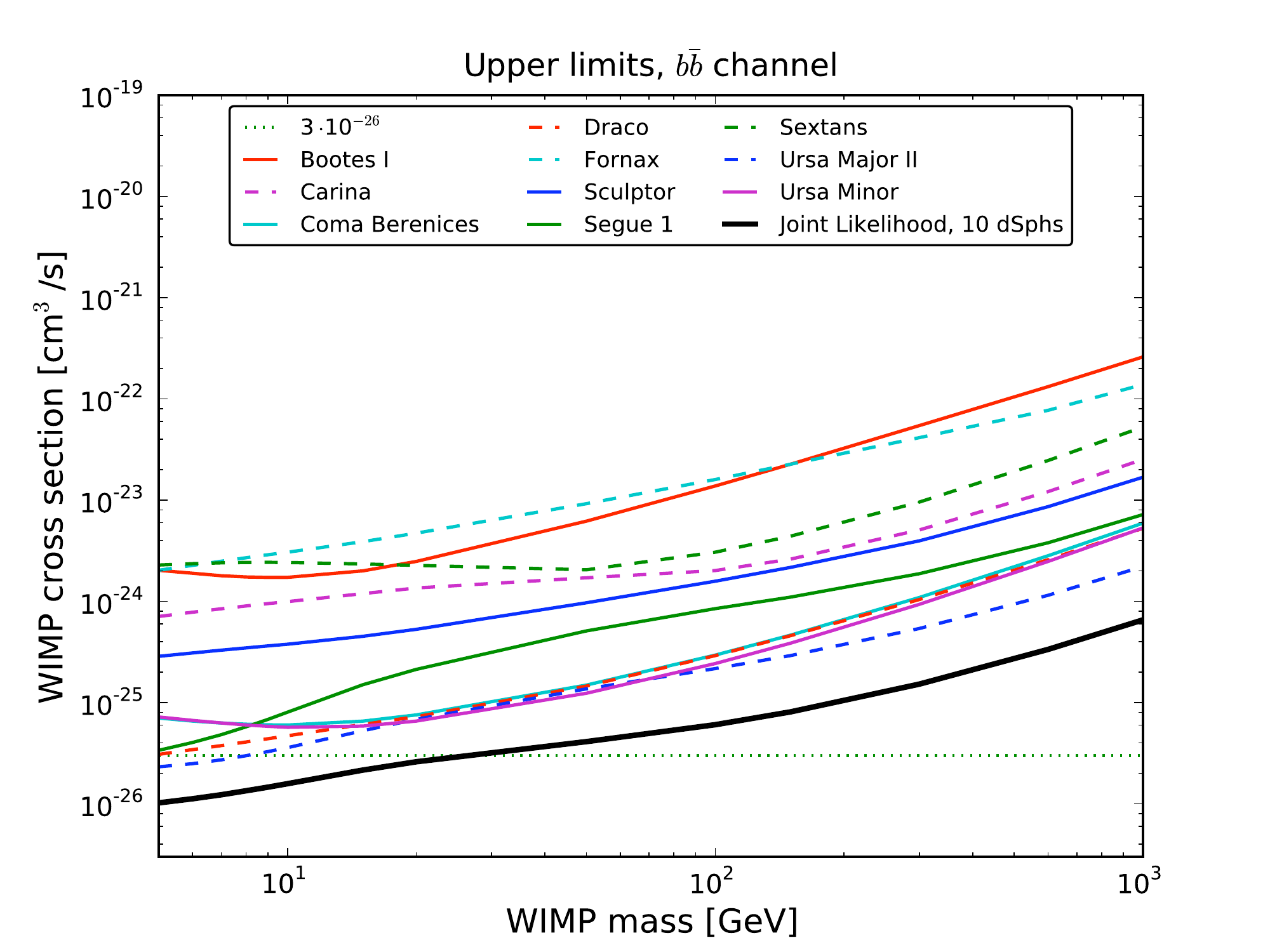} & 
\includegraphics[width=0.51\textwidth]{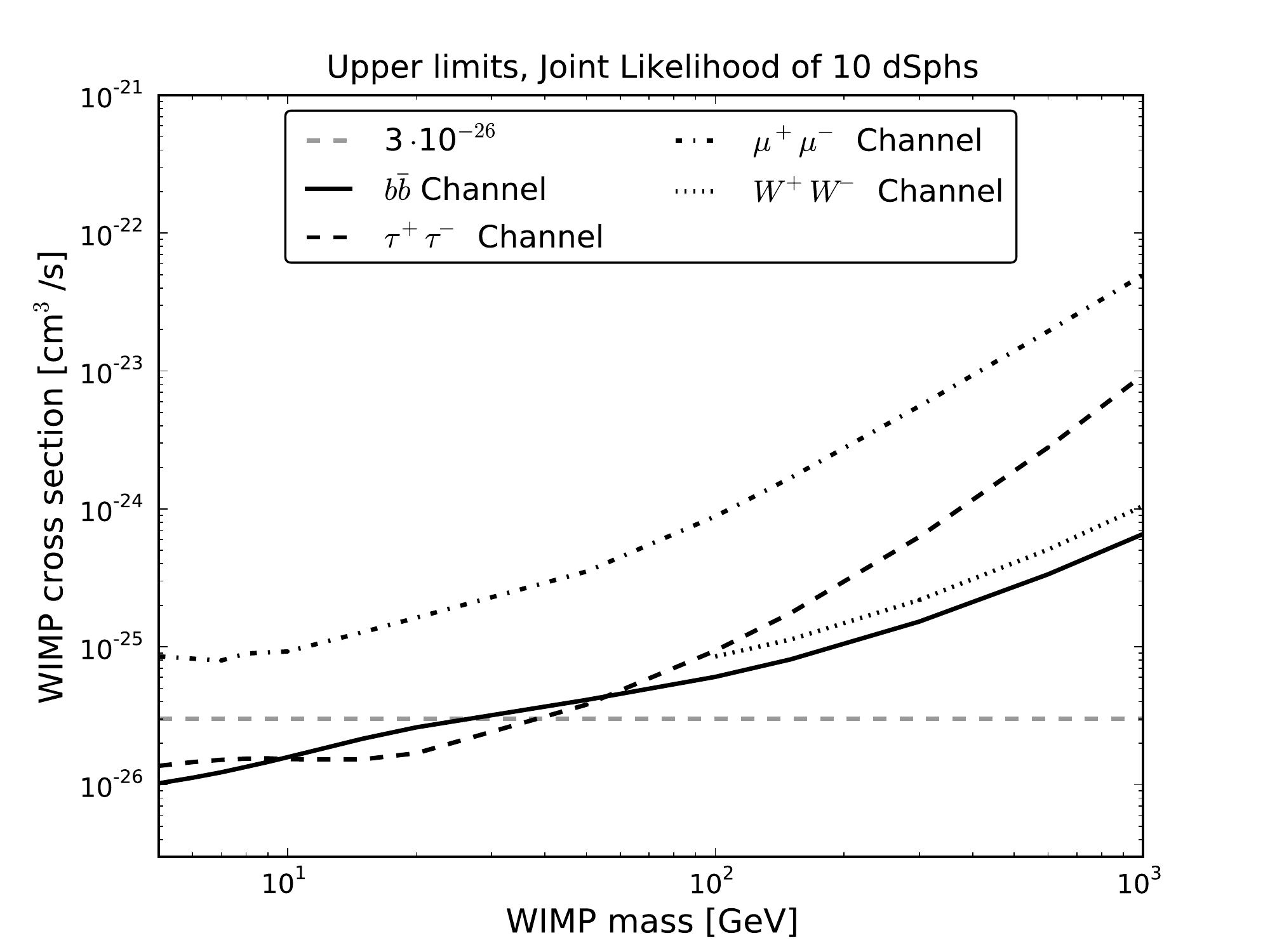} \\
\end{tabular}
\end{center}
\caption{{\it Left}: Limits on the annihilation cross section as a function of WIMP mass, assuming 100\% annihilation to $b \bar b$. Upper curves are limits from individual dSphs, and the lower solid curve is the combined limit. {\it Right}: Limits on the annihilation cross section as a function of WIMP mass for several different channels. From the Fermi-LAT Collaboration~\citep{Ackermann:2011wa}. 
}
\label{fig:fermi_dsph_limits}
\end{figure*}

 Using the $J$-values and their associated uncertainties, gamma-ray fluxes from each dSph are predicted using a model for the annihilation cross section and resulting gamma-ray spectrum. Analysis of two year Fermi-LAT data for all dSphs reveals no measured excess over the emission from diffuse Galactic backgrounds, extragalactic backgrounds, and nearby point sources~\citep{Ackermann:2011wa,GeringerSameth:2012sr}. With the null detection, a robust upper limit may be placed on the mass and the annihilation cross section. The procedure for determining the limit involves combining the posterior probability distributions of the $J$-value of each dSph with the likelihood for the gamma-ray data in the direction of each individual dSph (or a Region of Interest, ROI). In each ROI, the normalizations of the diffuse Galactic, diffuse extragalactic, and nearby points sources are separately fit to the data. While the $J$-values are distinct for each ROI, the parameters $\sigmavavg$ and $\mdm$ are shared between ROIs, thereby tying together different ROIs. The joint likelihood for the data, $D$, given $N$ ROIs, is 
\be
L(D | \vec p_w, \{ \vec p \}) = \prod_\imath^N  L_\imath^{\textrm{LAT}} (D | \vec p_w, \vec p_\imath) \nonumber \\
\times \frac{1}{\ln(10)J_\imath \sqrt{2\pi} \sigma_\imath} 
\exp \left[ - \frac{( \log_{10} (J_\imath) - \overline{ \log_{10} (J_\imath)} )^2}{2\sigma_\imath^2}\right],
\label{eq:joint_likelihood} 
\ee
where $L_\imath^{\textrm{LAT}}$ is the poisson likelihood for a ROI, $D$ is the binned gamma-ray data, $\vec p_w$ are the parameters that are independent of the ROI, in this case $\sigmavavg$ and $m$, and $\{ \vec p \}$ are the parameters that depend on the ROI, in this case the background normalizations. The $\sigma_\imath$ in this model are one sigma uncertainties on the $J$-values as derived from their posterior probability distributions. A profile likelihood method is used on Equation~\ref{eq:joint_likelihood}, which maximizes $L$ for all model parameters, and from this profile likelihood 95\% c.l. limits are defined. The above procedure for determining the upper limits on $\sigmavavg$ for a fixed mass was implemented by the LAT collaboration~\citep{Ackermann:2011wa}. Similar upper limits are obtained by using an ``on-off" source analysis for each dSph, which involves comparing the expected source region to a nearby region in which the background is smooth. In this case the background components do not explicitly have to be fit out~\citep{GeringerSameth:2011iw}. Resulting upper limits as a function of mass are shown in Figure~\ref{fig:fermi_dsph_limits}, for several different assumptions about the dominant annihilation channel for the dark matter.  

\par Analysis along the lines above has recently been implemented in a search for gamma-ray lines from dSphs~\citep{GeringerSameth:2012sr}. This is probably the most robust indirect detection signal imaginable because it would be a gamma-ray line emanating from a region where the dark matter distribution is well-understood. Unfortunately the predicted signal is well-below that expected from standard supersymmetric models discussed in Section~\ref{sec:indirect_detection_theory}. There is no evidence yet for gamma-ray lines from dSphs; the upper limits on the annihilation cross section into gamma-ray lines are $\sigmavavg \sim [10^{-27}$, $\sim 10^{-24}$] cm$^3$ s$^{-1}$ for masses of $[10,100]$ GeV, respectively. 

\par
\begin{figure*}
\begin{center}
\begin{tabular}{cc}
\includegraphics[width=0.55\textwidth]{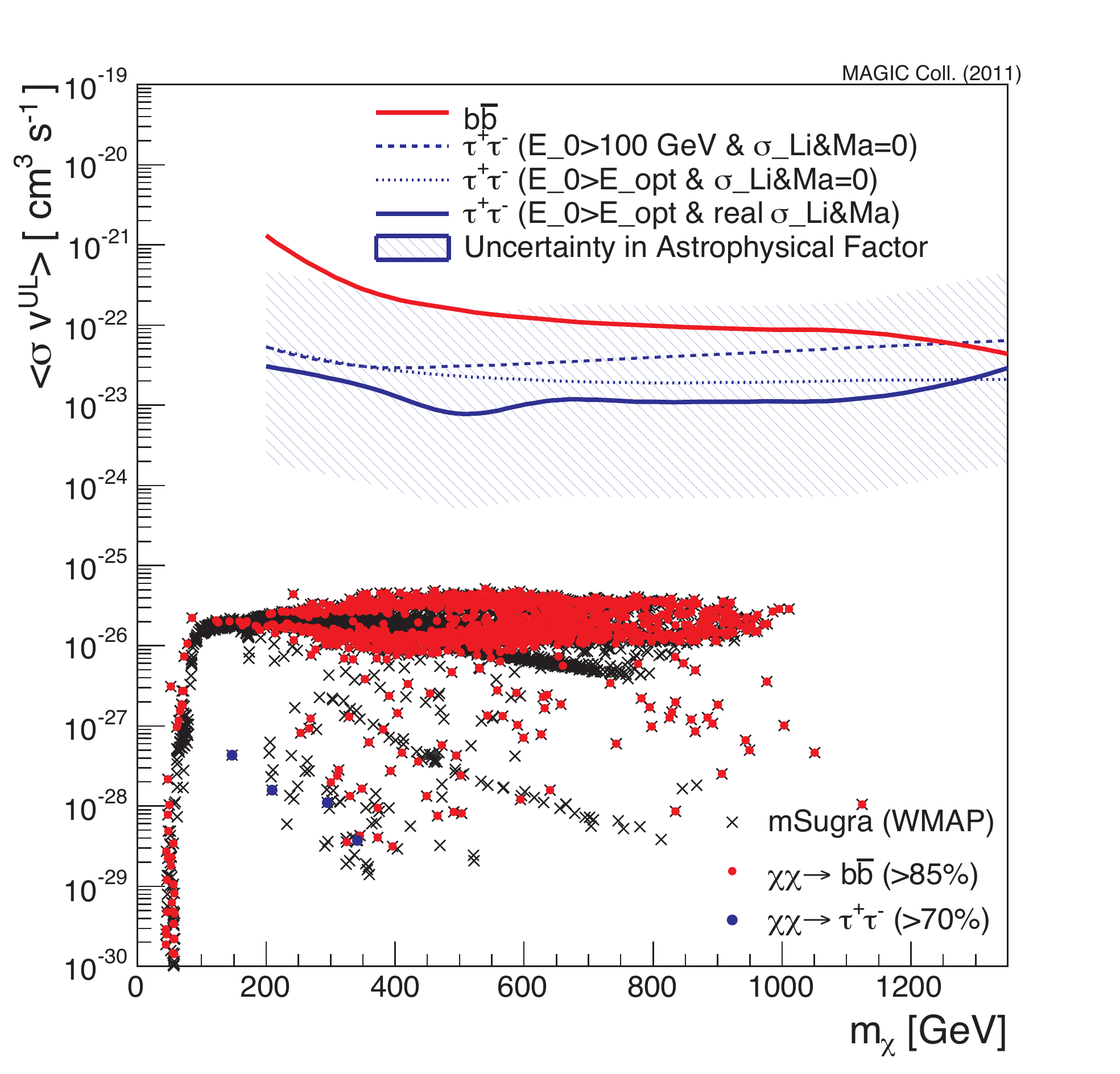} \\ 
\end{tabular}
\end{center}
\caption{Limits on the annihilation cross section as a function of WIMP mass, assuming 100\% annihilation to $b \bar b$ and $\tau \bar \tau$. From the MAGIC collaboration~\citep{Aleksic:2011jx}.   
}
\label{fig:magic_dsph_limits}
\end{figure*}

\begin{figure*}
\begin{center}
\begin{tabular}{cc}
\includegraphics[width=0.55\textwidth]{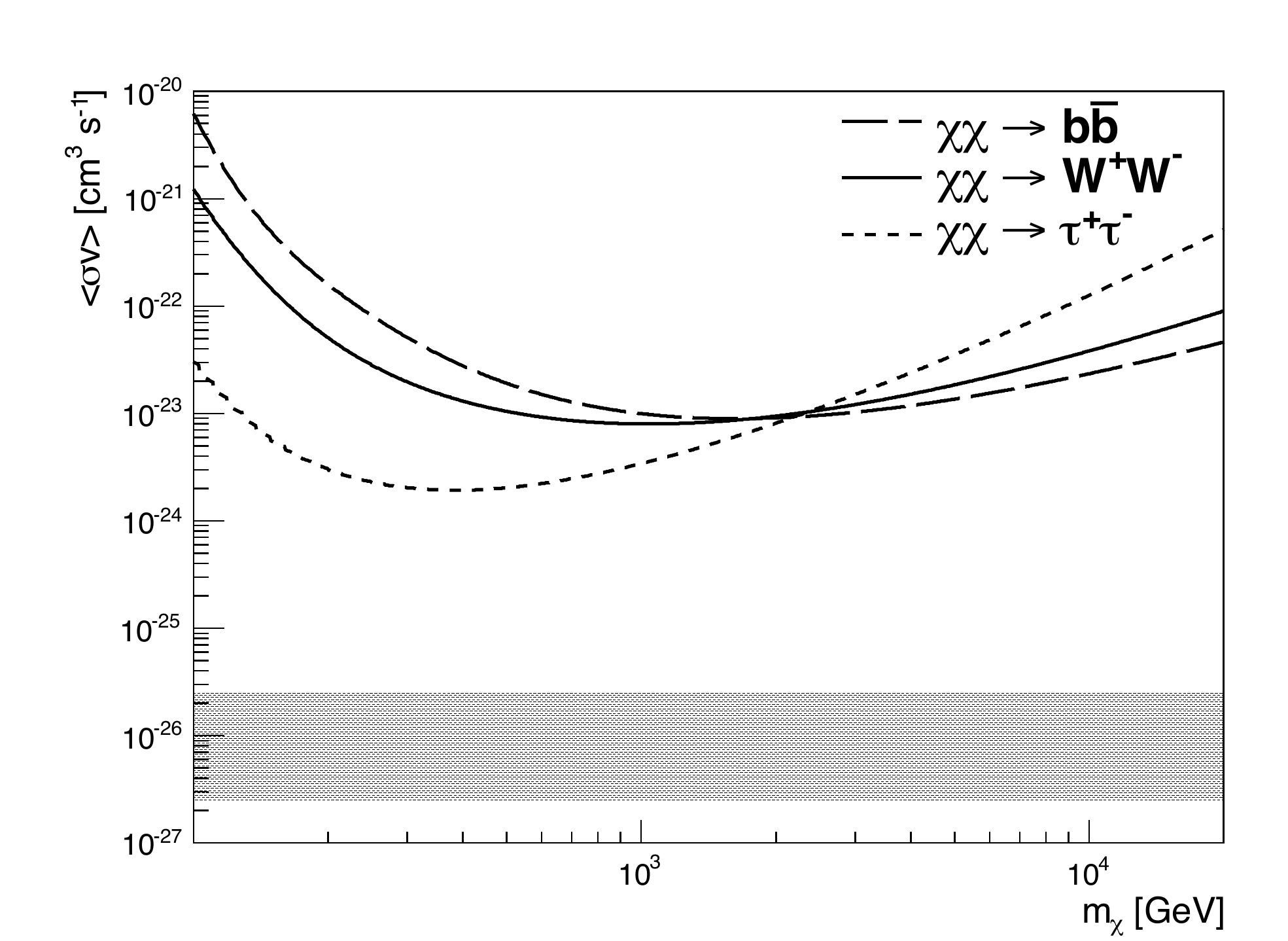} \\
\end{tabular}
\end{center}
\caption{Limits on the annihilation cross section as a function of WIMP mass, assuming 100\% annihilation to $b \bar b$ and $\tau \bar \tau$, and $W^{+} W^{-}$. From the Veritas Collaboration~\citep{Aliu:2012ga}. 
}
\label{fig:veritas_dsph_limits}
\end{figure*}

 Similar to the case of Galactic center observations, ACTs have also studied dSphs in search of a signal from dark matter. Because of the high energy and thus photon-limited regime in which ACTs operate, the dSphs are analyzed via an on-off strategy described above. Figures~\ref{fig:magic_dsph_limits} and~\ref{fig:veritas_dsph_limits} present the limits on $\sigmavavg$ from MAGIC~\citep{Aleksic:2011jx} and VERITAS~\citep{Rico:2011vs} observations of Segue 1. Figures~\ref{fig:magic_dsph_limits} and~\ref{fig:veritas_dsph_limits} clearly show that ACTs are more sensitive to the higher dark matter mass regime than the Fermi-LAT analysis, though because of smaller exposures the limits are weaker by a couple of orders of magnitude as compared to the Fermi-LAT limits. Similar to the Fermi-LAT analysis, $J$-values were determined from the kinematic data within the appropriate angular regime under the assumption of an Einasto dark matter profile. The indicated limits are at the 95\% c.l. Even though the limits from the ACTs are two to three orders of magnitude larger than the characteristic thermal relic scale, they do probe some of the parameter space for a dark matter explanation to recent cosmic ray data (see below). 

\par Forthcoming larger scale ACTs with angular resolution $\lesssim 0.1^\circ$ will be more sensitive to the dark matter distribution in the center of the dSphs. This is quantitatively shown in Figure~\ref{fig:J_pdfs}, which compares the integrated $J$-values within $0.1^\circ$ and $0.5^\circ$ for cored and cusped dark matter profiles. (See also~\cite{Charbonnier:2011ft}). The uncertainties in this figure are derived only from the kinematic data; before the launch of next generation ACTs it will also be important to improve the photometric data in the center of the dSphs. As was shown in Section~\ref{sec:satellites} there is a significant degeneracy between the central density of the stellar profile and the central density of the dark matter profile. 

\par In summary, indirect dark matter searches using dSphs are now at an advanced stage, and have delivered the most robust results on the dark matter mass and cross section. To the extent that is possible, they represent particle dark matter experiments in the sky. Gamma-ray observations of dSphs have been the first to achieve reliable sensitivity to thermal relic dark matter over a theoretically-motivated mass regime of $10-25$ GeV, and they will continue to push down limits to the thermal relic scale at larger masses. The sensitivity will continue to increase because of improved understanding of the systematics in the gamma-ray data, and also because new Milky Way satellites are sure to be detected in future surveys. Also in future analyses it will be important to model the satellites as extended sources, improving upon the point source modeling that has been used to derive the limits described above. 

\subsection{Diffuse Galactic and extragalactic searches}
\par Residual annihilation of dark matter particles throughout the Milky Way, and throughout external galaxies over all redshifts, provides an interesting signature that is complementary to the Galactic center and dSph searches. This is a truly diffuse signal that is spread out over many resolution elements of the detector, so by its nature it is more difficult to extract than searches for point sources. With its wide field-of-view, the Fermi-LAT is uniquely sensitive to a diffuse signal along these lines. Extracting it requires a detailed model for both the spectral and spatial distribution of the signal as well as the spectral and spatial resolution of the gamma-rays produced from cosmic ray sources. 

\par For diffuse searches, there are three relevant mechanisms of gamma-ray production. First, there is the standard prompt production that arises from hadronization, fragmentation, and subsequent $\pi_0$ decay, which as discussed above is strongly constrained for $b\bar b$ and $\tau \bar \tau$ channels from observations of dSphs. Second, diffuse radiation can also be used to search for gamma-ray lines that are produced directly through a $\gamma \gamma$ or $\gamma Z^0$ pair. Third, the inverse Compton component produced by the up-scattering of starlight by high energy electrons and positrons produced by the annihilating dark matter. Note that this third process is not relevant in the dSph searches because of the low density of starlight in the outer MW halo. 

\par The challenge in extracting the diffuse dark matter signal is to characterize the diffuse gamma-ray background from cosmic ray sources. The observed spectrum of the diffuse background radiation falls off like a power law over a wide energy regime, which is much different than the gamma-ray spectrum from dark matter. Several approaches for separating out the latter component have been considered in the literature. Using standard models for the Milky Way dark matter distribution and simply ascribing all of the background to other processes, the most conservative constraints on the annihilation cross section at $10$ GeV are a few times $10^{-24}$ cm$^3$ s$^{-1}$~\citep{Papucci:2009gd,Cirelli:2009dv}. Improving on this simple limit requires the development of more sophisticated theoretical models.~\cite{Baxter:2011rc} discuss an interesting technique to set a limit on the annihilation cross section that exploits the expected difference in the distributions between the signal and backgrounds in a ring of pixels around the Galactic center. They derive conservative upper limits of approximately $10^{-25}$ cm$^3$ s$^{-1}$ on the dark matter mass of approximately $10$ GeV for annihilation into $b\bar b$ and $\tau \bar \tau$ channels. Though this technique still is subject to uncertainties on the dark matter distribution it does provide an interesting method for reducing the Galactic emission from cosmic rays that peaks in the disk, without the need to understand details of the latter signal other than its spatial distribution. 

\par~\cite{Ackermann:2012rg} have recently performed a search for dark matter annihilation in the diffuse halo, modeling both the distribution of dark matter, and the properties of the cosmic ray source populations and their propagation through the Galaxy. A given model of the Galaxy is defined by the spectral index of the injection cosmic-ray energy spectrum, the scale-height of the diffusive cosmic ray halo, and the dust to HI ratio. After fixing a set of ``non-linear" parameters to define a model for the Galaxy, over twenty ``linear" parameters must still be determined that describe the cosmic ray source distribution, the normalization of the isotropic gamma-ray background in each bin, and the amplitude of the dark matter signal. Maximizing these linear model parameters allows for determination of the upper limits (or best-fitting) amplitude for the dark matter signal. The procedure is then repeated for a new set of non-linear parameters, and from a variety of these repeats different sets of constraints are placed on the linear parameters. Unfortunately, the large number of model parameters makes it difficult to marginalize over all of the parameters that describe the cosmic ray sources and their propagation. Given the complexity of the modeling, it is understandable that no conclusive dark matter signal can be extracted with the present data; at $\sim 10$ GeV ~\cite{Ackermann:2012rg} quote upper limits in the range $10^{-25} - 10^{-26}$ cm$^3$ s$^{-1}$ for $b\bar b$ and $\tau \bar \tau$ channels, where the uncertainty is driven by the assumed local dark matter density. 

\par The Fermi-LAT collaboration has reported two searches for gamma-ray lines throughout a large volume of the Galactic halo~\citep{Abdo:2010nc,Ackermann:2012qk}. In the latest results, which are based upon two years of data, the annihilation cross section upper limits vary from approximately $10^{-28}$ cm$^3$ s$^{-1}$ at 7 GeV to approximately $10^{-26}$ cm$^3$ s$^{-1}$ at 200 GeV. There are still factors of several uncertainties in these limits due to the uncertain dark matter distribution in the Milky Way (Figure~\ref{fig:fermi_lines}). 

\par Subtraction of the diffuse gamma-ray emission from Galactic sources leaves a residual, isotropic background. This background is believed to arise from a variety of unresolved sources, including active galactic nuclei, starburst galaxies, and shocks created during the formation of large scale structure. Over the energy range $200$ MeV - $100$ GeV, the spectrum of the isotropic emission is a power law that scales scales as $E^{-2.4}$~\citep{Abdo:2010nz}. Though the above sources are expected to contribute to this emission, there is no well-accepted model for the relative contributions of each of the sources~\citep{Fields:2010bw,Inoue:2011bm}. Similar to the case of Galactic emission, since the energy spectrum is a pure power law, it is clear that dark matter annihilation does not explain it over a large range of energies. The pertinent question is then whether dark matter can explain the background over a smaller energy window. 

\par The theoretical mass function of dark matter halos is now well-understood from N-body simulations, and from this mass function it is possible to predict the contribution to the extragalactic background from dark matter annihilation. Because the emission is from halos and subhalos over the entire mass range over all redshifts, the uncertainty on the boost factor from halo substructure is an important systematic in determining the background component from dark matter. Further, the uncertainty in the contributions to the background from non-dark matter sources preclude robust statements from being made on the mass and cross section at this time. However, it is possible to rule out some exotic models of the annihilation cross section using the isotropic background~\citep{Abdo:2010dk}.

\subsection{Galactic substructure}
\par As discussed in Section~\ref{sec:simulations}, N-body simulations of Galactic halos predict that the dark matter mass distribution in the halo is not smooth, but rather approximately $10-50\%$ of the mass is bound up in the form of subhalos with a mass function that may extend down to Earth masses or even below. Gamma-rays produced from dark matter annihilation in these subhalos may also be detectable, either from individual subhalo sources or from the cumulative effect from all such subhalos distributed throughout the Galaxy. 

\par
\begin{figure*}
\begin{center}
\begin{tabular}{c}
\includegraphics[width=0.70\textwidth]{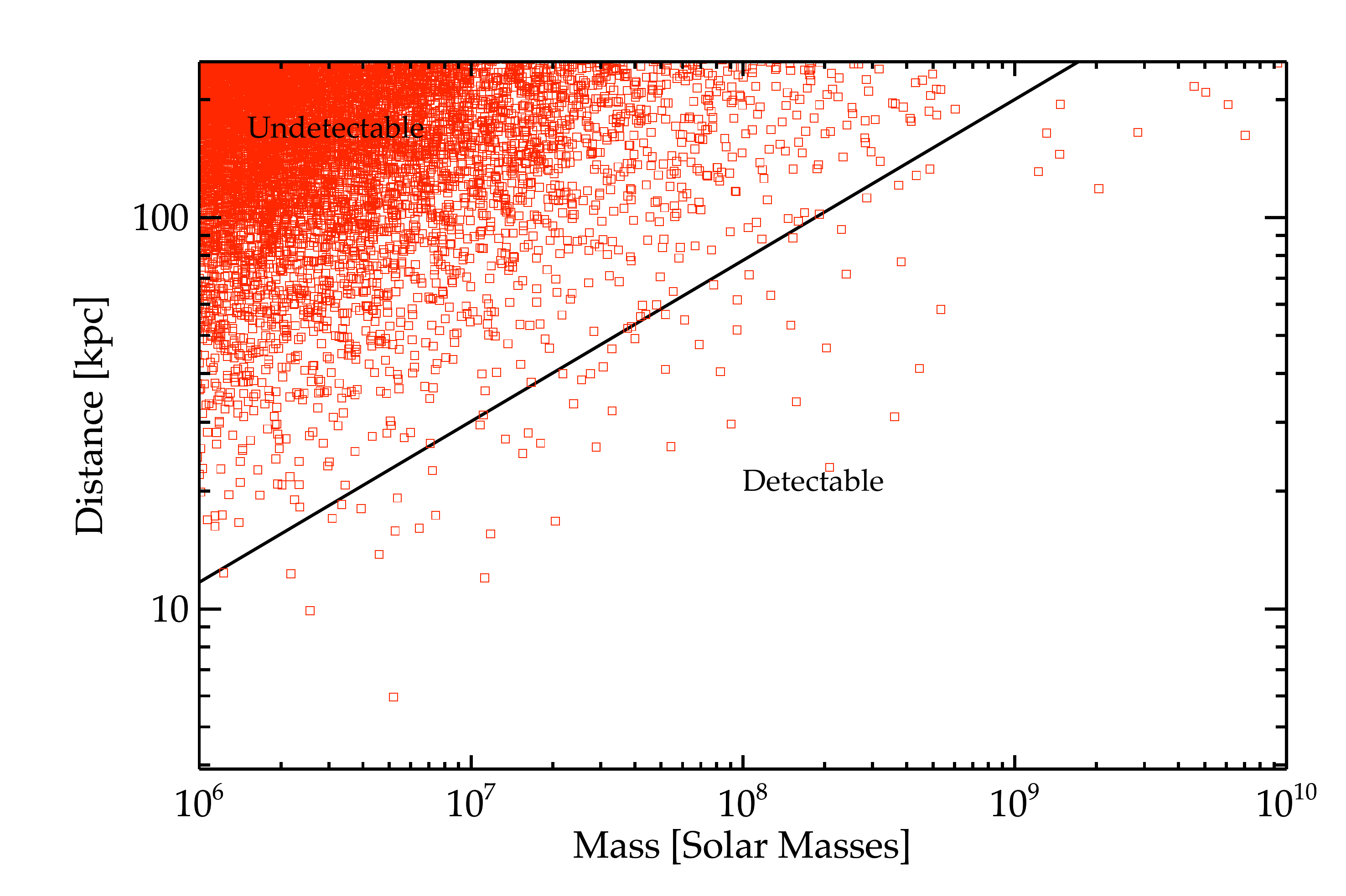} \\
\end{tabular}
\end{center}
\caption{Distance versus mass for subhalos in the Aquarius A-2 simulation~\citep{Springel:2008cc}. Above the diagonal line, the subhalos have a $J$-value that is less than Draco, so they would not be visible in the Fermi-LAT, given that no signal is yet observed from Draco. Adapted from~\cite{Ackermann:2012nb}. 
}
\label{fig:mass_distance} 
\end{figure*}

 Recall that in the LCDM model, the dSphs reside in a subset of the dark matter subhalos that are resolved by N-body simulations. When performing searches for dark matter annihilation from subhalos, the first question to ask is what fraction of these subhalos would be detectable at all given that no signal is yet observed from the dSphs, which have well-determined dark matter masses. Following~\cite{Ackermann:2012nb}, for the purpose of illustration, compare the potential signal from dark subhalos to the dSph Draco, which is at a distance of $80$ kpc, has an integrated mass within $300$ pc of $\sim 10^7$ $M_\odot$~\citep{Strigari:2008ib}, and a total halo mass in the range $\sim 10^8 - 10^9$ $M_\odot$. Figure~\ref{fig:mass_distance} shows the mass versus distance for subhalos from the Aquarius simulation, and on the diagonal line subhalos have the same $J$-value as Draco, assuming that the mass of Draco is $\sim 10^8$ $M_\odot$. Subhalos above the diagonal line would not be visible, while subhalos below the diagonal line do have higher $J$-values than Draco and may be detectable. It is interesting to note that a large majority of the dark subhalos that are detectable reside on the most massive end of the subhalo mass function. This behavior still holds even if the subhalo mass function continues as a power law $dN/dM \sim M^{-1.9}$ to below the resolution limit of modern simulations. In addition to these massive dark subhalos, if the mass function does continue down to approximately Earth mass scales, it still may also be possible that a dark subhalo is very local to the Sun, $\lesssim 1$ kpc, with emission detectable by the Fermi-LAT. 

\par Recall that the two-year Fermi-LAT point source catalog contains 1873 gamma-ray sources, over 30\% of which are unidentified at other wavelengths. So the first place to start the search for subhalos is to examine these unidentified sources. Based on their energy spectrum alone, it is possible to show that a fraction of these can be fit to dark matter models~\citep{Buckley:2010vg}, however further tests are required for a more robust identification. 

\par In order to robustly identify a gamma-ray sources as a subhalo, several criteria must be met. First, the source must have a characteristic spatial extension. It is most likely that either a massive subhalo in the outer part of the Galactic halo or a much lower mass subhalo that is very nearby will have a spatial extension that is resolvable by the Fermi-LAT. Second, if the source is a subhalo, its energy spectrum will deviate from a pure power law model. If the continuum gamma-ray spectrum is similar to a $b\bar b$ spectrum, simulations find that subhalo sources can be reliably distinguished from pure power law energy spectra~\citep{Ackermann:2012nb}. However, the energy spectra of pulsars typically have an exponential cut-off that mimics a $b \bar b$ spectrum, therefore based on the energy spectrum alone it is easy to confuse a potential subhalo source with a pulsar. Combining the extension and energy spectrum criteria with information on (lack of) variability, and cutting out sources near the Galactic plane where confusion is high, in the one year data no subhalo source could be uniquely identified~\citep{Ackermann:2012nb}. For assumptions on the distribution of subhalos, this lack of detection corresponds to an upper limit on $\sigmavavg$ of approximately $10^{-24}$ cm$^3$ s$^{-1}$ at $100$ GeV. 

\par The above analysis is based on the identification of individual subhalos as gamma-ray sources; since LCDM models extrapolate to predict up to $10^{16}$ subhalos in the Milky Way, it is important to ask whether the diffuse emission from these subhalos may be detectable as features in the gamma-ray background. Due to the spatial distribution and energy spectrum of the subhalos in the Galaxy, the probability distribution for the gamma-ray flux has a longer power law tail than a poisson distribution~\citep{Lee:2008fm}. Exploitation of the difference between this signal and background distribution may lead to a robust identification of the dark mater mass and cross section~\citep{Baxter:2010fr}. It is most likely that signals such as these could be extracted if a smooth emission component from the Galactic center or dSphs were discovered first. 

\subsection{Galaxy clusters}
\par Clusters of galaxies are also unique and distant targets from which a dark matter signal may be extracted. Gamma-ray photons from continuum emission, line emission, and inverse Compton emission are all produced through dark matter annihilation in clusters. Since clusters in many instances have well-determined mass distributions, in a manner that parallels the dSph analysis the gamma-ray emission from dark matter annihilation can be predicted starting from astrophysical data. 

\par The mass of a cluster is typically defined within a fixed overdensity, $\Delta$, with respect to the critical density, $\rho_c = 3H^2(z)/8\pi G$, as $M_\Delta = M(< r_\Delta)$ and $M_\Delta/(4/3) \pi r_\Delta = \Delta \times \rho_c$. The radius that encloses the given overdensity is $r_\Delta$. The most robust method to derive the mass of a cluster involves measuring the temperature profile from X-ray spectra, and the electron gas density profile from the X-ray luminosity. For nearby clusters with large values of $\Delta$, masses can then be robustly derived under the assumption of hydrostatic equilibrium, 
\be 
M(r) = - \frac{kT(r)r}{\mu m_p G} \left[ \frac{d \ln \rho(r)}{d \ln r} + \frac{d \ln T(r)}{d \ln r} \right], 
\label{eq:hydrostatic_eq_clusters}
\ee
where $T(r)$ is the temperature profile, and $\mu$ is the mean mass per particle in units of the proton mass. Numerical simulations find that the systematic uncertainty to due non-thermal pressure introduces a $\sim 10\%$ systematic in the mass determination~\citep{Nagai:2006sz}. For clusters without temperature or density profiles, the total mass of the cluster is derived from several observational proxies, which have been calibrated using both observations of low-redshisft clusters and numerical simulations. Three standard observational proxies are the average temperature, the mass of the hot gas, and the product of these two, which is the total thermal energy~\citep{Vikhlinin:2008cd}. The masses of the majority of the clusters are measured via these scaling relations; there are tens of clusters that have measured temperature and density profiles making it possible to use Equation~\ref{eq:hydrostatic_eq_clusters}.

\par Predictions for the gamma-ray signal from dark matter annihilation in clusters typically assume an NFW profile for the dark matter distribution. Recall that in the LCDM framework, the properties of an NFW dark matter halo are entirely specified once the total mass is determined. From the total mass estimate, $M_\Delta$, the concentration, $c_{vir}$, is determined from Equation~\ref{eq:concentration_mass}. From these quantities the virial radius, $R_{vir}$, is determined by standard formulae, e.g.~\cite{Maller:2004au}. The NFW scale radius is then $r_s = R_{vir}/c_{vir}$, and the scale density is determined from
\be 
\rho_s(c_{vir}) = \frac{M_{vir}}{4 \pi r_s^3} \left[ \ln (1+c_{vir}) - \frac{c_{vir}}{1+c_{vir}}\right]^{-1}. 
\label{eq:nfw_scale_density} 
\ee

\par Similar to the dSphs, the most important clusters for indirect dark matter detection are those that are the appropriate combination of the nearest and the most massive. Based solely on their derived mass distributions from the formalism outlined above and their distances, several recent studies have come to the agreement that the Fornax cluster is the brightest source of gamma-rays from dark matter annihilation~\citep{Ackermann:2010rg,Ando:2012vu,Pinzke:2011ek,Nezri:2012tu}. Coma and Virgo are also expected to be amongst the brightest gamma-ray sources. Though formalism above for determining cluster masses is well-motivated from a theoretical point of view, systematic uncertainties will be introduced if the mass profile deviates from an NFW model. This is particularly relevant for the gamma-ray analysis because a few of the most nearby clusters are predicted to be the brightest sources of gamma-rays. 

\par Unlike the case of dSphs, there is expected to be a significant increase in the gamma-ray signal from clusters due to dark matter substructure. This is pleasing from a detection perspective because it provides a well-motivated reason for clusters to have brighter emission than is deduced from the smooth dark matter component. However, in the case of a null-detection, it introduces a theoretical systematic that makes it difficult to interpret the meaning of an upper limit on the annihilation cross section. 

\par Numerical simulations provide the most reliable method for determining the gamma-ray emission from subhalos in clusters. From a sample of nine cluster-mass dark matter halos, which have individual particle masses of $\sim 10^6$ $M_\odot$ and identify subhalos down to a mass scale of $\sim 10^7$ $M_\odot$, ~\cite{Gao:2011rf} directly determine the overall boost factor and the surface density profile of the substructure component for clusters as a function of the cluster mass $M_{200}$. Due to the resolution limit of their simulations, only the gamma-ray luminosity near the cluster virial radius increases due to subhalos. Starting from a mass scale above which subhalos are resolved and extrapolating down to $10^{-6}$ $M_\odot$,~\cite{Gao:2011rf} determine a boost factor of
\be 
B(M_{200}) = 1.6 \times \left(\frac{M_{200}}{\textrm{M}_\odot}\right)^{0.39},
\label{eq:cluster_boost} 
\ee
and 
\be 
S_{sub}(R) = \frac{16 B(M_{200}) {\cal L} }{\pi \ln (17)} \frac{1}{r_{200}^2 + 16 R^2},
\label{eq:cluster_surface_brightness}
\ee
where $R$ is the projected radius and ${\cal L}$ is the luminosity of the smooth component of the cluster. Though it relies on an extrapolation, the increase in the luminosity over the smooth component in clusters is much greater than is expected in dSphs. 

\par Also unlike the case of dSphs, in clusters there is expected to be a significant flux of gamma-rays due to cosmic ray processes. For clusters of interest, the emission separates into component functions that depend on radius and energy, so that the source function decomposes as
\be
\frac{dN}{dt dV dE} = A(r) s(E).
\label{eq:cluster_source_function}
\ee
The spatial factor, $A(r)$, is proportional to the square of the gas density in the cluster, and the energy component, $s(E)$, varies as $dn/dE \sim E^{-2.3}$ for energies up to approximately $100$ GeV. Because the gamma-ray luminosity in clusters traces the gas density, it is more centrally-concentrated than the dark matter-induced signal, which is more extended because of the emission from subhalos. 

\par Theoretical calculations have shown that both the cosmic ray and the dark matter annihilation induced gamma-rays may be detectable by the Fermi-LAT~\citep{Pinzke:2011ek,Nezri:2012tu}. However, no gamma-ray emission has been conclusively associated with clusters in the Fermi-LAT one year data, assuming that they are point sources~\citep{Ackermann:2010rg}. Modeling clusters as extended sources, no gamma-ray emission was yet found in two year data~\citep{Han:2012uw}. At a dark matter mass of 10 GeV, the limits on the annihilation cross section deduced from these observations vary anywhere from $10^{-24}$ cm$^3$ s$^{-1}$ to below $10^{-26}$ cm$^3$ s$^{-1}$, depending on what is assumed for the emission from subhalos. 

\par HESS has made a pointed observation of the Fornax cluster, obtaining a total exposure time of 14.5 hours~\citep{Abramowski:2012au}. From a null detection, it has placed limits on $\sigmav$ for different assumptions of the dominant annihilation channel and the slope of the central dark matter profile. The limits obtained are very sensitive to both of these assumptions, in the latter case because the angular resolution of HESS probes the central region of the dark matter profile of the cluster. Depending on the assumption for the amount of substructure in the cluster, the limits on $\sigmavavg$ are in the range $10^{-21}$ cm$^3$ s$^{-1}$ to $10^{-23}$ cm$^3$ s$^{-1}$. 

\subsection{Summary of gamma-ray results} 

\par Table~\ref{tab:summary_cross_section} provides a summary of modern sensitivity limits for dark matter of mass $10$ and $100$ GeV from the various gamma-ray searches described above. 

\begin{deluxetable}{c c c c c c c c}
\tablecolumns{7}
\tablecaption{Summary of limits on the annihilation cross section times the relative velocity, $\sigmav$, in units cm$^3$ s$^{-1}$ for dark matter of mass 10 and 100 GeV for the different sources. For mass of 10 GeV, the limits are given for $100\%$ annihilation into $b \bar b$ and $\tau^{+} \tau^{-}$. For a mass of 100 GeV, the limits are given for $100\%$ annihilation into $b \bar b$, $\tau^{+} \tau^{-}$ and $W^{+} W^{-}$.
 \label{tab:summary_cross_section}}
\tablehead{
Source & \multicolumn{2}{c}{$10$ GeV} & \colhead{} & \multicolumn{3}{c}{$100$ GeV}\\
\cline{2-3} \cline{5-7}\\
\colhead{} & \colhead{$b \bar b$} & \colhead{$\tau \bar \tau$} & \colhead{} & \colhead{$b \bar b$} & \colhead{$\tau \bar \tau$} & \colhead{$W^{+} W^{-}$} & \colhead{}  }
\tablewidth{0pc}
\startdata
Dwarf spheroidals~\tablenotemark{a} & $1 \times 10^{-26}$ &$1 \times 10^{-26}$&& $7 \times 10^{-26}$&$1 \times 10^{-25}$&$1 \times 10^{-25}$\\
Diffuse Galactic halo~\tablenotemark{b}  &$1 \times 10^{-26}$ &$2 \times 10^{-26}$ && $1 \times 10^{-25}$&$1 \times 10^{-25}$&--\\
Diffuse extragalactic~\tablenotemark{c} &$2 \times 10^{-25}$&--&&$1\times10^{-24}$&--&--\\
Clusters~\tablenotemark{d}&$1 \times 10^{-25}$&$6 \times 10^{-24}$&&$1 \times 10^{-25}$&$3\times10^{-23}$&$1\times10^{-23}$
\enddata
\tablenotetext{a}{~\cite{Ackermann:2011wa,GeringerSameth:2012sr}}
\tablenotetext{b}{Results for NFW profile. Subject to the uncertainty in the local dark matter density. From~\cite{Ackermann:2012rg}} 
\tablenotetext{c}{~\cite{Abdo:2010dk}}
\tablenotetext{d}{Assumes just the emission from the smooth component and not subhalos. From~\cite{Han:2012uw,Ando:2012vu}}.
\end{deluxetable}

\par Because no clear dark matter signal has been detected yet in gamma-ray data, the above analysis has primarily focused on methodology for deriving limits on the annihilation cross section. Recalling the discussion in the beginning of this section it is always important to keep in mind that the nature of the signal we are searching for is not precisely predicted, so it is necessary to pay careful attention to details and residuals in data sets where a signal may be lurking. It must also be kept in mind that our knowledge of astrophysical systems is incomplete, and also our understanding of experimental and detector related issues is often incomplete. It is often tempting to associate our lack of understanding of conventional astrophysical processes or detector systematics with a dark matter signal. 

\par Though as discussed above the published Fermi-LAT results report null searches for gamma-ray lines, recently a feature has come to light in the data near the Galactic center where an excess over a pure power law model for the background is found~\citep{Bringmann:2012vr,Weniger:2012tx}. Even though the hint of a line-like signal is intriguing, there are certainly reasons to proceed with caution. First, the signal does not directly coincide with the Galactic center, so a mechanism must be invoked to displace the center of the dark matter halo from that of the baryons. Second, if the signal is in fact a result of dark matter, it should be much more spatially extended above and below the Galactic plane. Third, the power law parameterization of the background may not be appropriate for the small spatial regions that are being examined. Fourth, from a theoretical perspective, the line emission should generically be accompanied by continuum photons produced via hadronization and fragmentation. Since no continuum emission is found above the power law background~\citep{Cohen:2012me,Cholis:2012fb}, the ratio of the continuum to line signal must be much less than is generally found in particle dark matter models. A larger sample of data and cleaned data sets should be able to shed more light on this feature over the next few years. 

\subsection{Neutrino searches} 
\par Large-scale neutrino experiments, in particular Ice Cube and Super-Kamiokande, are now gaining sensitivity to study dark matter through indirect detection. Both of these experiments detect ``up-going" muon neutrinos that interact with nucleons in the Earth to produce muons that make Cerenkov light in the respective detectors. This mode of observation is used primarily to reduce the background from atmospheric muon neutrinos, which produce a background from ``down-going" muons above the horizon. The lower energy threshold for Ice Cube observations is 10 GeV, while Super-Kamiokande extends to a lower energy threshold allowing it to study lower mass dark matter. The angular resolution for up-going muon events is approximately one degree for these experiments. This section reports results from searches for neutrinos from the various sources discussed above in the context of gamma-ray searches. Searches for neutrinos from dark matter that gets captured by scattering off of nuclei in the Sun will be discussed in the section on direct dark matter limits (Section~\ref{sec:direct_detection_experiment}); these observations are most sensitive to the spin-dependent WIMP-nucleon cross section. 

\par Ice Cube has now reported limits on the annihilation cross section into several channels through observations of the diffuse Galactic halo with an exposure of $276$ days and a $22$-string configuration~\citep{Abbasi:2011eq}. For dark matter mass greater than $20$ GeV, the best-constrained channel is the direct annihilation of dark matter to neutrinos, for which the limits are $\sigmav \lesssim 10^{-22}$ cm$^3$ s$^{-1}$. For a $b \bar b$ annihilation channel, for a dark matter mass of $30$ GeV the upper limit is $\sigmav \simeq 10^{-19}$ cm$^3$ s$^{-1}$, while for a dark matter mass of $10$ TeV the upper limit is $\sigmav \simeq 10^{-19}$ cm$^3$ s$^{-1}$. For the other channels the limits are between the $b \bar b$ and direct neutrino limits. 

\par Ice Cube has very recently presented limits on the annihilation cross section into neutrinos from direct observations of the Galactic center~\citep{Abbasi:2012ws}. For direct annihilation into neutrinos, the limit on the annihilation cross section at approximately $200$ GeV is $\sim 10^{-23}$ cm$^3$ s$^{-1}$, which improves upon the aforementioned diffuse limits by about an order of magnitude. For the remaining channels, the limits are typically improved in comparison to the diffuse limits for WIMP masses less than a few hundred GeV. 

\subsection{Antimatter searches}
\par Positrons and anti-protons produced in dark matter annihilation will propagate to the Earth, and their detection is a complementary probe of dark matter in the Galaxy. There are now detailed theoretical treatments that analyze the propagation of positrons and anti-protons from their production sites, either from a dark matter source or astrophysical source such as pulsars, through the Galaxy. For non-dark matter sources, secondary production of positrons and anti-protons is probably the best established mechanism for describing their propagation. In this model cosmic ray nuclei interact inelastically with interstellar gas, producing charged pions that decay to positrons~\citep{Moskalenko:1997gh}. This process typically results in a positron fraction that declines with energy. 

\par Though not specifically designed for indirect dark matter detection, within the past few years it has been realized that antimatter experiments do have a unique position in Galactic searches for dark matter. The PAMELA experiment has found an interesting rise in the positron fraction at energies up to 100 GeV~\citep{Adriani:2008zr}, extending the previous results from the HEAT experiment~\citep{Barwick:1995gv}. This result has been corroborated by the measurements of the positron spectrum from the Fermi-LAT collaboration~\citep{Abdo:2009zk,FermiLAT:2011ab} and HESS~\citep{Aharonian:2009ah}. Measurements of the anti-proton flux by PAMELA on the other hand are in agreement with the predictions for local astrophysical sources~\citep{Adriani:2008zq}. Though the rise of the positron spectrum may be connected to interesting local astrophysical sources, it is unlikely to be due to dark matter because the required cross section is approximately two orders of magnitude larger than the thermal relic scale. Forthcoming results from the AMS-02 spectrometer are expected to shed more light on the PAMELA results (e.g.~\cite{Hooper:2012gq}). 

\subsection{Decaying dark matter and X-ray searches}
\par Up to this point the analysis in this section has focused on stable dark matter particles that annihilate into detectable standard model final states, restricting to energies in the gamma-ray band and above. A variety of models also predict that the dark matter is unstable, and a small fraction of the particles are decaying into detectable final states at the present epoch. For these searches the energy of the radiation produced can extend down to the X-ray energy band, and even below. 

\par Sterile neutrinos at the mass scale of approximately a few keV that are stable over the lifetime of the universe provide good candidates for dark matter~\citep{Abazajian:2012ys}. Because the sterile neutrino mixes with standard model neutrinos, it radiatively decays into a photon and a lighter neutrino. For a majorana neutrino, this decay rate is given by
\be
\Gamma \simeq 1.4 \times 10^{-30} \textrm{s}^{-1} \left[ \frac{\sin^2 2\theta} {10^{-7}} \right] \left[\frac{m_s}{1 \textrm{keV}}\right]^5, 
\label{eq:sterile_neutrino} 
\ee
where $m_s$ is the mass of the sterile neutrino and $\sin^2 2\theta$ is the mixing angle. This mixing angle is constrained to be $\lesssim 10^{-6}$ to avoid overproduction in the early universe. 

\par Beyond approximately $20$ keV, the null detection of M31 from XMM-Newton observations provides the best limit on the mass and mixing angle of sterile neutrinos~\citep{Watson:2006qb,Watson:2011dw}. Sterile neutrino masses greater than about $30$ keV can be ruled out from the non-observation of lines from the INTEGRAL satellite~\citep{Yuksel:2007xh}. 
Null reports have also been reported from searches for an X-ray signal from decaying dark matter from Galactic satellites~\citep{Loewenstein:2012px}. 

\subsection{Summary of results and interpretations}
 
\par As has been outlined through the course of this section, over the past several years indirect dark matter searches have made an extraordinary amount of progress. We have gone from wondering what contributions the Fermi-LAT could make to dark matter studies, to for the first time probing thermal relic scale cross sections. It is now appropriate to take a step back and consider some examples of what these results tell us about some theoretical models. 

\par In terms of the models presented in Equations~\ref{eq:sv_scalar_scal}-~\ref{eq:sigma_v-T}, because of the lack of velocity suppression in the annihilation cross section the indirect detection results are most sensitive to fermions with psuedoscalar, vector, and tensor interactions, and scalar dark matter with scalar, scalar-pseudoscalar, and vector-axialvector interactions.  In order to see how the results presented here constrain these dark matter models, it is informative to consider a specific example. For scalar dark matter consider the case of scalar interactions, with an annihilation cross section given by Equation~\ref{eq:sv_scalar_scal}. For universal couplings to fermions, the relic abundance is set by annihilation into all of the fermions. The relic abundance serves to set the couplings $F_s$. For the sake of clarity, say that the relic abundance is set by the annihilation into $b \bar b$, for which case the coupling is then fixed to $F_s \simeq 3 \times 10^{-4}$ GeV$^{-1}$. Including all fermionic annihilation channels decreases this by a factor of a few. It is clear that the Fermi-LAT dSph results now probe scalar couplings at this level in the mass range $10-25$ GeV. 

\par A second interesting example is fermionic dark matter particles with pseudoscalar couplings. This is particularly interesting because, as discussed in detail in the following two sections, these are most probably unconstrained by direct detection experiments (in the absence of light mediating particles), so indirect methods are the only ones that can probe dark matter with dominant pseudoscalar interactions. In contrast to the scalar dark matter case discussed above, the annihilation cross section in this case depends on the square of the dark matter mass (Equation~\ref{eq:sigma_v-P}). From the same arguments as above, for full annihilation into $b \bar b$, the coupling that the Fermi-LAT dSph results are probing in the mass range $10-25$ GeV is $\sim 5 \times 10^{-6}$ GeV$^{-2}$. This analysis clearly indicates that there is a complementarity between indirect and direct dark matter searches in probing different dark matter models.

\newpage 
\section{Direct Detection: Theoretical Developments} 
\label{sec:direct_detection_theory}
\par~~\cite{Goodman:1984dc} first suggested that if weakly interacting massive particles are the dominant component of mass of the Galactic halo, they could be detected by interactions with nuclei in low background underground detectors. Though the guiding theoretical framework is very much similar to what was originally discussed in the~\cite{Goodman:1984dc} paper, in the now almost thirty years subsequent to this work, a tremendous body of literature has been established that has refined and expanded upon direct dark matter detection theory. The theoretical work has led to a steady understanding that the theory of direct dark matter detection is very rich, both from the perspective of the dependence on forefront topics in modern Galactic astrophysics and on the dependence on particle dark matter models. 

\par Since the initial papers on direct detection, numerical simulations have dramatically-improved, and it has been realized that they play an important role in advancing our understanding of the dark matter distribution in galaxies like our Milky Way. Although they are still nowhere near the resolution limit that is needed to study the detailed phase space structure of the dark matter on scales that are relevant for direct detection, they do allow us to extend beyond the most naive models and provide clues as to how different phase space models affect direct detection. From the particle physics perspective, the flurry of experimental results, and hints of possible signals, has motivated a wide variety of theoretical models designed to accommodate them. Even if in the long run the experimental hints are unable to withstand further scrutiny, these models have led researchers down interesting theoretical paths that likely would not have been taken otherwise. It is certainly true that, in order to eventually extract necessary information from experimental data, we must have an in depth understanding of how to separate the dependence of the signal on Galactic model parameters and those parameters related to the nature of the interactions of the dark matter particle.  

\par There are several excellent reviews that lay out both the astrophysical and the particle physics aspects of the theoretical framework for direct detection~\citep{Jungman:1995df,Lewin:1995rx}. This section builds upon these reviews by highlighting the important theoretical progress that has occurred during the past several years in this field. This section begins by reviewing the ingredients that go into determining the flux and the detection rate of WIMPs at direct detection experiments. Following this is a discussion of basic models for the WIMP-nucleus elastic scattering cross section; modern experimental constraints on these models will be discussed in Section~\ref{sec:direct_detection_experiment}. The focus then moves to a discussion of how the detected rate of WIMPs depends on the Galactic halo model, accounting for the smooth dark matter distribution, the distribution of dark matter substructure, and modern estimates for the WIMP velocity distribution. Finally, recent theoretical work is discussed that attempts to understand how the properties of the WIMP can be extracted from experimental data sets by rigorous modeling of the dark matter distribution in the Galaxy. 

\subsection{WIMP detection methods}
\par The local flux of WIMPs is given by their number density in the Solar neighborhood times their velocity, $n v$. For target particles in an underground detector, the total cross sectional area seen by the WIMP is the number of targets, $N_T$, times the cross section per target, $\sigma$. The WIMP interaction rate is then $R = N_T \sigma n v$. 

\par The first part of the discussion in this section will take the observable signature of a WIMP to be the energy deposited to a nucleus in an underground detector, $E_r$. Define the local velocity distribution of WIMPs as $f(\vec v, \vec v_E)$, where the notation is now so that $\vec v$ represents the WIMP velocity relative to the Earth and $\vec v_E$ is the velocity of the Earth relative to the Galactic rest frame. The recoil momentum of the nucleus, in the lab frame, is $q = 2m_r v \cos \theta$, where $\theta$ is the angle between the incoming WIMP and the scattered nucleus, $m_N$ is the mass of the nucleus, $\mdm$ is the WIMP mass, and $m_r = m_N \mdm/(m_N + \mdm)$ is the WIMP-nucleus reduced mass. The recoiling energy of the nucleus is then $E_r = q^2/(2m_N) = 2 m_r^2 v^2 \cos^2 \theta/m_N$, and the minimum speed to scatter a nucleus to a recoil energy $E_r$ is 
\be 
v_{min} = \sqrt { \frac{m_N E_r}{2 m_r^2} }. 
\label{eq:vmin}
\ee
Note that Equation~\ref{eq:vmin} is strictly valid only for elastic scattering interactions; for inelastic scatterings, a term $\delta /  \sqrt{2m_N E_r}$ is added to Equation~\ref{eq:vmin}~\citep{TuckerSmith:2001hy}, with $\delta$ being the mass splitting for particles in the spectrum. 

\par The velocity of a WIMP in the Galaxy is of order  $v \sim 10^{-3} c$. For WIMP masses and nuclei that we are considering, the reduced mass is typically well approximated by $m_r \simeq 100$ GeV. For the above velocity and $m_r$, the momentum transfer is $q \simeq 100$ MeV. The energy imparted to the nucleon is then $E_r = |q|^2/2m_N \simeq (100 \, \textrm{MeV})^2/(2 \, A \, \textrm{GeV}) = 10^4 \, \textrm{keV}/(2 A)$, where $A$ is the mass number of the nucleus. Since we will be focusing on nuclei with $A \simeq 100$, the recoil energy of the nucleus is in the range of tens to hundreds of keV. 

\par With these definitions, the differential event rate as a function of nuclear recoil energy is then given as 
\be 
\frac{dR}{dE_r} = N_T \frac{\rho_\odot}{\mdm} \int_{v_{min}}^{v_{esc}} \frac{d\sigma(E_r,v)}{dE_r} v f(\vec v, \vec v_E) d^3 \vec v, 
\label{eq:WIMP_event_rate}
\ee
where $v_{esc}$ is the escape velocity in the Galactic reference frame, and $\rho_\odot = \rho(R_\odot)$ is the local mass density of WIMPs. 

\par The velocity of the Earth with respect to the Galactic rest frame, $\vec v_E$, is a sum of three separate components: the velocity of the LSR, the peculiar motion of the Sun with respect to the LSR, and the orbital motion of the Earth around the Sun. To calculate each of these components, consider a coordinate system so that $U$ points in the direction of the Galactic center, $V$ points in the direction of Galactic rotation, and $W$ towards the Galactic north pole. In these coordinates, the LSR velocity is $\vec v_{LSR} \simeq (0,220,0)$ km/s. The most recent determination of the mean velocity of the Sun with respect to the LSR gives $\vec v_\odot = (U,V,W) = (11.1,12.24,7.25) $ km/s~\citep{Schoenrich:2009bx}.

\par Because of the motion of the Earth about the Sun, the flux of WIMPs in the Earth rest frame varies periodically during the course of the year~\citep{Freese:1987wu,Freese:2012xd}. This effect manifests itself in a time dependence in $\vec v_E$ in the velocity distribution in Equation~\ref{eq:WIMP_event_rate}. Including both the motion of the Sun through the Galaxy and the Earth motion about the Sun, 
\be
\vec v_E(t) = \vec v_{LSR}  + \vec v_\odot + V_\oplus \left[ \cos \omega(t-t_1)\vec \epsilon_1 + \sin \omega(t-t_1) \vec \epsilon_2 \right],
\label{eq:annual_modulation}
\ee
where $V_\oplus = 29.8$ km/s is the orbital speed of the Earth around the Sun, $\omega = 2\pi$/year, $\vec \epsilon_1$ is the direction of the velocity of the Earth at a time $t$, and $\vec \epsilon_2$ is the direction of the velocity of the Earth at a time $t_1+0.25$ yr. 

\par The amplitude of the annual modulation signal is typically approximately 1-10\% of the averaged total event rate. It is a function of $v_{min}$, and depends strongly on the anisotropy of the velocity distribution. Due to the strong dependence on $v_{min}$, from a precise determination of this phase reversal of the signal it may be possible to strongly constrain the WIMP mass~\citep{Lewis:2003bv}.

\par The above analysis has focused on determining the event rate as a function of energy deposited in a detector from a WIMP-nucleus scattering event. It has ignored the direction of the recoiling nucleus, by integrating over all possible recoil directions. Understanding, and measuring, the distribution of nuclear recoil directions is also important, because it contains information both on the direction of the Earth through the Galactic halo and on the velocity distribution. Because of the former dependence on the direction of the Earth motion, it provides a clear confirmation that a signal determined from the energy spectrum is due to WIMPs and not from a various background component. Because of the latter dependence on the velocity distribution, the directional signal may contain information on the phase space properties of the dark matter halo (note that neither the outgoing WIMP momentum or its direction are detectable after a scattering event).

\par The simplest derivation of the directionally-dependent event rate spectrum neglects both the motion of the Sun with respect to the LSR, and the motion of the Earth around the Sun. Define $d\Omega = d\phi d(\cos \theta)$ as the solid angle associated with the direction of the recoiling nucleus, where as above $\theta$ is the angle between the direction of the incoming WIMP and the scattered nucleus, and $\hat v_R$ as the direction of the recoiling nucleus. From the dark matter velocity distribution $f(\vec v)$, the recoil rate as a function of both the energy and the direction is~\citep{Gondolo:2002np}
 \be 
 \frac{dR}{dE_r d \Omega_R} =  \frac{1}{2\pi} N_T \frac{\rho_\odot}{\mdm} \int_{v_{min}}^{v_{esc}} d^3 \vec v \, v^2 \, \frac{d \sigma}{d E_r} f(\vec v) \, \delta [(\vec v - \vec v_\odot) \cdot \hat v_R - v_{min}(E_r)].
 \label{eq:directional_rate}
 \ee
Here the explicit dependence of the cross section on the recoil energy and incoming velocity is indicated; simplified models for this cross section will be considered in more detail below. Note that for an isotropic velocity distribution, the event rate will be a maximum in the direction opposite to the motion of the LSR. 
 
\subsection{Scattering cross section models} 
\par The WIMP-nucleon scattering cross section fundamentally derives from the WIMP interactions with quarks. A theoretical model predicts specific interactions, which in general may reduce to functions of the WIMP spin, nuclear spin, the WIMP velocity, and the momentum transfer to the nucleus. Given the scale of the velocity and momentum transfer for WIMP-nucleon scattering, interactions that are not suppressed by powers of them will typically be the most relevant for direct detection. Spin-independent (SI) interactions are defined as those that do not depend on the nuclear spin, while spin-dependent (SD) interactions do depend on the nuclear spin. 

\par It is standard to define the total cross section in terms of the zero momentum transfer cross section, $\sigma_0$, as 
\be 
\sigma = \frac{\sigma_0}{q_{max}^2} \int d q^2 F^2(q)
\label{eq:dsigmadq2}
\ee
where $q_{max}^2 = 4 m_r^2 v^2$ is the maximum value of the momentum transfer. This assumes that the scattering is described by a contact interaction. The form factor is~\citep{Lewin:1995rx}, 
\be 
F^2(q) = \frac{9\left [\sin(qR_n) - qR_n \cos (qR_n) \right]^2}{(qR_n)^6},
\label{eq:form_factor}
\ee
where $R_n$ is the approximate nuclear radius, 
\be 
R_n \simeq \left[ 0.91 A^{1/3} + 0.3\right] \times 10^{-13} \textrm{cm}. 
\label{eq:radius}
\ee
All of the dependence on the momentum transfer is then contained in the form factor. 

\par Determining $\sigma_0$ is then a matter of specifying the nature of the WIMP and its interactions. Effective models in which a scalar WIMP couples to a fermion through scalar or vector interactions (Equations~\ref{eq:scalar:scal} and~\ref{eq:scalar:vect}) and have the right relic abundance of dark matter are already strongly constrained by direct detection experiments~\citep{Beltran:2008xg,Yu:2011by}. For Dirac and Majorana fermions that have scalar interactions with quarks (Equation~\ref{eq:fermion:scal}), the cross section is 
\be 
(\sigma_0)_S = C \frac{m_r^2}{\pi} \left[ Z \left(\frac{G_{S,p}}{\sqrt{2}}\right) + (A-Z)  \left(\frac{G_{S,n}}{\sqrt{2}}\right)  \right]^2,
\label{eq:fermion_scalar_cross_section} 
\ee
where $C = 1(4)$ for a Dirac (Majorana) particle. Here $Z$ is the number of protons, and $A$ is the mass number of the nucleus. This model gives SI interactions. The factors $G_{S,(p,n)}$ translate the coupling of the quarks to those of the protons and neutrons, 
\be
 \frac{G_{S,(p,n)}}{\sqrt{2}} = \sum_{q = u,d,s} \frac{m_{(p,n)}}{m_q}  \frac{G_{S,q}}{\sqrt{2}} f_{T_q}^{(p,n)} + \frac{2}{27} \left( 1- \sum_{q=u,d,s} f_{T_q}^{(p,n)} \right) \sum_{q=c,b,t}  \frac{G_{S,q}}{\sqrt{2}}  \frac{m_{(p,n)}}{m_q}. 
\label{eq:fpn_scalar} 
\ee
The various factors in Equation~\ref{eq:fpn_scalar} are $f_{Tu}^{(p)} = 0.020 \pm 0.004$, $f_{Td}^{(p)} = 0.026 \pm 0.005$, $f_{Tu}^{(n)} = 0.014 \pm 0.003$, $f_{Tu}^{(p)} = 0.036 \pm 0.008$, and $f_{Ts}^{(p,n)} = 0.118 \pm 0.062$~\citep{Jungman:1995df}. 

\par For a Dirac fermion with vector interactions to quarks (Equation~\ref{eq:fermion:vector}), the cross section is 
\be 
(\sigma_0)_V = \frac{m_r^2}{\pi} \left[ Z \left(\frac{G_{V,p}}{\sqrt{2}}\right) + (A-Z) \left(\frac{G_{S,n}}{\sqrt{2}}\right) \right]^2,
\label{eq:fermion_vector_cross_section} 
\ee
where 
\bea
G_{V,p} &=& 2 G_{V,u} + G_{V,d} \\
G_{V,n} &=& G_{V,u} + 2G_{V,d}
\eea
This model also gives SI interactions. Note that the vector current vanishes for Majorana fermions.


\par For a fermion with axial vector interactions to quarks (Equation~\ref{eq:fermion:axial}), the cross section is 
\be 
(\sigma_0)_A = C \frac{m_r^2}{\pi} \frac{J_A + 1}{J_A}\left[ \frac{G_{A,p}}{\sqrt{2}} S_p^A + \frac{G_{A,n}}{\sqrt{2}} S_n^A \right]^2 
\label{eq:axial_cross_section}, 
\ee
where C = 4 (16) for Dirac (Majorana) fermions, $J_A$ is the nuclear spin, and $S_p^A$ and $S_n^A$ are the expectation values for the spins of the proton and neutron, respectively. The couplings in Equation~\ref{eq:axial_cross_section} are given by $G_{A,{(p,n)}} = \sum_{q = u,d,s} G_{A,q} \Delta_q^{(p,n)}$, and $\Delta_u^p = 0.842$, $\Delta_d^p = -0.427$, $\Delta_s^p = -0.085$, $\Delta_u^n = \Delta_d^p$,  $\Delta_d^n = \Delta_u^p$, $\Delta_s^n = \Delta_s^p$~\citep{Jungman:1995df}. For a Dirac fermion, the tensor interaction has a similar form to Equation~\ref{eq:axial_cross_section} with $C = 16$. 

\par From Equations~\ref{eq:fermion_scalar_cross_section},~\ref{eq:fermion_vector_cross_section}, and~\ref{eq:fermion_vector_cross_section}, and assuming universal coupling to protons and neutrons, the cross section per nucleon is 
\be 
\sigma_N = \left( \frac{m_{r,\star}}{m_{r,N} A} \right)^2 \sigma_0, 
\label{eq:cross_section_per_nucleon} 
\ee
where $m_{r,N}$ is the reduced mass between the WIMP and the nucleus and $m_{r,\star}$ is the reduced masses between the WIMP and the nucleon. 

\par The interactions that lead to Equations~\ref{eq:fermion_scalar_cross_section},~\ref{eq:fermion_vector_cross_section}, and~\ref{eq:axial_cross_section} are typically the most important for direct detection. The remaining interaction terms are suppressed by at least one power of the momentum or the velocity (e.g.~\cite{Kurylov:2003ra} and~\cite{Fan:2010gt}). When the interaction is not suppressed at high momentum transfer, the SI interactions are enhanced relative to the SD interactions by the factors in the brackets proportional to $A^2$. 

\par On the most general grounds, a Dirac fermion that couples to the $Z^0$ through vector interactions (Equation~\ref{eq:fermion_vector_cross_section}) has a significant SI cross section that was large enough to be ruled out by the initial direct detection experiments nearly twenty years ago. This bound of course can be evaded for Majorana fermions since they do not have vector currents. It is straightforward to construct a Majorana fermion that has axial vector interactions and couples to the $Z^0$ and a scalar coupling to the Higgs; from the cross sections above these lead to SD and SI coupling respectively. For the later couplings, the scale for the scattering cross section is approximately a few times $10^{-44}$ cm$^2$; this scale is now being probed by direct detection experiments sensitive to SI couplings (Section~\ref{sec:direct_detection_experiment}). 

\par It is informative to compare to the predictions of different theoretical models. For example, in the scalar singlet model~\citep{McDonald:1993ex,Burgess:2000yq}, the scalar WIMP couples to the nucleons through a coupling to the Higgs. In this model the cross section for elastic scattering is a function of the coupling of the Higgs to the nucleus and the coupling of the dark matter to the Higgs. In supersymmetric models, there are two primary interactions that survive in the low energy limit. The dominant contribution to SI interactions between neutralinos and quarks arises from Higgs and squark exchanges. Neutralino SD interactions on the other hand have contributions from both $Z$ and squark exchanges~\citep{Jungman:1995df}. 

\subsection{Dependence on local dark matter density} 
\par The dependence of the WIMP rate on the local dark matter density is straightforward-- it is simply a normalizing factor that scales Equation~\ref{eq:WIMP_event_rate}. Recall from Section~\ref{sec:MW} that there has been a variety of recent determinations of the local dark matter density. Although it is typical for experiments to quote results assuming a canonical local dark matter density of $0.3$ GeV cm$^{-3}$, in Section~\ref{sec:MW} it was seen that the local dark matter density may be up to three times larger than the canonical value of $0.3$ GeV cm$^{-3}$, though these results still systematically depend on the inputs of the analysis and the specific stellar population that is utilized~\citep{Garbari:2011dh,Garbari:2012ff}. Note that even these estimates are still below the local baryonic material by up to a factor of several, and they still carry both significant systematic and statistical uncertainties. Various other analyses that add in constraints from the total Galactic potential find that this uncertainty is reduced, and the mean central value for the dark matter density is slightly larger~\citep{Catena:2009mf,McMillan:2011wd}. 

\par The local kinematic measurements described above are primarily sensitive to the ``smooth" distribution of dark matter in the Solar neighborhood. However, as discussed in Section~\ref{sec:simulations} from theoretical predictions of dark matter halo formation in cold dark matter cosmology, the distribution of dark matter in the Galactic halo is not smooth, so in principle it is possible that the Sun resides in either a significant local over or under density of dark matter, which may affect the implied constraint on the WIMP elastic scattering cross section. Numerical simulations~\citep{Vogelsberger:2008qb,Kuhlen:2009vh}, as well as analytic models~\citep{Kamionkowski:2008vw}, have begun to address this issue, finding that the probability for the Sun to reside in a significant over density or under density is small, of order $10^{-4}$\%. Further, there are predictions of a dark matter disk in the Galaxy, that may have its origin in the accretion of a massive satellite galaxy that was dragged into the disk by dynamical friction~\citep{Read:2008}. However, analysis of stellar kinematics that extend out beyond a few kpc place strong limits on a dark matter disk component in the Galaxy~\citep{Bidin:2010rj}. So in sum, our modern theoretical expectation is that the event rates in WIMP direct detection experiments will be dominated by the smooth component, i.e. the component that is not locked up into subhalos or streams. 

\subsection{Dependence on the velocity distribution} 
\par The dependence of the WIMP event rate on the velocity distribution is more phenomenologically-interesting than the dependence on the local dark matter density. However, because it cannot be directly inferred from stellar and gas kinematics like the local density is inferred, the velocity distribution is essentially impossible to deduce from astronomical observations. Indeed, probably the only method to measure the local velocity distribution is through a direct detection experiment itself. This means that our understanding of the impact of the velocity distribution on the WIMP event rate relies entirely on theoretical modeling. 

\par As was shown in Section~\ref{sec:simulations}, numerical simulations do provide a theoretical starting point for understanding the dark matter velocity distribution. From a theoretical perspective, the simulations provide a Monte Carlo realization of the velocity distribution in a given volume. However, the resolution of the distribution in velocity space is affected by both the finite number of particles and the finite size of the volume within which the velocity distribution is determined. To understand the effect of the finite number of particles, define $N^\prime$ as the number of particles within the volume of consideration, and $\sigma$ as the one-dimensional characteristic width of the distribution of velocities in the halo. The characteristic separation in velocity space then scales as $\Delta v \sim \sigma/(N^{\prime})^{1/3}$. From the definition of the velocity distribution function, the effect of the finite volume size scales as $\Delta v \sim \sigma (N^{\prime})^{1/3}$. So in order to accurately measure the velocity distribution from simulations, a compromise must be made. In position space, the highest resolution simulations of Milky Way mass halos have up to $10^5$ particles in a $2^3$ kpc$^3$ box~\citep{Vogelsberger:2008qb,Kuhlen:2009vh}, which implies a mean inter-particle separation of $\sim 50$ parsecs. For comparison, a terrestrial dark matter detector covers a distance of approximately $220$ km/s $\times 1$ yr $\simeq 2 \times 10^{-4}$ parsecs per year. So clearly we are a long way from resolving the distribution function on the scales that are relevant for direct dark matter detectors, with even the highest resolution cosmological simulations. 

\par From Equation~\ref{eq:WIMP_event_rate}, the WIMP event rate depends on the integral of the velocity distribution as $g(v_{min}) \equiv \int_{v_{min}}^{v_{esc}} \left[ f(\vec v, \vec v_E)/v \right] d^3 \vec v$. This scaling can be understood by noting that the scattering cross section depends on the velocity as $d\sigma /d E_r \sim 1/v^2$, and that the WIMP event rate is proportional to the mean WIMP speed. For the general case of anisotropic distributions, the function $g(v_{min})$ can be expressed as 
\be 
g(v_{min}) = \int_{v_{min}}^{v_{esc}} d v^\prime v^\prime \int d \Omega^\prime f(v^\prime, \Omega^\prime). 
\label{eq:gv} 
\ee
For a given model of the velocity distribution, the uncertainty in the WIMP event rate due to the unknown velocity distribution depends on this function. 

\par Because direct detect experiments at present still operate in the ``discovery" phase, and because the lack of resolution in simulations outlined above, in order to appropriately compare results experiments utilize a standard model for the velocity distribution. The velocity distribution function that the experiments use to interpret their reuslts is the standard halo model (SHM), which is an isotropic maxwellian distribution with a cut-off at the Galactic escape velocity, 
\be 
f(\vec v, \vec v_E) = k \exp \left[ - \frac{( \vec v + \vec v_E )^2}{v_0^2} \right] 
\label{eq:SHM} 
\ee
where the normalization is
\be 
k = (\pi v_0^2)^{3/2} \left[ \textrm{erf} \left(\frac{v_{esc}}{v_0}\right) - \frac{2}{\sqrt{\pi}}\frac{v_{esc}}{v_0} \exp \left( -\frac{v_{esc}^2}{v_0^2}\right) \right]. 
\label{eq:SHM_normalization} 
\ee
In the limit that $v_{esc} \rightarrow \infty$, the SHM reduces to the maxwell-boltzmann distribution in Equation~\ref{eq:maxwell_boltzmann}. From the form of Equations~\ref{eq:SHM} and~\ref{eq:SHM_normalization}, there is a link between $v_0$ and the velocity dispersion $\sigma$ associated to any of the components of the velocity. This link can be make more clear by examining the circular speed $v_0$ as derived from the spherical and isotropic jeans equation, 
\be 
\frac{1}{2} v_0^2 \left( \frac{d \ln \rho}{d \ln r} + \frac{d \ln \sigma^2}{d \ln r} \right)= v_c^2, 
\label{eq:spherical_jeans} 
\ee
where the circular velocity is $v_c = \sqrt{GM/r}$, $v_0 = \sqrt{2} \sigma$, and $\rho$ is the density. If the term in the brackets is Equation~\ref{eq:spherical_jeans} is unity, the one-dimensional velocity dispersion that appears in the SHM velocity distribution is related to the cirular velocity as $v_c^2 = 2 \sigma^2$. 

\par Although the SHM is a practical and versatile theoretical model for the dark matter velocity distribution, as was shown in Section~\ref{sec:simulations} the velocity distribution from numerical simulations deviates from the SHM, and may not ultimately be an adequate description of the velocity distribution in our Galactic halo. Recall from Section~\ref{sec:simulations} that simulations of dark matter halos (without the inclusion of baryonic physics) are converging on a form for the velocity distribution over a wide range of halo masses. In particular, the velocity distribution from simulations has a very high velocity tail that is steeper than a corresponding maxwell-boltzman distribution, a peak that is ``flattened" relative to the more peaked maxwell-boltzman distribution, and a larger fraction of particles at velocities less than the peak. This behavior is quantified by Equation~\ref{eq:universal_VDF}. 

\par
\begin{figure*}
\begin{center}
\begin{tabular}{cc}
{\includegraphics[width=0.45\textwidth]{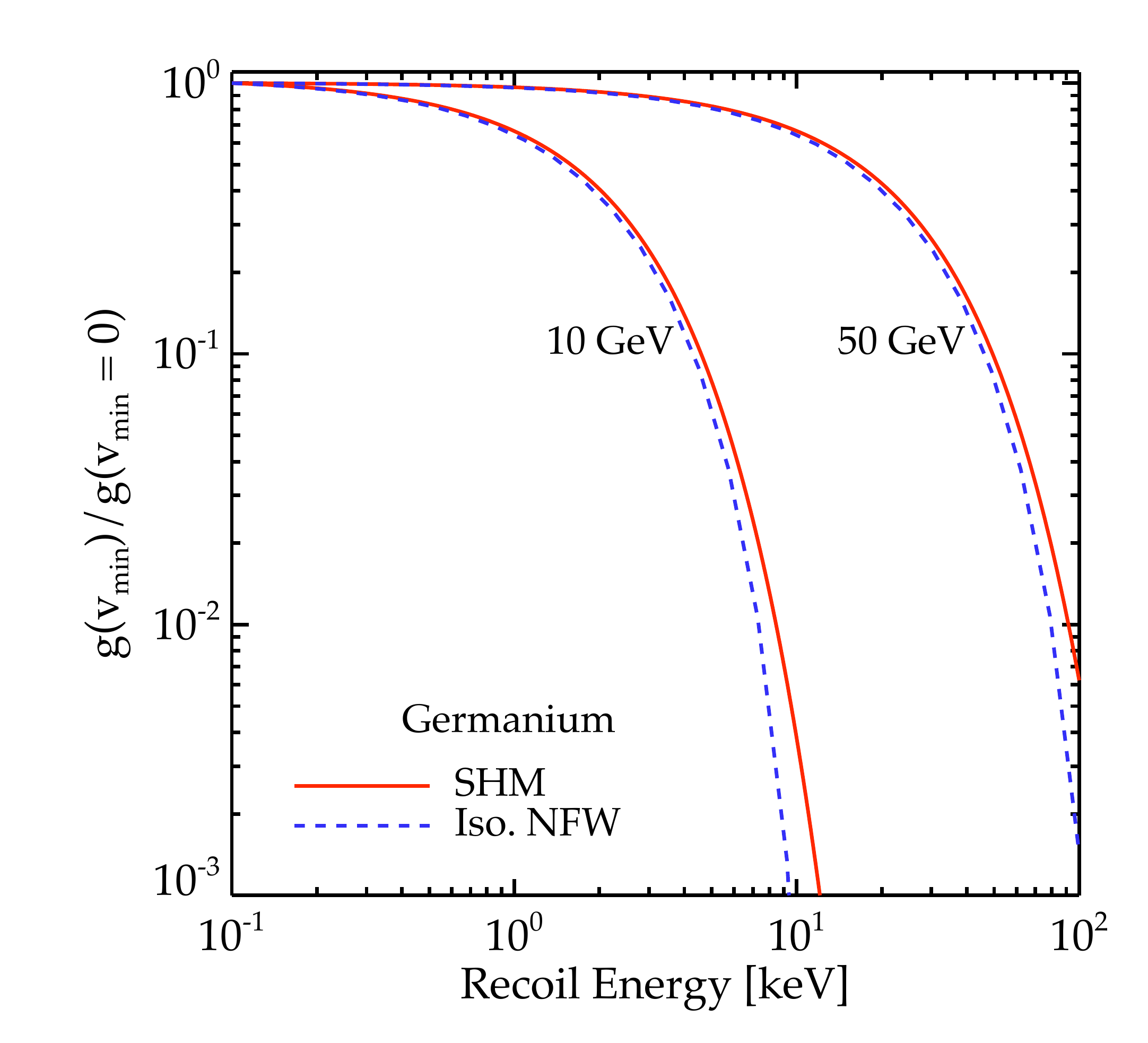}} & 
{\includegraphics[width=0.45\textwidth]{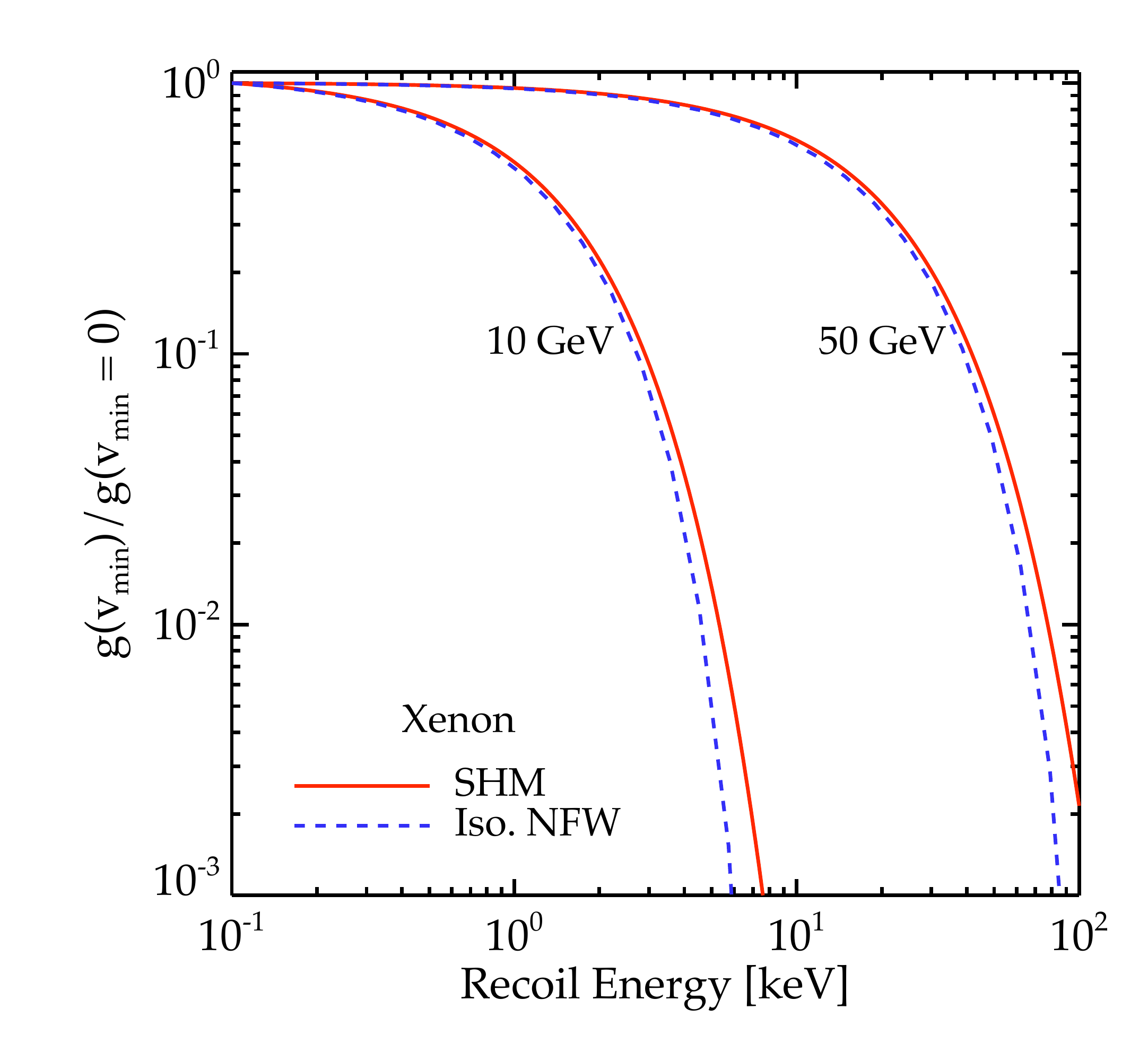}} \\
\end{tabular}
\end{center}
\caption{Integral in Equation~\ref{eq:gv}, normalized to the case with $v_{min} = 0$. The left panel is for a germanium target, and the right panel is for a xenon target. The solid red curve assumes the standard halo model, and the dashed blue curve assumes an  isotropic NFW model with a velocity distribution computed from Eddington's formula. For the NFW profile, the scale radius is $20$ kpc. For the inner curves the WIMP mass is $10$ GeV, and for the outer curves the WIMP mass is $50$ GeV. 
}
\label{fig:gv}
\end{figure*}

 The difference between different velocity distributions, and their impact on direct detection rates, is quantified by Equation~\ref{eq:gv}~\citep{Fox:2010bz}. Figure~\ref{fig:gv} shows this function for two different models, for a WIMP mass of $10$ and $100$ GeV for a germanium and xenon target. The first model is a standard halo model with $v_0 = 220$ km/s and $v_{esc} = 553$ km/s, and the second is an isotropic NFW model with a velocity distribution computed from Eddington's formula. For the NFW profile, the scale density is $8 \times 10^6$ M$_\odot$ kpc$^{-3}$ and a scale radius of $20$ kpc. Converted to nuclear recoil energy, the most significant difference between the two models is at high nuclear recoil energy. 

\subsection{Determining WIMP properties} 
\par Having developed an understanding of how the WIMP event rate depends on the scattering cross section, the local dark matter distribution, and the velocity distribution, it is now appropriate to address the important question of what WIMP properties and Galactic halo properties can ultimately be extracted from direct detection experiments. Because at this stage we only have upper limits on the dark matter cross section (as will be seen from the discussion in Section~\ref{sec:direct_detection_experiment}), at this point in time this question will be addressed, most optimistically, by next generation detectors. However, there is some sense of urgency to improve the theoretical projections, because this information will both help to set the program for future direct dark matter searches (Section~\ref{sec:future}) and help to clarify the need for complementarity between direct searches, indirect searches, and colliders. The three main signals in direct detection experiments, specifically the mean event rate, the annual modulation signal, and the direction of a nuclear recoil, all will play an important and complementary role in determining WIMP properties. 

\par To gain some intuition on how well the WIMP mass can be determined, it will be useful to start by appealing to some simple scalings. Consider the spin-independent zero momentum transfer cross section. In this case, for no form factor suppression the event rate scales simply as the inverse of the WIMP mass, so lighter WIMP have a larger flux than heavy WIMPs. However, the lighter the WIMP mass, the larger the value of $v_{min}$ because the reduced WIMP-nucleus mass factor. This latter point implies that the range of velocities that is integrated over in Equation~\ref{eq:WIMP_event_rate} is smaller for lighter WIMP masses, and that the flux is more sensitive to the shape of the high velocity tail of the WIMP distribution. It is easy to see in this case why the event rate is a sensitive function of the WIMP mass. For increasing WIMP masses, the suppression due to the nuclear form factor is more significant, so that the cross section is reduced from the zero momentum transfer value. In this case, for larger WIMP masses, ($\gtrsim 100$ GeV), the event rate is less sensitive to the WIMP mass. Understanding more precisely how the above competing factors play off of one another will help better understand how well a WIMP of a given mass may be constrained from an observed sample of events. 

\par Initial detailed studies that focused on reconstructing the WIMP mass and cross section from direct detection data were based on fixed models of the WIMP velocity distribution and local dark matter distribution~\citep{Green:2008rd}. These results showed that for an exposure of $3 \times 10^5$ kg-day ($\sim 0.8$ ton-yr), a 50 GeV mass WIMP with a spin-independent cross section of $10^{-44}$ cm$^2$ could be determined with an accuracy of $\sim 10\%$. The simple scaling arguments above are thus confirmed, in that for WIMPs $\gtrsim 100$ GeV, the constraints weaken so that only a lower bound can be placed on the WIMP mass. 

\par In order to understand the origin of these results and to make more rigorous predictions, it will be useful to develop a simple model to determine the constraints on the WIMP mass and cross section as a function of the most relevant experimental parameters. Begin first by considering the SHM velocity distribution. If future experiments have a large enough sample of events over their range of recoil energies, it will be possible to bin the events to obtain spectral information. From this type of data set, the probability to obtain a given number of counts in an energy bin can be approximated as a product of poisson distributions, 
\be 
P(\vec \Theta) = \prod_b \frac{N_R^{\hat N_b}}{\hat N_b !} \exp [-N_R], 
\label{eq:poisson_energy_bins} 
\ee
where $\hat N_b$ is the number of observed counts in each bin, and $N_R$ is the number of expected counts in the bin. The number of counts in each bin are determined from the theoretical model with parameters represented by $\vec \Theta$. 

\par When binning the experimental data and comparing to a theoretical model, it is important to account for the finite energy resolution of the detector. This requires smearing the true event rate spectrum according to
\be 
\frac{d\tilde R}{dE_r} = \int dE^\prime \frac{dR}{dE_r} (E^\prime) \frac{1}{\sqrt{2\pi} \epsilon} \exp \left[ -\frac{(E-E^\prime)^2}{2\epsilon^2}\right],
\label{eq:energy_resolution} 
\ee
where $dR/dE_r$ is the true event rate spectrum as determined from Equation~\ref{eq:WIMP_event_rate}, and $\epsilon$ is the energy-dependent energy resolution. The total number of events within a recoil energy range $(E_1, E_2)$ is 
\be 
N(E_1,E_2) = \int_{E_1}^{E_2} dE_r \epsilon_{eff} \frac{d \tilde R}{d E_r}, 
\label{eq:total_rate_smeared}
\ee
where $\epsilon_{eff}$ is the effective exposure as a function of energy. 

\par For the assumptions above from the poisson likelihood, we can define the elements of the inverse covariance matrix as
\be
F_{ab} = \sum_{\imath=1}^n \frac{N_{exp}}{R_\imath}  
         \frac{\partial R_\imath}{\partial \theta_a} 
         \frac{\partial R_\imath}{\partial \theta_b}, 
\label{eq:likelihood} 
\ee 
where $N_{exp}$ is the exposure of the experiment, i.e. the mass of the detector times the exposure time. From Equation~\ref{eq:likelihood}, the one-sigma uncertainty on parameter $a$ is ${\bf F}_{aa}^{-1}$, evaluated at the fiducial values for the parameters. Here $R_\imath$ is the predicted rate in the $\imath^{th}$ energy bin. Though for the analysis in this section backgrounds are not included in the analysis, these can be added as necessary (see Section~\ref{sec:future}). 

\par
\begin{figure*}
\begin{center}
\begin{tabular}{cc}
{\includegraphics[width=0.50\textwidth]{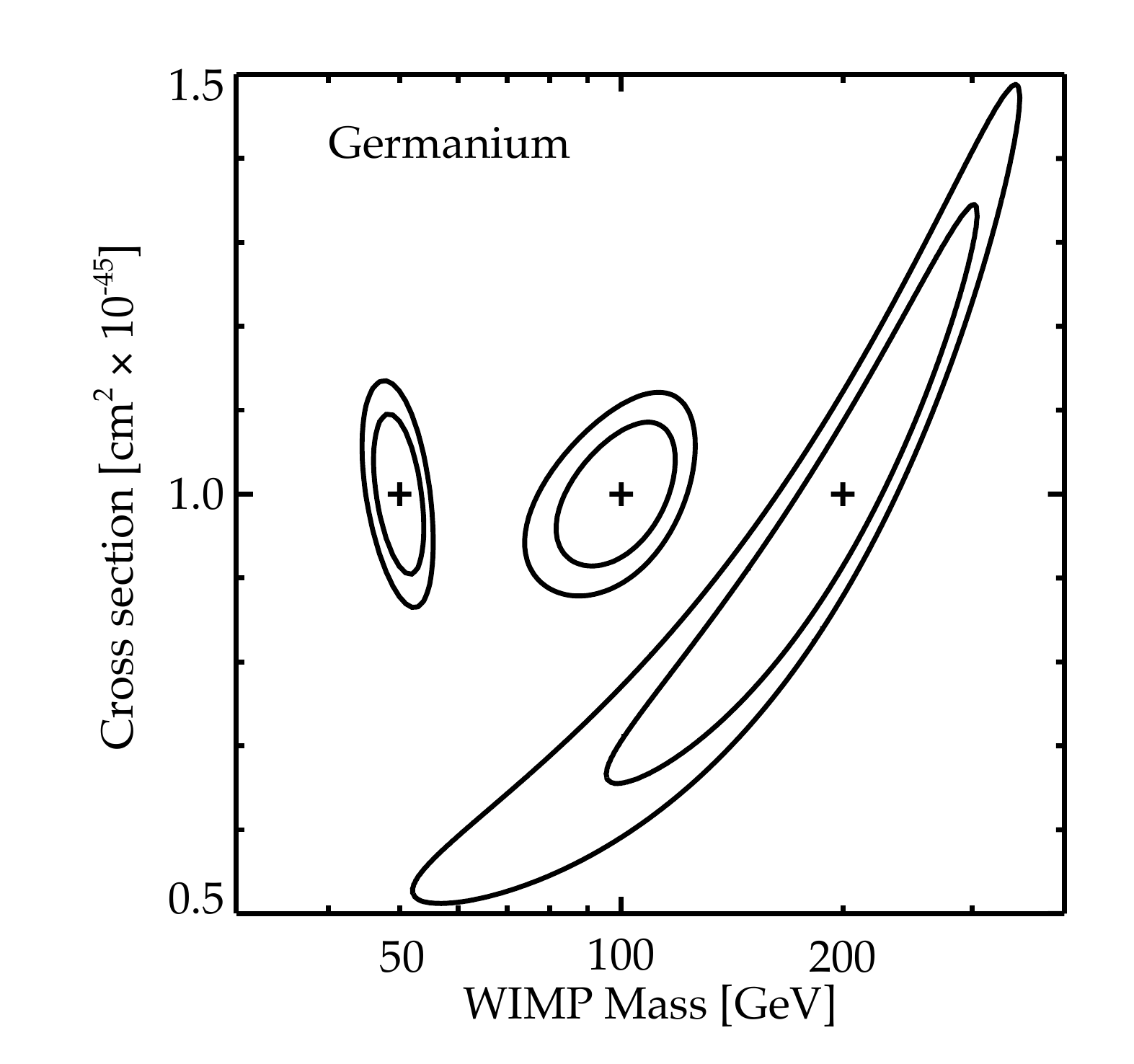}} & 
{\includegraphics[width=0.50\textwidth]{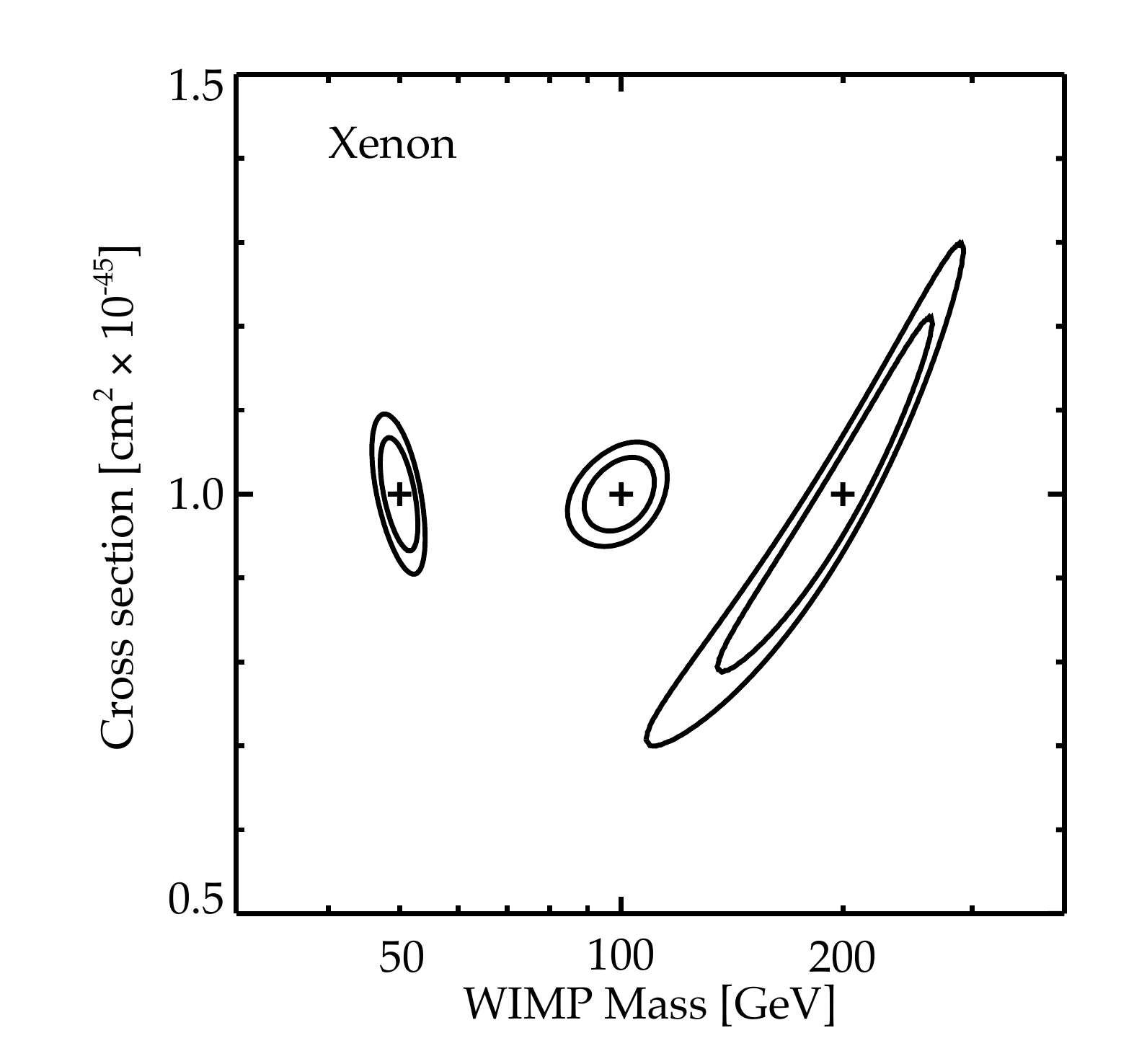}}
\end{tabular}
\end{center}
\caption{Projected constraint on the WIMP mass and spin-independent cross section assuming a 1 ton-yr exposure. The left panel shows the projections for a germanium target, and the right panel shows the projections for a xenon target. From left to right in each panel, the contours assume fiducial WIMP masses of $50$, $100$, and $200$ GeV. Both the local density of WIMPs and the velocity distribution are fixed.   
}
\label{fig:projection}
\end{figure*}

 Figure~\ref{fig:projection} shows the projected constraints on the WIMP mass and cross section, assuming the SHM velocity distribution with fixed parameters $V_c = 220$ km/s and $\rho_\odot = 0.3$ GeV cm$^{-3}$. The results in Figure~\ref{fig:projection} provide a simple way of assessing how well, in an ideal case, the WIMP mass and cross section can be determined. Because the SHM is probably only an approximation to the WIMP velocity distribution, and because a range of uncertain Galactic model parameters contribute to the determination of the WIMP event rate, when a detection is confirmed a more rigorous model for the Galactic dark matter distribution will be required. Using a discrete binned model for the WIMP velocity distribution,~\cite{Peter:2011eu} and~\cite{Kavanagh:2012nr} still show that there is a significant degeneracy in the reconstruction of the WIMP mass and cross section. Combining data from different nuclear targets does improve the constraining power on these quantities.

\par~\cite{Strigari:2009zb} introduced a Bayesian method to determine the WIMP mass and cross section using the general parameterization for the Milky Way dark matter density profile in Equation~\ref{eq:zhao}. Combining the uncertainties in the five parameters of the profile in Equation~\ref{eq:zhao} with uncertainties in the local dark matter density, the circular velocity, the dark matter halo mass, as well as the properties of the Milky Way disk, for a one ton-year exposure with liquid xenon, ~\cite{Strigari:2009zb} find that the mass of a fiducial $50$ GeV WIMP can be determined to a precision of less than approximately 25\%. For WIMPs greater than $100$ GeV, as above the mean event rate distribution will only be able to provide a lower bound on the WIMP mass. 

\par The~\cite{Strigari:2009zb} methodology developed a more robust connection between Galactic model parameters and the signal in direct dark matter detectors. This methodology can be improved upon by establishing a more theoretically-motivated connection between the dark matter density profile and the VDF. For isotropic models this can be readily accomplished through the use of Eddington's formula in Equation~\ref{eq:eddington}. In order to implement this method, the density profile parameters in Equation~\ref{eq:zhao} are constrained from the astrophysical data, and the constraints on these parameters are then propagated through to determine the isotropic velocity distribution, and from this the WIMP event rate distribution. This method can be extended to a few anisotropic models for which it is possible to obtain an analytic VDF from a density profile. 

\par Using a general parameterization for the isotropic WIMP VDF motivated by double power law density profile models,~\cite{Pato:2010zk} first examined the complementarity between ton-scale xenon, germanium, and argon targets. For assumed ton scale and greater exposures, the constraining power of argon, germanium, and xenon experiments were found to be able to reduce the uncertainty on the WIMP mass by a factor of two relative to utilizing just germanium and argon alone. Also,~\cite{Pato:2010zk} show that future experiments with different nuclear targets will be able to ``self-calibrate," so that uncertainties on astrophysical parameters will be able to largely normalize out. 

\par It is also interesting to considering what can be learned about particle physics, beyond simply the mass and the zero-momentum transfer cross section. With these motivations a useful general theoretical parameterization of the WIMP-nucleus cross section are effective field theory models~\citep{Fan:2010gt,Fitzpatrick:2012ix}. Phrasing the cross section in these terms clearly shows how different components of the cross section depend on the nuclear recoil energy. 

\par The aforementioned methods outlined for determining WIMP properties using the mean event rate spectrum are complemented by using information on the annual modulation. The phase of the annual modulation signal is a function of recoil energy, such that for typical WIMP masses for lower recoil energy, there is a complete phase reversal of the signal. The recoil energy at which the phase shifts depends on the WIMP mass. This fact has been utilized to determine how well detectors are able to constrain the WIMP mass from the annual modulation signal~\citep{Lewis:2003bv}. More detailed determination of the potential to extract the WIMP mass from the annual modulation signal using the methodology of~\cite{Strigari:2009zb} and~\cite{Pato:2010zk} represents the next step forward in these studies. Finally, using distribution function based models and future directional dark matter detectors,~\cite{Alves:2012ay} discuss the potential to reconstruct the WIMP distribution. These authors find that probably a sample of thousands of events is required in order to detect the presence of anisotropies in the dark matter halo.

\newpage 
\section{Direct Detection: Experimental Results} 
\label{sec:direct_detection_experiment}
\par The first direct dark matter detection experiments that reported significant scientific results utilized germanium crystals, and were largely dedicated to other scientific purposes such as double-beta decay. However at the time when Dirac neutrinos were strong candidates for particle dark matter, there was some sentiment that the detectors would be able to detect these particles in abundance. Null results were obtained, though in the process these experiments did rule out the first theoretically well-motivated dark matter candidate~\citep{Ahlen:1987mn,Caldwell:1988su}. As the technology has developed since these initial results, the field has been characterized by a continual stream of impressive results, and along with them a good deal of controversy. The first modern limits on spin-independent WIMP-nucleon cross section were reported by the Cryogenic Dark Matter Search (CDMS), which at the time operated at a shallow underground site at Stanford University. CDMS was able to rule out a cross section of a few times $10^{-42}$ cm$^2$ as $100$ GeV~\citep{Abusaidi:2000wg}. The improvement upon the initial germanium results was in large part due to improved background discrimination methods, in particular discrimination between electron and nuclear recoils. These limits were strong enough to essentially rule out an annular modulation signal that the DAMA experiment, which utilized sodium iodine crystals, claimed was due to dark matter~\citep{Bernabei:2000qi}. 

\par The past decade has been an even more extraordinary time for direct dark matter detection experiments. In addition to the development of larger and more sensitive germanium detectors, a variety of different technologies have been developed. Each of these technologies relies on different methods to eliminate backgrounds to extract the signal of a WIMP scattering off of a nucleus. 

\par This section reviews the results from modern direct dark matter detection experiments, focusing on those experiments that are now operating and have reported science results. The presentation of the results is broken up according to whether they were obtained from spin-independent searches or spin-dependent searches. Following this main line of discussion will be a short discussion of modern directional dark matter detection limits. For each of these three categories, the discussion is organized according to which experiment has the best limit at the time of the writing of this article. For much larger incarnations of these experiments, or for those that use technology that are currently not competitive with modern experiments, discussion is reserved until Section~\ref{sec:future}. After reviewing the experimental results, a connection is made to the both the theoretical models that were discussed in Section~\ref{sec:direct_detection_theory} and the results from indirect searches discussed in Section~\ref{sec:indirect_detection_experiment}. 

\subsection{Spin-independent limits} 
\par The organization of the spin-independent limits done according to which of the experiments have the best sensitivity. Note that some of the experiments discussed here also have reported competitive spin-dependent limits; where this is the case discussion of these results is reserved until the following subsection on spin dependent limits. 

\bigskip 

$\bullet$ {\it XENON100}-- The XENON100 experiment has been in operation in the Gran Sasso underground laboratory since 2009. It is the predecessor of the XENON10 experiment which operated in Gran Sasso from 2005-2007. It is a dual phase noble liquid detector that detects both primary scintillation light and secondary light proportional to the amount of ionization from particles that scatter in the detector volume. A recoiling particle induces ionization and excitation of Xe atoms. For a given amount of energy deposited into the detector, recoiling electrons produce a much larger amount of ionization than recoiling nuclei. In XENON100, an electric field is employed in the detector that drifts the ionized electrons into a region where the noble is in a gas phase, and there the electrons produce a signal proportional to the amount of ionization (this is the so-called ``$S_2$" signal). The primary scintillation signal (the so-called ``$S_1$" signal) results from the production of photons from the de-excitation of Xe atoms that have not been absorbed within the detector. Electron and nuclear recoils are then discriminated based upon the ratio of $S_2$/$S_1$. 

\par The low energy threshold of approximately $8$ keV for XENON100 is determined by photoelectron yield. The precise value of the WIMP mass that this corresponds to depends on the local WIMP velocity distribution (Sections~\ref{sec:simulations} and~\ref{sec:direct_detection_theory}). Because of the large atomic mass of Xe, XENON100 reaches its best sensitivity at a lower WIMP mass than detectors that rely on smaller atomic masses. The energy resolution of XENON100 is $\lesssim 1$ keV at $1$ keV and $\sim 3$ keV at $30$ keV~\citep{Aprile:2011dd}. 

\par
\begin{figure}
\begin{center}
\begin{tabular}{c}
\includegraphics[width=0.65\textwidth]{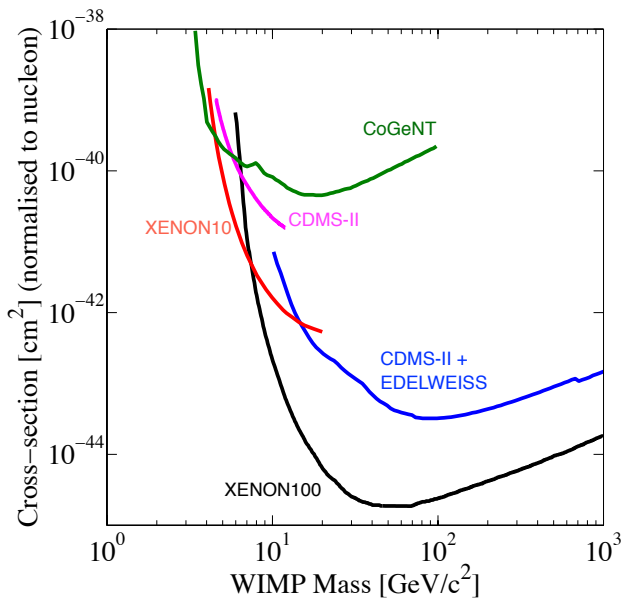} \\
\end{tabular}
\end{center}
\caption{Limits on the spin-independent WIMP-nucleon cross section as a function of mass from the experiments indicated.  Plot produced by DM tools (http://dmtools.brown.edu:8080/).
}
\label{fig:xenon100}
\end{figure}

 In July 2012, the XENON100 collaboration announced their results for 13 months of data between 2011-2012, with a total exposure of approximately 7665 kg-days. In the nuclear recoil energy range of $6.6-43.3$ keV there were two events that were observed, at recoil energies of $7.1$ and $7.8$ keV, which was consistent with the background expectation over this energy interval. From the lack of detected WIMP events over this energy interval, the deduced upper limit on the SI WIMP-nucleon cross section as a function of mass is given in Figure~\ref{fig:xenon100}. The XENON100 limits are the strongest limits on the SI cross section for masses greater than approximately $7$ GeV. XENON100 reaches its maximal sensitivity at approximately $40$ GeV, at which the limit on the cross section is $10^{-45}$ cm$^2$. 

\bigskip 

$\bullet$ {\it CDMS-II}-- The Cryogenic Dark Matter Search (CDMS-II) operated in the Soudan Mine from 2001-2008. CDMS-II uses germanium and silicon detectors that reject electron recoils using a combination of phonon light and ionization. In a cryogenic detector, the struck ion dissipates energy via scatterings with electrons and ions on the lattice. The dissipated energy can be detected by either phonon production, ionization, or scintillation. Phonons are produced as the kinetic energy of the recoil is converted into a collective excitation of the crystal. The shape and timing of the phonon pulses from ejection of electron events near the surface and the ionization yield allows for discrimination of bulk electron events from nuclear events within the detector. 

\par In December 2009, the CDMS-II collaboration announced the results of their data run from 2007-2008 with an exposure of $612$ kg-days. There were two candidate events detected in the nuclear recoil energy window of $10-100$ keV, at energies of $12.3$ and $15.5$ keV, which was statistically-consistent with the expected background over this energy interval. At this time of announcement this was the best limit for WIMP masses greater than approximately $40$ GeV, though now these results have been superseded by the aforementioned XENON100 results at all WIMP masses. 

\par In order to improve the sensitivity to WIMPs in the lowest mass regime that is probed by CDMS-II, the collaboration has presented an analysis with its threshold energy lowered to $2$ keV. Since the discrimination between nuclear and electron recoils significantly degrades at these lower energies, the limits on WIMP masses less than approximately $10$ GeV is degraded relative to WIMPs of larger mass. Even with the larger background acceptance, with an exposure of $241$ kg-days CDMS-II was able to improve the limits on the SI WIMP-nucleon cross section below a mass of approximatley $9$ GeV~\citep{Ahmed:2010wy}, and exclude cross section $\gtrsim 3 \times 10^{-40}$ cm$^2$ for WIMP masses greater than $5$ GeV. 

\bigskip 

$\bullet$ {\it EDELWEISS-II}-- EDELWEISS-II is a cryogenic germanium detector located in Laboratoire Souterrain de Modane. EDELWEISS-II has a different strategy for rejecting surface backrgound events relative to the method employed by CDMS-II. From an exposure of $384$ kg-day in 2009-2010, EDELWEISS-II detected five events above a recoil energy of $20$ keV, four of which have energies less than $25$ GeV, and one has an energy of $172$ GeV. This is also consistent with the expected number of background events~\citep{Armengaud:2011cy}. For WIMP masses larger than about $85$ GeV, EDELWEISS-II achieves a sensitivity that is similar to that of CDMS-II. EDELWEISS-II and CDMS-II have recently combined their data sets, achieving a total of $614$ kg-days of effective exposure, improving the respective individual limits by a factor of approximately $1.6$~\citep{Ahmed:2011gh}. 

\par EDELWEISS-II has also presented a low threshold analysis with a goal of improving the sensitivity to WIMPs with mass below $10$ GeV~\citep{Armengaud:2012kd}. From an exposure of $113$ kg-days and the observation of one event in the signal region, at $5$ GeV the limit on the cross section is approximately three times weaker than the corresponding limits from CDMS-II. 

\bigskip 

$\bullet$ {\it ZEPLIN-III}-- ZEPLIN-III is a liquid xenon experiment that operates in the Boulby Underground Laboratory. Similar to XENON100, it is a dual phase detector that relies on the detection of both ionization and scintillation light. From an exposure of $1344$ kg-days between 2010-2011, ZEPLIN-III observed a total of eight events in the nuclear recoil energy range of $7-29$ keV, again consistent with its expected backgrounds. Below approximately $70$ GeV, ZEPLIN-III achieves better sensitivity to the SI-cross section than CDMS-II or EDELWEISS-II do individually. 

\bigskip 

$\bullet$ {\it CRESST-II}-- The Cryogenic Rare Event Search with Superconducting Thermometers (CRESST) is located in Gran Sasso National Laboratory, using a scintillating material of Calcium Tungstate (CaWO$_4$). CRESST-II uses a combination of phonon and scintillation light to reduce backgrounds, and now has an exposure of 730 kg-days~\citep{Angloher:2011uu}. It has observed 67 events in its acceptance nuclear recoil energy range of $10-40$ keV. The indication from the analysis is that the known backgrounds are unable to account for these events observed in the signal region; from a maximum likelihood analysis, the CREST-II results can be made consistent with WIMPs with mass less than $25$ GeV. However, the cross sections which are able to explain the CRESST-II data are inconsistent with the CDMS-II and XENON100 limits described above. 

\bigskip 

$\bullet$ {\it DAMA}-- The DArk MAtter (DAMA) experiment is located in Gran Sasso and uses radiopure NaI (TI) scintillators as target detectors. DAMA does not distinguish between a WIMP signal and backgrounds on an individual basis, rather it infers the existence of WIMPs from single scatter interactions at low energy. 

\par The DAMA/NaI experiment has collected seven years of data between 1995-2002 with a mass of $100$ kg and a total exposure of $0.29$ ton-yr. Including the second phase of DAMA/LIBRA, the total mass is $250$ kg and the total exposure is $0.53$ ton-yr. In the $2-6$ keV energy interval, DAMA/NaI and DAMA/LiBRA report a $8.2\sigma$ detection of an annual modulation signal with both a phase and amplitude that is consistent with what is expected from a dark matter signal. No modulation is seen in the multiple scatter events, which can be interpreted as evidence that the signal is not from a modulating background. 

\par There have been numerous efforts to reconcile the observed DAMA signal with other null results, and windows have remained open for low-mass spin-independent interpretations~\citep{Gondolo:2005hh} until the recent announcement of the recent XENON100 results. There is manifest tension between these results, and it is not yet clear how to explain the DAMA results in light of all of the other dark matter limits. There is still the possibility that the reported results arise from an unknown experimental background; such possibilities are nicely laid out in~\cite{Schnee:2011ef}. 

\bigskip

$\bullet$ {\it CoGeNT}-- CoGeNT (short for Coherent Germanium Neutrino Technology) is a germanium detector cooled to liquid nitrogen temperatures of 77 K, with sensitivities to sub-keV nuclear recoils~\citep{Aalseth:2010vx} . These experiments only measure deposits of energy through ionization, and do not discriminate between electron and nuclear recoil backgrounds. 

\par In 2010, CoGeNT identified an excess of events at energies less than $3$ keVee, where here the energy units stand for equivalent electron energy. At the time it was determined that, if interpreted as a WIMP-induced spectrum, this excess is compatible with the DAMA signal of a $7$ GeV WIMP. As is the case of the DAMA signal, there is significant tension between this result and the recent results presented by the XENON100 collaboration. In 2011, CoGeNT presented $2.8\sigma$ evidence for an annual modulation signal that is consistent with a WIMP of mass $7$ GeV~\citep{Aalseth:2011wp}. CDMS-II searched for an annual modulation signal in its nuclear recoil energy range of $5-11.9$ keV and reported null results that appear in conflict with the CoGeNT and DAMA annual modulation signals~\citep{Ahmed:2012vq}. 

\bigskip 

$\bullet$ {\it TEXONO}-- TEXONO is a germanium detector similar to CoGeNT, which is located in the Kuo-Sheng Laboratory~\citep{Lin:2007ka}. From an exposure of $0.338$ kg-day, TEXONO has reported limits on the spin-independent cross section down to $2$ GeV, constraining the cross section to $\lesssim 10^{-38}$ cm$^2$ at this mass. 

\bigskip

$\bullet$ {\it XENON10}-- XENON10 has published a limit on low mass WIMPs that does not include the primary scintillation ``S1" signal~\citep{Angle:2011th}. Without this signal, it is unable to discriminate nuclear from electron recoils, so the sensitivity is reduced, however it does allow for nuclear recoil energy to be measured by the electron signal. This provides sensitivity down to approximately $1$ keV nuclear recoil energy. As shown in Figure~\ref{fig:xenon100}, this limit is in significant tension with the results that are interpreted as a WIMP detection at a mass near $10$ GeV. 

\subsection{Spin-dependent limits} 
\par
\begin{figure*}
\begin{center}
\begin{tabular}{c}
\includegraphics[width=0.65\textwidth]{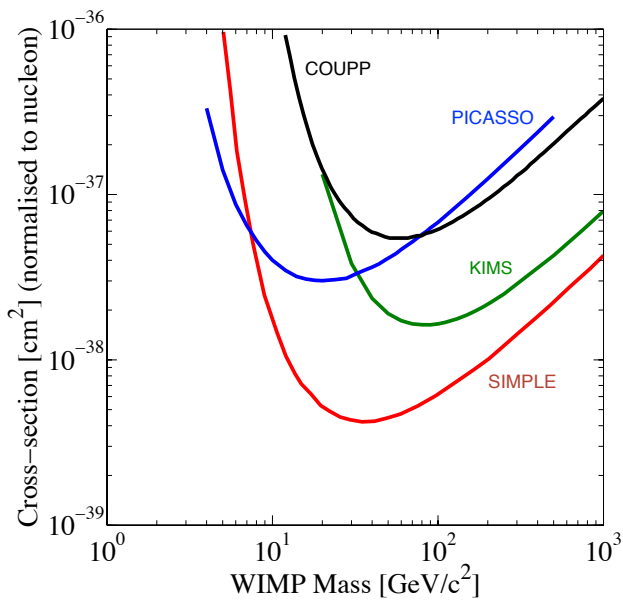} \\
\end{tabular}
\end{center}
\caption{Limits on the spin-dependent WIMP-proton cross section as a function of WIMP mass. Plot produced by DM tools (http://dmtools.brown.edu:8080/).
}
\label{fig:SD_proton_limits}
\end{figure*}

 This subsection now moves on to discuss spin-dependent limits. As above the results are organized based upon which experiments have the best limits as of the writing of this article. The updated limits discussed below are shown in Figures~\ref{fig:SD_neutron_limits} and~\ref{fig:SD_proton_limits}. 

\subsubsection{Direct detection} 

\par Begin by considering the so-called threshold detectors (SIMPLE, PICASSO, and COUPP); these operate by tuning the temperature and pressure of a superheated liquid to a point at which they are insensitive to electron recoils. Though these detectors provide limited information on the energy of an event, they do hold the promise of being readily scalable to larger volumes. 

\begin{figure*}
\begin{center}
\begin{tabular}{cc}
\includegraphics[width=0.65\textwidth]{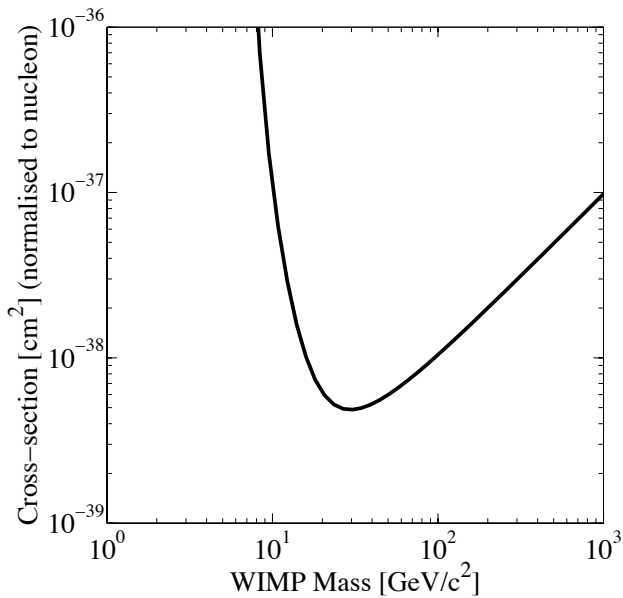} \\
\end{tabular}
\end{center}
\caption{Limits on the spin-dependent WIMP-neutron cross section as a function of mass, from the XENON10 Collaboration~\citep{Angle:2008we}. Plot produced by DM tools (http://dmtools.brown.edu:8080/).
}
\label{fig:SD_neutron_limits}
\end{figure*}

\bigskip  

$\bullet$ {\it SIMPLE}-- The Superheated Instrument for Massive ParticLe Experiments (SIMPLE)~\citep{Felizardo:2010mi} operates in the Low Noise Underground Laboratory in southern France.  It has sensitivity to spin-dependent WIMP-proton interactions through scattering on $^19$F, $^35$Cl, $^37$Cl,  and $^13$C. Combining the exposures for Stage 1 (13.47 kg-days) and Stage 2 (6.71 kg-days) of the SIMPLE experiments, it is able to exclude a WIMP-proton spin-dependent cross section $\gtrsim 5.7 \times 10^{-39}$ at $35$ GeV. 

\bigskip  

$\bullet$ {\it PICASSO}-- The Project in Canada to Search for Supersymmetric Objects (PICASSO)~\citep{Archambault:2012pm} is located at SNOLAB in Canada; it is sensitive to spin-dependent WIMP-proton interactions primarily through scattering on $^{19}$F. In 2012 PICASSO reported results from a total exposure of 114 kg-days, and from these results constrains the cross section to $\lesssim 3.2 \times 10^{-38}$ cm$^2$. 

\bigskip
  
 $\bullet$ {\it COUPP}-- The Chicagoland Observatory for Underground Particle Physics (COUPP)~\citep{Behnke:2010xt} operates in the MINOS tunnel at the Fermi National Accelerator Laboratory. It is sensitive to spin-dependent WIMP-proton interactions primarily through scattering on flourine. From a $28.1$ kg-day exposure in 2009, COUPP achieved a maximal sensitivity to the cross section at a WIMP mass of approximately $50$ GeV. 
 
\bigskip 

$\bullet$ {\it KIMS}-- The Korea Invisible Matter Search (KIMS) is located in the Yangyang Underground Laboratory. KIMS uses CsI detectors and a pulse shape discrimination method to statistically-discriminate between nuclear and electron recoil events. From an exposure of $\gtrsim 24524.$ kg-days between 2009-2010, KIMS has recently placed an upper limit on the spin-dependent WIMP-proton cross section that is more strict than any other experiment for WIMP masses $> 20$ GeV~\citep{Kim:2012rz}. 

\bigskip 

$\bullet$ {\it XENON10}-- The most stringent limits on the WIMP-neutron spin-dependent cross section come from the XENON10 experiment~\citep{Angle:2008we}. From an exposure of $136$ kg-days, XENON10 achieves a maximal sensitivity of  $\lesssim 10^{-38}$ cm$^2$ at approximately $25$ GeV (Figure~\ref{fig:SD_neutron_limits}).

\subsubsection{Neutrinos}
\par As discussed in Section~\ref{sec:indirect_detection_theory}, WIMPs get captured into the Sun and annihilate, and the only particles that can effectively escape to be detected are neutrinos. The neutrinos then interact in the Earth to create muon neutrinos which are then detectable via Cerenkov radiation. As a result neutrinos from the Sun are particularly effective at probing the spin-dependent WIMP-nucleon cross section. 

\par
\begin{figure*}
\begin{center}
\begin{tabular}{c}
\includegraphics[width=0.60\textwidth]{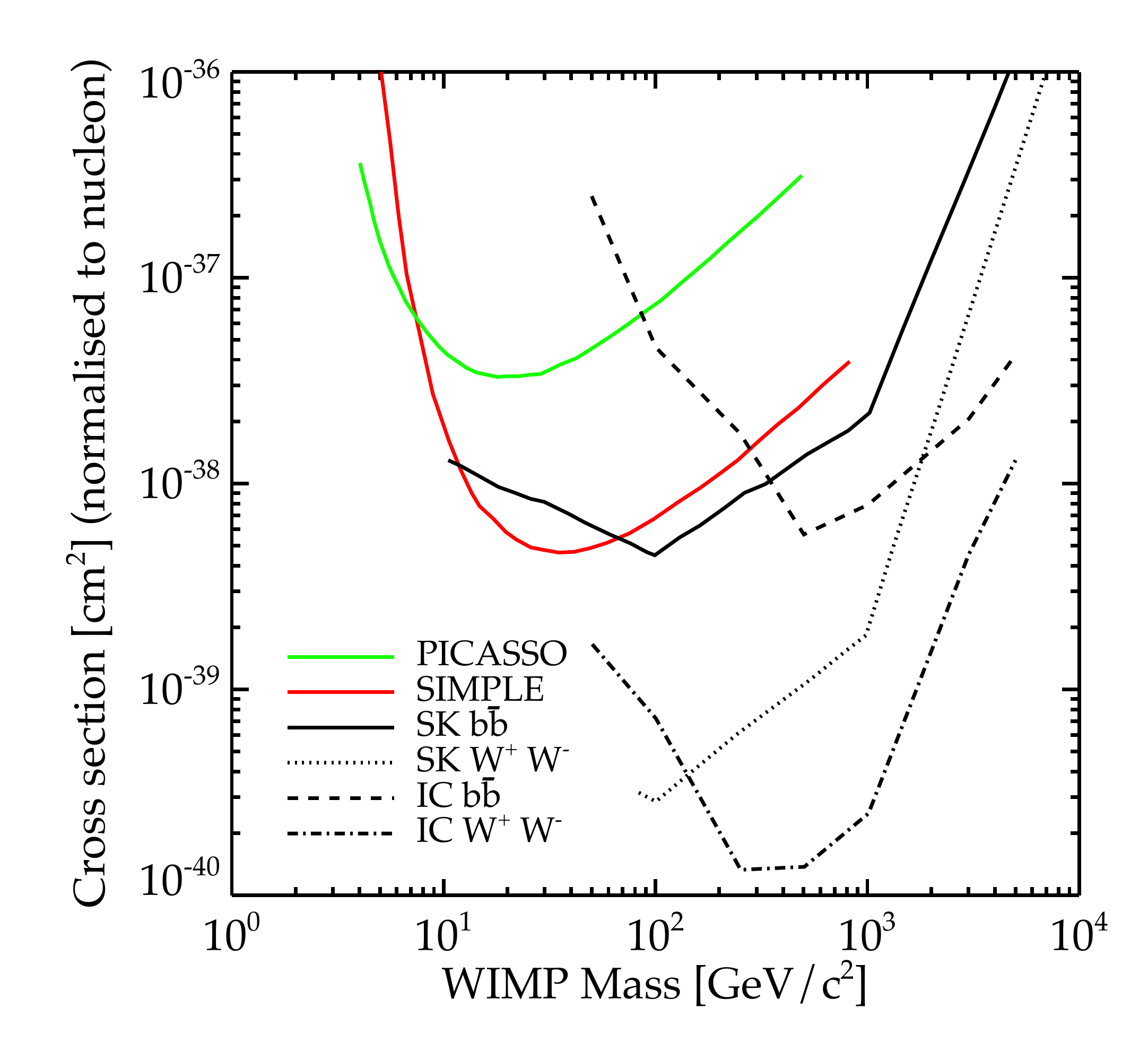} \\
\end{tabular}
\end{center}
\caption{Limits on the spin-dependent WIMP-proton cross section as a function of WIMP mass, including neutrino limits from Super-Kamiokande and Ice Cube. 
}
\label{fig:SD_neutrino_limits}
\end{figure*}

 Both Super-Kamiokande~\citep{Tanaka:2011uf} and Ice-Cube~\citep{IceCube:2011aj} have searched for neutrinos that come from dark matter capture and annihilation in the Sun. As in the case of diffuse neutrino searches discussed in Section~\ref{sec:indirect_detection_experiment}, neutrino experiments quote their limits for hard ($W^{+} W^{-}$) and soft ($b \bar b$) spectra for the sources of neutrinos. For Super-Kamiokande, the total exposure time for the search was approximately $3110$ days, while for Ice Cube the total exposure was $1065$ days. For both searches, no excess of events above backgrounds were reported, leading to flux upper limits of muon neutrinos from the Sun. For Super-Kamionkande, assuming a dark matter mass greater than $100$ GeV, the flux upper limits are approximately $6 \times 10^{-15}$ cm$^{-2}$ s$^{-1}$ for the soft channel, and approximately $4 \times 10^{-15}$ cm$^{-2}$ s$^{-1}$ for the hard channel. For the soft channel, the flux limits extend down to $20$ GeV, and are a factor of two larger than the limits above $100$ GeV. For Ice Cube, the reported limits cover a dark matter mass range of 50 to 50000 GeV, with the corresponding muon neutrino flux upper limits of approximately $10^4$ km$^{-2}$ yr$^{-1}$ to $10^2$ km$^{-2}$ yr$^{-1}$ over this range. The limits on the WIMP-proton spin-dependent cross section from Super-Kamiokande and Ice Cube are shown in Figure~\ref{fig:SD_neutrino_limits}, in comparison to the most stringent direct detection limits from Figure~\ref{fig:SD_proton_limits}. 

\subsection{Directional searches} 
\par Having surveyed the status of spin-independent and spin-dependent searches, now briefly consider the limits from dark matter detectors with directional sensitivity. The best limit on dark matter from this class of detector is from the Dark Matter Time Projection Chamber (DMTPC)~\citep{Ahlen:2010ub}. The DMTPC is a CF$_4$ gaseous detector, with a sensitivity to spin-dependent interaction of WIMPs off of flourine. The first results from the DMTPC~\citep{Ahlen:2010ub} utilize a 35.7 kg-day exposure, which is several orders of magnitude less than the exposure of the experiments reported in this section. They obtain an upper limit on the spin-dependent cross section of $\sim 10^{-32}$ cm$^{-2}$. The threshold energy of the DMTPC is much higher than the experiments described above, $\sim 80$ keV, so that the best sensitivity is obtained for WIMP masses $\sim 100$ GeV. 

\subsection{Summary of experimental results and interpretations}  

\par Taking a step back now and surveying the results presented in this section, it is clear that the direct dark matter searches are moving at a rapid pace. DAMA, CoGeNT, and CRESST have shown that their results cannot be explained by identified backgrounds. While each of these individually may be reconciled with a WIMP mass in the ballpark of $10$ GeV, there is no hint of a signal at this mass scale from either XENON100 or CDMS-II. While there have been models that have been put forward that attempt to reconcile these results~\citep{Feng:2011vu}, it is not clear that these satisfy all of the most recent constraints, even taking to account uncertainties in the dark matter local density and velocity distribution~\citep{Fox:2010bz,Frandsen:2011gi} 

\par The limits presented in this section have significantly improved constraints on dark matter with effective interactions~\citep{Beltran:2008xg,Zheng:2010js,Yu:2011by,Kumar:2011dr}. Scalar and fermionic WIMPs with mass $\lesssim 1$ TeV with scalar interactions and universal couplings with fermions are now strongly excluded by direct detection results. Note that fermionic WIMPs with scalar interactions have velocity-suppressed annihilation cross sections, so they are not detectable with Fermi or other indirect detection experiments, so direct detection is the only means to probe these models. Fermionic WIMPs with vector interactions (such as the heavy Dirac neutrino) are also strongly excluded by direct detection; these models do not have velocity-suppressed annihilation cross sections so they are also constrained by indirect detection results. Fermionic WIMPs with scalar interactions and universal couplings to fermions are now also strongly excluded; for mass-dependent couplings to fermions the limit on the WIMP mass is $\lesssim 200$ GeV. Note again that direct detection experiments are not sensitive to WIMPs with pseudo-scalar interactions.  
 
\par Direct detect searches, in particular the new XENON100 limits, are now reaching into the well-motivated focus point regime of the supersymmetric model parameter space. Though searching the entire supersymmetric parameter space for models that match relic abundance constraints and experimental constraints is challenging~\citep{Akrami:2009hp,Buchmueller:2011ab}, from general calculations of the neutralino-proton scattering cross section the spin-independent cross section is right where the XENON100 constraints are sensitive to for neutralino masses $\gtrsim 100$ GeV~\citep{Feng:2010ef}. This conclusion is further strengthened if the Higgs mass is $\sim 126$ GeV~\citep{Akula:2011aa}, and tells us that in the coming years it is very plausible that direct detection experiments will see hints of neutralino dark matter from supersymmetry.

\newpage 
\section{Future Directions and Complementarity with Other Searches}
\label{sec:future} 

\par This section reviews expected developments in dark matter searches over the next decade. Several issues are addressed, broken up into the following components: future developments in direct and indirect searches, complementarity with recent results from collider and cosmological searches, and finally forthcoming astronomical observations 

\subsection{Direct detection} 
\par There are a variety of direct dark matter searches that will improve on the limits presented in Section~\ref{sec:direct_detection_experiment} during the next decade. Several of these experiments are extensions of modern experiments, while others represent new technological directions.  

\bigskip 

$\bullet$ {\it Xenon}-- As discussed in Section~\ref{sec:direct_detection_experiment}, liquid xenon experiments are readily scalable, so that an order of magnitude increase in the fiducial volume is expected within the next few years. The next phase of the XENON100 experiment is Xenon 1 ton, which is fully funded and will be located in a tunnel in Gran Sasso near the XENON100 experiment. As its name indicates, the fiducial active volume of this experiment will be 1 ton. In addition to the increased volume, additional improvements are expected to increase the experimental sensitivity. These improvements include a water shield to be used as a muon veto to reduced backgrounds. Further there is expected to be a reduction of Kr and Rn backgrounds via the development of removal towers. Xenon 1 ton is currently in the planning and constructing phase, with first data expected in 2015. If the projected sensitivity is achieved, it is expected to cover a large portion of the so-called focus point region of supersymmetric parameter space~\citep{Feng:2000gh}. 

\par In addition to Xenon 1 ton, the Large Underground Xenon (LUX) experiment, which is located in the Homestake mine in South Dakota, is expected come online soon~\citep{McKinsey:2010zz}. Because it is liquid xenon, the principles are similar to those of the XENON100 experiment. LUX will have a 100 kg fiducial volume, but a background that is suppressed relative to XENON100 because of 200 kg of LXe that acts as self-shielding. Additionally, water shielding around the detectors act as a veto against muons. Data from its first science runs is expected in 2013. 

\bigskip 

$\bullet$ {\it SuperCDMS}-- SuperCDMS represents the next generation of the CDMS-II experiment. SuperCDMS will also be located in the Soudan mine, with a goal of eventually moving to the deeper SNOLAB facility in Sudbury, Ontario. The individual detectors of SuperCDMS will be increased in mass by greater than about a factor of two and, and it will have improved background discrimination relative to those of CDMS-II. The total target mass of the detectors is expected to be 100-200 kg~\footnote{http://cdms.berkeley.edu/}.

\bigskip 

$\bullet$ {\it Argon and Neon}-- Like liquid xenon detectors, liquid argon detectors are excellent scintillators and ionizers. Liquid argon detectors are easily scalable, though there is a significant contaminating radioactive background from $^{39}$Ar. There are several collaborations pursing a ton-scale liquid argon detector including Argon Dark Matter (ArDM)~\footnote{neutrino.ethz.ch/ArDM/} and the WIMP Argon Program (WArP)~\footnote{warp.lngs.infn.it/}

\par Like both xenon and argon, neon offers potential for good background rejection through pulse shape discrimination. Neon is being implemented as part of the DEAP/CLEAN program, with a $\sim 360$ kg prototype, mini-CLEAN, currently being built~\citep{Hime:2011ms}. Liquid neon has the additional benefit of being scalable to approximately 100 ton. 
\bigskip 

\par As a final note on direct dark matter detectors, recall that they were first conceived as Solar neutrino experiments~\citep{Drukier:1983gj}. This is because if a neutrino strikes a nucleus, a coherent recoil happens if the three-momentum imparted to the nucleus, $q$, obeys the inequality $q R_n \lesssim 1$, where $R_n$ is  the radius of the nucleus. For typical nuclear radii, coherent scattering leads to nuclear recoils in the range of a few keV for incoming neutrinos with energies in the range $\sim 1 - 100$ MeV. Though this neutrino-nucleus coherent scattering remains a fundamental prediction of the standard model, this process has yet to be detected. 

\par There are several sources of neutrinos that are expected to appear in next generation underground detectors that will serve as backgrounds to direct dark matter detection signals~\citep{Monroe:2007xp,Strigari:2009bq}. These include Solar neutrinos, atmospheric neutrinos, and diffuse supernova neutrinos~\citep{Horiuchi:2008jz,Beacom:2010kk}. The neutrinos produced from the $^8$B decay process are the most-easily detected component of the Solar neutrino flux. In the near term, direct detection experiments with sensitivities to nuclear recoil energies of a few keV or less will be most sensitive to this signal. 

\par
\begin{figure*}
\begin{center}
\begin{tabular}{cc}
\includegraphics[width=0.44\textwidth]{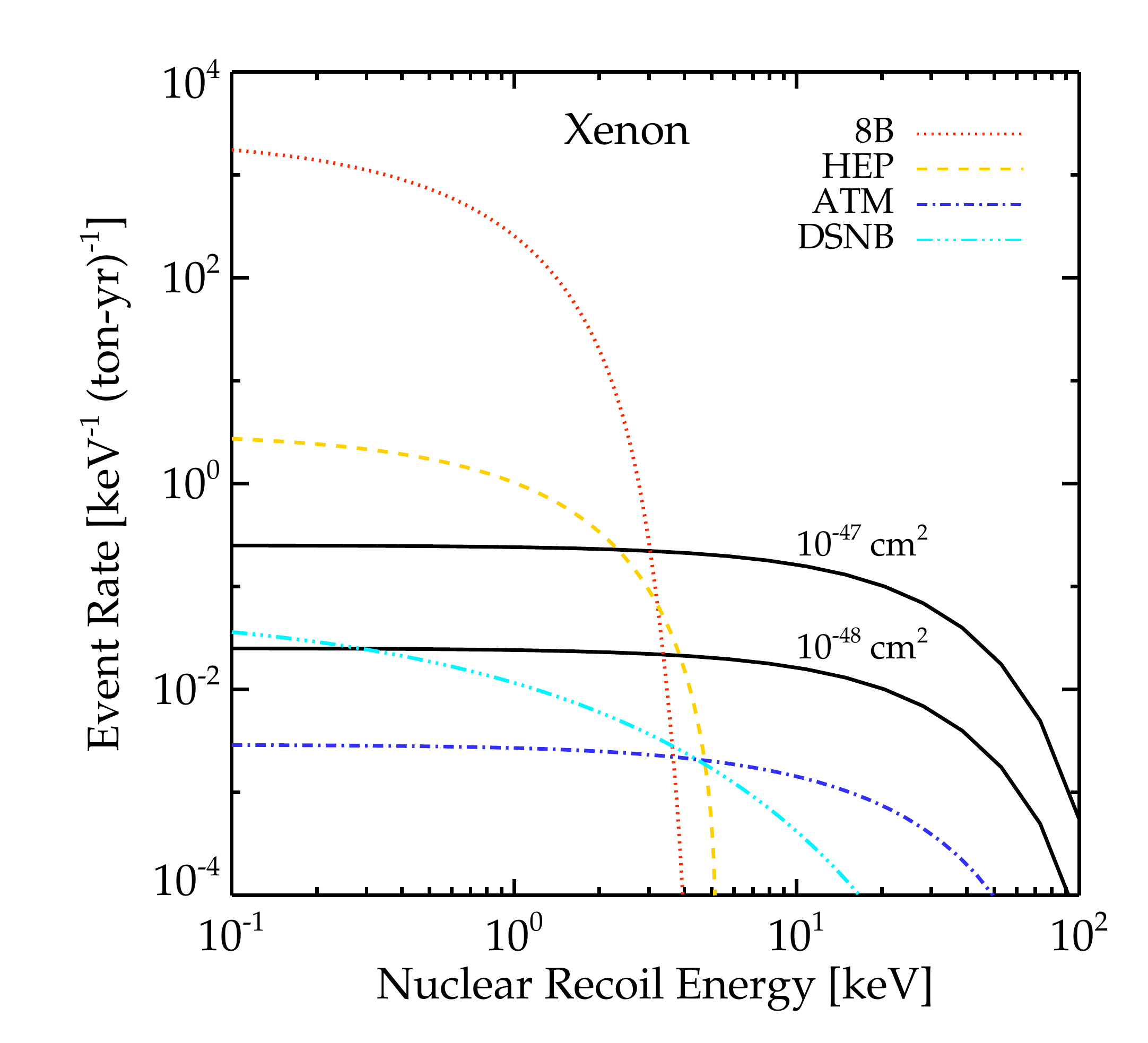} & 
\includegraphics[width=0.44\textwidth]{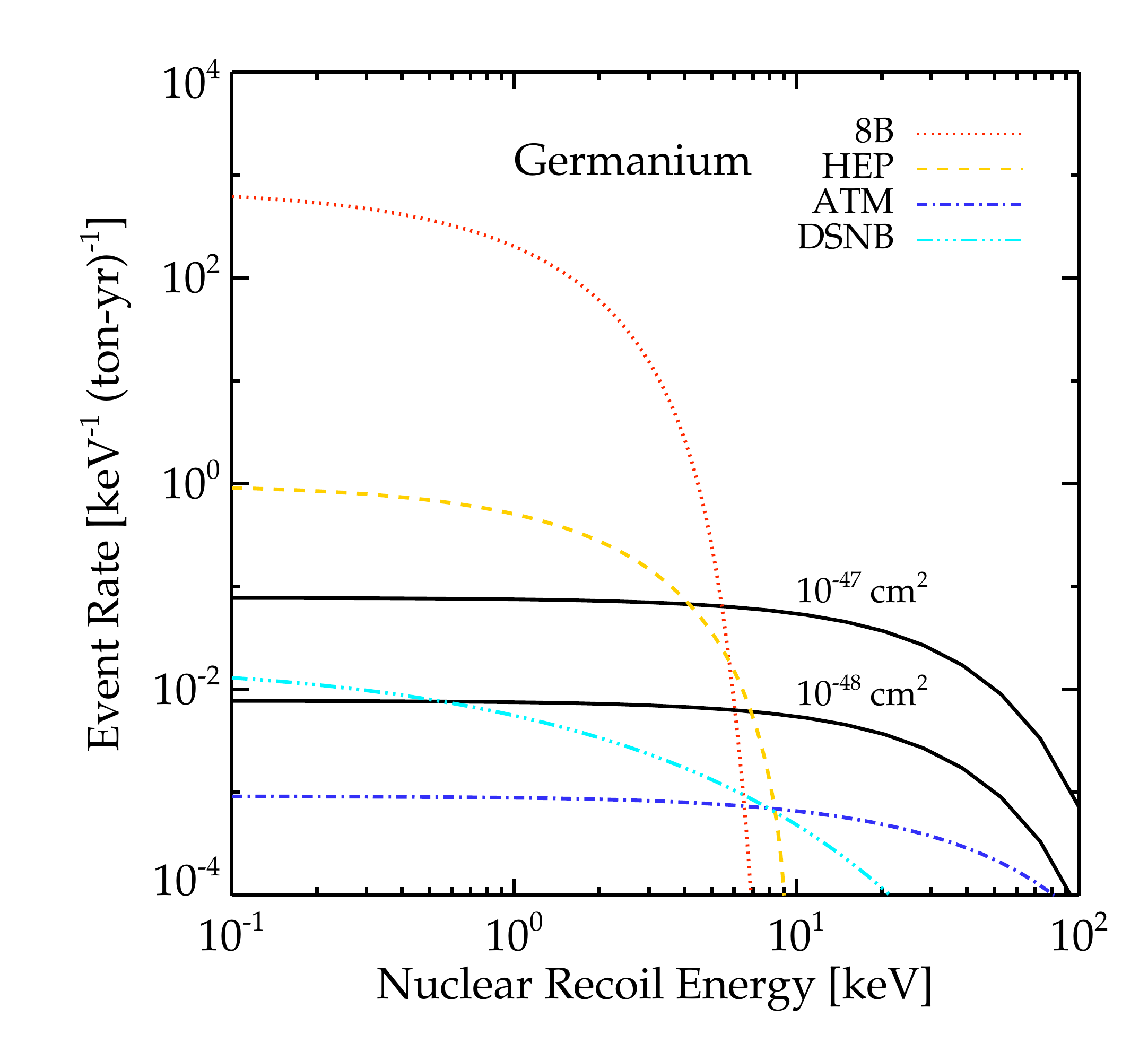}\\
\end{tabular}
\end{center}
\caption{Neutrino and WIMP coherent scattering rates in xenon (left) and germanium (right). The neutrino rates are determined as in~\cite{Strigari:2009bq}. The two WIMP spectra assume a mass of $100$ GeV and cross section per nucleon as indicated. 
}
\label{fig:neutrino_coherent}
\end{figure*}

 Figure~\ref{fig:neutrino_coherent} shows the expected neutrino fluxes in xenon and germanium detectors as a function of nuclear  energy. Over the entire nuclear recoil energy range, atmospheric and diffuse supernova neutrinos rival the WIMP event rate for a WIMP of mass $100$ GeV and cross section of $\lesssim 10^{-48}$ cm$^2$. Though the neutrino signal mimics the WIMP signal in many ways, some differences between the signals can be exploited to distinguish the neutrino signal from the WIMP signal. The most obvious method is to use directional dark matter searches, which would clearly indicate the presence of a WIMP signal due to the the motion of the Earth through the Galactic rest frame. By comparison both the diffuse supernova neutrino and atmospheric signal are completely isotropic. Even in the absence of a directional signal, it may be possible to perform a multi-component spectral fit to the neutrino spectra, since the Solar and atmospheric neutrino spectra are well understood from previous neutrino experiments. It is also interesting to consider the possibility for extracting new physics from the coherent neutrino signal, for example as a probe of non-standard neutrino interactions~\citep{Pospelov:2011ha,Harnik:2012ni}. Note that in their low threshold analysis, CDMS-II has reduced the energy threshold to $\sim 2$ keV, making them at this stage closest to measuring the coherent scattering from $^8$B Solar neutrinos. If the exposure in~\cite{Ahmed:2010wy} were extended down to $1$ keV, the  $^8$B event rate in CDMS-II would be one per ten years. 

\par Next generation dark matter detectors will also be sensitive to scattering of Solar neutrinos with electrons. Though the rate for this interaction is lower than the nuclear recoil rates in Figure~\ref{fig:neutrino_coherent}, this scattering on electrons is dominated by the neutrinos created directly in the proton-proton chain deep inside the Solar interior. Though in terms of number of neutrinos over all energies, the proton-proton component of the Solar neutrino flux dominates, neutrinos of this energy are too low to be detected with modern detectors. The neutrinos from the proton-proton chain would make a flat spectra of electron recoils that extends up to electron energies $\sim$ MeV~\citep{Malling:2011va}. 

\subsection{Indirect detection} 

\par Indirect dark matter detection experiments will also significantly improve their sensitivity to WIMPs over the course of the next several years. Focusing on gamma-ray searches, the progress can be characterized along two different fronts: more and improved data from the Fermi-LAT, and larger and more sensitive ground based ACTs. 

\bigskip 

$\bullet$ {\it Fermi-LAT}--  The Fermi-LAT will improve the results presented in section~\ref{sec:indirect_detection_experiment} for several reasons. For dSph searches along the lines of the current analysis, the limits will clearly improve linearly with exposure time in the high mass WIMP regime of a few hundred GeV. With a total of ten years of data, extremely sensitive limits or detections of dark matter will be obtained for the entire mass regime of $10-700$ GeV, and there is even reason to believe that better sensitivity will be achieved than can be currently projected. For example, galaxy surveys that are now coming online, such as Pan-STARRS, the Dark Energy Survey (DES), and farther into the future the Large Synoptic Survey Telescopes (LSST), are certain to discover more faint satellite galaxies. As new objects are detected, kinematic analysis of their constituent stars will provide their dark matter masses, and adding these objects to the Fermi-LAT analysis will certainly improve on the current limits. In addition, there will be improvements in sensitivities on all sources, and concurrent theoretical advances that will improve the understanding of their dark matter and astrophysical gamma-ray emission. Even extending beyond Fermi-LAT, Gamma-400, a Russian space-based mission that is designed to detect gamma rays in the energy range of $100$ MeV to $3$ TeV~\citep{Galper:2012ji}, will complement the above results. 

\bigskip 

$\bullet$ {\it ACTs}-- There is room for significant improvement in dark matter searches with ACTs. Modern experiments such as MAGIC, VERITAS, and HESS will continue to take data and improve the results presented in section~\ref{sec:indirect_detection_experiment}. The HESS experiment is being expanded with the addition of a fifth 600 m$^2$ telescope (HESS-II), which will both expand the effective area and reduce the energy threshold. Extending farther into the future, the proposed Cherenkov Telescope Array (CTA)~\footnote{www.cta-observatory.org/}, which is expected to begin operation near 2017, will extend WIMP dark matter limits to much higher masses, beyond the TeV scale, with better angular resolution than Fermi-LAT. CTA is expected to improve present limits from dSphs by at least an order of magnitude, and extend the Fermi-LAT studies of the Galactic center to a much higher energy regime~\citep{Doro:2012xx}. 

\bigskip 

\par Combining the satellite searches with the results from the Galactic center, line searches, and the other sources presented here through multiple wavelengths, the analysis will only improve over the next several years. In addition, we of course cannot anticipate surprising discoveries that will be made with more Fermi-LAT data, as well as forthcoming data from larger ACTs and neutrino experiments. Finally with regards to indirect detection experiments, new anti-matter detectors, in particular AMS-2, will begin reporting results, and test the increase in the positron fraction observed by PAMELA. 

\subsection{Complementarity with initial LHC searches} 

\par The effective model of dark matter interactions discussed in sections~\ref{sec:indirect_detection_theory} and~\ref{sec:direct_detection_theory} predicts interesting signals at colliders, which are complementary to indirect and direct searches. For couplings of dark matter to fermions, it is possible to translate the interactions in Equations~\ref{eq:scalar:v_Av}-\ref{eq:fermionic} to annihilation of fermions into dark matter and constrain the values of the respective couplings. Events in which a pair of dark matter particles are produced can be detected in colliders if a photon or a jet emanates from an incoming fermion.

\par
\begin{figure}
\begin{center}
\begin{tabular}{cc}
{\includegraphics[width=0.45\textwidth]{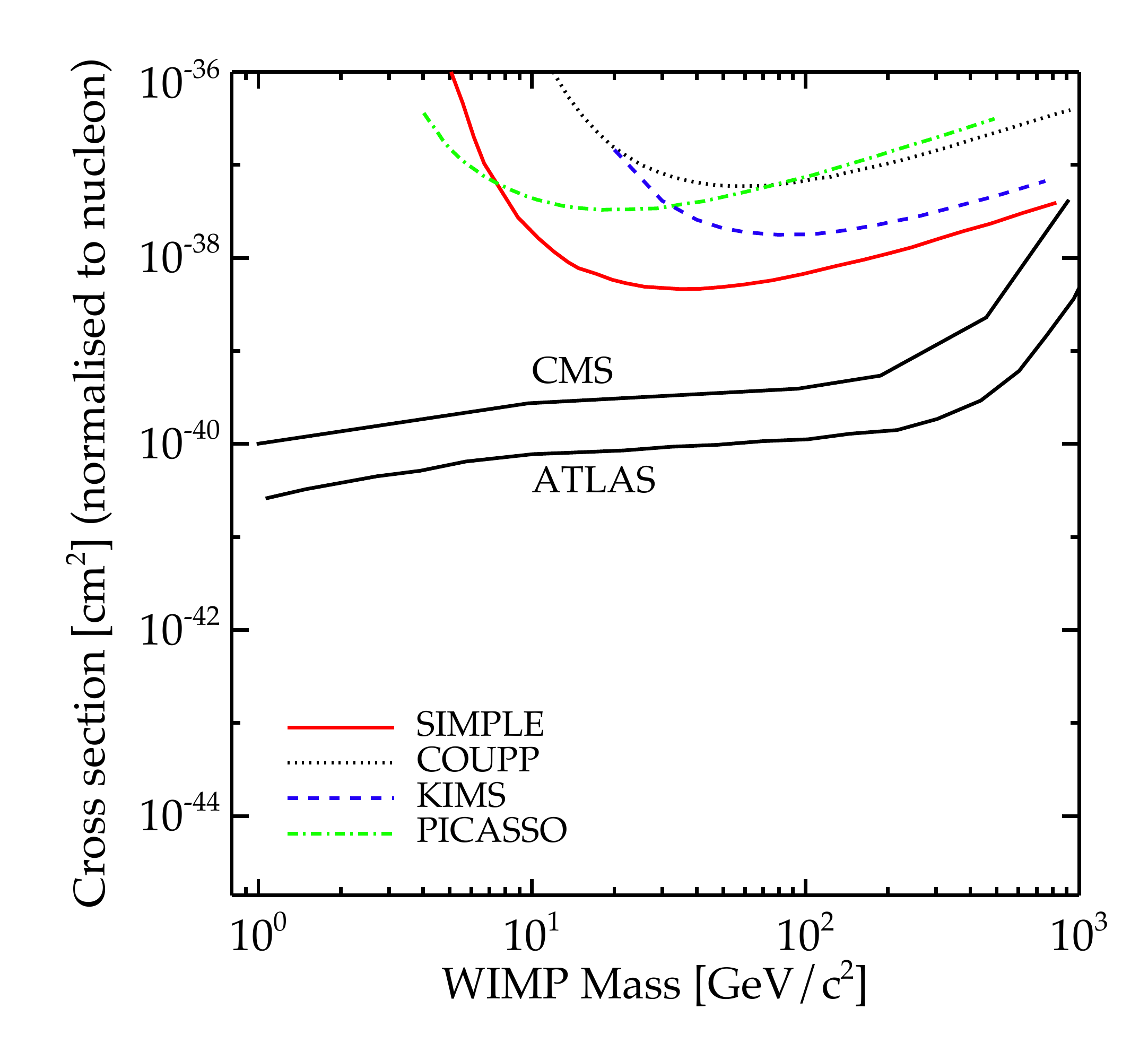}} & 
{\includegraphics[width=0.45\textwidth]{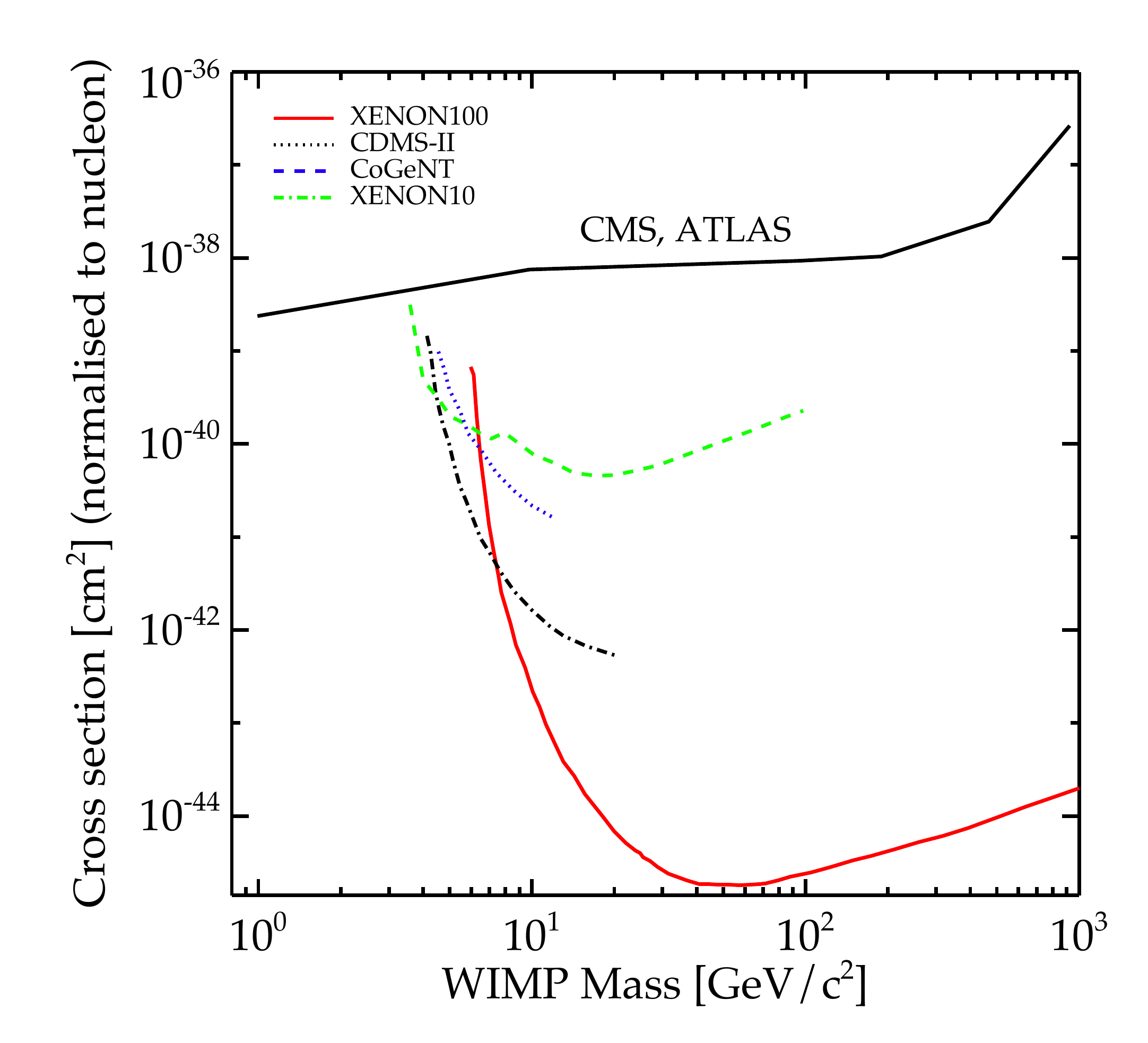}} \\
\end{tabular}
\end{center}
\caption{CMS and ATLAS limits on the WIMP-nucleon spin-dependent (left) and spin-independent cross section (right) from monophoton searches. Data from~\cite{Chatrchyan:2012tea} and~\cite{Aad:2012fw}. 
}
\label{fig:atlas}
\end{figure}

 Recently the CMS~\citep{Chatrchyan:2012tea} and ATLAS~\citep{Aad:2012fw} detectors have reported results on searches for so-called monophoton events. These events are characterized by a photon and missing transverse momentum in the final state; the missing transverse momentum in this case is the probe of new physics such as the production of dark matter (without the accompanying photon, the missing energy signal is unobservable at a hadron collider). CMS observed $73$ of these monophoton events, which was determined to be consistent with the background expected from standard model processes. ATLAS observed $116$ events, again consistent with its background expectations. The consistency with the standard model background implies an upper bound on the spin-independent and spin-dependent scattering elastic scattering cross sections. The limit on the former cross section is determined from scalar and vector couplings, while the later is deduced from axial-vector and tensor interactions. The upper limits are shown in Figure~\ref{fig:atlas}, in comparison to the limits from direct detection discussed in section~\ref{sec:direct_detection_theory}. Over the entire WIMP mass range, the limits from ATLAS and CMS are now more stringent than the spin-dependent limits deduced from direct detection. Above about $1$ TeV, the only more stringent constraint is the Ice Cube $W^+ W^-$ limit. For WIMP masses around $1$ GeV, the ATLAS and CMS limits are more strict than the spin-independent limits from direct detection. 

\par Limits on monophoton searches are also complementary to the Fermi-LAT limits from dSphs (Figure~\ref{fig:fermi_dsph_limits}). When translated into a limit on the annihilation cross section,  
the ATLAS and CMS limits on WIMPs with mass of $10$ GeV with vector current interactions described by the operator in Equation~\ref{eq:fermion:vector} are just as stringent as the Fermi-LAT limits. Further, the implied ATLAS and CMS upper limits on $10$ GeV mass WIMPs with axial vector interactions described by the operator in Equation~\ref{eq:fermion:axial} are an order of magnitude more stringent than the Fermi-LAT bounds~\citep{Aad:2012ky}. Above a WIMP mass of $100$ GeV the Fermi-LAT bound is more stringent than the ATLAS and CMS bounds. 

\par The aforementioned limits were obtained with $5$ fb$^{-1}$ of data for CMS and $4.6$ fb$^{-1}$ of data for ATLAS. With the increase of the luminosity at the LHC, more stringent limits on monophoton events will be set, and also the search will continue for supersymmetric dark matter through the detection of colored particles, quark jets, hard leptons, and missing transverse energy that signifies the presence of physics beyond the standard model. 

\subsection{Cosmological searches} 
\par In the early universe, WIMPs annihilate and inject energy into the photon-baryon plasma, ionize the gas and increase the number of free electrons. This process leaves an imprint on the CMB anisotropies, is most sensitive to WIMPs that do not annihilate largely into neutrinos, and is particularly sensitive to WIMPs that annihilate directly to electrons. Several authors have recently developed the theory for energy injection by WIMPs~\citep{Galli:2009zc,Slatyer:2009yq,Huetsi:2009ex,Finkbeiner:2011dx}; these authors show how the constraints on the WIMP mass and cross section are particularly sensitive to the fraction of the energy that is absorbed by the plasma. Using WMAP7 data, for direct annihilation into electrons the limits on the annihilation cross section reach the thermal relic scale at a WIMP mass of approximately $10$ GeV, and are a factor of a few weaker for direct annihilation into muons~\citep{Galli:2011rz}. Data from the Planck satellite mission, which will be released in 2013, is expected to further improve these limits by about an order of magnitude. 

\subsection{Astronomical observations} 
\par If WIMPs are not detected with the next decade, it is likely that there will be a paradigm shift in how dark matter is viewed in astrophysics, cosmology, and particle physics. In this circumstance of events, it will becomes more important to understand what astronomical surveys can teach us about the nature of dark matter. As a general point, over the course of the next several years, the improvement in the astronomical observations will be important from two perspectives. On the one hand, we will have both better measurements of astrophysical parameters as input to direct and indirect detection experiments. In addition, observations that are sensitive to deviations from the standard CDM model will become more relevant. 

\bigskip 

$\bullet$ {\it Milky Way surveys}-- The GAIA satellite is expected to launch in 2013. GAIA is the next generation large scale astrometric survey of the Milky Way which will significantly improve the corresponding measurements from the Hipparcos satellite. For stars brighter at 20th magnitude GAIA will have an astrometric accuracy of approximately 1 mas and a photometric accuracy of 60 mmag. From the GAIA data, it will be possible to re-analyze the local distribution of dark and luminous matter via the methods discussed in section~\ref{sec:MW} and to obtain better constraints on the local dark matter density. Further, GAIA will be able to probe the shape of the Milky Way dark matter halo from kinematic information on the population of stars in the outer halo, improving the measurements from current disk populations of stars discussed in section~\ref{sec:MW}. It will also be able to identify streams and other features in the phase space of stars. 

\bigskip 

$\bullet$ {\it Kinematics of Satellites}-- Larger samples of resolved stars in dSph satellite galaxies with kinematic information will continue to improve the measurements of their dark matter distributions. More line-of-sight velocities will be important to further test the initial results that have been obtained from multiple populations of stars in a small number of dSphs. Also, new ultra-faint satellites that will be discovered in future surveys will require kinematic follow up. It is very possible that a new survey will detect an object that is as massive and nearby as the SDSS discovery of Segue 1, so follow up on all satellites that are discovered will be important. Beyond measurements of line-of-sight velocities, in the future it will also be possible to obtain transverse velocities of bright stars in several dSphs. This will go a long way towards breaking the degeneracy between the velocity anisotropy and the dark matter mass profiles, and distinguishing between cold and warm dark matter models~\citep{Strigari:2007vn}. 

\bigskip 

$\bullet$ {\it Surveys for Satellite Galaxies}-- A variety of large-scale galaxy surveys are beginning to take data, or will take data, in the very near future: these include Pan-STARRS, Dark Energy Survey (DES), and even farther into the future the Large Synoptic Survey Telescope (LSST). Pan-STARRS will have a survey region that largely overlaps with the SDSS sky region, however, DES will cover about one-half of the Southern Sky that was not surveyed by SDSS, and LSST is scheduled to survey the entire southern sky down to a magnitude limit of $g = 24$. These surveys are each expected to make significant contributions to Galactic astrophysics and the studies of the dark matter distribution in the Milky Way. Though it is impossible at this stage to predict how many ultra-faint satellites will be detected in these surveys, only a very small fraction of the total Milky Way volume has been surveyed to the SDSS depth, so there is certainly a vast amount of discovery space available~\citep{Koposov:2007ni,Tollerud:2008ze,Willman:2009dv}. 

\par As discussed in section~\ref{sec:simulations}, surveys for faint satellites around external galaxies are now reaching a significant level of maturity. The results obtained to date will be improved by a combination of the aforementioned surveys. For example, it is likely that these surveys will identify more than $4000$ host galaxies with satellites to Fornax magnitude differences, approximately $1600$ galaxies with satellites as faint as Leo I, and approximately $100$ host galaxies with satellites as dim relative to their primary as Sculptor is to the Milky Way. A large statistical sample along these lines will also go a long way towards probing the subhalo population around other galaxies, and distinguishing between cold and warm dark matter models~\citep{Strigari:2011ps}. 

\bigskip 

$\bullet$ {\it Gravitational Lensing}-- As discussed in section~\ref{sec:simulations}, strong gravitational lensing is a powerful probe of the mass distribution of galaxies. Strong lensing is particularly interesting as a probe of dark matter substructure, providing the only method for detecting substructure outside of the Local Group. When a background quasar is lensed by a foreground galaxy, at optical and X-ray wavelengths the observed flux ratios are affected not only by dark matter substructure, but also by stars in the lensing galaxy. Because the contribution of the stars to the lensing signal varies, the contribution to the flux measurement from the stars is typically subtracted out by long term monitoring of the lightcurves. At radio wavelengths however, microlensing by stars is not important because the emission comes from a larger region within the source. Subtracting out this stellar emission through multi-wavelength studies of strong lensing systems is a promising technique for measuring the properties of dark matter substructure in these systems~\citep{Moustakas:2009na}. Time delay measurements of strong lensing systems may probe the substructure mass function down to mass scales $\lesssim 10^9$ $M_\odot$~\citep{Keeton:2008gq}. 

\par One can additionally ask whether dark subhalos in our own Milky Way Galaxy may cause a detectable gravitational lensing signal. The primary problem with lensing signatures in the Milky Way is that subhalos are predicted to be too diffuse-- only if subhalos happen to have inner slopes that are near that of a singular isothermal sphere will they generate a photometric microlensing signal. The situation is slightly better when considering an astrometric signal, however this measurement also requires slopes for subhalos steeper than are indicated in numerical simulations~\citep{Erickcek:2010fc}.

\newpage 
\section{Summary}
\label{sec:summary} 

\par Particle dark matter theories have in many ways become a victim of their own success. A myriad of models have been developed that explain a variety of observational phenomena, and because of this it is easy to see why researchers not directly tied to this field may have difficulty understanding the extent of the progress that has been made. The reality is of course that a significant amount of work remains to be done to solve the dark matter problem. As has been discussed in this article, the phenomenon of dark matter provides us with the clearest evidence for physics beyond the standard model, touching several different areas of modern scientific research, and many lingering questions remain as we try to bring together the various sets of observations from astronomy and particle physics. 

\par The process of obtaining a solution to the dark matter problem will likely continue to involve many more twists and turns. To appreciate the timescale that is required to fully understand the dark matter problem in an ideal scenario, we may compare to related problems in particle astrophysics. For example there was a twenty year interval between the time Pauli proposed the existence of the neutrino and its confirmed detection by Reines in 1954. Then it was not until the turn of the century that experiments were able to measure theoretical properties of the neutrinos such as their masses and mixing angles. Further, there was also a thirty year time window in between the initial discovery of the CMB to the time when the primary anisotropy was detected. Given its similarly broad implications for astrophysics and particle physics, the detection of dark matter will certainly generate an equivalent amount of interest, stoking an extended period of both experimental and theoretical research. 

\par It is clear that the phenomena of dark matter is one of the most profound mysteries in modern science. Solving this mystery will without a doubt led us down the path to new physics, and be one of the most significant scientific accomplishments in generations. One of the more exciting prospects in modern science is that, one way or another, in the very near future our prevailing view of dark matter will be affirmed, or there will be a paradigm shift in our view of it.

\section{Acknowledgements} 
\label{sec:acks} 
\par I would like to thank Ranjan Laha, Yao-Yuan Mao, and Kenny Chun Yu Ng for reading a draft of this article and verifying some of the calculations. I am also particularly grateful to Risa Wechsler for invaluable discussions as I was beginning this article. I additionally thank John Beacom, Marla Geha, Marc Kamionkowski, and Jason Kumar for providing valuable comments and suggestions.  
 
\newpage 


\end{document}